\documentclass[a4paper,11pt]{report}

\usepackage{amsmath, amssymb, graphicx}
\usepackage{epsfig}
\usepackage[hypertex]{hyperref}

\def\baselinestretch{1.2}

\def\Journal#1#2#3#4{{#1} {\bf #2}, #3 (#4)}

\begin{document}

\begin{titlepage}
\renewcommand{\baselinestretch}{1.7}\normalsize
\begin{center}
    {\large Yerevan Physics Institute\\ After A.I. Alikhanyan}

\parskip 3cm
 \textbf{\Large Lev Ananikyan}
\parskip 1cm

   \textbf{{\LARGE Spin Effects in Quantum Chromodynamics and\\
Recurrence Lattices with Multi-Site Exchanges}}

Thesis for acquiring the degree of candidate of
physical-mathematical
sciences\\
 in division 01.04.02 (Theoretical Physics)

\parskip 3cm
\begin{flushright}
\emph{\Large Scientific supervisor\\
Candidate of phys.-math. sciences\\
N. Ivanov}
\end{flushright}

\textbf{YEREVAN 2007}
\end{center}
\end{titlepage}



\tableofcontents
\listoffigures


\chapter{Introduction and Motivation}

\hspace{14pt} It is widely accepted that the notion of spin was
introduced by Uhlenbeck and Goudsmit to explain the data on the
energy levels of alcaline atoms. The full story is, however, more
intriguing and instructive. Pauli, in early 1925, decided that the
electron has an extra "classically undescribable" quantum number,
and formulated his famous exclusive principle, according to which
two electrons cannot be in the same state. This principle made it
possible to avoid the difficulties in interpretation of atomic
spectra and explain the counting of levels in weak and strong
magnetic fields.

Pauli was not, however, willing to make the big jump that the
electron has an intrinsic angular momentum $\frac{1}{2}\hbar$. It
was R.L. Kronig, who proposed this idea first. However, this idea
was not well received by Pauli, as well as in Copenhagen where
Kronig went visiting. There was also a problem about the spacing of
the levels which gave doubts to Kronig himself. Then in the fall of
1925, Uhlenbeck and Goudsmit, in Leiden, proposed the same idea
which they sent for publication to Naturwissenshaften
\cite{uhlenbeck}. After discussions with Lorentz, they tried to
withdraw their paper, but it was too late (fortunately) and it was
published! An excellent description of the history of spin and
statistics can be found in Ref.~\cite{history}.

Presently, spin is a powerful and elegant tool which plays a crucial role in both high energy
physics and statistical mechanics. Spin is one of the most fundamental properties of elementary
particles because it determines their symmetry behavior under space-time transformations.

High energy experiments with polarized beams or final-state spin effects provide often most deep
insights into the properties of elementary particles and their interactions. For example, the
world's best measurements of the weak mixing angle, $\sin \theta_W$, have been provided by the SLD
experiment at SLAC by using the left-right asymmetry in polarized $e^{+}e^{-}$ scattering, as well
as by the LEP results on the forward-backward asymmetry for $b$-quark final states \cite{data}.
(For a review, see Refs.~\cite{Ellis,Erler}.)

It is interesting to note that spin appeared in statistical mechanics also in 1925. The spin model
in a magnetic field was first solved in one dimension by E. Ising \cite{Ising} and for that reason
it now bears his name. In 1944, Onsager \cite{Onsager} first computed the free energy for the
two-dimensional Ising model. One of the most popular subjects of investigation in the modern
statistical physics are critical phenomena in spin systems. Present status of our knowledge about
the two- and three-dimensional equilibrium spin systems related to the Ising, Heisenberg and $O(N)$
universality classes is discussed in Ref.~\cite{Vicari}. Recent review on the spin glass models can
be found in Refs.~\cite{glass1,glass2}.

It is well known that there is a close relation between the Quantum Field Theory (QFT) and
Statistical Mechanics (SM). First, an external similarity there exist: the generating functional of
a QFT in the Euclidean formulation looks the same as the partition function of corresponding
statistical model. This similarity is, however, rather formal because both QFT and SM deal with an
infinite number of degrees of freedom, and further definitions are always needed to remove
corresponding divergences. A satisfactory understanding of the connection between QFT and SM was
reached only when the ideas of the scaling observed in investigation of critical behavior of SM
models were reconsidered in the general renormalization-group (RG) framework by Wilson
\cite{Wilson1,Wilson2}. Using the field-theoretical methods, it was possible to explain the
critical behavior of most of the systems and their universal features; for instance, why fluids and
uniaxial antiferromagnets behave quantitatively in an identical way at the critical point.

On the other hand, the RG theory of critical phenomena provides the natural framework for defining
quantum field theories at a nonperturbative level, i.e., beyond perturbation theory (see, e.g.,
Ref.~\cite{Zinn-Justin}). In particular, the Euclidean lattice formulation of gauge theories
proposed by Wilson \cite{Wilson3,Wilson4} provides a nonperturbative definition of Quantum
Chromodynamics (QCD), the theory of strong interactions of elementary particles. QCD is obtained as
the critical zero-temperature (zero-bare-coupling) limit of appropriate four-dimensional lattice
models and may therefore be considered as a particular four-dimensional universality class (see,
e.g., Refs.~\cite{Zinn-Justin,Creutz,Montvay}). Wilson's formulation represented a breakthrough in
the study of QCD, because it lent itself to nonperturbative computations using
statistical-mechanics techniques, for instance by means of Monte Carlo simulations (see, e.g.,
Ref.~\cite{Creutz2}).

In this thesis, we study some spin effects in QCD and recurrence lattices with multi-site
exchanges. In the framework of QCD, we consider the azimuthal asymmetries in heavy flavor
production in the lepton-nucleon deep inelastic scattering.

Investigation of the heavy flavor production plays a crucial role in QCD. This is because, for a
sufficiently heavy quark, the cross sections are calculable as a perturbation series in the running
coupling constant $\alpha_{s}$, evaluated at the quark mass. Thus, measurements of the heavy flavor
production provide an excellent testing ground for perturbative sector of QCD. Moreover, the charm
and bottom production is a good probe of the structure of the target hadron. In particular, the
heavy quark photoproduction is a viable way to measure the gluon structure functions (both
polarized and unpolarized), while the leptoproduction process is very sensitive at large Bjorken
$x$ to the intrinsic charm content of the target.

In the framework of perturbative QCD (pQCD), the basic spin-averaged characteristics of heavy
flavor hadro-, photo- and electroproduction are known exactly up to the next-to-leading order
(NLO). During the last fifteen years, these NLO results have been widely used for a
phenomenological description of available data. At the same time, the key question remains open:
How to test the applicability of QCD at fixed order to heavy quark production? The problem is
twofold. On the one hand, the NLO corrections are large; they increase the leading order (LO)
predictions for both charm and bottom production cross sections by approximately a factor of two.
For this reason, one could expect that higher-order corrections, as well as nonperturbative
contributions, can be essential, especially for the $c$-quark case. On the other hand, it is very
difficult to compare pQCD predictions for spin-averaged cross sections with experimental data
directly, without additional assumptions, because of a high sensitivity of the theoretical
calculations to standard uncertainties in the input QCD parameters. The total uncertainties
associated with the unknown values of the heavy quark mass, $m$, the factorization and
renormalization scales, $\mu _{F}$ and $\mu _{R}$, $\Lambda _{QCD}$ and the parton distribution
functions are so large that one can only estimate the order of magnitude of the pQCD predictions
for total cross sections at fixed target energies \cite{Mangano-N-R,Frixione-M-N-R}.

At not very high energies, the main reason for large NLO cross sections of heavy flavor production
in $\gamma g$ \cite{Ellis-Nason,Smith-Neerven}, $\gamma ^{*}g$ \cite{LRSN}, and $gg$
\cite{Nason-D-E-1,Nason-D-E-2,Nason-D-E-3,BKNS} collisions is the so-called threshold (or
soft-gluon) enhancement.  This strong logarithmic enhancement has universal nature in the
perturbation theory since it originates from incomplete cancellation of the soft and collinear
singularities between the loop and the bremsstrahlung contributions. Large leading and
next-to-leading threshold logarithms can be resummed to all orders of perturbative expansion using
the appropriate evolution equations \cite{Contopanagos-L-S,Laenen-O-S,Kidonakis-O-S}. Soft gluon
resummation of the threshold Sudakov logarithms indicates that the higher-order contributions to
 heavy flavor production are also sizeable. (For a review see
Refs.~\cite{Laenen-Moch,kid2,kid1}).

Since production cross sections are not perturbatively stable, it is of special interest to study
those (spin-dependent) observables that are well-defined in pQCD.  A nontrivial example of such an
observable was proposed in Refs.~\cite{we1,we2,we4,we3} where the azimuthal $\cos2\varphi$
asymmetry in heavy quark photo- and leptoproduction has been analyzed \footnote{The well-known
examples are the shapes of differential cross sections of heavy flavor production which are
sufficiently stable under radiative corrections.}. In particular, the Born level results have been
considered \cite{we1,we4} and the NLO soft-gluon corrections to the basic mechanism, photon-gluon
fusion (GF), have been calculated \cite{we2,we4}. It was shown that, contrary to the production
cross sections, the $\cos2\varphi$ asymmetry in heavy flavor photo- and leptoproduction is
quantitatively well defined in pQCD: the contribution of the dominant GF mechanism to the asymmetry
is stable, both parametrically and perturbatively. This fact provides the motivation for
investigation of the photon-(heavy) quark scattering (QS) contribution to the $\varphi$-dependent
lepton-hadron deep inelastic scattering (DIS).

In the present thesis, we calculate the azimuthal dependence of the next-to-leading order (NLO)
${\cal O}(\alpha_{em}\alpha _{s})$ heavy-flavor-initiated contributions to DIS. To our knowledge,
pQCD predictions for the $\varphi $-dependent $\gamma ^{* }Q$ cross sections in the case of
arbitrary values of the heavy quark mass $m$ and $Q^{2}$ are not available in the literature.
Moreover, there is a confusion among the existing results for azimuth-independent $\gamma ^{*}Q$
cross sections.

The NLO corrections to the $\varphi$-independent lepton-quark DIS have been calculated (for the
first time) a long time ago in Ref.~\cite{HM}, and have been re-calculated recently in \cite{KS}.
The authors of Ref.~\cite{KS} conclude that there are errors in the NLO expression for
$\sigma^{(2)}$ given in Ref.~\cite{HM} \footnote{For more details see PhD thesis \cite{KS-thesis},
pp.~158-160.}. We disagree with this conclusion. It will be shown below that a correct
interpretation of the notations for the production cross sections used in \cite{HM} leads to a
complete agreement between the results presented in Refs.~\cite{HM}, \cite{KS} and present thesis.

As to the $\varphi $-dependent $\gamma^{*}Q$ cross sections, our main result can be formulated as
follows. Contrary to the basic GF component, the QS mechanism is practically $\cos
2\varphi$-independent. This is due to the fact that the QS contribution to the $\cos 2\varphi $
asymmetry is absent (for the kinematic reason) at LO and is negligibly small (of the order of
$1\%$) at NLO. This fact indicates that the azimuthal distributions in charm leptoproduction could
be a good probe of the charm density in the proton.

Then we investigate the possibility of measuring the nonperturbative intrinsic charm (IC)
\footnote{The notion of the IC content of the proton has been introduced over 25 years ago in
Refs.~\cite{BHPS,BPS}. This  nonperturbative five-quark component, $\left\vert
uudc\bar{c}\right\rangle$, can be generated by $gg\rightarrow c\bar{c}$ fluctuations inside the
proton.} using the $\cos 2\varphi$ asymmetry. Our NLO analysis of the hadron level predictions
shows that the contributions of both GF and  IC components to the $\cos 2\varphi$ asymmetry in
charm leptoproduction are quantitatively well defined: they are stable, both parametrically and
perturbatively, and insensitive (at $Q^{2}> m^{2}$) to the gluon transverse motion in the proton.
At large Bjorken $x$, the $\cos 2\varphi$ asymmetry could be a sensitive probe of the intrinsic
charm content of the proton.

We have also considered the contribution to azimuthal distributions of the perturbative charm
density within the variable flavor number scheme (VFNS) \cite{ACOT,collins} \footnote{The VFNS is
an approach alternative to the traditional fixed flavor number scheme (FFNS) where only light
degrees of freedom ($u,d,s$ and $g$) are considered as active. Within the VFNS, potentially large
mass logarithms are resummed through the all orders into a heavy quark density which evolves with
$Q^{2}$ according to the standard evolution equation.}. Main result of our analysis is that the
charm densities of the recent CTEQ \cite{CTEQ6} and MRST \cite{MRST2004} sets of parton
distributions have a dramatic impact on the $\cos2\varphi$ asymmetry in the whole region of $x$
and, for this reason, can easily be measured.

Concerning the experimental aspects, azimuthal asymmetries in charm leptoproduction can, in
principle, be measured in the COMPASS experiment at CERN, as well as in future studies at the
proposed eRHIC \cite{eRHIC,EIC} and LHeC \cite{LHeC} colliders at BNL and CERN, correspondingly.

Another topic of our thesis are critical phenomena in spin systems
defined on the recurrence lattices with multi-site exchanges. It is
well established that the thermodynamic properties of a physical
system can be derived from a knowledge of the partition function.
 Since the discovery
of statistical mechanics, it has been a central theme to understand
the mechanism how the analytic partition function for a finite-size
system acquires a singularity in the thermodynamic limit when the
system undergoes a phase transition.
 The answer
to this question was given in $1952$ by Lee and Yang in their
seminal papers \cite{yl1,yl2}. It was shown that phase transitions
occur in the equilibrium systems in which the continuous
distribution of zeros of the partition function intersects the real
axis in the thermodynamic limit.
 For anti-ferromagnetic Potts models, by contrast, there are some
tantalizing conjectures concerning the critical loci, but many
aspects still remain obscure \cite{kas,kas2,salas}. Recently, the
Yang-Lee formalism has also been  applied for investigation of
nonequilibrium phase transitions \cite{priez}.

Presently, the investigation of the partition function zeros is a
powerful tool for studying phase transition and critical phenomena.
 Particularly, much attention is being attached to the study of
zeros of partition function of helix-coil transition of biological
macromolecules \cite{han,han2,han3,han4}.

 In this thesis, we investigate helix-coil phase  transition for polypeptides and proteins in
thermodynamic limit on recursive zigzag ladder with three-spin interaction. We use recursive
lattices because for the models formulated on them the exact recurrence relations for branches of
the partition function can be derived. For classical  hydrogen bond ($N-H\cdots O = C$), we have
got the Yang-Lee zeros corresponding to helix-coil phase transitions for polypeptides and proteins
in thermodynamic limit. We also take into account a non-classical helix-stabilizing term describing
a hydrogen bond of the type $C_\alpha - H\cdots O$. For this case we obtain folding and quasi
unfolding of the order parameter (degree of helicity $\Theta$). Applying multi-dimensional mapping
on zigzag ladder, we got Arnold tongues for non-classical helix-coil phase transition for neutral
points of mapping \footnote{Neutral points are defined by condition that eigenvalues of the mapping
Jacobian, $\lambda$, lie on the unit circle, $\lambda = e^{i\varphi}$.}.

There are two types of modulated phases: commensurate and incommensurate ones. For commensurate
phases, when $\varphi=\frac{p}{q} \pi$, the so-called Arnold tongues there exist. Typically, for
multi-dimensional maps, the border of such regions (Arnold tongues \cite{coom}) splits into two
branches in the parameter space.

Our main result is that we get two qualitatively different behaviors
for the degree of helicity that depend on input parameters.
 The first regime presents a
low-temperature helix structure which melts at higher temperatures.
We observe that the presence of a non-classical ($K_1$) interaction
sensibly enhances the melting temperature, and the transition is
smooth. In  second case, the presence of non-classical interaction
leads to a remarkably different low-temperature behavior. In this
regime, an quasi unfolding transition takes place for $T \rightarrow
0$ as well, like to cold denaturation \cite{Pri}.
 We point out that our results are
meaningful for long chains since, for such chains, a thermodynamic
limit is involved. Note that unfolding of biopolymer has also been
observed in phenomenological model \cite{bakk}, Monte Carlo
simulation \cite{collet}, Bethe approximation \cite{buzano}, and for
a short chain in Distance Constraint Model \cite{don}.

In this thesis, we also investigate  magnetic properties of the $^3$He.
 Investigation of magnetic phenomena and magnetic properties
of materials has a long history \cite{maghis}. The theory of magnetism and related problems
composes a fast and rather advanced field of research in the modern theory of condensed matter
physics intimately linked to many other fields of physics, mathematics, biophysics, chemistry and
materials science. These investigations have a wide application in various fields of electronics,
computer techniques e.t.c. The unexpected discovery of cooper-oxide high-$T_c$ superconductors in
1986 \cite{supcon} not only aroused hopes that one day we will have at our disposal materials which
exhibit superconductivity at room temperature, but also opened a new stage in the studies of
magnetic phenomena. This is because there is a strict evidence that two-dimensional
anti-ferromagnetism is one of the key components of high-temperature superconductivity. One of the
most remarkable achievements in this field is the progress in the studies of magnetism in solid
$^3$He.

Solid and fluid $^3$He  films absorbed on the surface of
         graphite have attracted extensive attention since (at low
         temperatures and high pressures, due to the light mass of helium
         atoms) it is a typical example of a two-dimensional frustrated quantum-spin system \cite{rog1,godfrin}.
         The nuclei of $^3$He are fermions with spin 1/2. It's reasonable to assume that solid
         $^3$He is a system of localized identical fermions. The microscopic theory of magnetism
         for such systems is based on the concept of
         the permutation of particles.
         In the films under consideration, the nuclei of  $^3$He form a system of quantum 1/2 spins on a
         triangular lattice. We know experimentally that the
         three-particle interactions dominate in this system. Transition from
         ferromagnetic behavior to antiferromagnetic one takes place when the
         coverage $($density$)$ of $^3$He atoms decreases. The explanation is suggested
         in terms of multiple$-$spin exchanges $($MSE$)$. In a dense clode-packed solid, ferromagnetic three-spin exchange
         is dominant \cite{Bernu}. At lower densities, ferromagnetic three-and-five spin exchanges compete with antiferromagnetic
         four-and-six spin exchanges and lead to a frustrated antiferromagnetic system.  The
         MSE
         produce frustration by themselves and a strong competition between
         odd$-$ and even$-$particle exchanges is also responsible for the
         frustration \cite{God}. For description of solid $^3$He films, one
         can use the dynamical system approach with MSE model that leads to appearance of various ordered phases
         and magnetization plateaus and one period
         doubling \cite{Bet,ana97}. The study of the above mentioned magnetization
         plateau is one of the main directions of
         present-day activity in the field of non-trivial quantum effects in condensed matter
         physics. Despite the purely quantum origin of this phenomenon,
         it was shown recently that magnetization plateau can appear in the Ising spin systems as
         well
          exhibiting in some cases fully qualitative correspondence with its Heisenberg
          counterpart \cite{ladderp,ladderp2,ladderp3,ladderp4}.

Using the dynamical system approach with MSE on the recurrent lattices (square, Husimi, hexagon),
we obtain magnetization curves
         with  plateaus (at $m=0, m=1/2, m=1/3$ and $m=2/3$) and one period
         doubling.

The next issue of our investigation is the so-called face-cubic model. We have considered a
 spin model with cubic symmetry defined on the Bethe lattice and containing both
linear and quadratic spin-spin interactions. An expression for the free energy per
      spin in the thermodynamic limit was obtained.
We have applied the methods of the
       dynamical systems theory or, more precisely, the theory of
       discrete mappings. In this technique, one exploits the
       self-similarity of the Bethe lattice and establishes a
       connection between the thermodynamic quantities defined for
       the lattices with different number of sites.

       We have identified the
       different thermodynamic phases of the system
       (disordered, partially ordered and completely ordered) in the
       ferromagnetic case ($J>0$, $K>0$) with
       different types of the fixed points of recurrent relation. Then
        we have obtained the phase diagrams of the model which
       are found to be different for $Q\leq 2$ and $Q > 2$. The
       case of  $Q\leq 2$ contains three tricritical points while only
       one tricritical and one triple points there exist at $Q > 2$.

Our results on the critical phenomena in spin systems defined on recurrence lattices with
multi-site exchanges are published in Refs.~\cite{arnold,Artuso,ar,vad,lev2,lev3}. Our studies of
the spin effects in QCD are presented in Refs.~\cite{we6,aniv}.

The thesis is organized as follows. In Chapter \ref{advreclatt}, the multi-dimensional mapping is
used for
 non-classical helix-coil phase
transition  of anti-ferromagnetic Potts model for biopolymer formulated on the recursive zigzag
ladder. Two qualitatively different behaviors for the degree of helicity are obtained.

 In Chapter \ref{chHellium}, three types of the recurrent
lattices with MSE are considered as approximation to solid $^3$He films. Using methods of the
dynamical systems theory, we've got magnetization plateaus, bifurcation points, one period doubling
 behavior and modulated phases at sufficiently high
 temperatures.

In Chapter \ref{FaceCubic}, we derive the system of recurrent relations for the face-cubic model on
the Bethe lattice. We identify different types of the  fixed points of the system of recurrent
relations with different physical phases.

In Chapter \ref{QCD}, we analyze the QS and GF parton level predictions for the $\varphi$-dependent
charm leptoproduction in the single-particle inclusive kinematics. Hadron level predictions for
azimuthal asymmetry are obtained. We consider the IC contributions to the asymmetry within the FFNS
and VFNS in a wide region of $x$ and $Q^{2}$.

Main observations and conclusions of this thesis are discussed in
Conclusion.

\chapter[Advantage of Recursive Lattices]{\label{advreclatt} Advantage of Recursive
 Lattices~\footnote{The results considered in this chapter are published in
Refs.~\cite{arnold,Artuso}.}}
\hspace{14pt}
 The advantage of recursive lattices is that for the models
formulated on them the exact recurrence relations for branches of
the partition function can be derived.
 Let us consider the recursive lattices which are
connected through the sites. As the first example of recursive
lattice is an usual chain. One can receive the exact recursion
relation for the partition function for Ising model. We divide  a
chain on two equal parts. The partition function can be written as
follows:
\begin{equation}
Z=\sum_{{\sigma}_0}\exp (\beta h {\sigma}_0 )\cdot g_n^2({\boldmath
\sigma}_0)
\end{equation}
where ${\sigma}_0$ is the center of the chain and $g_n(\sigma _0)$
is the contribution of each chain branch. $g_n(\sigma _0)$ can be
expressed trough $g_{n-1}(\sigma _1)$, that is, the contribution of
the same branch containing n-1 generations starting from the site
belonging to the first generation:
\begin{equation}
g_n(\sigma _0)=\sum _{\sigma _1}\exp {\left( J \sigma _0 \sigma _1 +
h \sigma _1 \right)}[g_{n-1}(\sigma _1)],
\end{equation}
where $\sigma _i$ takes values $\pm$ 1, \emph{J} is interaction
constant and $\emph{h}$ is the external magnetic field. We introduce
the following variable
\begin{equation}
x_n=\frac{g_n\left( +\right) }{g_n\left( -\right) },
\end{equation}
where we denote $g_n(\sigma _0)$ by $g_n\left( +\right)$  if the
spin  $\sigma _0$
 takes the value +1
and by  $g_n\left( -\right)$ if the spin  $\sigma _0$ takes the
value -1. For $x_n$ we can then obtain the recursion relation:
\begin{equation}
x_n=f\left( x_{n-1}\right).
\end{equation}
$f(x)$ is a ratio of two polynomials. We obtain one dimensional
 dynamic rational mapping.
We can get the magnetization of a central site through $x_n$.

Another example of recursive lattice is the Bethe one. Let us regard
the Potts model on this lattice with $\gamma$ coordination number.
The Hamiltonian can be written as:
\begin{equation}
{\mathcal{H}}= - J \sum _{<i,j>}\delta(\sigma _i, \sigma _j) - H
\sum _i \delta(\sigma _i, 1).
\end{equation}
where  $\delta(\sigma _i, \sigma _j) = 1$ for $\sigma _i = \sigma
_j$ and 0 otherwise, $\sigma _i$ takes the values 1,2,...Q, the
first sum is over the nearest-neighbor sites, and the second sum is
simply over all sites on the lattice. We use the notation K=J/kT and
h=H/kT. Cutting apart the Bette recursive lattice at the central
point we get $\gamma$ identical branches. As usual we can receive
one dimensional dynamic rational mapping for partition function. The
same ideas can be used as for the recursive chain. Denoting
$g_n(\sigma _0)$ the contribution of each lattice branch one can
receive the recursive dynamic relation. Introducing the notation
\begin{equation}
x_n=\frac{g_n\left( \sigma  \neq 1\right) }{g_n\left( \sigma =
1\right) }
\end{equation}
one can obtain the Potts-Bethe map
\begin{equation}
x_n=f\left( x_{n-1},K,h\right), \qquad f\left( x,K,h\right)
=\frac{e^h +(e^K+Q -2)x^{\gamma -1}}{%
e^{K+h}+(Q-1)x^{\gamma -1}}.
\end{equation}
The magnetization of the central site for the Bethe lattice can be
written as
\begin{equation}
M_n= <\delta(\sigma _0, 1)>=\frac {e^h}{e^h + (Q-1)x_n^ \gamma}
\end{equation}
The situation changes drastically for $Q < 2$ with antiferromagnetic
interactions. The plot of the M (magnetization) versus h (external
magnetic field) has a bifurcation point and chaotic behavior at low
temperatures \cite{hu}.

An other example is Husimi lattice. It can be regarded as recursive
lattice. The three-site antiferromagnetic Ising model on Husimi
lattice is investigated in an external magnetic field using the
dynamic system approach.
 Making the same procedure for Husimi recursive lattice one can obtain one dimensional
 rational  recursive relation for partition function. The full bifurcation diagram, including chaos,
 of the magnetization was exhibited. It is shown that this system displays in the chaotic region
 a phase transition at a positive "temperatures" whereas in a class of maps close to $x\rightarrow 4x(1-x)$, the
 phase transitions occure at negative "temperatures".
 The Frobenius-Peron recursion equation was numerically solved and the density
 of the invariant measure was obtained \cite{mon91}.

 The ladders \cite{ladder,ladder01,ladder02} also can be regarded as a recursive lattice.
 They are connected through the bonds and  have  multi dimensional rational mapping for
 partition function. A zigzag ladder with axial next-nearest-neighbor Ising model has attracted
 many investigators on account of the fact that it is a particularly simple model exhibiting quasi
 specially modulated phases that can be either commensurate or incommensurate with
 the underlying lattice \cite{selke}. Using the dynamic approach
 one can receive three dimensional rational mapping for partition function.

\section{Arnold Tongues in Ising and Potts Models}
\hspace{14pt}Let us regard the anti-ferromagnetic  Ising and Potts
models on the recursive Bethe lattice connected through sites. For
Ising model the partition function can be written as:
\begin{equation}
\label{Z} Z=\sum_{\{\sigma_0\}}\exp\{h\sigma_0\} g^q_n(\sigma_0),
\end{equation}
where $\sigma_0$ is the central spin, $g_n(\sigma_0)$ - the
contribution of each lattice branch, $h$ - magnetic field  and $q$ -
coordination number\cite{bax}. $g_n(\sigma_0)$ is obviously
expressed through $g_{n-1}(\sigma_1)$:
\begin{equation}
g_n(\sigma_0)=\sum_{\sigma_1}\exp\{{\frac{h \sigma_1 - \sigma_0
\sigma_1}{T}}\}g^2_{n-1}(\sigma_1) \label{gn}
\end{equation}
for $q=3$ and interaction between the spins is constant  $J=-1$.
Introducing the notation
\begin{equation}
\label{xni} x_n=\frac{g_n(+)}{g_n(-)}
\end{equation}
the recursion relation (\ref{gn}) can be rewritten in the form
\begin{equation}
x_n=f(x_{n-1},T,h).
\end{equation}
As is known,   if the  derivative of $f(x,T,h)$ is equal to $-1$ we
have a bifurcation point, corresponding to the second order phase
transition for anti-ferromagnetic model. We defined
$v=e^{-\frac{2}{T}}$ and after a simple calculation we have got the
following system of equations:
\begin{equation}
\left\{ \begin{array}{c}
 x=\displaystyle{\frac{v^h + vx^2}{v^{h+1} + x^2}}\\
 \\
 \displaystyle{\frac{2vx - 2x^2}{v^{h+1} + x^2}=-1}
\end{array} \right.
\end{equation}
Eliminating $x$ we obtain the following equation:
\begin{equation}
4v^2(v^{h+1} +1)(v^h +v)=v^h(1-v^2)^2.
\end{equation}
Solving this equation we get:
\begin{equation}
\begin{array}{l}
-\frac{2h}{T}=-3\ln2+\frac{6}{T} +\ln\Big\{1-6v^2-3v^4
\pm\sqrt{(1-6v^2-3v^4)^2-64v^6}\Big\}
\end{array}
\end{equation}
This equation define Arnold tongues between paramagnetic and
modulated phases with winding number $w=1/2$.

\begin{figure}
\begin{center}
 \psfig{figure=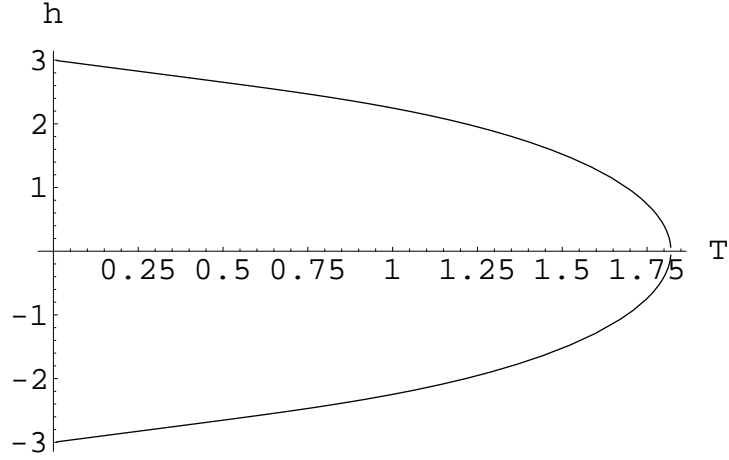,width=10.0cm}
 \caption[Arnold tongue for  anti-ferromagnetic Ising model on
recursive Bethe lattice]{\small Arnold tongue for
anti-ferromagnetic Ising model on recursive Bethe lattice with
coordination number q=3. \label{toung1}}
\end{center}
\end{figure}
 The Arnold tongue begins at the temperature of
$T=\frac{2}{\ln 3}$, when the external magnetic field $h=0$, and
ends ($T=0$) at $h=\pm3$ (see figure \ref{toung1}).
\begin{figure}
\begin{center}
\psfig{figure=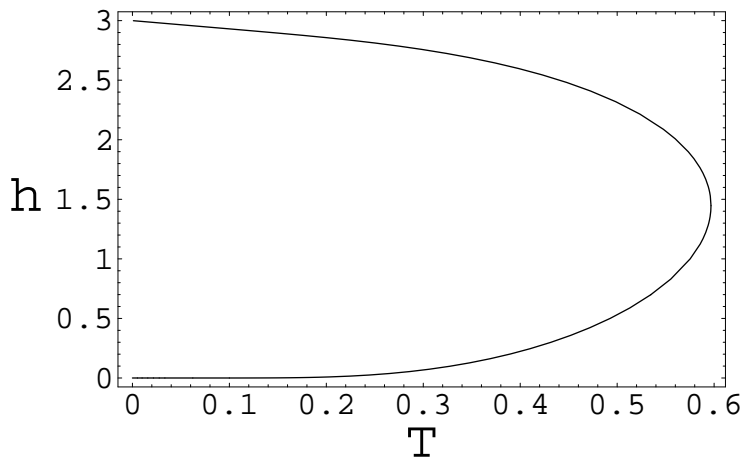,width=10.0cm} \caption[Arnold tongue for
anti-ferromagnetic Potts model on recursive Bethe lattice]{\small
Arnold tongue for anti-ferromagnetic Potts model on recursive Bethe
lattice with coordination number q=3. \label{toung2}}
\end{center}
\end{figure}

The same procedure we can perform for anti-ferromagnetic Potts model
on recursive Bethe lattice with Hamiltonian
\begin{equation}
-\beta H = -\sum_{<i,j>}\delta(\sigma_i,\sigma_j) +
h\sum_i\delta(\sigma_i,1),
\end{equation}
where $\sigma_i$ takes the values $1,2,3$. Introducing the notation
\begin{equation}
\label{xnp} x_n=\frac{g_n(*)}{g_n(1)},
\end{equation}
where $g_n(1)$ is the branch of partition function with central spin
$\sigma=1$ and $g_n(*)$ is the branch of partition function with
central spin $\sigma\neq1$. For the coordination number of the Bethe
lattice  $q=3$ we obtain following system of equations
\begin{equation}
\left\{ \begin{array}{c}
 x=\displaystyle{\frac{z^{-h} + (z+1)x^2}{z^{-h+1} + 2x^2}}\\
 \\
 \displaystyle{\frac{2(z+1)x - 4x^2}{z^{-h+1} + 2x^2}=-1}
\end{array} \right.
\end{equation}
here again the  derivative of $f(x,T,h)$ is equal to $-1$ which
corresponds to the  second order phase transition of
anti-ferromagnetic model and where $z=e^{-\frac{1}{T}}$. The Arnold
tongue begins at $z=\frac{1}{6}(\sqrt{17}-3)$, when external
magnetic field $h=1.5$, and ends ($T=0$) at $h=0$ and $h=3$ (see
figure \ref{toung2}).

\section{Multi-dimensional Mapping for the Biological Macromolecules}
\hspace{14pt}The structure of a protein is completely encoded in the
amino-acid sequence \cite{Anf}. Understanding of the folding and
unfolding processes of proteins (hetero-polymers) and polypeptides
(homo-polymers) is one of the current challenges in molecular
biophysics.  A lot of effort has been devoted to clear up the
mystery of protein or polypeptide folding nature by using lattice
models \cite{Dill}. These simple lattice models single out the
formation of a helix structure in protein as the basic mechanism to
be understood. Thermodynamics of homo and hetero-polymers folding
has been investigated in this perspective by introducing a variety
of different lattices: chains \cite{one,one2,one3}, square lattices
\cite{two,Salvi}, and cubic ones \cite{three,three2}. Off-lattice
models have been discussed by Irback \emph{et al.} and Klimov and
Thirumalai \cite{off_l,off_l2}. Chaotic behavior in off-lattice
models of hetero-polymers (proteins) and folding and unfolding  have
been analyzed in two-dimensional systems  by means of Monte Carlo
simulations \cite{off_lat}. The theory of finite-size scaling of
helix-coil transition was studied by Okamoto and Hansmann
\cite{han,han2,han3,han4} by multi-canonical simulation. They have
chosen three types of polypeptides with aliphatic neutral amino
acids (alanine, valine, and glycine). It was shown that
$\alpha$-helix formation in short peptide systems  agrees with
experimental results \cite{Bal}. But proteins are composed of
different types of monomers. Hydrophobic monomers, such as leucine
or proline, try to hide their surfaces from the solvent. The
simplest protein theoretical model divides the amino acids into two
categories: hydrophobic (H) and polar (P) surrounded by the solvent
\cite{Maritan,Maritan2}. Kamtekar et al. \cite{Kam} made experiments
with a variety of hydrophobic (H) and polar (P) amino acids in
hetero-polymers and showed that a simple code of polar (P) and
nonpolar hydrophobic (H) residues arranged in an appropriate order
could drive polypeptide chains to collapse into globular
$\alpha$-helical folds.By using a simple HP lattice model
\cite{Dokh,Dokh2} a theory explained the experimental phenomenon of
cold denaturation (unfolding) on real proteins \cite{Pri}.
 The study of
relaxation processes in biopolymer is of particular significance
since the functional abilities of thes molecules are related to the
dynamical properties \cite{zhou}.

We point out that different theoretical models were proposed to
study both unfolding and folding of  proteins \cite{bakk}.

From a statistical mechanics perspective different approaches can be
attempted to investigate the nature of the helix-coil phase
transition: from the analysis of Yang-Lee zeroes \cite{yl1,yl2}, to
multicanonical Monte Carlo simulation for finite samples
\cite{han,han2,han3,han4,two,Salvi,three,off_lat,Dokh,Dokh2}.

In this thesis we study in thermodynamic limit a model for the helix
structure of proteins and polypeptides, where we take into account
both the \emph{classical} hydrogen bond \cite{Sch} between three
$\alpha$-carbons by using $CO$ and $NH$ H-bond connection, and the
\emph{non-classical} H-bonds \cite{Shi} in every $C_\alpha - H$. The
classical ($\alpha - $ helix) H-bond is formed in the following way:
three neighboring angle pairs
[$C_\alpha(\varphi_i,\psi_i),C_\alpha(\varphi_{i+1},\psi_{i+1})$ and
$C_\alpha(\varphi_{i+2},\psi_{i+2})$]
form a H-bond when rotations are such that the distance between H
$[N(i-1)-H]$ and O $[O=C(i+3)]$ becomes less than 2 {\AA} (fig.1).
The hydrogen bond is a unique phenomenon in structural chemistry and
biology. Its functional importance stems from both thermodynamic and
kinetic reasons. In supermolecular chemistry, the hydrogen bound is
able to control and direct the structures of molecular assemblies
because it is sufficiently strong and sufficiently directional. The
subject of hydrogen bonding is of major interest and remains
relevant with each new phase in the kaleidoscope of chemical and
biological research (see references in \cite{hboundbook}).
Non-classical $C-H \dots O$ bonds have been recognized to play an
important role in biological macromolecules (see \cite{TIBS}), and
they were for instance observed
between water and amino acids alanine [$C_\alpha - H\cdots OH_{2}$]
\cite{Kar} or between two helices of collagen \cite{bella}.

Traditionally, the transition from random coiled conformation to the
helical state in DNA, RNA or proteins are described in the framework
of Zimm-Bragg \cite{zimbrag} type Ising model.
 But this type of one-dimensional
model cannot account for non-trivial topology of hydrogen bonds
\cite{Sch}.
\begin{figure}[h!]
\begin{center}
\psfig{figure=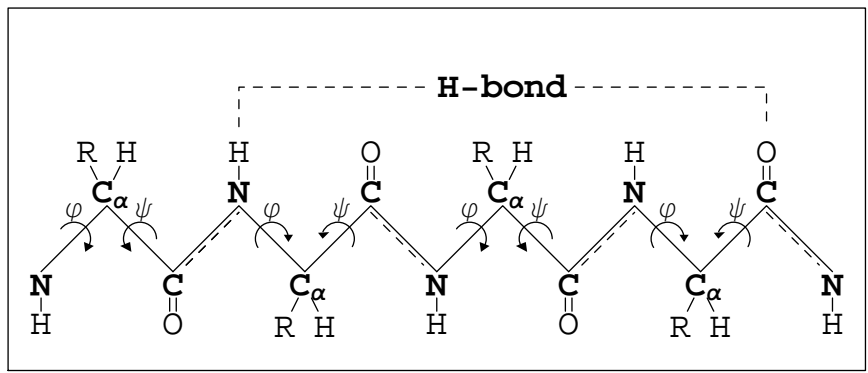,width=12cm}
\end{center}
\caption[The backbone of polypeptide or protein]{\small The backbone
of polypeptide or protein. The classical H-bond interaction between
N-H and C=O is pointed out by dashed line. \label{hbound}}
\vspace{0.2cm}
\end{figure}
\begin{figure}[h!]
\begin{center}
\psfig{figure=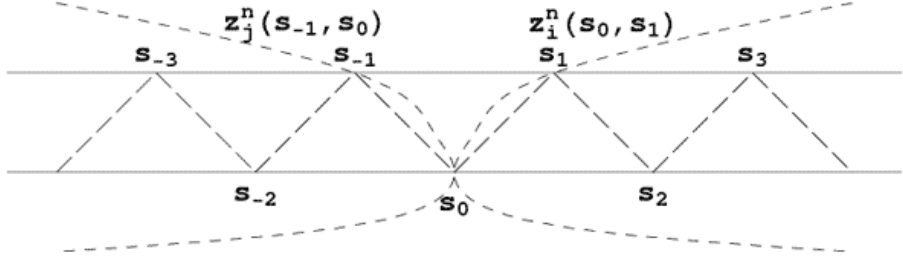,width=12cm}
\end{center}
\caption[The zigzag ladder]{\small The zigzag ladder. 3-site Potts
H-bond interaction is marked by solid line. \label{zigzag}}
\vspace{0.2cm}
\end{figure}

We show the backbone chain of the polypeptide molecule in  fig.
\ref{hbound}. {\bf R} (amino acid residue) denotes the side chain.
 Because of
the planar structure of the amide group, almost the whole
conformational flexibility of the polypeptide backbone chain is
determined by the rotation angles around the single bonds
$N-C_{\alpha}$ and $C_{\alpha}-C$ which are usually denoted as
$\varphi$ and $\psi$ respectively.

We now formulate our model: first of all the angle pairs
$(\varphi_{i},\psi_{i})$ are discretized \cite{ra}, the possible $Q$
values are labeled by a discrete variable $s_i$.When three
successive rotation pairs (spins) are zero
($s_i=s_{i+1}=s_{i+2}=0$), an H-bond appears which leads to some
energy gain. When one of three neighboring spins is not
zero($s_i=*$) an interaction with solvent is taken into account.
This leads to a three-site interaction Potts model \cite{hay1} on a
zigzag ladder (see fig. \ref{zigzag})
 The Hamiltonian of the system is written as
\begin{eqnarray}
\label{Ham}
 H=-J\sum_{\Delta^i}\delta(s_{i-1},0)\delta(s_{i}
,0)\delta(s_{i+1},0)\\ \nonumber
 -
 K\sum_{\Delta^i}[1-\delta(s_{i-1},0)\delta(s_{i},0)\delta(s_{i+1},0)]\\\nonumber
 -K_1\sum_i\delta(s_{i},0),
\end{eqnarray}
where  $J$ is  the energy of  hydrogen bond, $K$ is the energy of
protein-solvent hydrogen bond, $s_i$ denotes the Potts variable at
the site $i$ and takes the values $0,1,2,\cdots,Q-1$, $K_1$ is the
energy of non-classical H-bond,
and $\Delta _i$ label each triangle in Fig. \ref{zigzag}.

The model we thus introduced is indeed quite a simplified one, but
it allows to discuss how non-classical bonds compete with classical
hydrogen interaction in an idealized setting.
We will take advantage of the recursive nature of the zigzag ladder:
this makes it possible to derive exact recursion relations for
branches of the partition functions, and in this way statistical
properties in the thermodynamic limit may be investigated by
dynamical systems techniques \cite{arnold}. In their simplest
realization recursive relations yield one dimensional mappings
\cite{bax,ana97}.

By cutting the zigzag ladder in the central triangle
($s_{-1},s_{0},s_{1}$) one gets the partition function associated to
the hamiltonian ({\ref{Ham})
\begin{eqnarray}
\label{partfunc}
Z\sim\sum_{\{s_{-1},s_{0},s_{1}\}}[e^{-\frac{H_0}{T}}Z^{(n)}(s_{-1},s_{0})Z^{(n)}(s_{0},s_{1})],
\end{eqnarray}
where
\begin{eqnarray}
\label{H0} H_0=-(J-K)\delta(s_{-1},0)\delta(s_{0} ,0)\delta(s_{1},0)
-K_1\Bigl(\delta(s_{-1},0)+\delta(s_{0} ,0)+\delta(s_{1},0)\Bigr),
\end{eqnarray}
T is temperature (room temperature is $T=0.6\frac{Kcal}{mol}$),
$s_{-1}, s_{0},s_{1}$ are spins of central triangle,
$Z^{(n)}(s_{-1},s_{0})$ and $Z^{(n)}(s_{0},s_{1})$ are the parts of
partition function corresponding to two branches, $n$ is generation
of recursive lattice (see Fig.\ref{zigzag}). By introducing the
following notation
\begin{eqnarray}
\label{z1z4} Z^{(n)}(0,0)=Z^{(n)}_1;\quad Z^{(n)}(0,*)=Z^{(n)}_2;\nonumber \\
Z^{(n)}(*,0)=Z^{(n)}_3; \quad Z^{(n)}(*,*)=Z^{(n)}_4.
\end{eqnarray}
and
\begin{eqnarray}
\label{gammaz} \gamma=\exp{\frac{J-K}{T}};\quad
z=\exp{\frac{K_1}{T}},
\end{eqnarray}
(\ref{partfunc}) can be rewritten as:
\begin{eqnarray}
\label{partition}Z\sim \gamma z^3 [Z^{(n)}_1]^2+2(Q-1)z^2 Z^{(n)}_1
Z^{(n)}_2  + (Q-1)^2 z [Z^{(n)}_2]^2 +(Q-1)z^2 [Z^{(n)}_3]^2\\
\nonumber + 2(Q-1)^2zZ^{(n)}_3 Z^{(n)}_4 +(Q-1)^3[Z^{(n)}_4]^2
\end{eqnarray}
By applying the "cutting" procedure to an $n$th generation branch
one can derive the recurrence relations for $Z^{(n)}_1, Z^{(n)}_2,
Z^{(n)}_3, Z^{(n)}_4$,
\begin{eqnarray}
\label{recrelb} Z^{(n)}_1=\gamma z Z^{(n-1)}_1 +Z^{(n-1)}_2;
Z^{(n)}_2=z Z^{(n-1)}_3 +Z^{(n-1)}_4\\ \nonumber Z^{(n)}_3=z
Z^{(n-1)}_1 +Z^{(n-1)}_2; Z^{(n)}_4= zZ^{(n-1)}_3 +Z^{(n-1)}_4
\end{eqnarray}
If we notice  that $Z^{(n)}_2=Z^{(n)}_4$, and introduce the notation
\begin{eqnarray}
x_n=\frac{Z^{(n)}_1}{Z^{(n)}_4}; \quad
y_n=\frac{Z^{(n)}_3}{Z^{(n)}_4}, \label{x_y}
\end{eqnarray}
we can obtain a two-dimensional mapping from (\ref{recrelb}),
\begin{eqnarray}
 \label{f1f2}
x_n=f_1(x_{n-1},y_{n-1}),\quad f_1(x,y)=\frac{\gamma z x +Q-1}{z y +
Q-1}; \\ \nonumber y_n=f_2(x_{n-1},y_{n-1}),\quad f_2(x,y)=\frac{ z
x +Q-1}{z y + Q-1}.
\end{eqnarray}

\section{Helicity and Arnold Tongues for the Macromolecules}

\hspace{14pt}The (dimensionless) order parameter or helicity defined
as
\begin{equation}
\label{Theta} \Theta=\frac{Q^3\hat\Theta-1}{Q^3-1},
\end{equation}
where
$\hat\Theta=\langle\delta(s_{-1},0)\delta(s_{0},0)\delta(s_{1},0)\rangle$
(when recursion relations admit a stable fixed point the order
parameter is independent of the triangle we consider). Since our
procedure implicitly involves a thermodynamical limit, its
biological significance is motivated by the existence of long chains
of proteins, like collagen, that may exist in the form of three
intertwined peptide chains, each containing a thousand of amino
acids (we also remark that the importance of \emph{non-classical} H
bonds in collagen has been pointed out in \cite{bella}).

Thus our task is that of investigating the asymptotic behavior of
recursion relations (\ref{f1f2}), this allow to characterize the
macroscopic  order parameter $\Theta$ as a function of the physical
parameters T, $J$ (energy of \emph{classical} H bond), $K$ (energy
of protein-solvent H bond) and $K_1$ (energy of \emph{non-classical}
H bond).

We observe that in the whole range of the parameters triplet ($J, K,
K_1)$ the recursion relations (\ref{f1f2}) admit a single (real)
fixed point ($\tilde{x},\tilde{y}$). An investigation of the
Jacobian matrix of the transformation at ($\tilde{x},\tilde{y}$)
moreover indicates that such a fixed point is always stable: thus we
do not get any phase diagram marked by the stability border for the
fixed point like in mean field Ising models with competing
interactions on hierarchical lattices \cite{yokoi}. We also point
out that we did not observe in our tests any other dynamically
relevant attracting structure: under iteration of (\ref{f1f2})
generic initial conditions collapse to the stable fixed point. (We
mention that for physical values of microscopic parameters there
exist complex fixed points ($\tilde{x},\tilde{y}$) at which the
absolute value of eigenvalue of Jacobian equals to one. In this case
the order parameter ($\Theta$) would be complex too).
\begin{figure}[h!]
\begin{center}
\psfig{figure=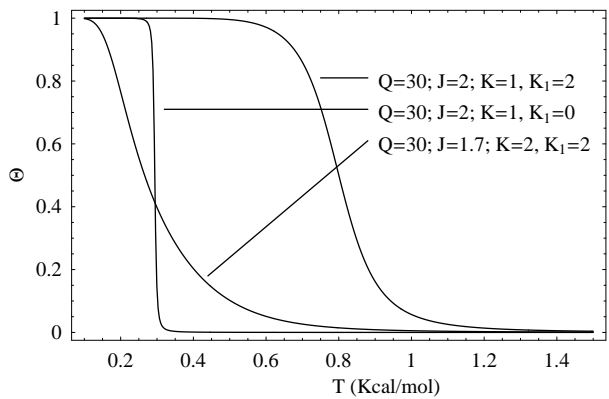,width=8.5cm}
\end{center}
\caption[The order parameter $\Theta $ as a function of
temperature]{\small The order parameter $\Theta$ as a function of
temperature (T) for different values$(Q,J,K,K_1)$.\label{HelB}}
\vspace{0.2cm}
\end{figure}

Once ($\tilde{x},\tilde{y}$) is determined, the degree of helicity
(\ref{Theta}) can be computed. Our main result is that we get two
possible qualitatively different behaviors. The first regime
presents a low-temperature helix structure, which melts at higher
temperatures (see fig.\ref{HelB}), a qualitative feature that may be
observed if the non classical interaction is absent. By looking at
two of the curves in fig.\ref{HelB} we observe that the presence of
a non-classical ($K_1$) interaction sensibly enhances the melting
temperature, and that, coherently with the dynamical analysis of
recursion relation,  the transition is smooth. For other parameter
values, the presence of non-classical interaction leads to a
remarkably different low temperature behavior, with an quasi
unfolding transition also for $T \rightarrow 0$, akin to cold
denaturation \cite{Pri}, see fig.\ref{HelA}. Real unfolding behavior
is when order parameter's peak is near one. In fig.\ref{HelA} for
parameters: $Q=4, J=0.8, K=2$ and $K_1=3.5$ we have a strange
situation. At low temperature we have a coil. The protein changes
conformation upon heating. Some of $[N(i-1)-H]$ and $0=C(i+3)$ in
average are near and they form H-bond. In these parameters the
degree of helicity (order parameter) $\Theta$ becomes larger. We
call that quasi helix. At higher temperature the protein becomes
coil again.
\begin{figure}[h!]
\begin{center}
\vspace{0.3cm} \psfig{figure=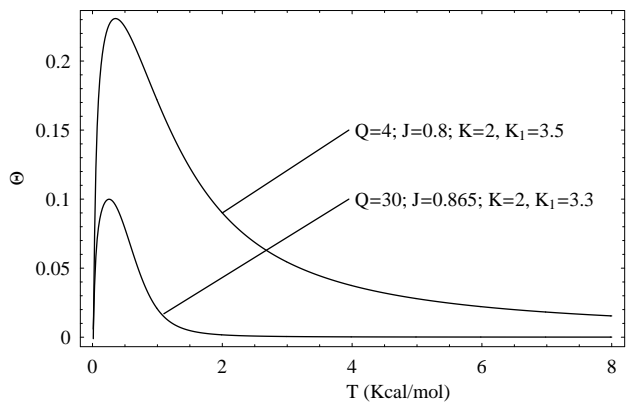,width=8.5cm}
\end{center}
\caption[T-dependence of the order parameter $\Theta $ with
non-classical interaction]{\small Starting from low temperature and
upon heating, the protein changes conformation from coil to quasi
helix, and then at still higher temperature becomes coil
(disordered).\label{HelA}} \vspace{0.2cm}
\end{figure}

Using the theory of dynamical systems for two-dimensional mapping,
we have obtained the separating line, which divides the coil
(paramagnetic, disordered) phase from helix (modulated, ordered) one
(see figure \ref{line}).Two example of Arnold tongues for
non-classical helix-stabilizing interaction with $Q=50$ for
$\varphi=\frac{5}{6}\pi$ $w=\frac{5}{12}$ and
$\varphi=\frac{3}{4}\pi$ $w=\frac{3}{8}$ are shown on  figures
\ref{tongue_m1}, \ref{tongue_m2}.
\begin{figure}
\begin{center}
\psfig{figure=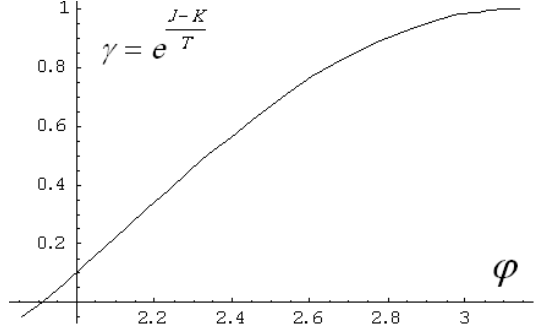,width=7cm}
\end{center}
\caption[The line separating coil and helix phases]{\small The line
separating coil (paramagnetic, disordered) and helix (modulated,
ordered)phases.\label{line}} \vspace{0.2cm}
\end{figure}
\begin{figure}
\begin{center}
\psfig{figure=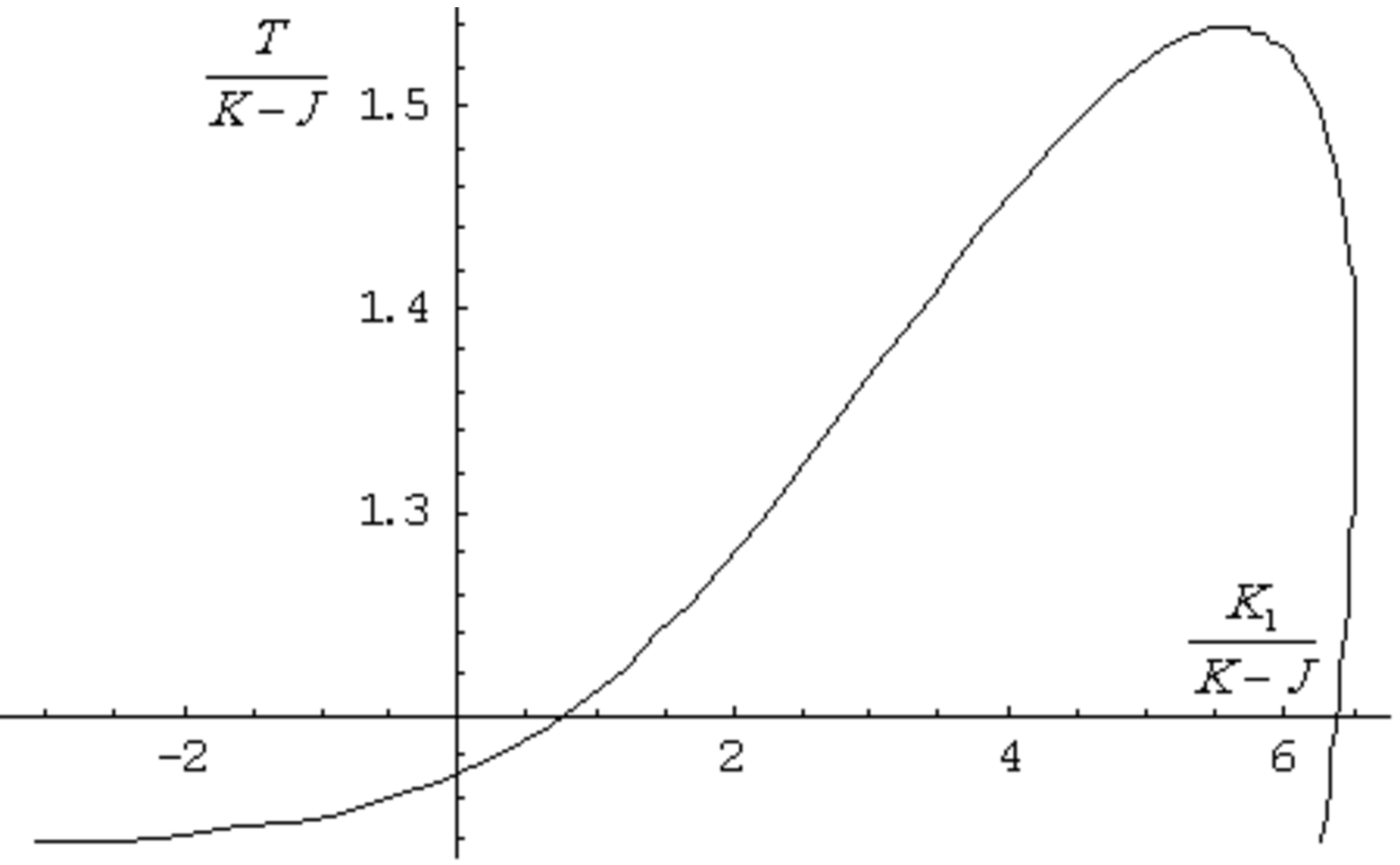,width=7cm}
\end{center}
\caption[Arnold tongue with winding number
$w=\frac{5}{12},\varphi=\frac{5}{6}\pi$ and $Q=50$]{\small Arnold
tongue with winding number $w=\frac{5}{12},\varphi=\frac{5}{6}\pi$
and $Q=50$ for non-classical  helix-stabilizing  interaction.
\label{tongue_m1}} \vspace{0.2cm}
\end{figure}
\begin{figure}[h]
\begin{center}
\psfig{figure=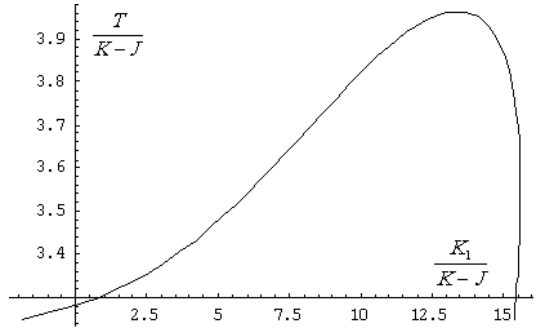,width=7cm}
\end{center}
\caption[Arnold tongue with winding number
$w=\frac{3}{8},\varphi=\frac{3}{4}\pi$ and $Q=50$]{\small Arnold
tongue with winding number $w=\frac{3}{8},\varphi=\frac{3}{4}\pi$
and $Q=50$ for non-classical helix-stabilizing  interaction.
\label{tongue_m2}} \vspace{0.2cm}
\end{figure}
 We point out that our result are
meaningful for long chains, since a thermodynamic limit in the
statistical model is involved. We notice however that unfolding of
biopolymer has been observed in phenomenological model \cite{bakk},
Monte Carlo simulation \cite{collet}, Bethe approximation
\cite{buzano}, and for a short chain in Distance Constraint Model
\cite{don}.

\section{Yang-Lee Zeroes for the Biological Macromolecules}
\hspace{14pt}When we take into account only classical hydrogen bound
the Hamiltonian of the system is written as
\begin{eqnarray}
\label{Ham2}
 -\beta H=J\sum_{\Delta^i}\delta(s_{i-1},0)\delta(s_{i}
,0)\delta(s_{i+1},0)
 + K\sum_{\Delta^i}[1-\delta(s_{i-1},0)\delta(s_{i},0)\delta(s_{i+1},0)],
\end{eqnarray}
%
%
%
We obtain again the two dimensional rational mapping relation for
$x_n$ and $y_n$
\begin{eqnarray}
\label{xnyn}
x_n=f_{1}(x_{n-1},y_{n-1}) \nonumber \\
y_n=f_{2} (x_{n-1},y_{n-1}),
\end{eqnarray}
where
\begin{eqnarray}
\label{xnyn1}
f_{1}(x,y)=\frac{\gamma x+(Q-1)}{y+(Q-1)} \nonumber \\
f_{2} (x,y)=\frac{x+(Q-1)}{y+(Q-1)}.
\end{eqnarray}
In case of multi-dimensional rational mapping the fixed point
$x^*,y^*$ is attracting when the eigenvalues of Jacobian $|\lambda|
<1$, repelling, when $|\lambda| >1$, and neutral, when $|\lambda|
=1$. So the system undergoes a phase transition when
\begin{equation}
\label{jac} \left|
\begin{array}{lr}
\frac{\partial f_1}{\partial x}-\exp{(\imath \varphi)}&\frac{\partial f_1}{\partial y} \\
\frac{\partial f_2}{\partial x}&\frac{\partial f_2}{\partial
y}-\exp{(\imath \varphi)}
\end{array}
\right|=0
\end{equation}

After eliminating $x$ and $y$ from (\ref{xnyn1}) and (\ref{jac}) we
obtain the following equation for the partition function zeros
\begin{eqnarray}
\label{reim}  b_0+b_1\cos(\varphi)+b_2\cos^2(\varphi)+
b_3\cos^3(\varphi)[\cos(3\varphi)+ \imath\sin(3\varphi)]=0,
\end{eqnarray}
where
\begin{eqnarray}
\label{b0b3}
b_0&=&Q^2(\gamma-1)+(\gamma-1)\gamma^2+Q^3(\gamma+1)+Q\gamma(\gamma^2+\gamma-2)\nonumber \\
b_1&=&2\{Q^3+2Q(\gamma-1)-Q^2(\gamma^2-1)+\gamma(\gamma^2+\gamma-2)\} \nonumber \\
b_2&=&-4(Q-1)(\gamma-1)(2+Q+\gamma) \nonumber \\
b_3&=&-8(Q-1)(\gamma-1).
\end{eqnarray}

One can solve (\ref{reim}) for $\mu$ and find the Yang- Lee zeros of
partition function with different parameters $Q$, $J$ and $T$ like
in  Refs.\cite{sl}. These parameters are different for each
polypeptides and proteins.
\begin{figure}[h]
\begin{center}
\psfig{figure=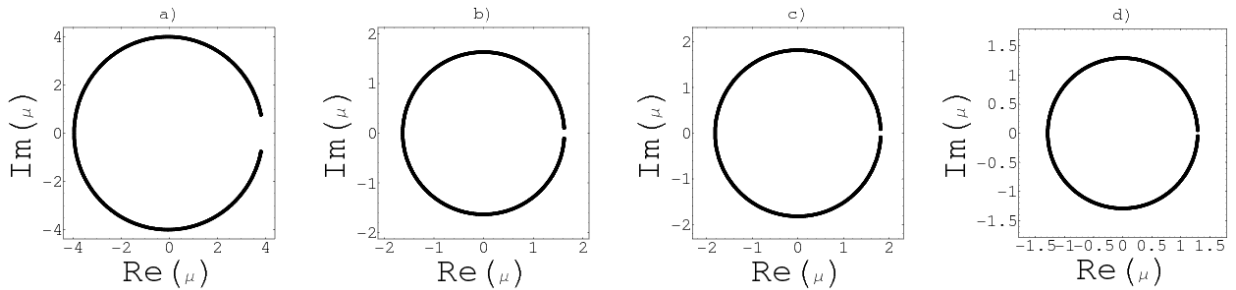,height=1.5in}
\end{center}
\caption[The Yang-Lee zeros for several values of $Q,J,T$ and
$K$]{\small \label{yanlee}The Yang-Lee zeros: a) Q=9, J=2.0, T=0.56,
K=0.76; b) Q=30, J=2.1, T=0.54, K=0.26; c) Q=40,
 J=2.4, T=0.56, K=0.334; d) Q=50, J=2.5, T=0.6, K=0.1523.}
\end{figure} After making discrete values of $Q$ and
comparing with Ramachardan and Shceraga\cite{ra} we confirm  that
the circle in classical helix-coil transition  does not cut the real
axis. So we have not a real phase transition in polypeptides
(proteins). According to phenomenological theory of Zimm-Bragg or
Lifson-Roig there is only pseudo phase transition in 2-site (Ising)
model. Our results describe the  microscopic theory of helix-coil
transition of polypeptides or proteins  with non-trivial topology of
hydrogen bonds and find Yang-Lee zeros of pseudo phase transition.
Yang-Lee zeroes of helix-coil transition for polyalanine, polyvaline
and polyglysine was regarded \cite{han}. The authors made Monte
Carlo simulation technique and considered polypeptide chain up to
$N=30$ monomers and determine the (pseudo) critical temperatures of
the helix-coli transition in all-atom model of polypeptides.

\chapter[Fluid and Solid $^3$He]{Fluid and Solid $^3$He~\footnote{
The results considered in this chapter are published in
Refs.~\cite{ar,lev2,lev3}.} \label{chHellium}}

\hspace{14pt}As mentioned in introduction, fluid and solid   $^3$He films absorbed on the surface
of graphite have attracted extensive attention, since it is a typical example of a two-dimensional
frustrated quantum-spin system \cite{rog1,godfrin}. The first and second layer of the nuclei of
$^3$He forms a system of quantum one-half spins on a triangular lattice. The third layer forms a
Kagome one \cite{rog2}. Many experimental \cite{exp} and theoretical \cite{theor} studies suggest
that the exchange of more than two particles are dominated in these systems. For such systems a
change from ferromagnetic behavior to anti-ferromagnetic takes place. Spin ladder anti-ferromagnets
have been attracting extensive interest because they have a spin gap. A special type of frustration
due to cyclic exchange interactions was recently found to be important in the spin ladder material
La$_x$Ca$_{14-x}$Cu$_{24}$O$_{41}$ \cite{ladder}. It is experimentally also known that a many-body
exchange interaction cannot be neglected especially in $^3$He on graphite \cite{ladder2}.

Last decade the investigation of magnetization plateaus in a strong
magnetic field has taken on special significance. The magnetization
plateaus are famous for the fact that they are an example of
essentially macroscopic quantum phenomenon. For the first time, Hida
has  theoretically predicted an appearance of the magnetization
plateau  for the ferromagnetic-ferromagnetic-antiferromagnetic
Heisenberg chain of 3CuCl$_2 \cdot$2 dioxane compound, which consist
of the antiferromagnetic coupled ferromagnetic trimers \cite{hida}.
The values of magnetization at which the plateaus appear are
quantized to fraction values of the saturation magnetization. The
theoretical explanation of this fact was been given in 1997 by
Oshikawa, Yamanaka and Affleck \cite{pl}. These magnetization
plateaus were observed as a simple origin in the Ising limit
\cite{pl2}. Geometric frustrated quantum magnets are a class of
magnetic materials with various unusual properties at low
temperature and high pressure. Due to strong frustration and quantum
effects, these materials  may be in principle considered as a source
of new strongly correlated physics. The most of studied geometric
frustrated quantum magnets are the Kagome and pyrochlore lattices of
antiferromagnetic coupled nearest neighbor spins. As mentioned
above, the third layer of the nuclei of $^3$He forms a Kagome
lattice. Usually, the antiferromagnetic Kagome nets are investigated
using numerical simulations \cite{kag}. We propose a dynamic
approach based on exact recursive relations for partition functions.
Our method makes possible to research  magnetization plateaus,
bifurcation points and period doubling  in anti-ferromagnetic case
at low temperatures and high pressures.
\section{Ising Model Approach to the Solid $^3$He System on the
Square Recursive Lattice}\label{square}
 \hspace{14pt}The most general expression for the
Hamiltonian with multi spin-exchanges on a triangular lattice is
\begin{equation}
{\mathcal{H}}={\mathcal{H}}_{Ph}+{\mathcal{H}}_{ex}+{\mathcal{H}}_Z.
\end{equation}

The term ${\mathcal{H}}_{Ph}$ describes the phonon contribution and
is not essential. ${\mathcal{H}}_{ex}$ responses for two-, three-,
and four exchange interactions. ${\mathcal{H}}_Z$ term is
responsible for magnetism in solid $^3$He and is given by Zeeman
Hamiltonian
\begin{equation}
{\mathcal{H}}_Z=-\sum_i\frac \gamma 2\hbar \mbox{\boldmath $H$}
\cdot \mbox{\boldmath $\sigma$}_i \label{Zeemanp}
\end{equation}
where $\gamma $ is gyromagnetic ratio of the $^3$He nucleus.

One can write down exchange Hamiltonian for the first and second
layers of planar solid $^3$He in the following way:

\begin{equation}
{\mathcal {H}}_{ex}=J_2\sum_{pairs}\left( P_2+P_2^{-1}\right)
-J_3\sum_{triangles}\left( P_3+P_3^{-1}\right)
+J_4\sum_{rectangles}\left( P_4+P_4^{-1}\right), \label{ham4}
\end{equation}
where sum in first term is going over all pairs of particles, in
second term over all triangles and in third term over all rectangles
consisting of two triangles.

         The expression of a pair transposition operator $P_{ij}$ has been given by Dirac,
         \begin{equation}\label{P2}
           P_{ij}=\frac{1}{2}(1+\bf\sigma_{\it i}\bf\sigma_{\it j}),
         \end{equation}
         where $\bf\sigma_{\it i}$ is the Pauli matrix, acting on the spin at the position number $i$.
         The operator $P_n^{-1}$ in general works in entirely different way, but in case of $n=2$
         the pair transplonation operators are equal $(P_{ij}^{-1}=P_{ij}^1)$,
         that we can't write in case of $n=3$.

         For $n=3$ we have
         \begin{equation}
           P_{ijk}=\frac{1}{4}(1+\bf\sigma_{\it i}\bf\sigma_{\it j})(1+\bf\sigma_{\it i}\bf\sigma_{\it k}),
         \end{equation}
         and
         \begin{equation}
           P_{ijk}^{-1}=\frac{1}{4}(1+\bf\sigma_{\it i}\bf\sigma_{\it k})(1+\bf\sigma_{\it i}\bf\sigma_{\it j}).
         \end{equation}

         Using the identity
         \begin{equation}
           (\bf\sigma_{\it i}\bf\sigma_{\it j})(\bf\sigma_{\it i}\bf\sigma_{\it k})=(\bf\sigma_{\it j}\bf\sigma_
           {\it k})+\bf\sigma_{\it i}[\bf\sigma_{\it j}\times\bf\sigma_{\it k}],
         \end{equation}
         we can write the former expression as
         \begin{equation}
           P_{ijk}=\frac{1}{4}(1+\bf\sigma_{\it i}\bf\sigma_{\it j}+\bf\sigma_{\it j}\bf\sigma_{\it k}+\bf\sigma_
           {\it k}\bf\sigma_{\it i}+\bf\sigma_{\it i}[\bf\sigma_{\it j}\times\bf\sigma_{\it k}])
         \end{equation}
         and
         \begin{equation}
           P_{ijk}=\frac{1}{4}(1+\bf\sigma_{\it i}\bf\sigma_{\it j}+\bf\sigma_{\it j}\bf\sigma_{\it k}+\bf\sigma_
           {\it k}\bf\sigma_{\it i}+\bf\sigma_{\it i}[\bf\sigma_{\it j}\times\bf\sigma_{\it k}])
         \end{equation}
         hence
         \begin{equation}\label{P3}
           P_{ijk}+P_{ijk}^{-1}=\frac{1}{2}(1+\bf\sigma_{\it i}\bf\sigma_{\it j}+\bf\sigma_{\it j}\bf\sigma_
           {\it k}+\bf\sigma_{\it k}\bf\sigma_{\it i}).
         \end{equation}

The four-spin permutation operators can be written as:
\begin{equation}
P_{ijkl}=P_{ijk}\cdot P_{il},  \label{4p}
\end{equation}
\begin{equation}
P_{ijkl}+\left( P_{ijkl}\right) ^{-1}=\frac 14\left( 1+\sum_{\mu
<\nu } (\mbox{\boldmath $\sigma$}_\mu \cdot \mbox{\boldmath
$\sigma$}_\nu)+G_{ijkl}\right) , \label{P4}
\end{equation}
where the sum is taken over six distinct pairs ($\mu \nu $) among
the four particles (ijkl), and
\begin{equation}
G_{ijkl}=\left( \mbox{\boldmath $\sigma$}_i \cdot \mbox{\boldmath
$\sigma$}_j\right) \left( \mbox{\boldmath $\sigma$}_l \cdot
\mbox{\boldmath $\sigma$}_k\right) +\left( \mbox{\boldmath
$\sigma$}_i \cdot \mbox{\boldmath $\sigma$}_l\right) \left(
\mbox{\boldmath $\sigma$}_j \cdot \mbox{\boldmath $\sigma$}_k\right)
-\left( \mbox{\boldmath $\sigma$}_i \cdot \mbox{\boldmath
$\sigma$}_k\right) \left( \mbox{\boldmath $\sigma$}_j \cdot
\mbox{\boldmath $\sigma$}_l\right) .  \label{g}
\end{equation}
So, in terms of Pauli matrices, we have:
\begin{eqnarray}
{\mathcal{H}}_{ex} &=&\frac{J_2}2\sum_{\left\langle i,j\right\rangle
}
\left( 1+%
\mbox{\boldmath $\sigma$}_i \cdot \mbox{\boldmath $\sigma$}_j\right)
-
\frac{J_3}%
2\sum_{\left\langle i,j,k\right\rangle }\left( 1+\mbox{\boldmath
$\sigma$}_i
 \cdot %
\mbox{\boldmath $\sigma$}_j+\mbox{\boldmath $\sigma$}_j \cdot
\mbox{\boldmath
 $\sigma$}_k
+\mbox{\boldmath $\sigma$}_k \cdot\mbox{\boldmath $\sigma$}_i\right)
+
\label{ex4} \\
&&+\frac{J_4}4\sum_{\left\langle i,j,k,l\right\rangle
}(1+\mbox{\boldmath $\sigma$}_i
 \cdot \mbox{\boldmath $\sigma$}_j+\mbox{\boldmath $\sigma$}_i \cdot %
\mbox{\boldmath $\sigma$}_k+\mbox{\boldmath $\sigma$}_i \cdot
\mbox{\boldmath $\sigma$}_l +\mbox{\boldmath $\sigma$}_j \cdot
\mbox{\boldmath $\sigma$}_k+\mbox{\boldmath
 $\sigma$}_j
 \cdot \mbox{\boldmath $\sigma$}_l+\mbox{\boldmath $\sigma$}_k \cdot %
\mbox{\boldmath $\sigma$}_l  \nonumber \\
&&+\left( \mbox{\boldmath $\sigma$}_i \cdot \mbox{\boldmath
$\sigma$}_j\right) \left( \mbox{\boldmath $\sigma$}_l \cdot
\mbox{\boldmath $\sigma$}_k\right) +\left( \mbox{\boldmath
$\sigma$}_i \cdot \mbox{\boldmath $\sigma$}_l\right) \left(
\mbox{\boldmath $\sigma$}_j \cdot \mbox{\boldmath $\sigma$}_k\right)
-\left( \mbox{\boldmath $\sigma$}_i \cdot \mbox{\boldmath
$\sigma$}_k\right) \left( \mbox{\boldmath $\sigma$}_j \cdot
\mbox{\boldmath $\sigma$}_l\right) ).
 \nonumber
\end{eqnarray}

For the first and second solid layers, we consider a recursive
lattice, instead of the periodic triangular one. This lattice is
given in Fig.\ref{latt1}. \setlength{\unitlength}{2mm}

\begin{center}
\newsavebox{\sqa} \savebox{\sqa}(2,4){%
\begin{picture}(2,4)
\put(1,0){\line(1,2){1}} \put(1,0){\line(-1,2){1}}
\put(1,0){\line(0,1){4}} \put(1,4){\line(1,-2){1}}
\put(1,4){\line(-1,-2){1}}

\end{picture}}

\newsavebox{\sqb} \savebox{\sqb}(4,2){%
\begin{picture}(4,2)
\put(2,0){\line(2,1){2}} \put(2,0){\line(-2,1){2}}
\put(2,2){\line(2,-1){2}} \put(2,2){\line(-2,-1){2}}
\put(0,1){\line(1,0){4}}
\end{picture}}

\newsavebox{\sqc} \savebox{\sqc}(6,10){%
\begin{picture}(6,10)
\put(3,0){\line(0,1){10}} \put(3,10){\line(-3,-5){3}}
\put(0,5){\line(3,-5){3}} \put(3,0){\line(3,5){3}}
\put(6,5){\line(-3,5){3}}

\end{picture}}

\newsavebox{\sqd} \savebox{\sqd}(10,6){%
\begin{picture}(10,6)
\put(0,3){\line(1,0){10}} \put(10,3){\line(-5,-3){5}}
\put(5,0){\line(-5,3){5}} \put(0,3){\line(5,3){5}}
\put(5,6){\line(5,-3){5}}

\end{picture}}

\begin{figure}[h!]
\begin{center}
\begin{picture}(20,40)(10,0)
\put(15,15){\usebox{\sqc}}

\put(15,25){\usebox{\sqc}} \put(15,5){\usebox{\sqc}}
\put(21,17){\usebox{\sqd}} \put(5,17){\usebox{\sqd}}
\put(17,35){\usebox{\sqa}} \put(17,1){\usebox{\sqa}}
\put(25,23){\usebox{\sqa}} \put(25,13){\usebox{\sqa}}
\put(9,23){\usebox{\sqa}} \put(9,13){\usebox{\sqa}}
\put(31,19){\usebox{\sqb}} \put(1,19){\usebox{\sqb}}
\put(11,9){\usebox{\sqb}} \put(21,9){\usebox{\sqb}}
\put(21,29){\usebox{\sqb}} \put(11,29){\usebox{\sqb}}
\put(14.4,14){$S_{0}^{(3)}$} \put(13.5,21.6){$S_{0}^{(4)}$}
\put(18.4,24.7){$S_{0}^{(1)}$} \put(19.9,16.8){$S_{0}^{(2)}$}
\put(13.5,31){$S_{1}$} \put(20.5,31){$S_{1}$} \put(13.5,8){$S_{1}$}
\put(20.5,8){$S_{1}$} \put(15.5,34.6){$S_{1}$}
\put(15.5,4.6){$S_{1}$} \put(4,20.8){$S_{1}$} \put(30,20.8){$S_{1}$}
\put(7.5,16){$S_{1}$} \put(7.5,23){$S_{1}$} \put(27,16){$S_{1}$}
\put(27,23){$S_{1}$}
\end{picture}
\end{center}
\caption[The recursive Bethe-type lattice of $4$-polygons with additional inner bond] {\small {The
recursive Bethe-type lattice of $4$-polygons with additional inner bond. $S_{0}^{(i)}$ are the spin
variables of $0$-th shell, $S_{1}$ of the first shell. }} \label {latt1}
\end{figure}
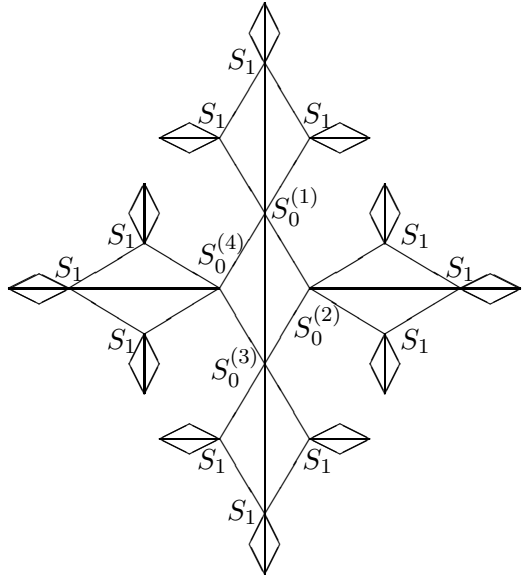
\end{center}

We can attach to each site of the central plaquette a new one.
Carrying out this procedure successively for each new
 shell, we can obtain a recursive
lattice which actually is Bethe-type lattice of square plaquettes
with additional inner links . It is evident that, for each
plaquette, the coordination numbers of its sites are 6 or 5.
Introducing the following parameters,
\begin{eqnarray}
\alpha _1&=&\beta \left( \frac{J_3}2-\frac{J_4}4-\frac{J_2}2\right)\\
\alpha _2&=&\beta \left( J_3-\frac{J_4}4-\frac{J_2}2\right)\\
\alpha_3&=&-\frac{\beta J_4}4, \label{parametres} \\
h&=&\beta \frac{\gamma \hbar H}2
\end{eqnarray}
the Hamiltonian can be rewritten as:
\begin{eqnarray}
-\beta \mathcal{H} &=&\sum_{\diamondsuit }\Bigl\{\alpha _1 \left(
\mbox{\boldmath $S$}_i\cdot%
\mbox{\boldmath $S$}_j+\mbox{\boldmath $S$}_j \cdot \mbox{\boldmath $S$}_k+%
\mbox{\boldmath $S$}_k \cdot \mbox{\boldmath $S$}_l+\mbox{\boldmath
$S$}_l \cdot \mbox{\boldmath $S$}_i \right)
+\alpha_2(\mbox{\boldmath $S$}_i \cdot
\mbox{\boldmath $S$}_k)  \label{mainhamil} \\
&+&\alpha _3 \bigl( \mbox{\boldmath $S$}_j \cdot \mbox{\boldmath
$S$}_l+\left( \mbox{\boldmath $S$}_i \cdot \mbox{\boldmath $S$}_j
\right)
\left( \mbox{\boldmath $S$}_k \cdot %
\mbox{\boldmath $S$}_l\right) +\left( \mbox{\boldmath $S$}_l
\cdot \mbox{\boldmath $S$}_i%
\right) \left( \mbox{\boldmath $S$}_j \cdot \mbox{\boldmath
$S$}_k\right)\nonumber\\
&& -\left( \mbox{\boldmath $S$}_i \cdot \mbox{\boldmath
$S$}_k\right)
\left( \mbox{\boldmath $S$}_l \cdot %
\mbox{\boldmath $S$}_j\right) \bigr)  \nonumber \\
&+&h\left( S_i^z+S_j^z+S_k^z+S_l^z\right) \Bigr\}.  \nonumber
\end{eqnarray}
If we use the multisite interaction Ising model, Eq.
(\ref{mainhamil}) takes the following form:
\begin{eqnarray}
-\beta {\mathcal{H}} &=& \sum_{\diamondsuit} \{\alpha _1 \left(
s_is_j+s_js_k+s_ks_l+s_ls_i\right)
+\alpha _2s_is_k+\alpha _3\left( s_js_l+s_is_js_ks_l\right)  \nonumber \\%
[0.02in] &&+h\left( s_i+s_j+s_k+s_l\right) \},
\end{eqnarray}
where $s_{i}$ takes values $\pm 1$ .

The partition function takes the form \cite{ar}:
\begin{eqnarray}
Z &=&\sum_{\left\{ S_0\right\} }\exp \Bigl\{\alpha _1\left(
s_0^{\left( 1\right) }s_0^{\left( 2\right) }+s_0^{\left( 2\right)
}s_0^{\left( 3\right) }+s_0^{\left( 3\right) }s_0^{\left( 4\right)
}+s_0^{\left( 4\right) }s_0^{\left( 1\right) }\right) +\alpha
_2s_0^{\left( 1\right) }s_0^{\left(
3\right) }+  \nonumber \\
&&+\alpha _3\left( s_0^{\left( 2\right) }s_0^{\left( 4\right)
}+s_0^{\left( 1\right) }s_0^{\left( 2\right) }s_0^{\left( 3\right)
}s_0^{\left( 4\right) }\right) +h\left( s_0^{\left( 1\right)
}+s_0^{\left( 2\right) }+s_0^{\left(
3\right) }+s_0^{\left( 4\right) }\right) \Bigr\}\times  \nonumber \\
&&\times g_N\left( s_0^{\left( 1\right) }\right) g_N\left(
s_0^{\left( 2\right) }\right) g_N\left( s_0^{\left( 3\right)
}\right) g_N\left( s_0^{\left( 4\right) }\right) ,
\end{eqnarray}
where $s_0^{\left( a\right) }$ are spins of central plaquette,
$g_N\left( s_0^{\left( a\right) }\right) $denotes contribution of
branch at a-th site of central plaquette and $N$ is the number of
generations ($N\rightarrow \infty $ case corresponds to the
thermodynamic limit and neglecting the surface effects).
 For both values ($\pm 1$) of $s_0^{\left(
1\right) }$, one can easily calculate:
\begin{eqnarray}
\label{gn(+)}
 g_N(+) & = &a^4bc^2d^3g_{N-1}^3(+) + 2bc^{-2}dg_{N-1}^2(+)g_{N-1}(-)\\
 &+& b^{-1}dg_{N-1}^2(+) g_{N-1}(-) + a^{-4}bc^2d^{-1}g_{N-1}(+) g_{N-1}^2(-)\nonumber \\
 &+&2b^{-1}d^{-1}g_{N-1}(+)g_{N-1}^2(-)  + b^{-1}d^{-3}g_{N-1}^3(-)\nonumber ,
\end{eqnarray}
\begin{eqnarray}
\label{gn(-)} g_N(-)& = &b^{-1}d^3g_{N-1}^3(+) + 2b^{-1}dg_{N-1}^2(+)g_{N-1}(-) \\
&+&a^{-4}bc^2dg_{N-1}^2(+) g_{N-1}(-) + b^{-1}d^{-1}g_{N-1}(+)g_{N-1}^2(-)\nonumber \\
&+&2bc^{-2}d^{-1}g_{N-1}(+)g_{N-1}^2(-)+a^4bc^2d^{-3}g_{N-1}^3(-)\nonumber,
\end{eqnarray}
where the following notations has been introduced:
\begin{equation}
a=\exp \alpha _1,\qquad b=\exp \alpha _2,\qquad c=\exp \alpha _3,
\qquad d=\exp h.  \label{abcd}
\end{equation}

Using Eq. (\ref{gn(+)}) and Eq. (\ref{gn(-)}), one can obtain the
recursion relation for variable $x_N=\frac{g_N\left( +\right)
}{g_N\left( -\right) }$:
\begin{equation}
x_N=f\left( x_{N-1}\right), \qquad f\left( x\right)
=\frac{A{\mu}^3x^3+(2B+1)
{\mu}^2x^2+(C+2){\mu}x+1}{%
{\mu}^3x^3+(C+2){\mu}^2x^2+(2B+1){\mu}x+A}.  \label{recrel}
\end{equation}
Here
\begin{eqnarray}
A&=&\exp(\beta(4J_3-2J_4-3J_2))\\
B&=&\exp(\beta(2J_3-J_2))\\
 C&=&\exp({\beta}J_2)\\
 \mu&=&\exp(2h)=\exp(\beta \gamma \hbar H).
\end{eqnarray}
\section{Magnetic Properties of the Antiferromagnetic Model on the Square Lattice}
\hspace{14pt} The  recursion relation (\ref{recrel}) plays the
central role in our further investigations. Because it provides  all
the thermodynamic properties of the system. In particular,
analogously, recursion relation for the magnetization per site,
 \begin{eqnarray}
 m=\frac{\sum_{(s)}s_ie^{-{\beta}{\mathcal{H}}}}
{\sum_{(s)}e^{-{\beta}{\mathcal{H}}}}, \label{mag}
 \end{eqnarray}
 can be derived.
\begin{figure}
\begin{center}
\begin{tabular}{cc}
{\small (a)}& {\small (b)}\\
\epsfig{file=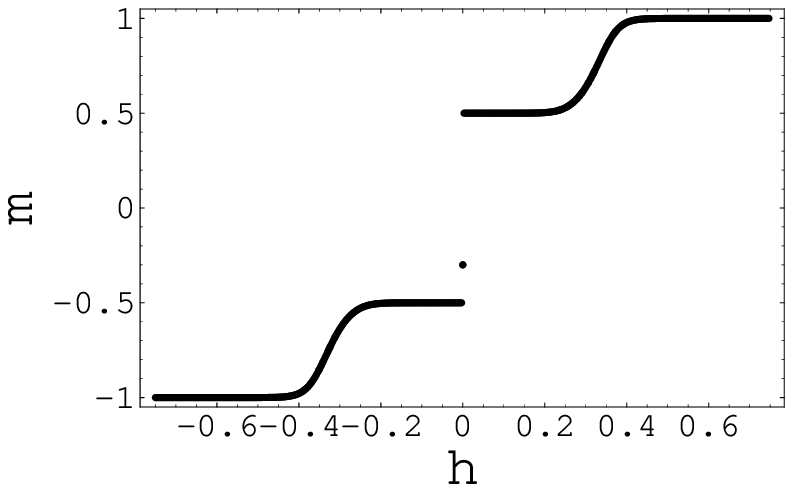,width=200pt} & \epsfig{file=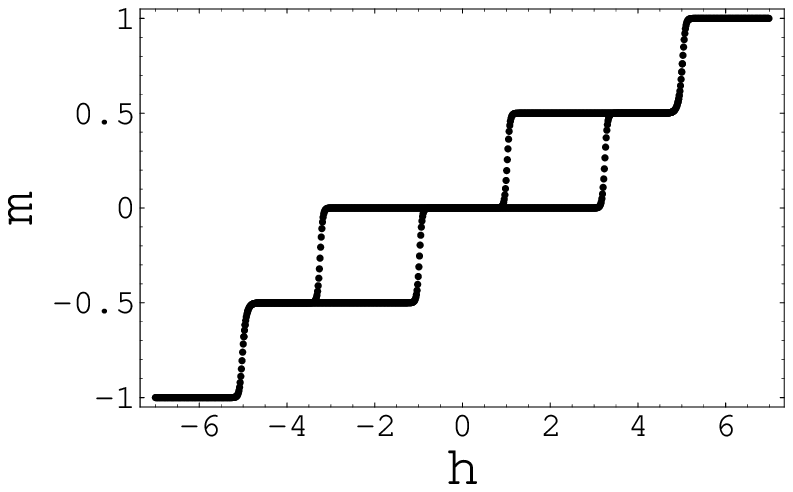,width=200pt}\\
\end{tabular}
\caption[Magnetization plateau (a) and bifurcation points and period doubling (b)] {\small {
Magnetization plateau for temperature $T=0.04mK$ (a); bifurcation points
 and period doubling  for temperature $T=0.06mK$ (b).}}
 \label{latt2}
 \end{center}
\end{figure}
Let us describe  the magnetic behavior of our model in a strong
magnetic field. For the $2D$ $^3$He films recent experimental
measurements predict the following relations between the exchange
energies $J_n$ on the regular triangular lattice:
\begin{equation}
J_3 > J_2 > J_4. \label{param}
\end{equation}
The main features of the resulting magnetic behavior of the system
 under consideration are caused by the interplay
 between ferromagnetic ($J_3$) and antiferromagnetic ($J_2$ and $J_4$)
interactions. Also we use the well-known relations between exchange
parameters $(J_2, J_3, J_4$), which have been estimated from
susceptibility and specific-heat  data in the low density region
\cite{Bernu}:
   \begin{eqnarray}
   J=J_2-2J_3 \approx - 3 mK, \qquad       K=J_4 \cong 1.873 mK. \label{par}
   \end{eqnarray}

We have taken the following exchange parameters: $J_2=1.75 mK$,
$J_3=2.35 mK$ and $J_4=1.5 mK$. At  $T=0.04 mk$, we get the
magnetization plateau [see Fig. \ref{latt2}(a)]. The plateaus at
$m=0$ and $m/m_{sat}=1/2$, bifurcation points and period doubling
take place  for the values $J_2=2 mK$, $J_3=0.5 mK$ and $J_4=0.5 mK$
at $T=0.06 mk$ [see Fig. \ref{latt2}(b)].

\section{Recursive Approximation to Kagome Lattice}
\hspace{14pt}Since the density ratio  of the third layer of $^3He$
is less then the first and second ones, it can use a Kagome
lattice\cite{rog2,kag}. For using the dynamic system approach it is
necessary to approximate the Kagome lattice by recursive one. We use
the Husimi lattice with two triangles coming out from one site (see
Fig.\ref{kag}).

\begin{figure}
\begin{center}
\epsfig{file=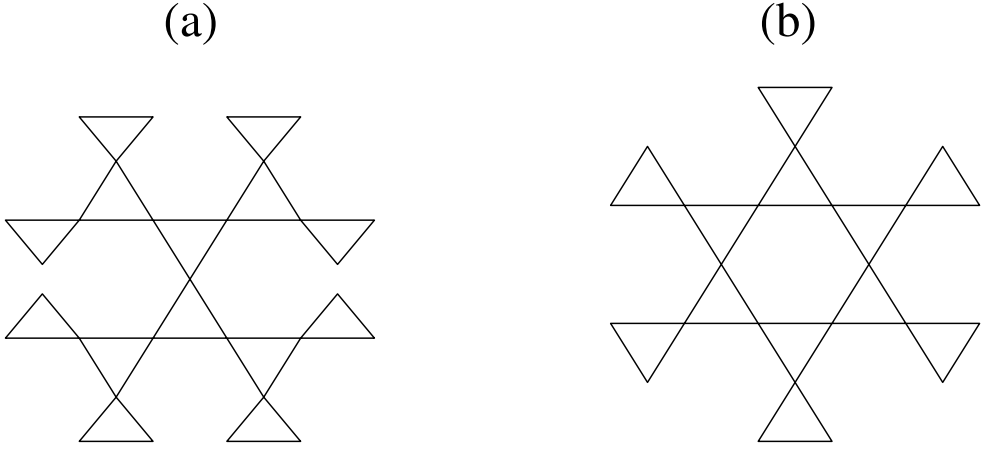,width=13.0cm} \caption {\small {A recursive approximation to the Kagome (b)
lattice by Husimi (a) one}}
 \label{kag}
 \end{center}
\end{figure}

In this case
\begin{equation}
{\mathcal {H}}_{ex}=J_2\sum_{pairs}\left( P_2+P_2^{-1}\right)
-J_3\sum_{triangles}\left( P_3+P_3^{-1}\right) \label{K2}
\end{equation}
and
\begin{equation}
{\mathcal{H}}_Z=-\sum_i\frac \gamma 2\hbar \mbox{\boldmath $H$}
\cdot \mbox{\boldmath $\sigma$}_i \label{Zeeman}
\end{equation}
where $\gamma $ is the gyromagnetic ratio of the $^3$He nucleus. The
two-,and three-spin exchanges are given by Eq.(\ref{P2}) and
Eq.(\ref{P3}). We have done further approximation passing to
classical Ising model as in the $^3He$ solid case \cite{ar}.

The partition function for recursive lattice is written in the form:
\begin{equation}
Z=\sum_{\sigma_0}\exp (\beta h {\sigma}_0)\cdot g_n^2({ \sigma}_0).
\end{equation}
Here $\sigma_i$ takes the values $\pm 1$, $\sigma_0$ is the central
spin and $g_n({\sigma}_0)$ denotes the contribution of each branch
of the partition function.

Introducing, as in previous section, $g_n(+)$, $g_n(-)$ and
$x_n=g_n(+)/g_n(-)$ one can receive the exact one dimensional
rational mapping for partition function
\begin{equation}
x_n=f\left( x_{n-1}\right), \qquad f\left( x,\mu,z\right)
=\frac{z{\mu}^2x^2+ 2{\mu}x+1}{%
{\mu}^2x^2+2{\mu}x+z},  \label{Krecrel}
\end{equation}
where $z=e^{-4 \beta J}, \mu=e^{2\beta h}$ and $J=\frac
{(J_2-J_3)}{2}$.

\begin{figure}[h]
\begin{center}
\begin{tabular}{cc}
{\small (a)}& {\small (b)}\\
\epsfig{file=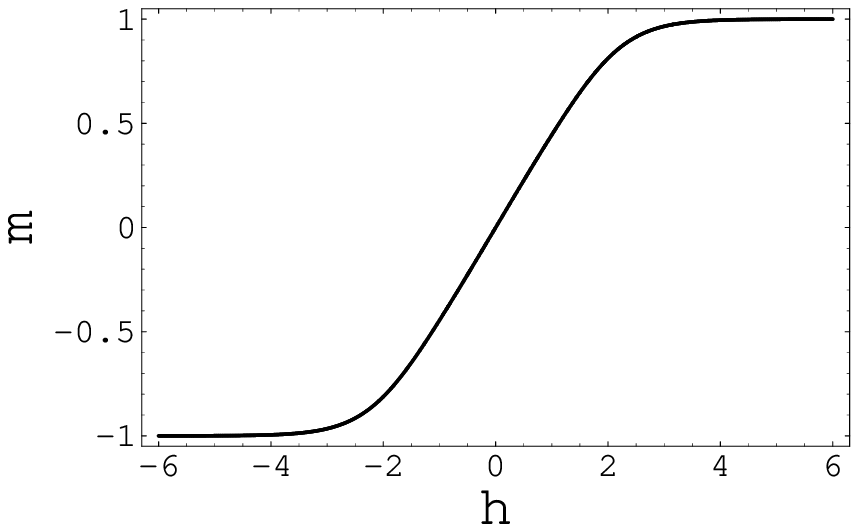,width=7.0cm} & \epsfig{file=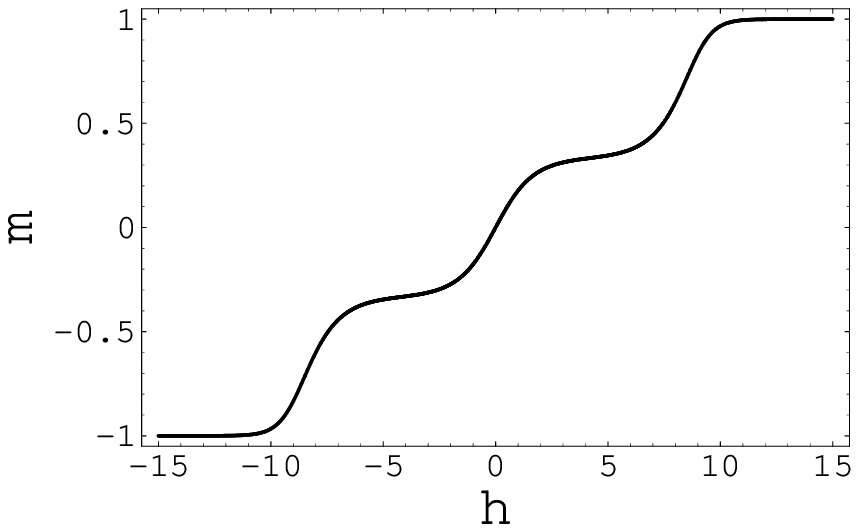,width=7.0cm}\\
{\small (c)}& {\small (d)}\\
\epsfig{file=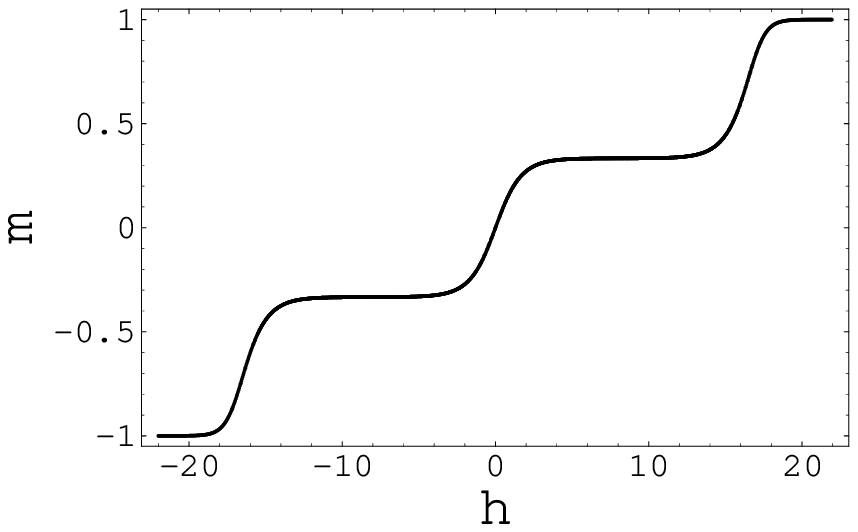,width=7.0cm}& \epsfig{file=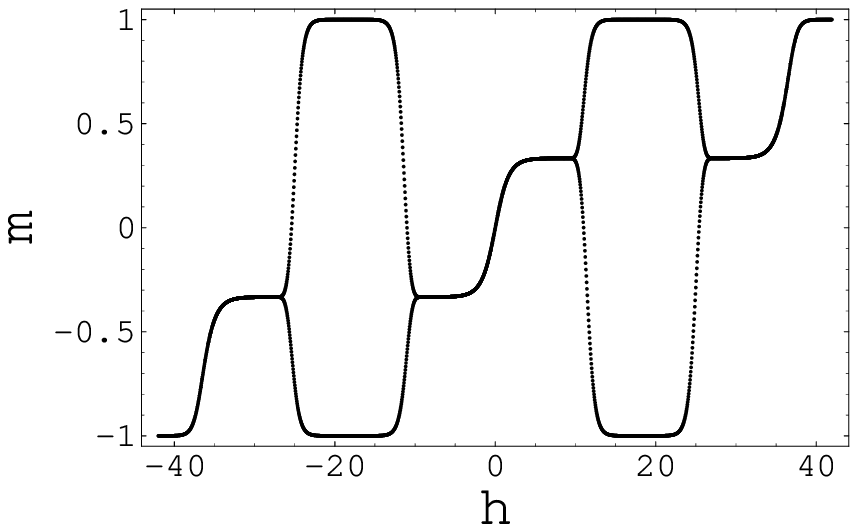,width=7.0cm}\\
\end{tabular}
\caption[Magnetization processes for several values of $J$ at $T=1mK$] {\small {Magnetization
processes for the values of exchange constants J
 at temperature $T=1mK$: (a) J=0.5mK, (b) J=4mK, (c)J=8mK and (d) J=18mK}}
 \label{plots}
 \end{center}
\end{figure}
The magnetization $m$ Eq.(\ref{mag}) as a function  of $x$,
temperature and external magnetic field can be written as:
\begin{equation}
m=\frac {\mu x^2 -1}{\mu x^2 +1}.
\end{equation}

 As an example, we take the temperature $T=1mK$ and obtain
 the figures of \emph{m}(magnetization) versus  \emph{h}(external magnetic field)(see Fig.(\ref{plots})).
  We take $J>0$ since the antiferromagnetic  pair exchange  interaction  may be larger than
the ferromagnetic three-spin interaction on a Kagome-type lattice.

At J=0.5mK and 4mK, we have a usual behavior for $m$ (see
Fig.(\ref{plots}a) and Fig.(\ref{plots}b)). The magnetization
plateau (m=1/3) takes place at J=8mK, Fig.(\ref{plots}c). At higher
values of J, J=18, bifurcation points and one-period doubling point
occur (see Fig.(\ref{plots}d))

\section{Hexagonal Recursive Lattice as an Approximation of the Triangular One}
\hspace{14pt}One can write down the exchange Hamiltonian for planar
solid $\ ^3$He in the following way:
           \begin{eqnarray}
            H_{ex}  &=& J_2\sum_{pairs}(P_2+P_2^{-1})-J_3\sum_{triangles}(P_3+P_3^{-1})
              +J_4\sum_{rectangles}(P_4+P_4^{-1})\nonumber\\
              &-&J_5\sum_{pentagons}(P_5+P_5^{-1})
             +J_6\sum_{hexagons}(P_6+P_6^{-1}),
           \end{eqnarray}
         where the sum in the first term is going over all pairs of particles, in the second term over
         all triangles and so on $($see figure \ref{fig:FIGURE_2}$)$.

%
         Using the same technique as in Sec.\ref{square} we can derive expressions for fith and sixth exchange ineractions

         \begin{equation}
           P_{ijklm}+P_{ijklm}^{-1}=\frac{1}{8} (1+\sum_{\mu < \nu } (\bf\sigma_{\it \mu}\bf\sigma_{\it \nu})+\sum_
           {\mu < \nu <\lambda <\rho}G_{ \mu \nu \lambda \rho})
         \end{equation}

         \begin{equation}
           P_{ijklmn}+P_{ijklmn}^{-1}=\frac{1}{16} (1+\sum_{\mu < \nu } (\bf\sigma_{\it \mu}\bf\sigma_{\it \nu})+
           \sum_{\mu < \nu <\lambda <\rho}G_{ \mu \nu \lambda \rho} + S_{ijklmn})
         \end{equation}
         where $S_{ijklmn}$ is
           \begin{eqnarray}
             S_{ijklmn} &=&  \bigl[(\bf\sigma_{\it i}\bf\sigma_{\it j})(\bf\sigma_{\it k}\bf\sigma_{\it l})(\bf\sigma_
               {\it m}\bf\sigma_{\it n})+(\bf\sigma_{\it j}\bf\sigma_{\it k})(\bf\sigma_{\it l}\bf\sigma_{\it m})
               (\bf\sigma_{\it n}\bf\sigma_{\it i})\bigr]\\
             &-&\bigl[(\bf\sigma_{\it i}\bf\sigma_{\it j})(\bf\sigma_{\it k}\bf\sigma_{\it m})(\bf\sigma_{\it l}
               \bf\sigma_{\it n})+(\bf\sigma_{\it j}\bf\sigma_{\it k})(\bf\sigma_{\it l}\bf\sigma_{\it n})
               (\bf\sigma_{\it m}\bf\sigma_{\it i}) \nonumber\\
               &+&(\bf\sigma_{\it k}\bf\sigma_{\it l})(\bf\sigma_{\it m}
               \bf\sigma_{\it i})(\bf\sigma_{\it n}\bf\sigma_{\it j})+(\bf\sigma_{\it l}\bf\sigma_{\it m})
               (\bf\sigma_{\it n}\bf\sigma_{\it j})(\bf\sigma_{\it i}\bf\sigma_{\it k})+(\bf\sigma_{\it m}
               \bf\sigma_{\it n}) \nonumber\\
               \nonumber
                &+&(\bf\sigma_{\it i}\bf\sigma_{\it k})(\bf\sigma_{\it j}\bf\sigma_{\it l})+
               (\bf\sigma_{\it n}\bf\sigma_{\it i})(\bf\sigma_{\it j}(\bf\sigma_{\it n}\bf\sigma_{\it k})
               \sigma_{\it m})\bigr]+\bigl[(\bf\sigma_{\it i}\bf\sigma_{\it l})(\bf\sigma_{\it j}\bf\sigma_{\it n})
               (\bf\sigma_{\it k}\bf\sigma_{\it m}) \\
               \nonumber
                &+&(\bf\sigma_{\it j}\bf\sigma_{\it m})(\bf\sigma_{\it k}
               \bf\sigma_{\it i})(\bf\sigma_{\it l}\bf\sigma_{\it n})+(\bf\sigma_{\it k}\bf\sigma_{\it n})
               (\bf\sigma_{\it l}\bf\sigma_{\it j})(\bf\sigma_{\it m}\bf\sigma_{\it i})\bigr]
             + \bigl[(\bf\sigma_{\it i}\bf\sigma_{\it j})(\bf\sigma_{\it k}\bf\sigma_{\it n})(\bf\sigma_{\it l}
               \bf\sigma_{\it m}) \\
               \nonumber
                &+&(\bf\sigma_{\it j}\bf\sigma_{\it k})(\bf\sigma_{\it l}\bf\sigma_{\it i})
               (\bf\sigma_{\it m}\bf\sigma_{\it n})+(\bf\sigma_{\it k}\bf\sigma_{\it l})(\bf\sigma_{\it m}
               \bf\sigma_{\it j})(\bf\sigma_{\it n}\bf\sigma_{\it i})\bigr]\\
               &-&\bigl[(\bf\sigma_{\it i}\bf\sigma_{\it l})
               (\bf\sigma_{\it j}\bf\sigma_{\it m})(\bf\sigma_{\it k}
               \bf\sigma_{\it n})\bigr]\nonumber
           \end{eqnarray}

          \begin{figure}[h]
           \begin{center}
           \psfig{figure=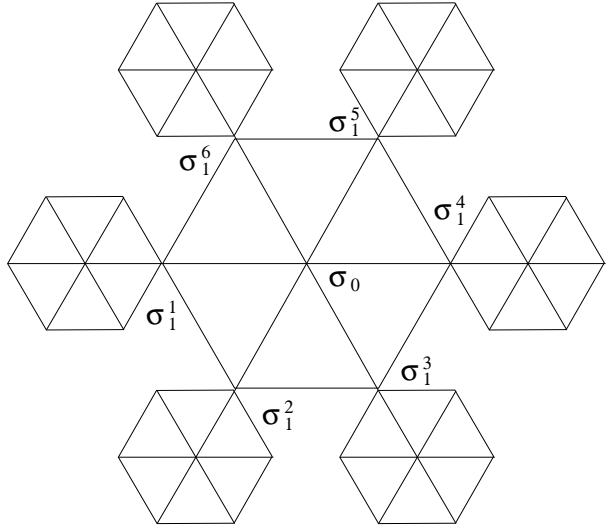,width=8.0cm}
           \caption{\label{fig:FIGURE_2} \small The hexagon recursive lattice.}
           \end{center}
         \end{figure}
         Now let us consider the recursive lattices which are connected through
         the sites for getting one$-$dimensional rational mapping which is
         close to triangular lattice. The central hexagon is approached to
         the triangular lattice by connection of center to vertices of
         hexagon, so we get six triangles in the central hexagon. The
         recursive lattice is constructed by adding next shells
         $($branches$)$ of hexagons on vertices of previous one $($see
         figure \ref{fig:FIGURE_2}$)$. Next shells are constructed in the same way. For
         hexagonal recursive lattice we obtain the exact recursive
         relations for partition function. It's important to mention that
         this recursive relation is dynamic.

           By now we considered Heisenberg model for description of our recursive
         lattice. We can assume that in the strong external magnetic field all
         atom' spins will be directed mostly by certain z axis. So it will be
         possible to calculate the eigen-values of spin$-$operators. One may consider
         the variables $s_i$ as classical vectors as well as Pauli
         matrices. If we use the multisite
         interaction Ising model, we can take +1 and -1 instead of $\sigma_i$.
         In the rest of the paper all constance will be taken in temperature's
         scaling $(J_2=$$\bf J_2$$/T, J_3=$$\bf J_3$$/T, J_4=$$\bf J_4$$/T,
         J_5=$$\bf J_5$$/T, J_6=$$\bf J_6$$/T,$ $h=$$H$$/T, )$.

         When the lattice cut apart from the central hexagon, it separates
         into six identical branches. So we can first realize a summation
         over all spin configurations on each branch, getting the same
         result every time, and then the sum over spins of the central
         hexagon.
         Here follows, that we can represent the partition function
         as
           \begin{eqnarray}
             Z=\sum_{s_0s_1s_2s_3s_4s_5s_6}e^{-h(s_0+s_1+s_
               2+s_3+s_4+s_5+s_6)+H_{ex}}\\
               \times g_n(s_1)g_n(s_2)g_n(s_3)g_n(s_4)
             g_n(s_5)g_n(s_6)
           \end{eqnarray}
         from which the expression takes the form
           \begin{eqnarray}
             Z & = &
             e^{-5h-6J_2-2J_6}g(+)^6+e^{-7h-12J_2+12J_3-12J_4+12J_5-2J_6}g(+)^6\\\nonumber
             &+&6e^{-3h-5J_2}g(+)^5g(-) + 9e^{-h-4J_2}g(+)^4g(-)^2 + 6e^{-h-6J_2+2J_3}g(+)^4g(-)^2 \\\nonumber
             &+&3e^{-3h-6J_2+4J_3}g(+)^4g(-)^2 + 6e^{-3h-6J_2+4J_3-2J_4}g(+)^4g(-)^2 \\\nonumber
             &+&6e^{-3h-8J_2+6J_3-4J_4+2J-5}g(+)^4g(-)^2 + 2e^{-h-3J_2}g(+)^3g(-)^3\\\nonumber
             &+&2e^{h-3J_2}g(+)^3g(-)^3 + 12e^{-h-5J_2+2J_3}g(+)^3g(-)^3 + 12e^{h-5J_2+2J_3}g(+)^3g(-)^3\\\nonumber
             &+&6e^{-h-7J_2+4J_3-2J_4}g(+)^3g(-)^3+6e^{h-7J_2+4J_3-2J_4}g(+)^3g(-)^3 \\\nonumber
             &+&9e^{h-4J_2}g(+)^2g(-)^4+6e^{h-6J_2+2J_3}g(+)^2g(-)^4+3e^{3h-6J_2+4J_3}g(+)^2g(-)^4 \\\nonumber
             &+&6e^{3h-6J_2+4J_3-2J_4}g(+)^2g(-)^4+6e^{3h-8J_2+6J_3-4J_4+2J_5}g(+)^2g(-)^4\\\nonumber
             &+&6e^{3h-5J_2}g(+)g(-)^5+6e^{5h-9J_2+8J_3-6J_4+4J_5}g(+)g(-)^5+e^{5h-6J_2-2J_6}g(-)^6\\\nonumber
             &+&e^{7h-12J_2+12J_3-12J_4+12J_5-2J_6}g(-)^6\nonumber
           \end{eqnarray}
         where
         \begin{eqnarray}
           \nonumber
           g_n(s_i)&=&\sum_{s_1^i}e^{-h(s_i+s_1+s_2+s_3+s_4+s_5+s_6)+H_{ex}}\\
          &&\times
          g_{n-1}(s^{(2)}_1)g_{n-1}(s^{(2)}_2)g_{n-1}(s^{(2)}_3)g_{n-1}(s^{(2)}_4)g_{n-1}(s^{(2)}_5).
           \end{eqnarray}

         We can obtain the recurrent relation for the variable $x_n=\frac{g_n(+)}{g_n(-)}$: $$x_n=f(x_{n-1}),$$
           \begin{eqnarray}
             f(X) & = & \Bigl[e^{-4h+7J_2}+e^{-6h+3J_2+8J_3-6J_4+4J_5}
               +(3e^{-2h+8J_2}+2e^{-2h+6J_2+2J_3} \\
               &&
               +e^{-4h+6J_2+4J_3}+2e^{-4h+6J_2+4J_3-2J_4}+2e^{-4h+4J_2+6J_3-4J_4+2J_5})X\nonumber\\
               \nonumber
               &+&(e^{9J_2} + e^{-2h+9J_2}+6e^{7J_2+2J_3}+6e^{-2h+7J_2+2J_3}\\
               &&+3e^{5J_2+4J_3-2J_4}+3e^{-2h+5J_2+4J_3-2J_4})X^2\nonumber \\
               \nonumber
               &+&(6e^{8J_2}+4e^{6J_2+2J_3}+2e^{2h+6J_2+4J_3}\\
               &&+4e^{2h+6J_2+4J_3-2J_4}+4e^{2h+4J_2+6J_3-4J_4+2J_5})X^3\nonumber\\
               &+&(5e^{2h+7J_2}+5e^{4h+3J_2+8J_3-6J_4+4J_5})X^4 \nonumber\\
               \nonumber
               &+&(e^{4h+6J_2-2J_6}+e^{6h+12J_3-12J_4+12J_5-2J_6})X^5\Bigr]\Big/\Bigl[(e^{4h+7J_2}+e^{6h+3J_2+8J_3-6J_4+4J_5})X^5 \\
               \nonumber
               &+&(3e^{2h+8J_2}+2e^{2h+6J_2+2J_3}+e^{4h+6J_2+4J_3} \\
               \nonumber
               &&+2e^{4h+6J_2+4J_3-2J_4} +2e^{4h+4J_2+6J_3-4J_4+2J_5})X^4\\
               &+&(e^{9J_2}+e^{2h+9J_2} + 6e^{7J_2+2J_3} + 6e^{2h+7J_2+2J_3}\nonumber\\
               &&+3e^{5J_2+4J_3-2J_4} + 3e^{2h+5J_2+4J_3-2J_4})X^3\nonumber\\
               &+&(6e^{8J_2} +4e^{6J_2+2J_3}+2e^{-2h+6J_2+4J_3}+4e^{-2h+6J_2+4J_3-2J_4}\nonumber\\
               &&+4e^{-2h+4J_2+6J_3-4J_4+2J_5})X^2 + (5e^{-2h+7J_2}+5e^{-4h+3J_2+8J_3-6J_4+4J_5})X\nonumber\\
               &+&e^{-4h+6J_2-2J_6}+e^{-6h+12J_3-12J_4+12J_5-2J_6}\Bigr]\nonumber
           \end{eqnarray}

         This recurrent relation plays a crucial role in our further investigations,
         because it's getting possible obtain all thermodynamic quantities
         of the considered system.
\newpage
\section[Magnetic Properties of the Antiferromagnetic Model on the Hexagonal Lattice]{Magnetic Properties of the Antiferromagnetic Model on the Hexagonal Recursive Lattice}
\hspace{14pt}Using the technique of dynamic system       theory,
applied to one$-$dimensional rational mapping, we can get
magnetization relations per
         site. As mentioned above for an homogeneous lattice the magnetization
         function is given by the formula

         \begin{equation}
           m=\frac{\sum_{(s)}s_ie^{-\beta H}}{\sum_{(s)}e^{-\beta H}}
         \end{equation}
         where $s_i$ is the spin on an arbitrary site. In our case, the
         lattice is homogeneous, so on central hexagon we divided the
         tree of hexagons on six branches.

         We could obtain the magnetization value of sublattice composed by
         hexagon central vertices.
           \begin{eqnarray}
             M_0=\frac{1}{Z}\sum_{s_0s_1s_2s_3s_4s_5s_6}
             s_0e^{-h(s_0+s_1+s_2+s_3+s_4+s_5+s_6)+H_{ex}}
              g_n(s_1)g_n(s_2)g_n(s_3)g_n(s_4)g_n(s_5)g_n(s_6)
           \end{eqnarray}
         and for magnetization sublattice composed by corner verticies.
           \begin{eqnarray}
             M_1=\frac{1}{Z}\sum_{s_0s_1s_2s_3s_4s_5s_6}s_1e^{-h(s_0+s_1+s_2+s_3+s_4+s_5+s_6)+H_{ex}}
             g_n(s_1)
               g_n(s_2)g_n(s_3)g_n(s_4)g_n(s_5)g_n(s_6)
           \end{eqnarray}
         So from this expressions finally we could get
           \begin{eqnarray}
             M_0 & = &
             \bigl[(e^{-5h-6J_2-2J_6}-e^{-7h-12J_2+12J_3-12J_4+12J_5-2J_6})\\
             \nonumber
               &+&(6e^{-3h-5J_2}-6e^{-5h-9J_2+8J_3-6J_4+4J_5})X \\
               \nonumber
               &+&(9e^{-h-4J_2}+6e^{-h-6J_2+2J_3}-3e^{-3h-6J_2+4J_3} \\
               \nonumber
               &&-6e^{-3h-6J_2+4J_3-2J_4}-6e^{-3h-8J_2+6J_3-4J_4+2J_5})X^2\\
               \nonumber
               &+&(2e^{h-3J_2}-2e^{-h-3J_2}+12e^{h-5J_2+2J_3} -12e^{-h-5J_2+2J_3} \\
               \nonumber
               &&+6e^{h-7J_2+4J_3-2J_4}-6e^{-h-7J_2+4J_3-2J_4})X^3 \\
               \nonumber
               &+&(9e^{h-4J_2} +6e^{h-6J_2+2J_3}+3e^{3h-6J_2+4J_3}\\
               \nonumber
               &&+6e^{3h-6J_2+4J_3-2J_4}+6e^{3h-8J_2+6J_3-4J_4+2J_5})X^4 \\
               \nonumber
               &+&(6e^{3h-5J_2}+6e^{5h-9J_2+8J_3-6J_4+4J_5})X^5\\
               \nonumber
               &+&(e^{5h-6J_2-2J_6}+e^{7h-12J_2+12J_3-12J_4+12J_5-2J_6})X^6\bigr]\Big/\\
               \nonumber
               &&\bigl[(e^{-5h-6J_2-2J_6}+e^{-7h-12J_2+12J_3-12J_4+12J_5-2J_6}) \\
             \nonumber
             &+&(6e^{-3h-5J_2}+6e^{-5h-9J_2+8J_3-6J_4+4J_5})X \\
               \nonumber
               &+&(9e^{-h-4J_2}+6e^{-h-6J_2+2J_3}+3e^{-3h-6J_2+4J_3}+6e^{-3h-6J_2+4J_3-2J_4} \\
               \nonumber
               &&+6e^{-3h-8J_2+6J_3-4J_4+2J_5})X^2\\
               \nonumber
               &+&(2e^{-h-3J_2}+2e^{h-3J_2}+12e^{-h-5J_2+2J_3} +12e^{h-5J_2+2J_3}\\
               \nonumber
               &&+6e^{-h-7J_2+4J_3-2J_4}+6e^{h-7J_2+4J_3-2J_4})X^3\\
               \nonumber
               &+&(9e^{h-4J_2}+6e^{h-6J_2+2J_3}+3e^{3h-6J_2+4J_3}\\
               \nonumber
               &&+6e^{3h-6J_2+4J_3-2J_4}+6e^{3h-8J_2+6J_3-4J_4+2J_5})X^4 \\
               \nonumber
               &+&(6e^{3h-5J_2}+6e^{5h-9J_2+8J_3-6J_4+4J_5})X^5\\
               \nonumber
               &+&(e^{5h-6J_2-2J_6}+e^{7h-12J_2+12J_3-12J_4+12J_5-2J_6})X^6\bigr]
           \end{eqnarray}
         and
           \begin{eqnarray}
             M_1 & = &\bigl[-e^{-5h-6J_2-2J_6}-e^{-7h-12J_2+12J_3-12J_4+12J_5-2J_6}\\
               &-&(4e^{-3h-5J_2}+4e^{-5h-9J_2+8J_3-6J_4+4J_5})X \nonumber\\
               \nonumber
               &-&(3e^{-h-4J_2}+2e^{-h-6J_2+2J_3}+e^{-3h-6J_2+4J_3}\\
               &&+2e^{-3h-6J_2+4J_3-2J_4}+2e^{-3h-8J_2+6J_3-4J_4+2J_5})X^2\nonumber\\
               &+&(3e^{h-4J_2}+2e^{h-6J_2+2J_3}+e^{3h-6J_2+4J_3}\nonumber \\
               \nonumber
               &&
               +2e^{3h-6J_2+4J_3-2J_4}+2e^{3h-8J_2+6J_3-4J_4+2J_5})X^4\\
               \nonumber
               &+&(4e^{3h-5J_2}+4e^{5h-9J_2+8J_3-6J_4+4J_5})X^5 \\
               \nonumber
               &+&(e^{5h-6J_2-2J_6}+e^{7h-12J_2+12J_3-12J_4+12J_5-2J_6})X^6\bigr]\Big/\nonumber\\
             &&\bigl[(e^{-5h-6J_2-2J_6}+e^{-7h-12J_2+12J_3-12J_4+12J_5-2J_6})\nonumber \\
               \nonumber
               &&
               +(6e^{-3h-5J_2}+6e^{-5h-9J_2+8J_3-6J_4+4J_5})X\\
               &+&(9e^{-h-4J_2}+6e^{-h-6J_2+2J_3}+3e^{-3h-6J_2+4J_3}\nonumber \\
               \nonumber
               &&
               +6e^{-3h-6J_2+4J_3-2J_4}+6e^{-3h-8J_2+6J_3-4J_4+2J_5})X^2\\
               &+&(2e^{-h-3J_2}+2e^{h-3J_2}+12e^{-h-5J_2+2J_3} +12e^{h-5J_2+2J_3}\nonumber \\
               \nonumber
               &&+6e^{-h-7J_2+4J_3-2J_4} +6e^{h-7J_2+4J_3-2J_4})X^3 \\
               \nonumber
               &+&(9e^{h-4J_2}+6e^{h-6J_2+2J_3}+3e^{3h-6J_2+4J_3} \\
               \nonumber
               &&
               +6e^{3h-6J_2+4J_3-2J_4}+6e^{3h-8J_2+6J_3-4J_4+2J_5})X^4\\
               &+&(6e^{3h-5J_2}+6e^{5h-9J_2+8J_3-6J_4+4J_5})X^5\nonumber \\
               \nonumber
               &+&(e^{5h-6J_2-2J_6}+e^{7h-12J_2+12J_3-12J_4+12J_5-2J_6})X^6\bigr]
           \end{eqnarray}

         Having these dynamical expressions one can draw the plots of
         magnetization vs. external magnetic field for sublattices in
         various temperatures. To do that one should fix the value of the
         dimensionless magnetic field $\bf h$ $($for a given temperatures
         and exchange parameters$)$ and implement the simple iteration from
         the recursion relation for $f(x)$, beginning with some initial
         $x_0$. For achieving to the thermodynamical limit we have to
         apply infinite amount of iterations $(n\rightarrow\infty)$.

From experimental measurements and
         theoretical calculations we could conclude that the relations
         between the $J_n$ exchange energies on the regular triangular
         lattice are\cite{rog1}\cite{Bernu}\cite{godfrin}:
         \begin{equation}
           J_3>J_2>J_4\geq J_6>J_5
         \end{equation}

         It's important to mention that the values of pure $ J_n$ are not
         observable in the experiment measurements, because each n-spin
         exchanges makes also a contribution to a few $(n-1)$-spin
         exchanges, and thus there are some effective exchanges parameters,
         which are certain contributions of $ J_n$ $($such as $ J=J_2-2J_3$
         and $ K=J_4-2J_5)$ and can be directly obtained from
         experiments \cite{rog1,godfrin,Bernu,pl}.

         The exchange parameters depend on the particle density and on the
         type of the lattice, particularly on its coordination number and
         dimensionality. For instance, it was found that for 2D triangular
         lattice at high densities the exchange $ J$ is dominant, mainly
         due to the three-spin exchange, but the ratio $ |K/J|$
         increases rapidly with the lowering of the particle density, and
         below other exchanges become important also. The magnetic
         properties of the system $($ferromagnetic or antiferromagnetic$)$
         depend on which exchange interactions are dominant at the present
         value of the particle density.

         From the arguments stated above one can conclude that there is a
         large freedom in the choice of concrete values of the exchange
         parameters $ J_2$, $ J_3$, $ J_4$, $ J_5$ and $
         J_6$. Moreover, it is quite difficult to identify our model,
         having so many assumptions, with some concrete value of particle
         density.

         The simplest case of pure ferromagnetic behavior $(J_3=2.5, J_5=0.5,
         J_2=J_4=J_6=0)$ is presented on figure \ref{figure2}. Figure \ref{figure2}a represents the magnetization for the central
         vertex of hexagon recursive lattice with temperature $T=$0.1, figure \ref{figure2}b the magnetization for the corner
         vertex lattice with temperature $T=0.1$ and figure \ref{figure2}c the average
         magnetization for the lattice with temperature $T=0.1$. At relatively high
         temperatures the magnetization curve has a smooth monotone form of
         Langevin type with a rather large value of the saturation field.
         Decreasing the temperature, the curve becomes more steep and the
         value of the saturation field decreases. Further decreasing the
         temperature leads to the magnetization diagram.
         We can see smoothly increasing curves, which are expected and
         corresponding to only ferromagnetic interactions.

         The physical case $(J=3mK)$ with ferromagnetic and antiferromagnetic interactions
         $(J_2=2, J_3=2.5, J_4=1.8, J_5=0.5, J_6=1)$ for
         enough high temperature $T=10$ is represented on
         figure \ref{figure3}. On plots we can
         see the paramagnetic limit as expected in high temperatures.
         \begin{figure}
           \begin{center}
             \psfig{figure=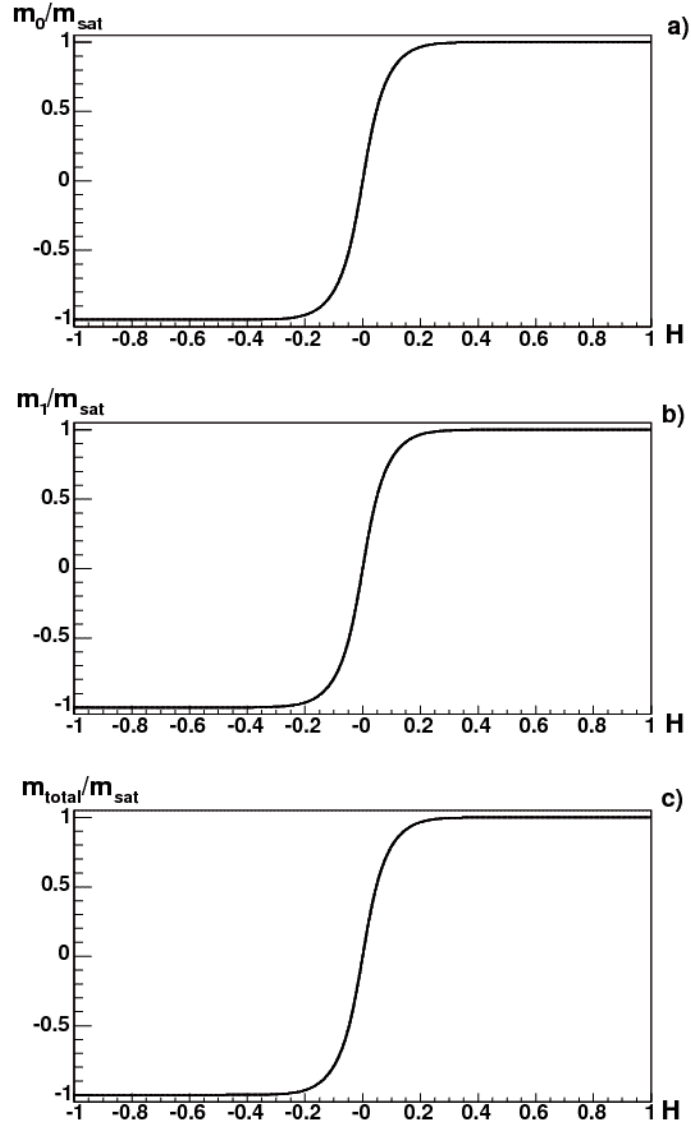,width=9.0cm}
           \end{center}
           \caption[Magnetization vs. magnetic field for $J_3=2.5, J_5=0.5,
             J_2=J_4=J_6=0$]{\small{ $J_3=2.5, J_5=0.5,
             J_2=J_4=J_6=0$ pure ferromagnetic case. a$)$ is the magnetization for the central
             vertex lattice with temperature $T=0.1$, b$)$ is the magnetization for the corner
             vertex lattice with temperature $T=0.1$, c$)$ is the average
             magnetization for the lattice with temperature $T=0.1$} \label{figure2}} \vspace{0.2cm}
         \end{figure}

         \begin{figure}
           \begin{center}
             \psfig{figure=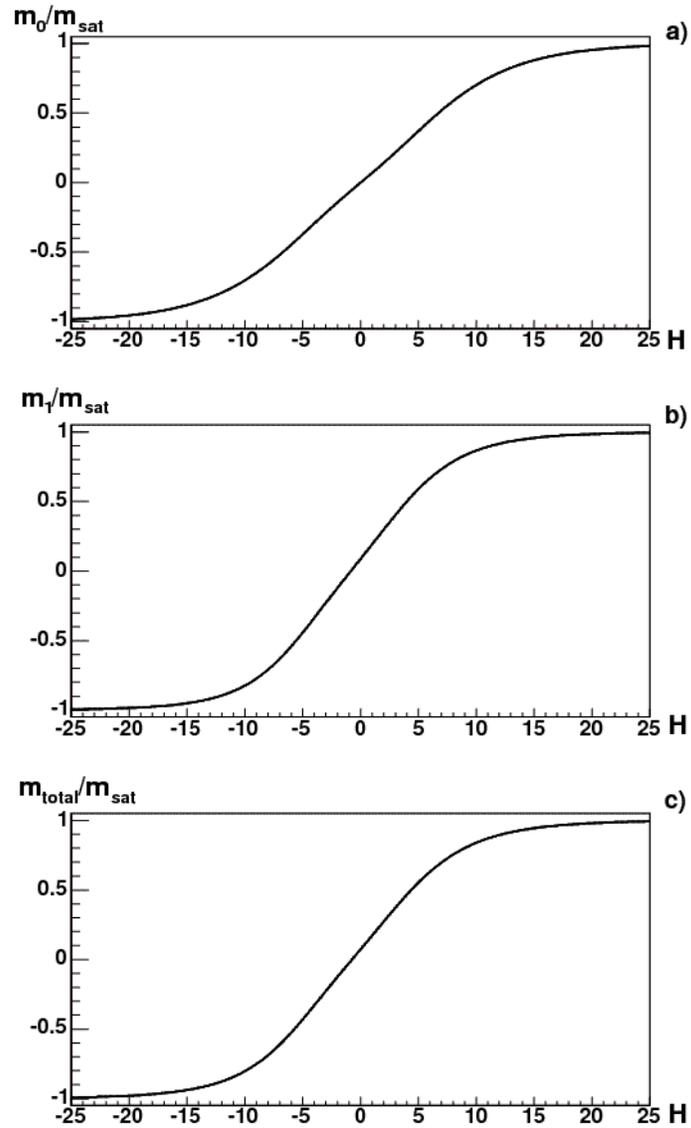,width=9.0cm}
           \end{center}
           \caption[Magnetization vs. magnetic field for $J_2=2, J_3=2.5, J_4=1.8, J_5=0.5,
             J_6=1$]{\small{ $J_2=2, J_3=2.5, J_4=1.8, J_5=0.5,
             J_6=1$ paramagnetic limit at enough high temperature. a$)$ is the magnetization for the central
             vertex lattice with temperature $T=10$, b$)$ is the magnetization for the corner
             vertex lattice with temperature $T=10$, c$)$ is the average
             magnetization for the lattice with temperature $T=10$}.\label{figure3}}\vspace{0.2cm}
         \end{figure}

         There is represented the depends between external magnetic field
         and magnetization on figure \ref{figure4} $($a$)$ is the magnetization for the central
         vertex lattice with temperature $T=0.1$, (b) is the magnetization for the corner
         vertex lattice with temperature $T=0.1$, (c) is the average
         magnetization for the lattice with temperature
         $T=0.1$ for physical case with $J=3mK$ $(J_2=2$, $J_3=2.5$, $J_4=1.8$, $J_5=0.5$,
         $J_6=1)$. The hatch line represents the magnetization value with
         $m/m_{sat}=2/3$. On figure \ref{figure4} we can see bifurcation points in high magnetic
         fields. This fact is making clear the advantage of represented model
         in high external magnetic field.There is a one period doubling
         behavior between bifurcation points, so that the contribution of
         antiferromagnetic interaction is visible.

         \begin{figure}
           \begin{center}
             \psfig{figure=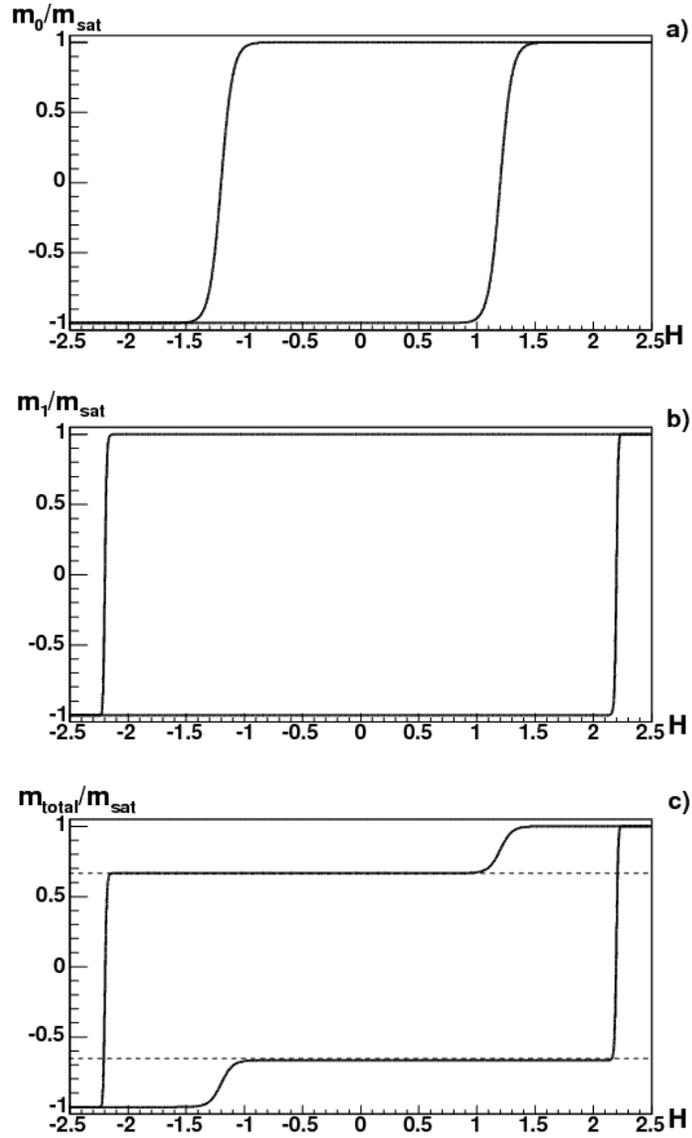,width=9.0cm}
           \end{center}
           \caption[Magnetization vs. magnetic field for $J_2=2, J_3=2.5, J_4=1.8, J_5=0.5,
             J_6=1$ with $J=3mK$]{\small { $J_2=2$, $J_3=2.5$, $J_4=1.8$, $J_5=0.5$,
             $J_6=1$ physical $J=3mK$  case with ferromagnetic and antiferromagnetic interactions
             contributions. There are bifurcation points and one period
             doubling between bifurcation points. a) is the magnetization for the central
             vertex lattice with temperature $T=0.1$, b$)$ is the magnetization for the corner
             vertex lattice with temperature $T=0.1$, c$)$ is the average
             magnetization for the lattice with temperature $T=0.1$.}\label{figure4}}\vspace{0.2cm}
         \end{figure}

         The magnetization plateau on $m/m_{sat}=2/3$ on figure \ref{figure5}c $($a$)$ is the magnetization for the central
         vertex lattice with temperature $T=0.1$, (b) is the magnetization for the corner
         vertex lattice with temperature $T=0.1$, (c) is the average
         magnetization for the lattice with temperature $T=0.1)$ with $(J_2=2, J_3=2.05, J_4=0, J_5=0,
         J_6=0)$. As in previous example we can see bifurcation points in high
         magnetic fields and a one period doubling behavior between bifurcation
         points.
         \begin{figure}
           \begin{center}
             \psfig{figure=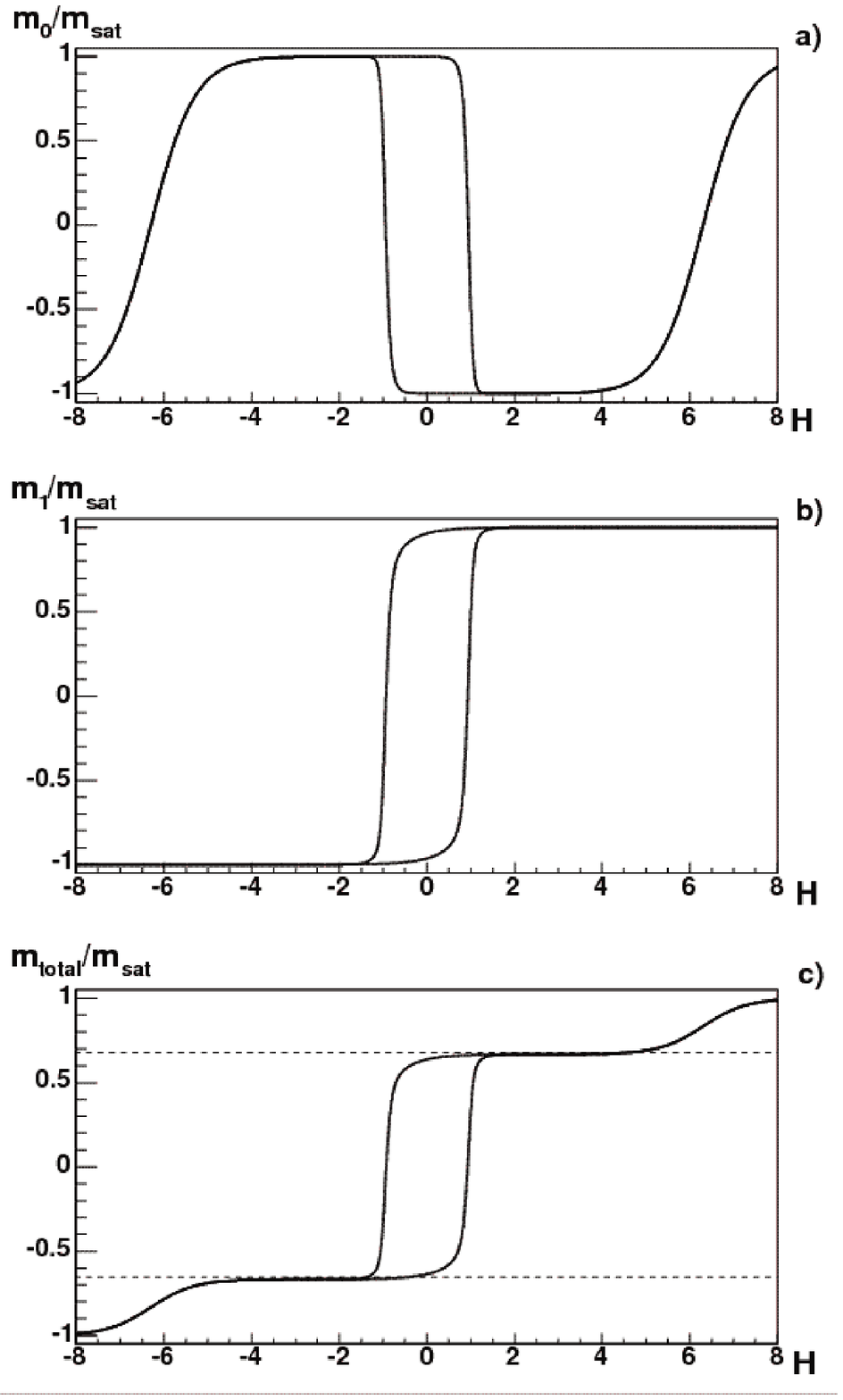,width=9.0cm}
           \end{center}
           \caption[Magnetization vs. magnetic field for $J_2=2, J_3=2.05, J_4=J_5=J_6=0$]
           {\small{ $J_2=2, J_3=2.05, J_4=0, J_5=0,
             J_6=0$ case with ferromagnetic and antiferromagnetic interactions
             contributions. There are bifurcation points, a one period
             doubling between bifurcation points and magnetization plateau at
             $m/m_{sat}=2/3$. a$)$
             is the magnetization for the central
             vertex lattice with temperature $T=0.1$, b$)$ is the magnetization for the corner
             vertex lattice with temperature $T=0.1$, c$)$ is the average
             magnetization for the lattice with temperature $T=0.1$}.\label{figure5}}\vspace{0.2cm}
         \end{figure}

         From figures \ref{figure5}a, \ref{figure5}b and \ref{figure5}c we can conclude that in represented
         recursive lattice it's possible to separate two possible kinds of lattice's
         structure with four sublattices (shells of the considered hexagon tree) each one represented on figure \ref{figure6} and \ref{figure7}
         which are changing with each other the directions of their's spins positions during
         the process of interaction. That phenomena is called modulated phases
         process.
         The Roman numbering is corresponding to each sublattice and the bottom
         side of each figure represents one period of
         structure. We can conclude
         from doubling of plots on figures  \ref{figure5}a and \ref{figure5}b the Heisenberg interaction between spins of
         hexagonal recursive lattice are going on opposite phases depending on the fact of
         dominating either ferromagnetic or antiferromagnetic essence of interaction.

         Supporting on results reported above we are able to conclude that the whole lattice
         maybe constructed in two ways of recursively repeating shells
         represented on bottom side of figure \ref{figure6} and figure \ref{figure7}. The possible
         not monosemantic behavior of lattice structure in the thermodynamic
         limit may come from difference of initial value of ratio $x_0=g(+)/g(-)$.

         \begin{figure}
           \begin{center}
             \psfig{figure=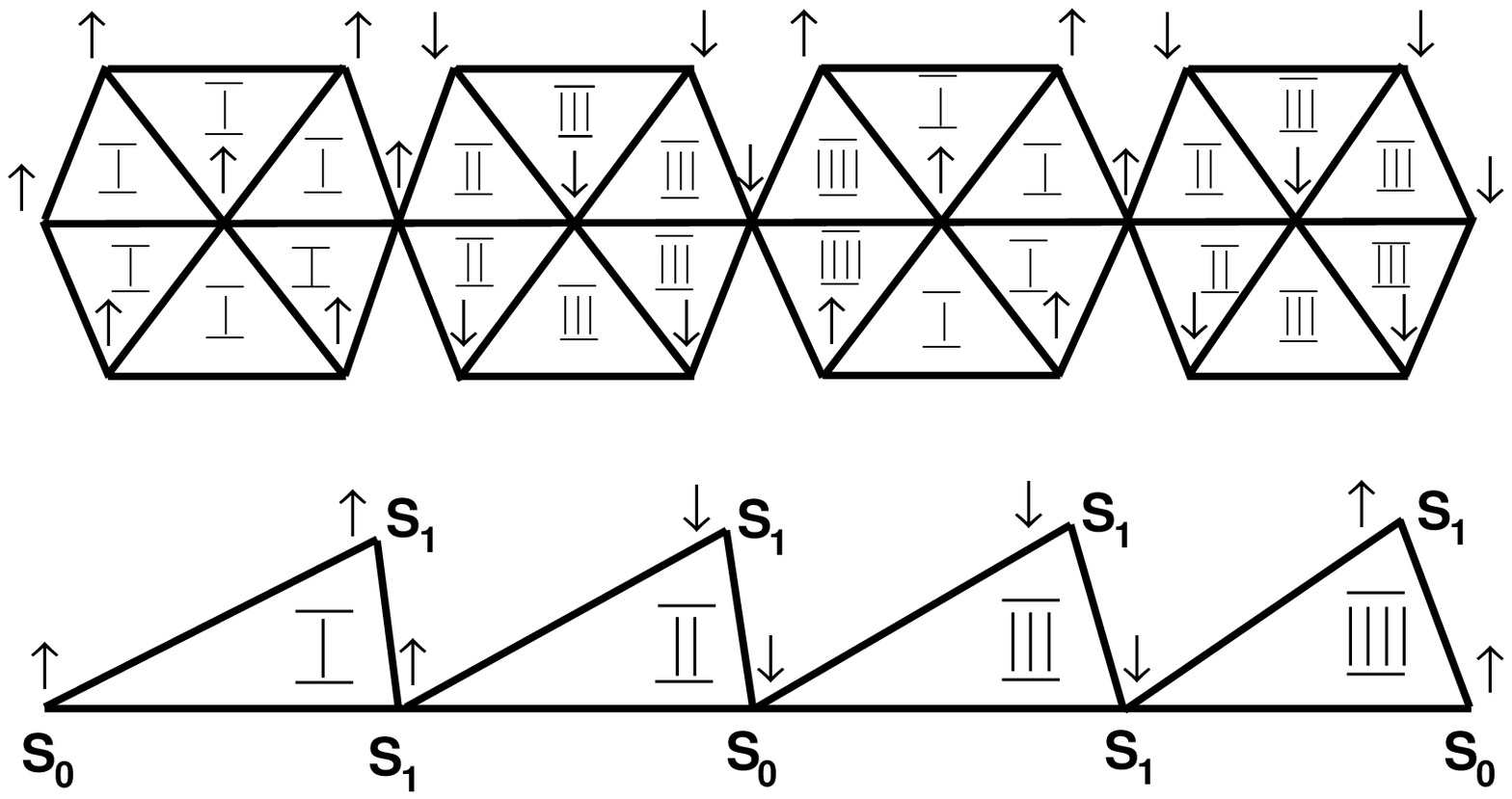,width=8.0cm}
           \end{center}
           \caption[First of the two possible kinds of lattice's
             structure with four sublattices]{\small First of the two possible kinds of lattice's
             structure with four sublattices (shells of the considered hexagon tree).\label{figure6}}
         \end{figure}

         \begin{figure}[h]
           \begin{center}
             \psfig{figure=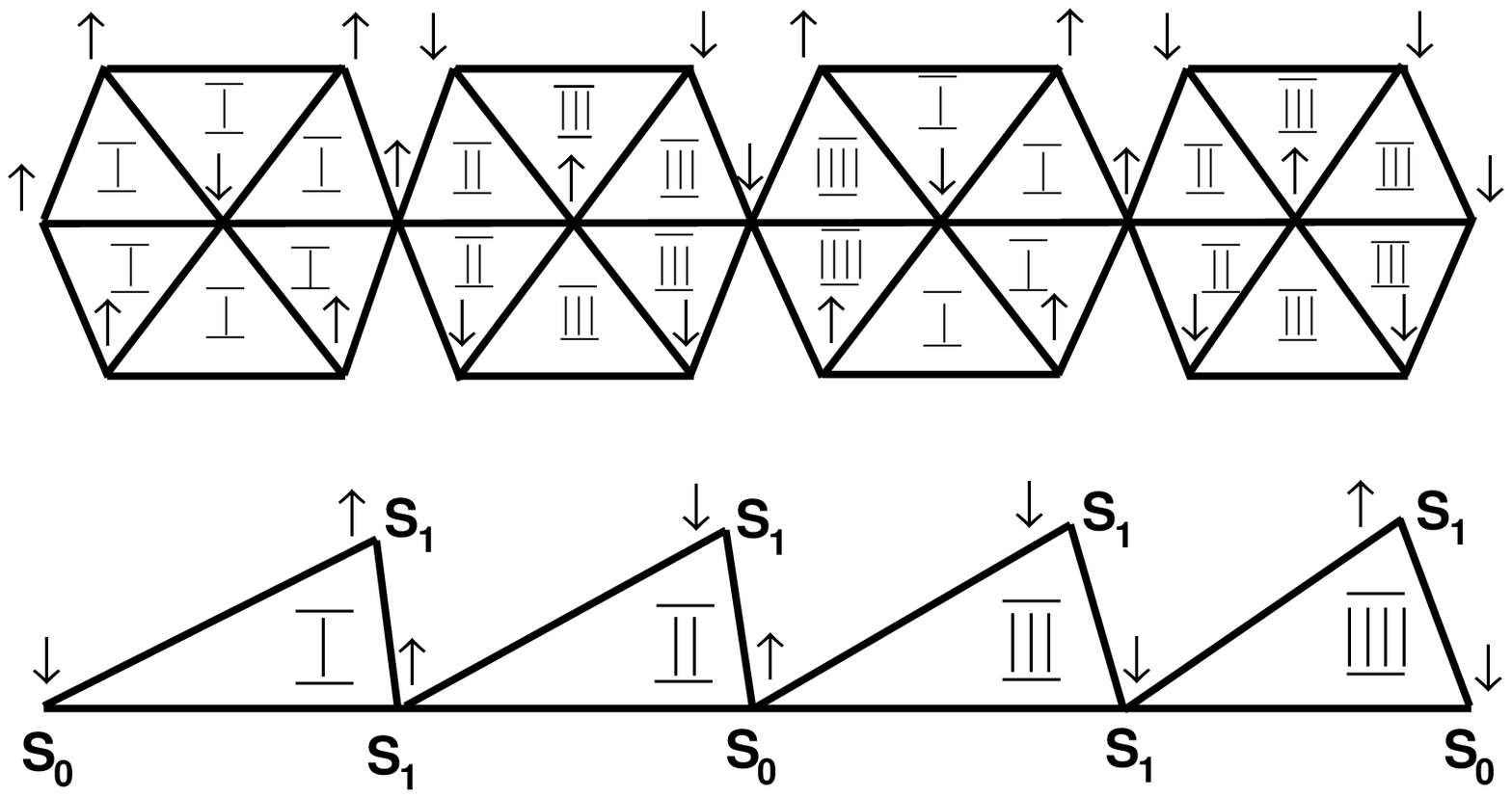,width=8.0cm}
           \end{center}
           \caption[Second of the two possible kinds of lattice's
             structure with four sublattices]{\small Second of the two possible kinds of lattice's
             structure with four sublattices (shells of the considered hexagon tree).\label{figure7}}
         \end{figure}

         In the following plots $(J_2=2, J_3=2.3, J_4=1, J_5=0.1, J_6=0.5)$
         we can see bifurcation points and one period doubling phenomena
         between bifurcation points represented
         on figure \ref{figure8} $($a$)$ is the magnetization for the central
         vertex lattice with temperature $T=0.8$, b$)$ is the magnetization for the corner
         vertex lattice with temperature $T=0.8$, c$)$ is the average
         magnetization for the lattice with temperature $T=0.8)$.

         \begin{figure}
           \begin{center}
             \psfig{figure=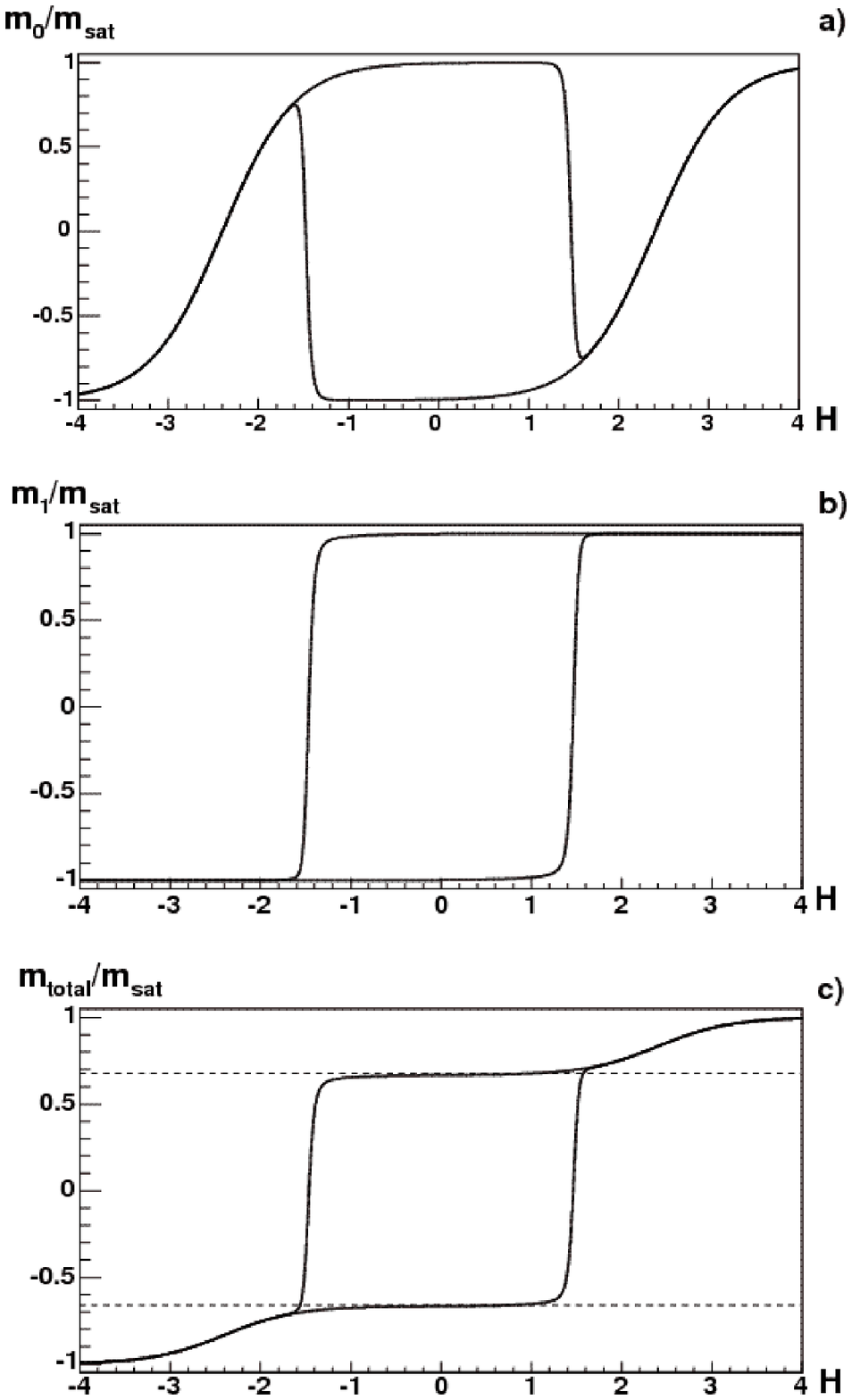,width=9.0cm}
           \end{center}
           \caption[The $J_2=2, J_3=2.3, J_4=1, J_5=0.1,
             J_6=0.5$ physical case]{\small  $J_2=2, J_3=2.3, J_4=1, J_5=0.1,
             J_6=0.5$ physical case with ferromagnetic and antiferromagnetic interactions
             contributions. There are bifurcation points and one period
             doubling between bifurcation points. a$)$ is the magnetization for the central
             vertex lattice with temperature $T=0.8$, b$)$ is the magnetization for the corner
             vertex lattice with temperature $T=0.8$, c$)$ is the average
             magnetization for the lattice with temperature
             $T=0.8$.\label{figure8}}
         \end{figure}

         Thus we've got magnetization plateaus, bifurcation points, doubling
         behavior, modulated phases for multiple exchange
         model where is taken into account both ferromagnetic
         and antiferromagnetic interactions. When we are
         changing interaction parameters or as the same the
         density in the physical limits got from
         experiment\cite{Bernu} $($close to $J=3mK)$ we see magnetization plateaus, bifurcation points, doubling
         behavior, modulated phases at low temperatures.
         $($fig. \ref{figure4},\ref{figure5},\ref{figure8}$)$.

\chapter[Face-cubic Model]{\label{FaceCubic} Face-cubic Model~
\footnote{The results considered in this chapter are
published in Ref.~\cite{vad}.}}

\hspace{14pt}Cubic symmetry plays a significant role in many types
of phase
 transitions and critical phenomena. Still in 70-s various models
possessing cubic symmetry have been introduced to determine the
nature of the displacive phase transitions in perovskites
\cite{aha74,bru75}. Magnetic phenomena in cubic crystals are also
affected by the lattice structure.
  For instance, in the crystalline solids
 with cubic--symmetric lattices (Fe, Ni, etc.) it leads to the
 modification of the magnetic exchange interaction giving rise to
 additional contributions to the conventional O(N)-symmetric Heisenberg Hamiltonian.
 The simplest contribution of the underlying cubic symmetry is the
 single-ion anisotropy of the form ${\sum}_i \mathbf{s}_i^4$.
 During the last three decades different aspects of this
 issue have been the subject of various investigations.

    In the framework of the field--theoretical approach to critical
phenomena modelling of effects of cubic symmetry is usually
performed in
 terms of continuous-spin Landau-Ginzburg effective Hamiltonian with cubic
 anisotropy, i.e. the $\phi^4$ - theory with an additional
 cubic term which breaks explicitly the O(N) invariance to a residual discrete cubic symmetry \cite{aha76} - \cite{cal04}:
 \begin{eqnarray}
 - \beta{\mathcal{H}}= \int d^Dx
 \left\{\frac{1}{2}\left(\partial_{\mu} \phi \right)^2+
 \frac{1}{2}a\phi^2+\frac{1}{4}b\phi^4+
 v \sum_{\alpha=1}^{N} (\phi_{\alpha})^4\right\},
 \end{eqnarray}

 where $\phi(x)= (\phi_1,\phi_2,\ldots ,\phi_N) $ is a continuous N-component local vector order
 parameter, $\phi^4 = (\phi^2)^2$ and $\beta=1/kT$ is inverse temperature.
 The spin orientations, orthogonal to faces of an $N$-dimensional
 hypercube will be favorable for $v<0$, whereas positive values of
 $v$ favor the orientations toward the corners.

    This model received numerous important applications, among
    which are such as the oxygen ordering in YBaCu$_3$O$_{6+x}$
    \cite{bar89}, one of the most studied high-T$_c$
    superconductor; the buckling instability of confined colloid
    crystal layer \cite{cho93}; the micellar binary solution of
    water and amphiphile \cite{bek00} etc. The two--dimensional
    case has been recently intensively studied within the framework
    of renormalization group technique in the space of fixed
    dimensionality up to five--loop approximation
    \cite{cal02,cal04}. The model was found to have four
    fixed points: the Gaussian one, the Ising one with N decoupled
    components, the O(N)--symmetric and the cubic fixed point.
    Analogous calculations were made for $D=3$, as well revealing
    some peculiar type of cubic fixed point for $N>2$
    corresponding to specific anisotropic mode of the critical
    behavior \cite{kle97}-\cite{cal02a}. The surface critical
    behavior of the model was also studied \cite{usa04}. It has been
    recently shown that the behavior of the spherical many--spin
    magnetic nanoparticle with surface anisotropy may be modelled
    by an effective single macro--spin with cubic anisotropy terms
    in the effective energy \cite{kac06}.

        Another class of lattice cubic--symmetric spin models was
        introduced by Kim, Levy and Uffer in Ref. \cite{kim75} to
        explain the tricritical--like behavior of cubic rare--earth
        compounds, particularly holmium antimonide, HoSb
        \cite{kim75}-\cite{nie83}. Projecting the pair exchange
        interaction onto the sixfold degenerated ground-state
        manifold of Ho$^{3+}$ ion they arrived at the following
        effective Hamiltonian:
\begin{eqnarray}
-\beta {\mathcal{H}} = \sum_{\langle
 i,j\rangle}J_{i  j}{\mathbf{S}}_i{\mathbf{S}}_j, \label{1.2}
\end{eqnarray}
where the classical spin variables ${\mathbf{S}}_i$ are the unit
vectors restricted to have orientations, orthogonal to the faces of
the cube and the sum is going over all the pairs of
nearest--neighbor sites. Generalization for $Q$ component spin is
obvious: each spin can assume $2Q$ orientations:
\begin{eqnarray}
{\mathbf{S}}_i \in \{(\pm1, 0, \ldots, 0), (0, \pm1, \ldots, 0)
\ldots (0, \ldots, 0, \pm1)\}. \label{1.3}
\end{eqnarray}
This model is known as the Face-cubic model. It is connected to the
continuous cubic model of Eq. (\ref{1}) via the limit of strong
anisotropy $(|v|\gg |b|)$. In a similar way one can also consider
 quadrupolar pair interaction terms and obtain the following
Hamiltonian provided all interactions are homogenous \cite{kim76,
aha77}:
\begin{eqnarray}
-\beta {\mathcal{H}}_{FC}=J\sum_{\langle i,j\rangle}
({\mathbf{S}}_i{\mathbf{S}}_j)+K\sum_{\langle i,j\rangle}
({\mathbf{S}}_i{\mathbf{S}}_j)^2. \label{1.4}
\end{eqnarray}
It is easy to see that Hamiltonian (\ref{1.4}) can be represented in
terms of two sets of discrete variables: a Potts--like determining
which component of ${\mathbf{S}}_i$ is non-zero and an Ising--like,
corresponding to the sign of the component. Indeed,
\begin{eqnarray}
{\mathbf{S}}_i{\mathbf{S}}_j=\sigma_i \sigma_j \delta_{\alpha_i
\alpha_j}, \label{1.5}
\end{eqnarray}
where $\sigma_i = \pm1$ and $\alpha_i = 1,2,...Q$. Therefore we have
 \begin{eqnarray}
 - \beta{\mathcal{H}}_{FC}=J\sum_{\langle i,j\rangle}\sigma_i \sigma_j
 \delta_{\alpha_i, \alpha_j}+K\sum_{\langle i,j\rangle }\delta_{\alpha_i,
 \alpha_j}. \label{1.6}
\end{eqnarray}
Formally we can enlarge the Hamiltonian (\ref{1.4}) up to the
interaction terms of higher power of the
$({\mathbf{S}}_i{\mathbf{S}}_j)$:
\begin{eqnarray}
-\beta {\mathcal{H}}_{FC}^{(L)}=\sum_{\langle i,j\rangle}
\sum_{n=0}^L J_n ({\mathbf{S}}_i{\mathbf{S}}_j)^n .\label{1.7}
\end{eqnarray}
For arbitrary finite $L$ this expression leads to the same
Hamiltonian that of Eq. (\ref{1.6}) with $J=\sum_{k=0}^L J_{2k+1}$
and $K=\sum_{k=0}^L J_{2k}$. At $K=0$ the Hamiltonian (\ref{1.7})
reduced to the $2Q$ -- state Potts model; $J=0$ corresponds to two
decoupled $Q$ -- state Potts model, and at $Q=2$ one obtains the
Ashkin--Teller model. Variational renormalization--group study of
the pure and diluted $Q$ -- component face--cubic model in two
dimensions has revealed existence of four competing possible types
of critical behavior corresponding to $Q$ -- state Potts model, $2Q$
-- state Potts model, Ising model and special "cubic" fixed point
\cite{rie81, nie83}. It was also found that at $Q<Q_c=2$ the
transitions are continuous and critical behavior of the discrete
face--cubic model belongs to the O(N)-model universality class. For
$Q=2$ the Ashkin--Teller--like behavior occurs and for $Q>2$
transitions are of first order. There are also early results
obtained within the mean--field (MF) theory \cite{kim75} and using
the Bethte--Peirels (BP) approximations and high--temperature series
\cite{kim75d} for the case when only dipolar pair interactions are
included $(K=0)$. In the BP approximation the critical value of spin
component $Q_c$, above which transitions are of the first order was
found to be given by
\begin{eqnarray}
Q_c=1+\frac{2}{3}q \left[ \left(1+\frac{6}{q} \right)^{1/2}-1
\right], \label{1.8}
\end{eqnarray}
where $q$ is the coordination number of the lattice. The limit $ q
\to \infty$ corresponds to the MF solution and gives $Q_c=3$. In
Ref. \cite{kim75d} the high--temperature series for $Q$ -- component
face--cubic model on three dimensional fcc lattice were constructed
up to 5-th order, from which authors obtained $Q_c=2.35 \pm 0.2$,
whereas Eq. (\ref{1.8}) gives us $Q_c \simeq 2.8$ Moreover, although
the MF solution of cubic model predicted the tricritical like
behavior \cite{kim75}, the BP approximation showed that it is not
the case \cite{kim75d}. Only inclusion of single--ion--anisotropy
terms, quadrupolar pair interactions and crystal fields may drive
the system tricritical \cite{kim76}.
\section{Recursive Methods, Cayley Tree and Bethe Lattice}
\hspace{14pt}Among the vast variety of statistical mechanics lattice
models
 with strong local interactions only very limited amount allows
 exact solutions. These exact solutions are known only for
 low--dimensional systems, more precisely for $d=1$ and $d=2$;
 most of exact solutions for two--dimensional systems are known
 only in the absence of the external field and/or for the special
 choice of model parameters \cite{bax}. Well known conventional
 approximate methods like MF theory and BP approximation in
 general can provide only more or less qualitatively satisfactory
 picture and in some cases they just fail. That is why the
 quest for the alternative approaches, which can provide more
 reliable results for the thermodynamics of lattice models is very
 important.
  \begin{figure}
           \begin{center}
           \epsfig{file=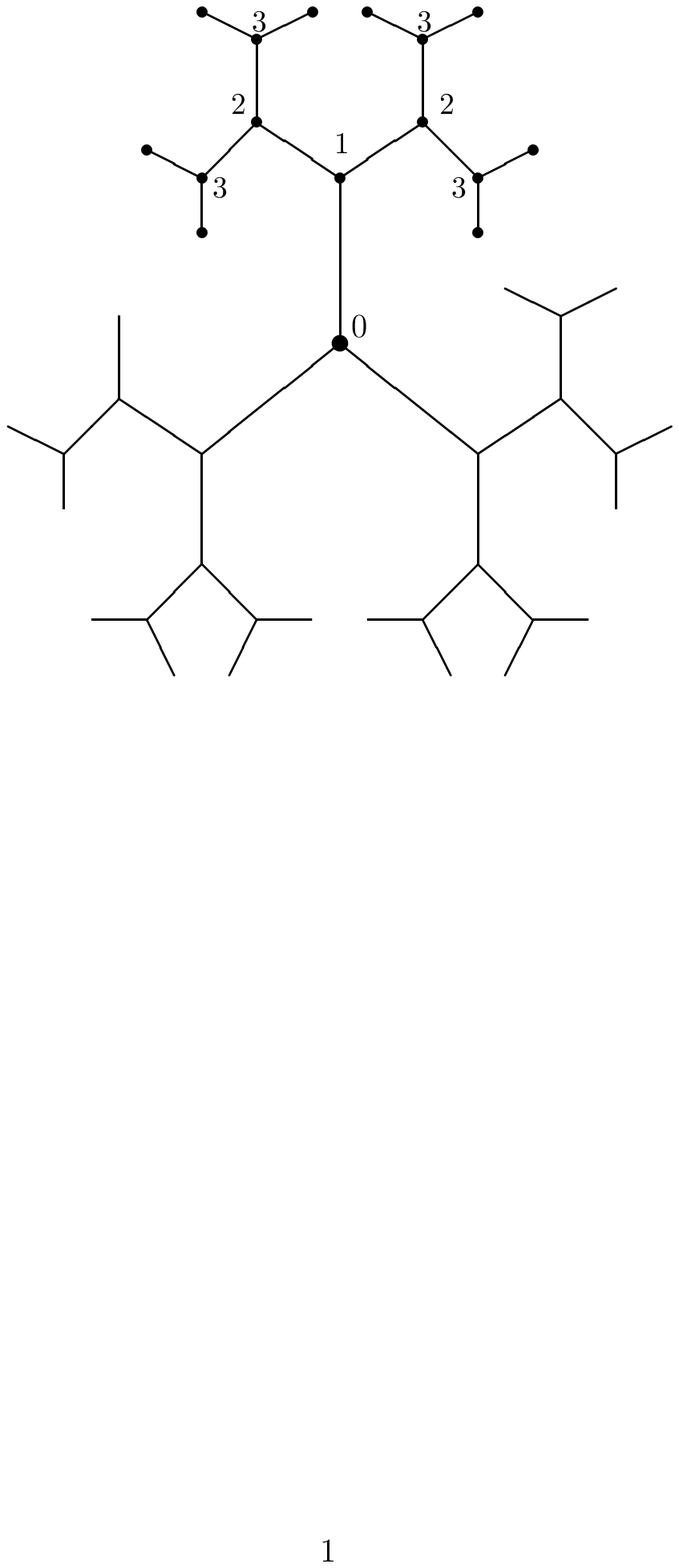, width=10.0cm}
           \caption{\label{fig1}  The Cayley tree with coordination number $q=3$ and 3 shells.}
           \end{center}
         \end{figure}

  The recursive lattices are of twofold interest. On the one hand,
  the models, defined on a recursive lattices can be considered as
  an independent original area of research in which the powerful
  methods of dynamical systems theory and fractal geometry are
  successfully exploited for determining the thermodynamical
  properties of the statistical mechanics models. On the other
  hand, recursive lattices provide a
  specific kind of approximate treatment of many--particle systems physics.
  In the heart of these approximations not the simplification
  of the character or/and strength of the interaction between the
  system elements, but the modification of the topology of the
  underlying lattice lies. This modification in
  most cases consist in the replacement  of the regular periodic
  "physical" lattice by a recursive one, constructed by the certain algorithm
  and possessing the self--similarity,  maintaining all
  interaction unchanged. For the large class of lattice models
  with commuting variables ("classical models") this approach
  leads
  to the exact solution of the statical mechanics problem in
  terms of the theory of dynamical systems or, more precisely, in terms
  of discrete maps. The solution formally is rather similar to
 that which corresponds to conventional BP approximation but, in
 contrast to the latter, it is the exact solution for the lattice
 endowed with the recursive structure provided the boundary sites
 are properly taken into consideration. It was argued that in some
 cases, particularly in the models where multi--site interactions
 are presented \cite{ana97,mon91,ar} the recursive lattice
 approximation gives more reliable results than conventional BP \cite{hu}.

  The simplest example of recursive lattices is a Cayley tree,
  the self-similar graph, which contains no cycles. In order to
  construct a Cayley tree one should start with a central site
  $O$. At the first step this site should be connected by links
  with $q$ others, which constitute the first shell or the first
  generation of the Cayley tree. The second shell of the Cayley
  tree is constructed by repeating this procedure for all sites of
  the first shell: each of the $q$ sites belonging to the first
  shell connects to the $q-1$ new sites. Thus, the second shell
  contains $N_2=q(q-1)$ sites. Accomplishing this construction for
  n steps one arrives at the recursive connected graph, which
  contains no cycles and is called the Cayley tree with coordination
  number $q$ and $n$ generations (shells) (Fig. \ref{fig1}). As the
  number of sites in $k$-th shell is $q(q-1)^{k-1}$ the total
  number of sites then
  \begin{eqnarray}
N_n=1+\sum_{k=1}^n q \left(q-1 \right)^{k-1}=\frac{q \left(q-1
\right)^n-2}{q-2}. \label{Nn}
  \end{eqnarray}
  Thus, each site of the Cayley tree has coordination number
  equal to $q$ except the sites on the last shell which have only
  one neighbor. The peculiar property of the Caylet tree is the large
  amount of boundary sites which even in the thermodynamical limit
  $n \to \infty$ comprise a finite fraction of the total number of
  sites
  \begin{equation}
  \lim_{n \to \infty}\frac{q\left( q-1 \right)^{n-1}}{N_n}=\lim_{n \to
  \infty}\frac{\left( q-2 \right) q \left( q-1 \right)^{n-1} }{q \left(
  q-1 \right)^n-2}=\frac{q-2}{q-1}. \label{fra}
  \end{equation}
  This feature causes anomalous properties of the statistical
  mechanics systems on the full Cayley tree. For instance, in
  Ising model  there is no zero field magnetization, whereas the
  derivations of free energy with respect to the external field
  exhibit singular behavior \cite{egg74} - \cite{sto03}.

   The Bethe lattice is an object intimately linked to the
   Cayley tree but, in contrast to the latter, it is devoid of the
   complications which arise from the boundary sites. When one
   deals with the Bethe lattice the general structure and topology
   of the Cayley tree are preserved whereas only contributions from
   the bulk sites lying deep inside the Cayley tree are taken into
   consideration. In this regard the Bethe lattice is the interior
   of the Cayley tree \cite{bax, che74}. Dealing with the Bethe
   lattice we suppose the underlying Cayley tree to be large
   enough to achieve thermodynamical limit and consider only the
   sites situated far away from a boundary. One can regard these
   sites as a uniform lattice with coordination number $q$.

    An undeniable advantage of the Bethe lattice is the
    possibility of exact solutions for many types of statistical
    mechanics problems. A large amount of problems of the theory
    of magnetism, macromolecule physics, lattice gauge theory,
    self--organized criticality, dynamical mean--field theory (DMFT) and general questions of
    statistical mechanics like zeroes of partition functions have
    been successfully considered on Bethe or Bethe-like lattices,
    revealing many interesting exact results and deep connections
    between the theory of dynamical systems and statistical
    mechanics \cite{egg74} - \cite{ala98}.

  \section{Face-cubic Model on Planar Graphs}
\hspace{14pt}It is possible to calculate exactly the partition
function of the $FC_Q$-model on Cayley tree in case of the absence
of external field. Let us consider the partition function of the
model without external field on arbitrary planar graph $G$
\begin{eqnarray}
{\mathcal{Z}}_G=\sum_{\{ \sigma \}} \sum_{\{ \alpha \}}
\prod_{\langle i,j\rangle \in G}\exp \left\{ (J\sigma_i \sigma_j +
K)\delta_{\alpha_i, \alpha_j} \right\}. \label{2.2}
\end{eqnarray}
As in the case of Potts model \cite{wu} one can represent it in the
following way:
\begin{eqnarray}
{\mathcal{Z}}_G=\sum_{\{ \sigma \}} \sum_{\{ \alpha \}}
\prod_{\langle i,j\rangle \in G}(1+U_{\sigma_i
\sigma_j}\delta_{\alpha_i, \alpha_j}), \label{2.3}
\end{eqnarray}
where $U_{\sigma_i \sigma_j}=e^{J\sigma_i \sigma_j+K}-1$. Then
summing out over the all $\alpha_i$ we obtain that
\begin{eqnarray}
{\mathcal{Z}}_G=\sum_{\{ \sigma \}}\sum_{G^{\prime}\subseteq G}
Q^{R_0(G^{\prime})}\prod_{\langle i,j\rangle \in
G^{\prime}}U_{\sigma_i \sigma_j}, \label{2.4}
\end{eqnarray}
where the second sum is going over all spanning graphs $G^{\prime}$
of the underlying planar graph $G$ and $R_0(G^{\prime})$ is the 0-th
Betti's number of $G^{\prime}$ which coincides with the number of
connected components of $G^{\prime}$ and the product is over all
links $\langle i,j\rangle$ belonging to $G^{\prime}$. Using the
identity $\sigma_i \sigma_j = 2 \delta_{\sigma_i \sigma_j}-1$ one
can represent Eq. (\ref{2.4}) in terms of double power series
associated with the underlying lattice provided latter is a planar
graph:
\begin{eqnarray}\nonumber
{\mathcal{Z}}_G&=&\sum_{G^{\prime}\subseteq
G}Q^{R_0(G^{\prime})}u^{e(G^{\prime})}\sum_{\{ \sigma
\}}\prod_{\langle i,j\rangle \in G^{\prime}}(1+v\delta_{\sigma_i
\sigma_j}), \\ \label{2.5} u&=&e^{K-J}-1, \nonumber \\
v&=&\frac{e^{2J}-1}{1-e^{J-K}},
\end{eqnarray}
where $e(G^{\prime})$ is the edges number in the graph $G^{\prime}$
and all $\sigma_i$ assume values 0 or 1. So, for each spanning
subgraph $G^{\prime}$ of planar graph $G$ we obtain a partition
function of 2-state Potts model:
\begin{eqnarray}\nonumber
{\mathcal{Z}}_G&=&\sum_{G^{\prime}\subseteq
G}Q^{R_0(G^{\prime})}u^{e(G^{\prime})}{\mathcal{Z}}_{G^{\prime}}^{Potts}(v)
\\ &=&\sum_{G^{\prime}\subseteq
G}Q^{R_0(G^{\prime})}u^{e(G^{\prime})}\sum_{G^{\prime
\prime}\subseteq G^{\prime}}2^{R_0(G^{\prime \prime})}v^{e(G^{\prime
\prime})}, \label{2.6}
\end{eqnarray}
where the second sum is going over all subgraph of $G^{\prime}$
 Thus, we have succeeded in representing the partition function of the $FC_Q$ -
 model given by Hamiltonian (\ref{1.6}) in terms of double power
 series associated with the underlying lattice provided the latter
 is a planar graph.
 As in case of ordinary spin Potts model one can make
 $1/Q$-expansions for $FC_Q$-model with the aid of Eq.
 (\ref{2.6}). For instance, for the two--dimensional square lattice one
 obtains
 \begin{eqnarray}
{\mathcal{Z}}&=&(2Q)^N
\{1+2N\frac{u}{Q}(\frac{v}{2}+1)+N(2N-1)(\frac{u}{Q})^2(\frac{v}{2}+1)^2\\
&+&C_{2N}^3 (\frac{u}{Q})^3(\frac{v}{2}+1)^3 + (C_{2N}^4 -
N)(\frac{u}{Q})^4(\frac{v}{2}+1)^4\nonumber\\ \nonumber
  &+& N\frac{u^4}{Q^3}((\frac{v}{2}+1)^4 + (\frac{v}{2})^4)+ (C_{2N}^5 - N(2N-4))(\frac{u}{Q})^5 (\frac{v}{2}+1)^5\\
  &+& N(2N-4)\frac{u^5}{Q^4}((\frac{v}{2}+1)^5+(\frac{v}{2})^4+(\frac{v}{2})^5)\nonumber \\ \nonumber
  &+&2N\frac{u^7}{Q^5}((\frac{v}{2}+1)^7+2((\frac{v}{2})^4+(\frac{v}{2})^5)+3(\frac{v}{2})^6)+ \ldots \}.
 \end{eqnarray}

 Now we can use the well known relations
 between the topological invariants of graphs \cite{graph} to
 rewrite Eq. (\ref{2.6}) in another way suitable for our further
 purposes. Applying the Euler theorem to planar graph $G$ we get
 \begin{eqnarray}
 R_0(G)-R_1(G)=n-e(G), \label{2.7}
 \end{eqnarray}
 where the first Betti's number $R_1(G)$ coincides with the number
 of independent cycles of $G$ and $n$ is the number of sites.
 Therefore
 \begin{equation}
{\mathcal{Z}}_G=(2Q)^n\sum_{G^{\prime}\subseteq
G}Q^{R_1(G^{\prime})}(\frac{u}{Q})^{e(G^{\prime})}\sum_{G^{\prime
\prime}\subseteq G^{\prime}}2^{R_1(G^{\prime
\prime})}(\frac{v}{2})^{e(G^{\prime \prime})}. \label{2.8}
 \end{equation}

 Partition function for $FC_Q$-model in the form of Eq.
 (\ref{2.8}) can easily be calculated in case of the so-called
 forests, the planar graphs, which contain no cycles. Connected
 part of forest is called tree. Let us suppose that we have a tree
 $T$ with $n$ sites and $L_n$ edges. Then Eq.(\ref{2.8}) takes the
 form
 \begin{eqnarray}\nonumber
{\mathcal{Z}}_T&=&(2Q)^n\sum_{G^{\prime}\subseteq
T}(\frac{u}{Q})^{e(G^{\prime})}\sum_{G^{\prime \prime}\subseteq
G^{\prime}}(\frac{v}{2})^{e(G^{\prime \prime})} \\
&=&(2Q)^n\sum_{k=0}^{L_n}C^k_{L_n}(\frac{u}{Q})^k\sum_{l=0}^kC^l_k(\frac{v}{2})^l,
\label{2.9}
 \end{eqnarray}
 where $C^l_k=\frac{k!}{l! (k-l)!}$ are the binomial coefficients
 and the sum is going over all subgraphs with given volume $k$.  With the aid of Newton's binom we obtain
 \begin{eqnarray}
{\mathcal{Z}}_T=(2Q)^n \left( \frac{u}{Q} \left( \frac{v}{2}+1
\right)+1 \right)^{L_n}. \label{2.10}
 \end{eqnarray}
If $T$ is the Cayley tree then $L_n=n-1$ and we will have
\begin{eqnarray}
{\mathcal{Z}}_{Cayley}=2Q \left(u \left( v+2 \right)+2Q
\right)^{n-1}. \label{2.11}
\end{eqnarray}
Thus, the free energy per site for $FC_Q$-model on Cayley tree is
\begin{eqnarray}\nonumber
f=-k_BT \lim_{n \to \infty}\frac{\log{\mathcal{Z}}_{Cayley}}{n}=
-k_BT\log\left(u \left( v+2 \right)+2Q \right)\\
=-k_BT\log2-k_BT\log\left(e^K\cosh J+Q-1 \right).
\end{eqnarray}

 It is noteworthy that this result is in full agreement
with that obtained by Aharony for one--dimensional model by
transfer--matrix technique \cite{aha77}. Indeed, a one--dimensional
chain can be regarded as a "tree" in the sense mentioned above.
Thus, the free energy of the $FC_Q$ model on a Cayley tree in a
thermodynamic limit is continuous for all $T$. Formally it coincides
with that for one--dimensional systems where the continuity is
really the case. But, as is known, the BP approximations for the
lattice model with an arbitrary number of component solves exactly
the problem on a Cayley tree \cite{whe70}. Therefore the model on a
Cayley tree must possess a finite critical temperature as predicted
by BP approximation \cite{kim75d}. The origin of this peculiar
properties of the Cayley tree was revealed by Eggarter \cite{egg74}
on the example of Ising model. He argued that the equivalence of all
sites in the thermodynamic limit, which is one of the key points of
BP approximation, breaks down for the sites of the Cayley tree which
are situated close to the surface. As appears from Eq. (\ref{fra})
these sites comprise rather large fraction of the sites of the
Cayley tree and, thus, they determine the behavior of all
thermodynamic quantities to a great extent.

 However, even though no phase transitions in the thermodynamic
 limit occur at the value of critical temperature $T_c$ predicted
 by the BP method the value of the order parameter (magnetization
 etc.) in the interior of the lattice, i. e. for the region far
 from the surface, which is generally called "Bethe lattice", will undergo jumps from zero to its BP value
 when the temperature is below $T_c$. In order to
 illustrate it we will use the technique of dynamical system
 theory, which becomes a powerful tool for investigating various
 physical problems on recursive lattices.

 \section{Recursive Method for Face Cubic model on Bethe lattice.}
  \hspace{14pt}Many statistical systems defined on the recursive lattices are famous
 for the possibility of exact solution in terms of dynamical
 system theory. In the heart of these exact solutions the
 self-similarity of recursive lattices lies. For instance, if we cut
 apart the Cayley tree at the central site it will give q identical
 branches each of which is the same Cayley tree with the number of
 generation decreased by 1. Using this fact one can establish
 the connection between the partition function of a model defined
 on the Cayley tree containing n shells with the partition
 function of the same model defined on the Cayley tree containing
 n-1 shells. Therefore, the thermodynamical problem is
 reformulated in terms of discrete maps, given by recurrent
 relations. Let us introduce a symmetry breaking field into
 Hamiltonian (\ref{1.6}). The field can be considered as a uniform
 magnetic field pointing along the first coordinate axis
 ${\mathbf{H}}=\left(H,0,...0 \right)$. The corresponding
 interaction Hamiltonian is of the conventional Zeeman type:
 \begin{eqnarray}
 {\mathcal{H}}_Z=-{\mathbf{H}}\sum_i {\mathbf{S}}_i, \label{zee}
 \end{eqnarray}
 which leads to the following form of the $FC_Q$-model Hamiltonian
 in the field:
 \begin{equation}
 - \beta{\mathcal{H}}_{FC}=J\sum_{\langle i,j\rangle}\sigma_i \sigma_j
 \delta_{\alpha_i, \alpha_j}+K\sum_{\langle i,j\rangle }\delta_{\alpha_i,
 \alpha_j}+h \sum_i \sigma_i \delta_{\alpha_i , 1}, \label{3.0}
 \end{equation}
 where $h= \beta H$. Hereafter we pass from the Cayley tree to the Bethe
 lattice having in mind one of the ways mentioned in Section II.
 Thus, we should no longer care about boundary sites and boundary
 conditions.
 According to that one can represent the partition function
 of the  $FC_Q$-model on the Bethe lattice in the following form:
 \begin{eqnarray}
 Z=\sum_{(\sigma_0,\alpha_0)}e^{h\sigma_0\delta_{\alpha_0,1}}[g_n(\sigma_0,
 \alpha_0)]^q, \label{3.1}
 \end{eqnarray}
 where $\sigma_0$ and $\alpha_0$ are the variables of the spin in
 the central site and $g_n(\sigma_0,
 \alpha_0)$ refers to a partition function of individual branch:
 \begin{eqnarray}
 g_n(\sigma_0, \alpha_0)=\sum_{\sigma\neq\sigma_0}\sum_{\alpha\neq\alpha_0}
 \exp \Bigl((J\sigma_0\sigma_1+K)\delta_{\alpha_0, \alpha_1}+
 \sum_{\langle i,j\rangle}(J\sigma_i\sigma_j+K )\delta_{\alpha_i, \alpha_j}+
 \sum_i \sigma_i \delta_{\alpha_i, 1} \Bigr). \label{3.2}\nonumber
 \end{eqnarray}

 Each branch, in its turn, can be cut at the site which was
 previously connected to the central one. This will give us q-1
 identical branches being the Bethe lattices with n-1 generations.
 Thus, the connection between $g_n$ and $g_{n-1}$ is
 \begin{eqnarray}
g_n(\sigma_0,
 \alpha_0)=\sum_{(\sigma_1, \alpha_1)}\exp\left(\left(J\sigma_0\sigma_1+K \right)\delta_{\alpha_0,
 \alpha_1}+ h\sigma_1\delta_{\alpha_1,1}\right)[g_{n-1}(\sigma_1,
 \alpha_1)]^{q-1}. \label{3.3}
  \end{eqnarray}
 Here we obtain the system of $2Q$ recursion relations but, in
 fact, only three of them are independent as all $g_n(\sigma,
 \alpha)$ for $\alpha\neq1$ are identical. So, from (\ref{3.3}) we
 have
 \begin{eqnarray}
 g_n(+, 1)=e^{J+K+h}[g_{n-1}(+,1)]^{q-1}+e^{-J+K-h}[g_{n-1}(-,1)]^{q-1}
 +2(Q-1)[g_{n-1}(\pm,*)]^{q-1}, \\
g_n(-,1)=e^{-J+K+h}[g_{n-1}(+,1)]^{q-1}+e^{J+K-h}[g_{n-1}(-,1)]^{q-1}
+2(Q-1)[g_{n-1}(\pm,*)]^{q-1},  \\
 g_n(\pm,*)=e^{h}[g_{n-1}(+,1)]^{q-1}+e^{-h}[g_{n-1}(-,1)]^{q-1}+(e^{J+K}+e^{-J+K}
 +2(Q-2))[g_{n-1}(\pm,*)]^{q-1},
 \label{ggg}
 \end{eqnarray}
where $g_n(\pm,*)$ stand for any $(2Q-2)$ partition functions
corresponding to the individual branch with $\vec{S}_1$ whose
direction is non collinear with the first coordinate axes.
Introducing the variables
\begin{eqnarray}\label{xyg}
x_n=\frac{g_n(+,1)}{g_n(\pm,*)},
\\\nonumber
y_n=\frac{g_n(-, 1)}{g_n(\pm,*)},
\end{eqnarray}
we obtain the system of two recurrent relations
\begin{eqnarray}\label{rr1}
x_n=f_1(x_{n-1}, y_{n-1}), \\\nonumber  y_n=f_2(x_{n-1}, y_{n-1}),
\end{eqnarray}
with
\begin{eqnarray}
\label{rrpuk} f_1(x,y)=\frac{P_1(x,y)}{R(x,y)}=\frac{a\mu x^{q-1}+b
\mu^{-1}y^{q-1}+2(Q-1)}{\mu x^{q-1}+\mu^{-1}y^{q-1}+a+b+2(Q-2)},
\\ \nonumber f_2(x,y)=\frac{P_2(x,y)}{R(x,y)}=\frac{b\mu x^{q-1}+a
\mu^{-1}y^{q-1}+2(Q-1)}{\mu x^{q-1}+\mu^{-1}y^{q-1}+a+b+2(Q-2)},
\end{eqnarray}
where the following notations are adopted
\begin{eqnarray}
a&=&\exp(J+K),\\ \nonumber b&=&\exp(-J+K), \\ \nonumber \mu
&=&\exp(h). \label{33dva}
\end{eqnarray}
In this approach statistical averages of all physical quantities can
be expressed in terms of $x$ and $y$ variables defined by Eq.
(\ref{rr1}). For instance, when the total number of spins is $n$ the
magnetization along the first coordination axis which is the thermal
average of the following form:
\begin{eqnarray}
m=\frac{1}{n}\sum_{i=1}^n\langle
S_i^{(1)}\rangle=\frac{1}{n}\sum_{i=1}^n\langle \sigma_i
\delta(\alpha_i, 1 )\rangle, \label{3.10}
\end{eqnarray}
in terms of $x$ and $y$ are expressed as
\begin{eqnarray}
m=\frac{\mu x^q - \mu^{-1} y^q}{\mu x^q +\mu^{-1}y^q+2(Q-1)},
\label{m}
\end{eqnarray}
provided all sites of the lattice are equivalent. It is easy to see
that this quantity plays the role of an order parameter of Ising
universality class. Another order parameter belonging to the
$Q$--state Potts universality class is
\begin{eqnarray}
p=\frac{1}{n}\sum_{i=1}^n\langle \delta(\alpha_i, 1 )\rangle,
\label{p} \label{3.12}
\end{eqnarray}
which also can be regarded as a "quadrupolar" moment $\langle {
S_i^{(1)}}^2 \rangle$. Thus
\begin{eqnarray}
p=\frac{\mu x^q + \mu^{-1} y^q}{\mu x^q
+\mu^{-1}y^q+2(Q-1)}.\label{3.13}
\end{eqnarray}

  The order parameters $m$ and $p$ define the three possible
  phases of the $FC_Q$ model in case of the ferromagnetic
  couplings, i.e. $J>0, K>0$, provided external magnetic field is vanished:\\
   \\
   (a) disordered (paramagnetic) phase:  $m=0$, $p=1/Q$  \\
   \\
   (b) ferromagneticaly ordered phase: $m\neq0$ , $p\neq1/Q$ \\
   \\
   (c) partially ordered (quadrupolar) phase: $m=0$ , $p\neq1/Q$

 \section{Investigations of phase transitions in terms of dynamical systems theory}
 \hspace{14pt}In order to obtain physical results one must implement an
 iterative procedure for the RR
 (\ref{rrpuk}). Namely, starting from the random initial
 conditions $(x_0, y_0)$ one uses the simple iterations scheme and
 examines the behavior of physical quantities after a large number
 of iterations \cite{van81}. In a simplest case  the iterative
 sequence $\{x_n, y_n\}$ converges to a fixed point $(x^*, y^*)$,
 which is defined by
 \begin{eqnarray}
 \left\{ \begin{array}{rl}
 & x^*=f_1(x^*,y^*) \\
 & y^*=f_2(x^*,y^*)
 \end{array} \right.
 \end{eqnarray}
 This situation is inherent to the ferromagnetic case when both $J$
 and $K$ are positive.
 As can easily be seen from Eqs. (\ref{3.10}) - (\ref{3.13}), when the
 magnetization and quadrupolar moment of the system take values
 $m$ and $p$ respectively, then the corresponding fixed point of
 the RR (\ref{rrpuk}) is the following,
 provided $h=0$:
 \begin{eqnarray}
 \left\{ \begin{array}{rl}
 & x^*=\left( \frac{(Q-1)(p+m)}{1-p} \right)^{1/q}\\
 & y^*=\left( \frac{(Q-1)(p-m)}{1-p} \right)^{1/q}
 \end{array} \right.
 \end{eqnarray}
 Thus, according to the properties of different phases of the
 ferromagnetic
 $FC_Q$ model one can establish the connection between them and the
 classification of the possible types of fixed points:\\
   \\
   (a) disordered (paramagnetic) phase:  $x^*=y^*=1$  \\
   \\
   (b) ferromagneticaly ordered phase: $x^* \neq y^* \neq 1$ \\
   \\
   (c) partially ordered (quadrupolar) phase: $x^*=y^* \neq 1$\\
   \\
 The condition
 $x^*=y^*$ leads to $g_n(+,1)=g_n(-,1)$, which means that the probability
 of spin "up" is equal to the probability of spin "down" which is really the
  case in paramagnetic phase.
Applying the simple iterative scheme to Eqs. (\ref{rr1}), (\ref{m})
and (\ref{3.13}) one can obtain the plots of magnetization processes
($m$ vs. $H$ at fixed values of $T$) as well as the zero field
magnetization and quadrupolar moment of the system.
\begin{figure}
\begin{center}
\begin{tabular}{cc}
\epsfig{file=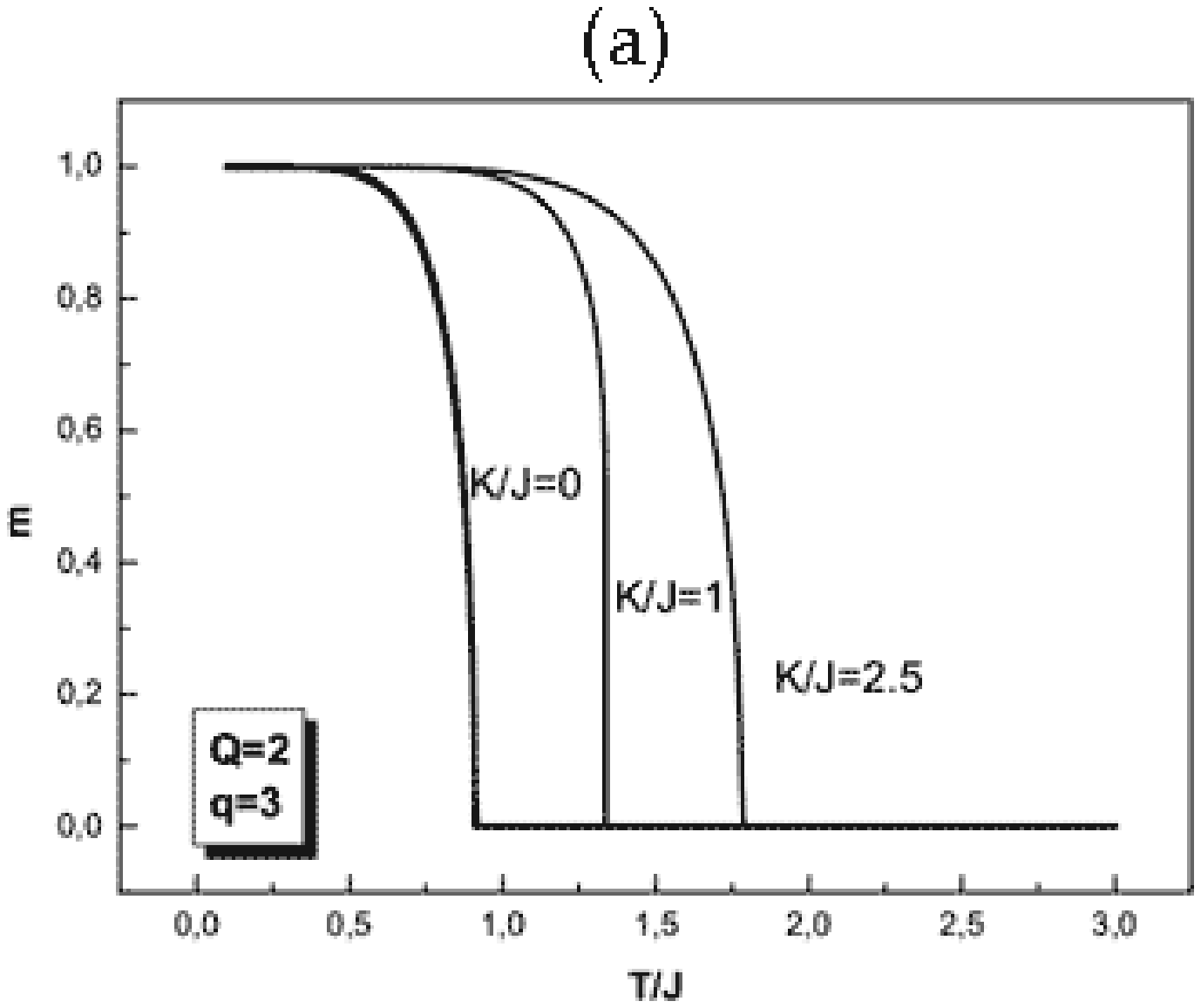,width=9cm}\\
\\
\epsfig{file=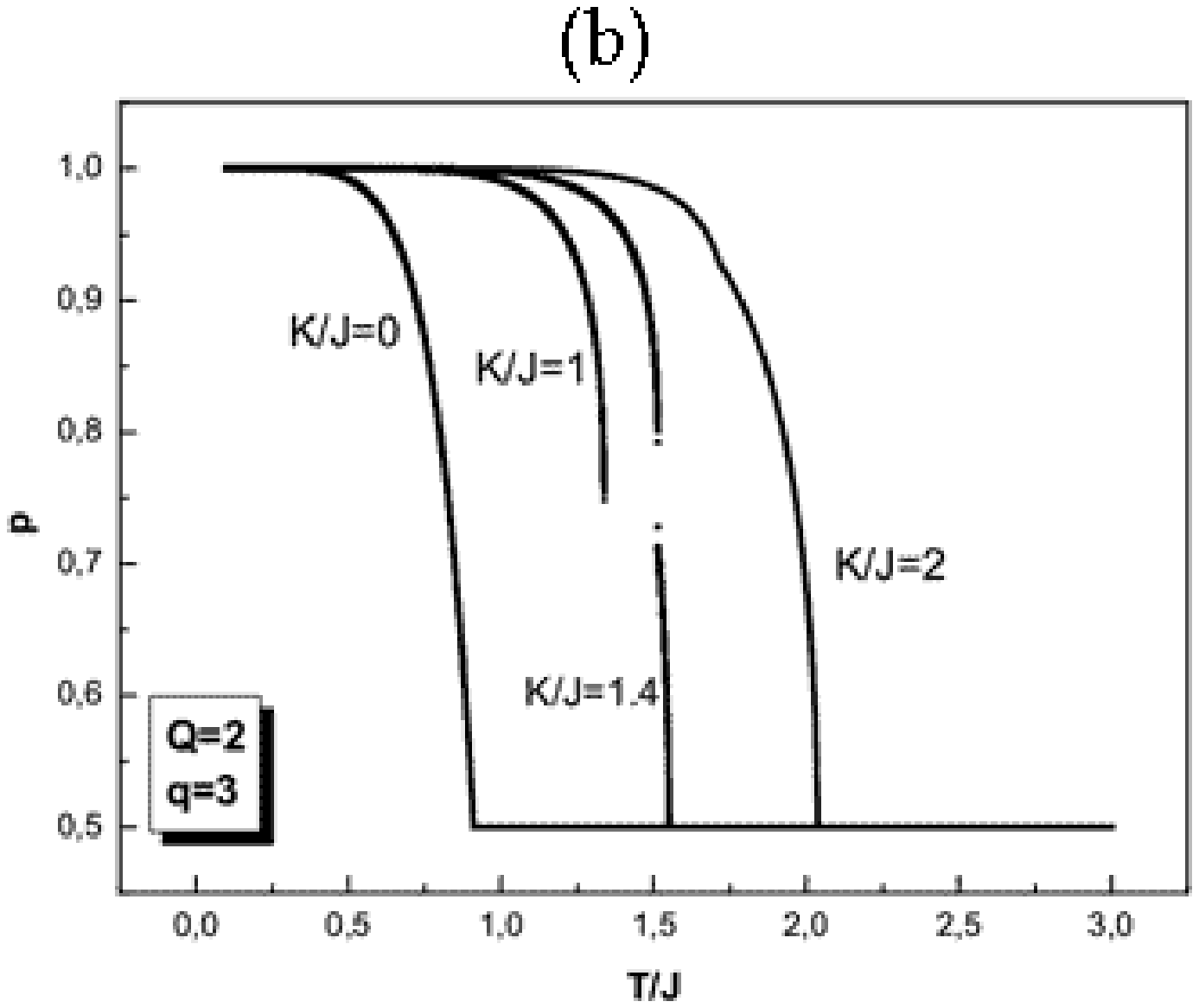,width=9cm}
\end{tabular}
\caption[The temperature behavior of the order parameters for $Q=2$
and $q=3$] {\small The temperature behavior of the order parameters
for $Q=2$ and $q=3$. The transitions from disordered to
ferromagnetic phase is of second order for all range of $K$ .(a)
Zero field magnetization for $K/J=0; 1$ and 2.5; (b) Quadrupolar
moment in zero field for $K/J=0;1;1.4$ and 2. For $K/J=1$ one can
see the first order transition between disordered and ferromagnetic
phases, whereas at $K/J=1.4$ one can see two subsequent phase
transitions, the continuous transition from disordered to
quadrupolar phase and very close to that the first order transition
to the ferromagnetic phase. With the further increase of $K/J$ the
second transition becomes of the second order.}
 \label{fig2}
 \end{center}
\end{figure}

 In Fig. \ref{fig2} one can see the temperature dependencies
 of the order parameters of the $FC_Q$-model on Bethe lattice with
 coordination number $q=3$ and $Q=2$. As is obviously seen from
 Fig. \ref{fig2}(a) in this case the zero field magnetization is
 always continuous, whereas the behavior of $p$ is quite different.
 Depending on the value of ratio $K/J$ the $p(T)$ curve could be
 continuous or can include one first order transition
 point. In Fig. \ref{fig2}(b) one can see that at small values of $K/J$ only the one phase
 transition occurs from disordered phase to ferromagnetic phase omitting the partially ordered phase.
This transition is of the second order until $K/J$ reaches some
intermediate value above which one can see a discontinuity in the
order parameter $p$ at critical temperature corresponding to the
transition between disordered and ferromagnetic phases.
 At large values of $K/J$ the system undergoes successively two
 phase transitions at temperatures $T_q(K)$ and $T_f(K)$, the
 larger one corresponds to the transition between disordered and
 quandrupolar phases, the lower one -- to the transitions from
 quandrupolar to ferromagnetic. Within this region of the values
 of $K$ one can see the second order transition from disordered
 phase into "quadrupolar" phase and the further first order transition
 to the ferromagnetic phase. However, beginning with some value of
 $K$ the transition between quadrupolar and ferromagnetic phases
 becomes continuous.
\begin{figure}
\begin{center}
\epsfig{file=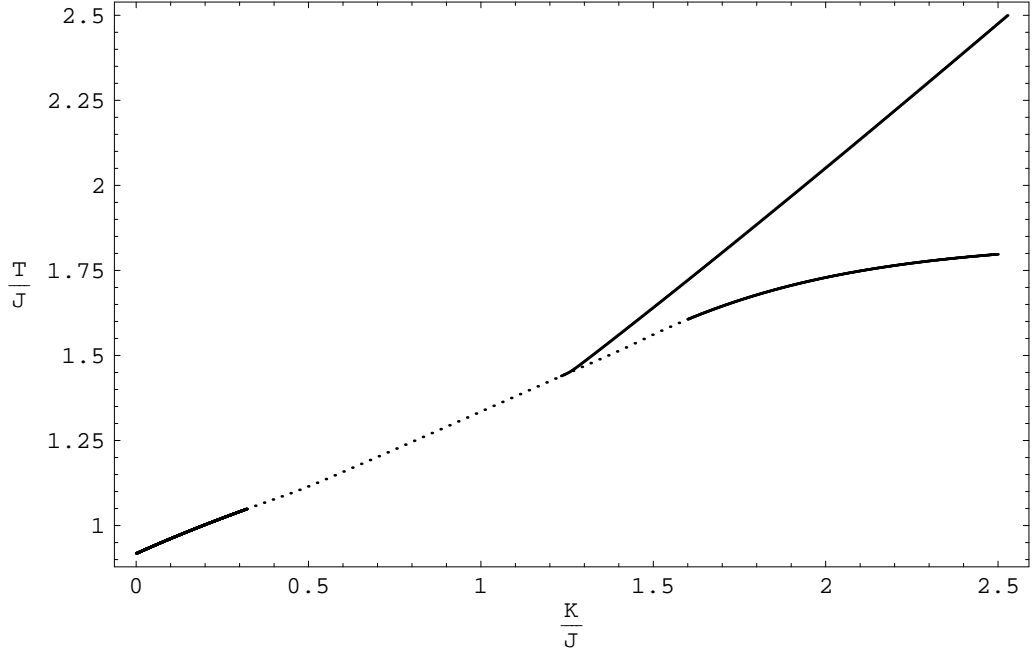,width=14cm}
   \end{center}
  \caption[The phase diagram of the model for $Q=2$ and $q=3$]
  {\small The phase diagram of the model for $Q=2$ and $q=3$. Solid
line corresponds to the second order transitions, dotted - to the first order transitions. For
small values of the ratio $K/J$ one can see only single phase transition line between disordered
hight--temperature phase and completely ordered ferromagnetic phase, whereas for $K/J\geq 1.233$
there is the partially ordered "quadrupolar" phase between them. The tricritical points are
$K/J\approx 0.32;1.233$ and 1.6. \label{fig.3} }
\end{figure}

  In order to complete the picture of the phase structure of the model
  under consideration at $Q=2$ we plot a phase diagram by
  separating the regions of the fixed points of different kinds of the
  RR
  from Eq. (\ref{rrpuk}) in the $(K/J, T/J)$-plane. One can see in
   Fig. \ref{fig.3} that in case of $Q=2$, $q=3$ at $K=0$ the
  phase transition from disordered to ferromagnetic phase is of the
  second order. This feature maintains up to $K/J\approx 0.32$,
  where one can notice the tricritical point separating the region
  of continuous transitions from the region of the first order
  transitions. The line of the second order phase transitions between
  disordered and quadrupolar phases merges the line of
  transitions between disordered and ferromagnetic phase at $K/J\approx
  1.233$, the latter, in its turn, again becomes of the second order at
  $K/J\approx1.6$ and for large values of $K/J$ it becomes parallel
  to the $K/J$ axis which means that the transition temperature
  between quadrupolar and ferromagnetic phases is unaffected by
  the value of $K$ beginning with $K/J\approx4$.
\begin{figure}
\begin{center}
\begin{tabular}{cc}
\includegraphics[width =9cm]{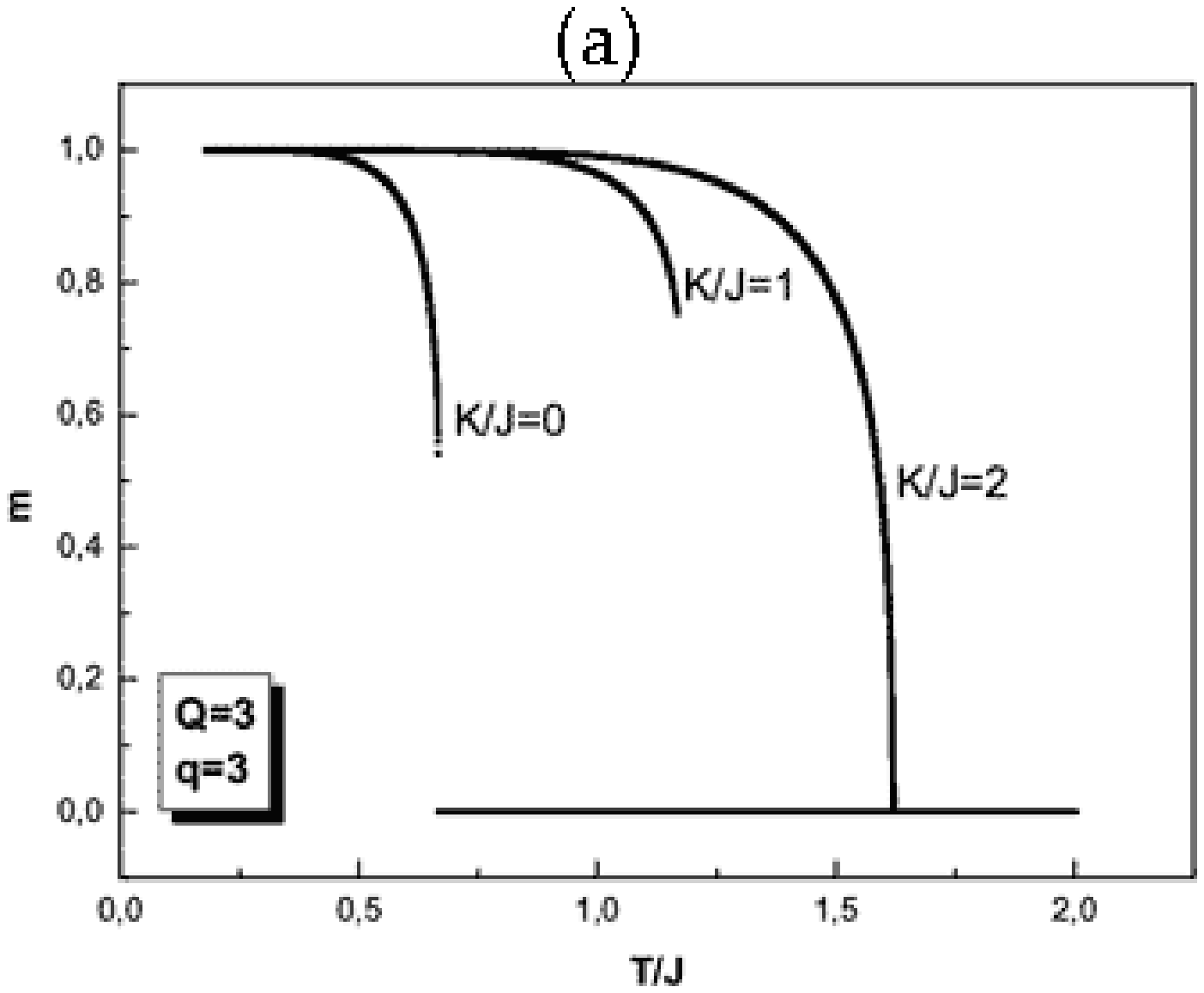}\\
\\
\includegraphics[width =9cm]{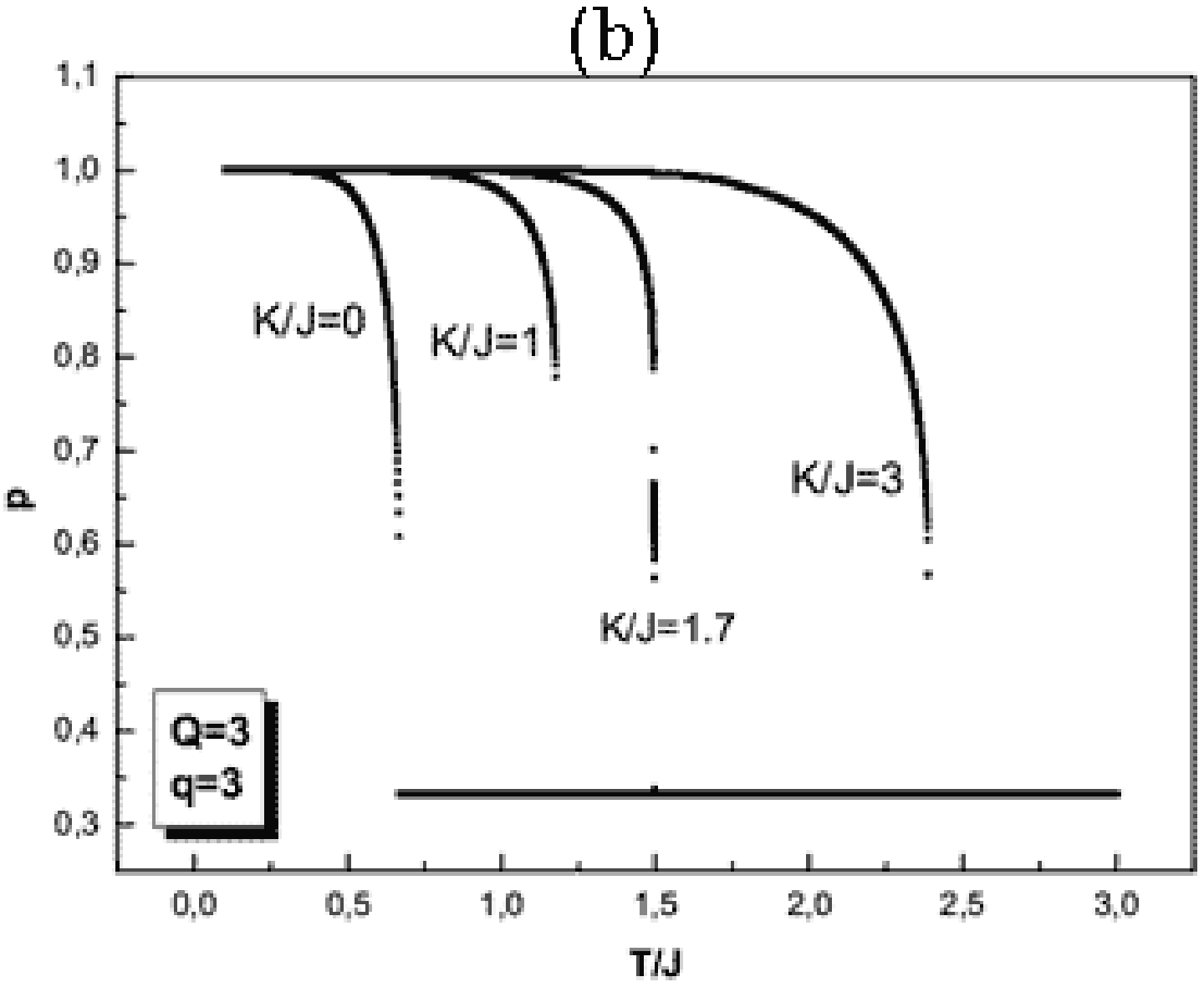}
\end{tabular}
\caption[The temperature behavior of the order parameters for $Q=3$
and $q=3$] {\small The temperature behavior of the order parameters
for $Q=3$ and $q=3$. Here magnetization of the system undergoes a
jump for small values of $K/J$ and becomes continuous beginning with
$K/J\approx 1.84$. Another order parameter always remains
discontinuous at the transition points. In the plots $p(T)$ for
$K/J=1.7$ one can see that the transitions between "quadrupolar" and
ferromagnetic phases is of the first order, whereas with the further
increase of $K/J$ in becomes continuous. }
 \label{fig.4}
 \end{center}
\end{figure}

\begin{figure}
\begin{center}
\centerline{\includegraphics[width=14cm]{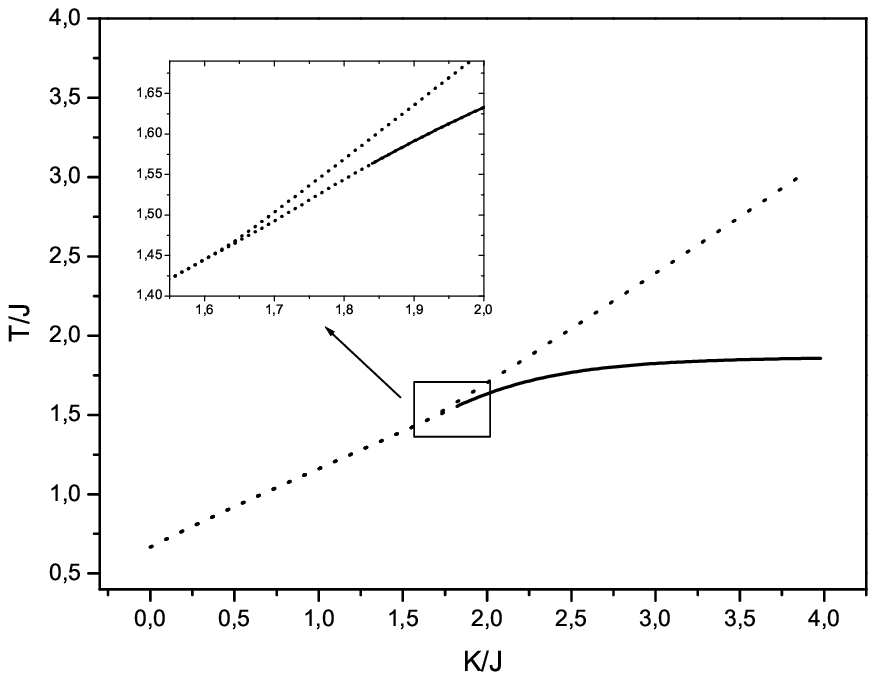}} \caption[The
phase diagram of the model for $Q=3$ and $q=3$]{\small The phase
diagram of the model for $Q=3$ and $q=3$. Solid line corresponds to
the second order transitions, dotted - to the first order
transitions. The inset shows the merge point of the three first
order lines (triple point): disordered--ferromagnetic,
disordered--"quadrupolar" and "quadrupolar"--ferromagnetic. In this
point at $K/J\approx 1.67$ these three phase are in the equilibrium.
The tricritical point on the line between "quadrupolar" and
ferromagnetic phases is situated rather close to the triple point.
The distance between them tends to zero with the increase of $Q$.}
\label{fig.5}
\end{center}
\end{figure}

   The situation is quite different for $Q>2$. In Fig.
   \ref{fig.4} the plots of zero field magnetization and
   quadrupolar moment for $Q=3$ and $q=3$ are presented. In this
   case both $m$ and $p$ are discontinuous at the critical
   temperatures. Moreover, this feature is preserved for $p$ for
   all values of $K/J$, whereas $m$ becomes continuous after
   $K/J\approx1.84$. the corresponding phase diagram is presented
   in Fig. \ref{fig.5}. Here one can see a triple point at
   $K/J\approx1.67$ where the merging of the three lines of the
   first order transitions takes place. There is also a
   tricritical point at $K/J\approx1.84$ which separates the
   region of the first and the second order phase transitions on the
   line between quadrupolar and ferromagnetic phases. The line, as
   in case of $Q=2$, goes to plateau beginning with $K/J \approx 5$,
   so the transition temperature again does not depend on $K$ for
   $K \geq 5J$.
     In general, for $Q>3$ the pase diagrams of the model under
     consideration have the same topology as in case of $Q=3$. The
     main feature is the decrease of the distance between triple
     and tricritical points with the increase of the $Q$. So, for $Q=30$
     one can obtain that this distance is of $10^{-2}$ order with $K/J$.
     Apparently, they have never shrunk to a single point at
     finite values of $Q$, but it could be the case when
     $Q\rightarrow\infty$.
      The value of critical spin component number $Q_c$ at which
      the phase transitions become of the first order at $K=0$ in
      the Face-cubic model on Bethe lattice is exactly equal to
      3, which coincides with the mean--field results at
      $q\rightarrow\infty$, but, in contrast to that, on Bethe
      lattice this value is unaffected by the value of
      coordination number $q$. Obviously, this is the direct
      consequence of the fact, that the Bethe--Pierels
      approximation become exact on the Bethe lattice.

\chapter[Azimuthal Asymmetries in DIS as a Probe
of Intrinsic Charm Content of the Proton]{\label{QCD}Azimuthal
Asymmetries in DIS as a Probe of Intrinsic Charm Content of the
Proton~\footnote{The results considered in this chapter are
published in Refs.~\cite{we6,aniv}.}}

 \hspace{14pt}The notion of the
intrinsic charm (IC) content of the proton has been introduced over
25 years ago in Refs~\cite{BHPS,BPS}. It was shown that, in the
light-cone Fock space picture \cite{brod1,brod2}, it is natural to
expect a five-quark state contribution to the proton wave function.
The probability to find in a nucleon the five-quark component
$\left\vert uudc\bar{c}\right\rangle$ is of higher twist since it
scales as $1/m^{2}$ where $m$ is the $c$-quark mass \cite{polyakov}.
This component can be generated by $gg\rightarrow c\bar{c}$
fluctuations inside the proton where the gluons are coupled to
different valence quarks. Since all of the quarks tend to travel
coherently at same rapidity in the $\left\vert
uudc\bar{c}\right\rangle $ bound state, the heaviest constituents
carry the largest momentum fraction. For this reason, one would
expect that the intrinsic charm component to be dominate the $c$
-quark production cross sections at sufficiently large Bjorken $x$.
So, the original concept of the charm density in the proton
\cite{BHPS,BPS} has nonperturbative nature and will be referred to
in the present paper as nonperturbative IC.

A decade ago another point of view on the charm content of the
proton has been proposed in the framework of the variable flavor
number scheme (VFNS) \cite{ACOT,collins}. The VFNS is an approach
alternative to the traditional fixed flavor number scheme (FFNS)
where only light degrees of freedom ($u,d,s$ and $g$) are considered
as active. It is well known that a heavy quark production cross
section contains potentially large logarithms of the type
$\alpha_{s}\ln\left( Q^{2}/m^{2}\right)$ whose contribution
dominates at high energies, $Q^{2}\rightarrow\infty$. Within the
VFNS, these mass logarithms are resummed through the all orders into
a heavy quark density which evolves with $Q^{2}$ according to the
standard DGLAP \cite{grib-lip,dokshitzer,alt-par} evolution
equation. Hence the VFN schemes introduce the parton distribution
functions (PDFs) for the heavy quarks and change the number of
active flavors by one unit when a heavy quark threshold is crossed.
We can say that the charm density arises within the VFNS
perturbatively via the $g\rightarrow c\bar{c}$ evolution and will
call it the perturbative IC.

Presently, both perturbative and nonperturbative IC are widely used
for a phenomenological description of available data. (A recent
review of the theory and experimental constraints on the charm quark
distribution can be found in Refs.~\cite{pumplin,brod-higgs}. See
also Appendix~\ref{exp} in the present thesis). In particular,
practically all the recent versions of the CTEQ \cite{CTEQ6} and
MRST \cite{MRST2004} sets of PDFs are based on the VFN schemes and
contain a charm density. At the same time, the key question remains
open: How to measure the intrinsic charm content of the proton? As
mentioned in Introduction, the main theoretical problem is that
production cross sections are not perturbatively stable.

In the present thesis we investigate the possibility to measure the
charm content of the proton using the azimuthal asymmetries in heavy
quark leptoproduction. For this purpose we calculate the IC
contribution to azimuth-dependent process:
\begin{equation}
l(\ell )+N(p)\rightarrow l(\ell -q)+Q(p_{Q})+X[\overline{Q}](p_{X}).
\label{1}
\end{equation}
Neglecting the contribution of $Z-$boson as well as the target mass
effects, the cross section of the reaction (\ref{1}) for unpolarized
initial states  may be written as
\begin{eqnarray}\label{2}
\frac{\text{d}^{3}\sigma
_{lN}}{\text{d}x\text{d}Q^{2}\text{d}\varphi }=\frac{\alpha
_{em}}{(2\pi
)^{2}}\frac{1}{xQ^{2}}\frac{y^2}{1-\varepsilon}\Bigl[\sigma _{T}(
x,Q^{2})+ \varepsilon\sigma_{L}( x,Q^{2})+ \varepsilon\sigma_{A}(
x,Q^{2})\cos2\varphi\\\nonumber
+2\sqrt{\varepsilon(1+\varepsilon)}\sigma_{I}( x,Q^{2})\cos
\varphi\Bigr],
\end{eqnarray}
The quantity $\hat{\varepsilon}$ measures the degree of the
longitudinal polarization of the virtual photon in the Breit frame
\cite{dombey},
\begin{equation}\label{3}
\varepsilon=\frac{2(1-y)}{1+(1-y)^2},
\end{equation}
and the kinematic variables are defined by
\begin{eqnarray}
\bar{S}=\left( \ell +p\right) ^{2},\qquad &Q^{2}=-q^{2},\qquad &x=\frac{Q^{2}}{%
2p\cdot q},  \nonumber \\
y=\frac{p\cdot q}{p\cdot \ell },\qquad &Q^{2}=xy\bar{S},\qquad &\rho =\frac{4m^{2}%
}{\bar{S}}.  \label{4}
\end{eqnarray}
The cross sections $\sigma _{i}$ $(i=T,L,A,I)$ in Eq.~(\ref{2}) are
related to the structure functions $F_{i}(x,Q^{2})$ as follows:
\begin{eqnarray}
F_{i}(x,Q^{2}) &=&\frac{Q^{2}}{8\pi^{2}\alpha _{em}x}\,\sigma_{i}(x,Q^{2}), \qquad (i=T,L,A,I)\nonumber \\
F_{2}(x,Q^{2}) &=&\frac{Q^{2}}{4\pi^{2}\alpha
_{em}}\,\sigma_{2}(x,Q^{2}),\label{6}
\end{eqnarray}
where $F_{2}=2x(F_{T}+F_{L})$ and
$\sigma_{2}=\sigma_{T}+\sigma_{L}$.
\begin{figure}
\begin{center}
\mbox{\epsfig{file=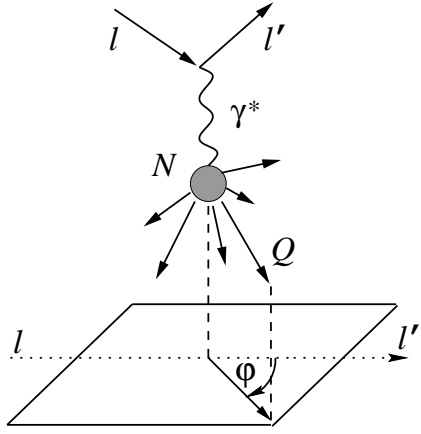,width=200pt}} \caption[Definition of
the azimuthal angle $\varphi$ in the nucleon rest
frame.]{\label{Fg.1}\small Definition of the azimuthal angle
$\varphi$ in the nucleon rest frame.}
\end{center}
\end{figure}
In Eq.~(\ref{2}), $\sigma _{T}\,(\sigma _{L})$ is the usual $\gamma
^{*}N$ cross section describing heavy quark production by a
transverse (longitudinal) virtual photon. The third cross section,
$\sigma _{A}$, comes about from interference between transverse
states and is responsible for the $\cos2\varphi$ asymmetry which
occurs in real photoproduction using linearly polarized photons
\cite{we1,we2,we3}. The fourth cross section, $\sigma _{I}$,
originates from interference between longitudinal and transverse
components \cite{dombey}. In the nucleon rest frame, the azimuth
$\varphi $ is the angle between the lepton scattering plane and the
heavy quark production plane, defined by the exchanged photon and
the detected quark $Q$ (see Fig.~\ref{Fg.1}). The covariant
definition of $\varphi $ is
\begin{eqnarray}
\cos \varphi &=&\frac{r\cdot n}{\sqrt{-r^{2}}\sqrt{-n^{2}}},\qquad
\qquad
\sin \varphi =\frac{Q^{2}\sqrt{1/x^{2}+4m_{N}^{2}/Q^{2}}}{2\sqrt{-r^{2}}%
\sqrt{-n^{2}}}~n\cdot \ell ,  \label{7} \\
r^{\mu } &=&\varepsilon ^{\mu \nu \alpha \beta }p_{\nu }q_{\alpha
}\ell _{\beta },\qquad \qquad \quad n^{\mu }=\varepsilon ^{\mu \nu
\alpha \beta }q_{\nu }p_{\alpha }p_{Q\beta }.  \label{8}
\end{eqnarray}
In Eqs.~(\ref{4}) and (\ref{7}), $m$ and $m_{N}$ are the masses of
the heavy quark and the target, respectively. Usually, the azimuthal
asymmetry associated with the $\cos 2\varphi $ distribution,
$A_{2\varphi}(\rho ,x,Q^{2})$, is defined by
\begin{eqnarray}
A_{2\varphi}(\rho ,x,Q^{2})&=&2\langle \cos 2\varphi \rangle(\rho,x,Q^{2})\nonumber \\
&=&\frac{\text{d}^{3} \sigma _{lN}(\varphi =0)+\text{d}^{3}\sigma
_{lN}(\varphi =\pi )-2\text{d}^{3}\sigma _{lN}(\varphi =\pi /2)}{
\text{d}^{3}\sigma _{lN}(\varphi =0)+\text{d}^{3}\sigma
_{lN}(\varphi =\pi
)+2\text{d}^{3}\sigma _{lN}(\varphi =\pi /2)} \nonumber\\
&=&\frac{\varepsilon \,\sigma _{A}( x,Q^{2}) }{\sigma _{T}( x,Q^{2})
+\varepsilon \,\sigma _{L}( x,Q^{2})
}=A(x,Q^{2})\frac{\varepsilon+\varepsilon R(x,Q^{2})}{1+\varepsilon
R(x,Q^{2})}, \label{9}
\end{eqnarray}
where $\text{d}^{3}\sigma _{lN}(\varphi )\equiv
{{\displaystyle{\text{d}^{3}\sigma _{lN} \over
\text{d}x\text{d}Q^{2}\text{d}\varphi }} }( \rho ,x,Q^{2},\varphi)$
and the mean value of $\cos n\varphi$ is
\begin{equation}
\langle \cos n\varphi \rangle (\rho ,x,Q^{2})=
\frac{\int\limits_{0}^{2\pi }\text{d}\varphi \cos n\varphi
{\displaystyle {\text{d}^{3}\sigma _{lN} \over
\text{d}x\text{d}Q^{2}\text{d}\varphi }} ( \rho ,x,Q^{2},\varphi )
}{\int\limits_{0}^{2\pi }\text{d}\varphi {\displaystyle
{\text{d}^{3}\sigma _{lN} \over
\text{d}x\text{d}Q^{2}\text{d}\varphi }} ( \rho ,x,Q^{2},\varphi )
}.  \label{10}
\end{equation}
In Eq.~(\ref{9}), the quantities $R(x,Q^{2})$ and $A(x,Q^{2})$ are
defined as
\begin{eqnarray}
R(x,Q^{2})&=&\frac{\sigma_{L}}{\sigma_{T}}(x,Q^{2})=\frac{F_{L}}{F_{T}}(x,Q^{2}), \label{11}\\
A(x,Q^{2})&=&\frac{\sigma_{A}}{\sigma_{2}}(x,Q^{2})=2x\frac{F_{A}}{F_{2}}(x,Q^{2}).
\label{12}
\end{eqnarray}
Likewise, we can define the azimuthal asymmetry associated with the $\cos \varphi$ distribution,\\
$A_{\varphi}(\rho,x,Q^{2})$:
\begin{eqnarray}\label{13}
A_{\varphi}(\rho ,x,Q^{2})&=&2\langle \cos \varphi \rangle(\rho
,x,Q^{2})\\
&=&\frac{2\text{d}^{3} \sigma _{lN}(\varphi =0)-2\text{d}^{3}\sigma
_{lN}(\varphi =\pi )}{ \text{d}^{3}\sigma _{lN}(\varphi
=0)+\text{d}^{3}\sigma _{lN}(\varphi =\pi
)+2\text{d}^{3}\sigma _{lN}(\varphi =\pi /2)} \nonumber\\
&=&\frac{2\sqrt{\varepsilon(1+\varepsilon)}\,\sigma _{I}( x,Q^{2})
}{\sigma _{T}( x,Q^{2})
+\varepsilon\,\sigma_{L}(x,Q^{2})}=A_{I}(x,Q^{2})\sqrt{\varepsilon(1+\varepsilon)/2}
\frac{1+R(x,Q^{2})}{1+\varepsilon R(x,Q^{2})}, \nonumber
\end{eqnarray}
where
\begin{equation}\label{14}
A_{I}(x,Q^{2})=2\sqrt{2}\frac{\frac{}{}\sigma_{I}}{\sigma_{2}}(x,Q^{2})=
4\sqrt{2}\,x\frac{F_{I}}{F_{2}}(x,Q^{2}).
\end{equation}
Remember that $y\ll 1$ in most of the experimentally reachable
kinematic range. Taking also into account that
$\varepsilon=1+{\cal{O}}(y^{2})$, we find:
\begin{equation}\label{15}
A_{2\varphi}(\rho ,x,Q^{2})=A(x,Q^{2})+{\cal{O}}(y^{2}), \qquad
\qquad A_{\varphi}(\rho ,x,Q^{2})=A_{I}(x,Q^{2})+{\cal{O}}(y^{2}).
\end{equation}
So, like the $\sigma _{2}(x,Q^{2})$ cross section in the
$\varphi$-independent case, it is the parameters $A(x,Q^{2})$ and
$A_{I}(x,Q^{2})$ that can effectively be measured in the
azimuth-dependent production.

In this paper we concentrate on the azimuthal asymmetry $A(x,Q^{2})$
associated with the $\cos2\varphi$-distribution. We have calculated
the IC contribution to the asymmetry which is described at the
parton level by the photon-quark scattering (QS) mechanism given in
Fig.~\ref{Fg.2}. Our main result can be formulated as follows:
\begin{itemize}
\item[$\star$] Contrary to the basic GF component, the IC mechanism is practically
$\cos2\varphi$-independent. This is due to the fact that the QS
contribution to the $\sigma_{A}(x,Q^{2})$ cross section is absent
(for the kinematic reason) at LO and is negligibly small (of the
order of $1\%$) at NLO.
\end{itemize}
As to the $\varphi$-independent cross sections, our parton level
calculations have been compared with the previous results for the IC
contribution to $\sigma_{2}(x,Q^{2})$ and $\sigma_{L}(x,Q^{2})$
presented in Refs.~\cite{HM,KS}. Apart from two trivial misprints
uncovered in \cite{HM} for $\sigma_{L}(x,Q^{2})$, a complete
agreement between all the considered results is found.

Since the GF and QS mechanisms have strongly different
$\cos2\varphi$-distributions, we investigate the possibility to
discriminate between their contributions using the azimuthal
asymmetry $A(x,Q^{2})$. We analyze separately the nonperturbative IC
in the framework of the FFNS and the perturbative IC within the
VFNS.

The following properties of the nonperturbative IC contribution to
the azimuthal asymmetry within the FFNS are found:
\begin{itemize}
\item  The nonperturbative IC is practically invisible at low $x$, but affects essentially the GF
predictions at large $x$. The dominance of the
$\cos2\varphi$-independent IC component at large $x$ leads to a more
rapid (in comparison with the GF predictions) decreasing of
$A(x,Q^{2})$ with
growth of $x$.%
\item  Contrary to the production cross sections, the $\cos 2\varphi$ asymmetry in charm  azimuthal
distributions is practically insensitive to radiative corrections at
$Q^{2}\sim m^{2}$. Perturbative stability of the combined GF+QS
result for $A(x,Q^{2})$ is mainly due to the
cancellation of large NLO corrections in Eq.~(\ref{12}).%
\item  pQCD predictions for the $\cos 2\varphi$ asymmetry are parametrically stable; the GF+QS
contribution to $A(x,Q^{2})$ is practically insensitive to most of
the standard uncertainties in
the QCD input parameters: $\mu _{R}$, $\mu _{F}$, $\Lambda _{QCD}$ and PDFs.%
\item  Nonperturbative corrections to the charm azimuthal asymmetry due to the gluon transverse
motion in the target are of the order of $20\%$ at $Q^{2}\leq m^{2}$
and rapidly vanish at $Q^{2}> m^{2}$.
\end{itemize}
We conclude that the contributions of both GF and IC components to
the $\cos 2\varphi$ asymmetry in charm leptoproduction are
quantitatively well defined in the FFNS: they are stable, both
parametrically and perturbatively, and insensitive (at $Q^{2}>
m^{2}$) to the gluon transverse motion in the proton. At large
Bjorken $x$, the $A(x,Q^{2})$ asymmetry could be a sensitive probe
of the nonperturbative IC.

The perturbative IC has been considered within the VFNS proposed in
Refs.~\cite{ACOT,collins}. The following features of the azimuthal
asymmetry should be emphasized:
\begin{itemize}
\item[$\ast$]  Contrary to the nonperturbative IC component, the perturbative one is significant
practically at all values of Bjorken $x$ and $Q^{2}>m^{2}$.%
\item[$\ast$]  The charm densities of the recent CTEQ and MRST sets of PDFs  lead to a sizeable
reduction (by about 1/3) of the GF predictions for the
$\cos2\varphi$ asymmetry.
\end{itemize}
We conclude that impact of the perturbative IC on the $\cos2\varphi$
asymmetry is sizeable in the whole region of $x$ and, for this
reason, can easily be detected.

Concerning the experimental aspects, azimuthal asymmetries in charm
leptoproduction can, in principle, be measured in the COMPASS
experiment at CERN, as well as in future studies at the proposed
eRHIC \cite{eRHIC,EIC} and LHeC \cite{LHeC} colliders at BNL and
CERN, correspondingly.

The paper is organized as follows. In Section~\ref{part} we analyze
the QS and GF parton level predictions for the $\varphi$-dependent
charm leptoproduction in the single-particle inclusive kinematics.
In particular, we discuss our results for the NLO QS cross sections
and compare them with available calculations. Hadron level
predictions for $A(x,Q^{2})$ are given in Section~\ref{hadr}. We
consider the IC contributions to the asymmetry within the FFNS and
VFNS in a wide region of $x$ and $Q^{2}$. Some details of our
calculations of the QS cross sections are presented in
Appendix~\ref{virt}. An overview of the soft-gluon resummation for
the photon-gluon fusion mechanism is given in Appendix~\ref{soft}.
Some experimental facts in favor of the nonperturbative IC are
briefly listed in Appendix~\ref{exp}.

\section{\label{part} Partonic Cross Sections}
\subsection{Quark Scattering Mechanism}
\begin{figure}
\begin{center}
\mbox{\epsfig{file=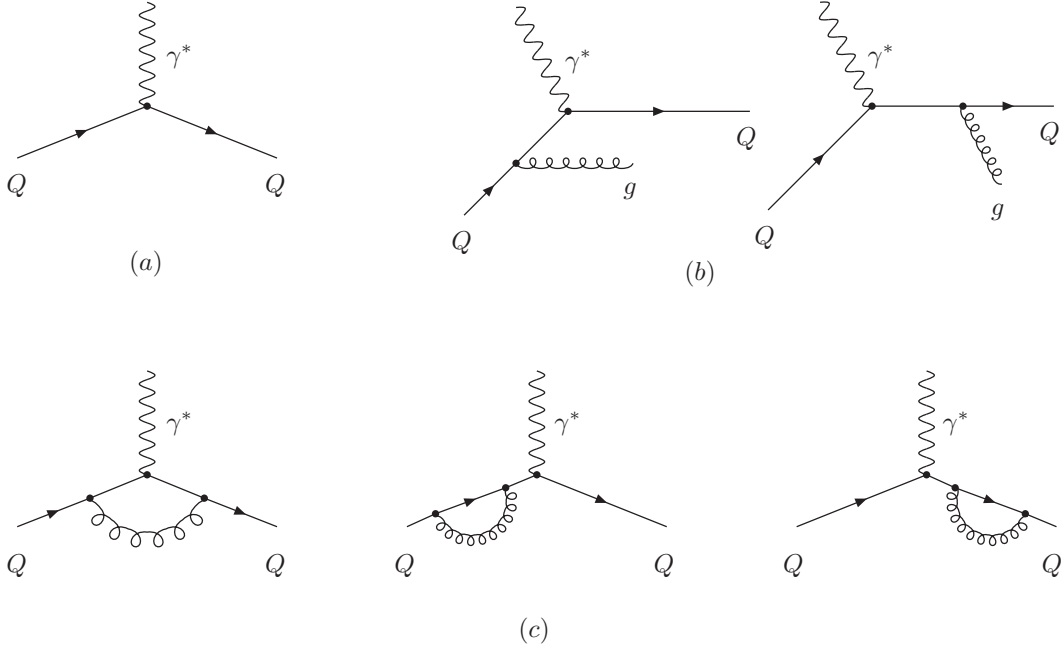,width=400pt}}
\end{center}
\caption[The LO (a) and NLO (b and c) photon-quark scattering
diagrams.]{\label{Fg.2}\small The LO (a) and NLO (b and c)
photon-quark scattering diagrams.}
\end{figure}
\hspace{14pt}The momentum assignment of the deep inelastic
lepton-quark scattering will be denoted as
\begin{equation}
l(\ell )+Q(k_{Q})\rightarrow l(\ell -q)+Q(p_{Q})+X(p_{X}).
\label{17}
\end{equation}
Taking into account the target mass effects, the corresponding
partonic cross section can be written as follows \cite{we6}
\begin{eqnarray}\label{18}
\frac{\text{d}^{3}\hat{\sigma}_{lQ}}{\text{d}z\text{d}Q^{2}\text{d}\varphi
}=\frac{\alpha _{em}}{(2\pi )^{2}}\frac{y^2}{z
Q^{2}}\frac{\sqrt{1+4\lambda
z^{2}}}{1-\hat{\varepsilon}}\Bigl[\hat{\sigma}_{2,Q}(z,\lambda)-
(1-\hat{\varepsilon})\hat{\sigma}_{L,Q}(z,\lambda)\\
+\hat{\varepsilon}\hat{\sigma}_{A,Q}(z,\lambda)\cos 2\varphi+
2\sqrt{\hat{\varepsilon}(1+\hat{\varepsilon})}\hat{\sigma}_{I,Q}(z,\lambda)\cos
\varphi\Bigr].\nonumber
\end{eqnarray}
In Eq.~(\ref{18}), we use the following definition of partonic
kinematic variables:
\begin{equation}\label{19}
y=\frac{q\cdot k_{Q}}{ \ell\cdot k_{Q} },\qquad \qquad
z=\frac{Q^{2}}{2q\cdot k_{Q}},\qquad \qquad\lambda
=\frac{m^{2}}{Q^{2}}.
\end{equation}
In the massive case, the (virtual) photon polarization parameter,
$\hat{\varepsilon}$, has the form \cite{we6}
\begin{equation}\label{20}
\hat{\varepsilon}=\frac{2(1-y-\lambda
z^{2}y^{2})}{1+(1-y)^2+2\lambda z^{2}y^{2}}.
\end{equation}
At leading order, ${\cal O}(\alpha _{em})$, the only quark
scattering subprocess is
\begin{equation}
\gamma ^{*}(q)+Q(k_{Q})\rightarrow Q(p_{Q}).  \label{21}
\end{equation}
The $\gamma ^{*}Q$ cross sections, $\hat{\sigma}_{k,Q}^{(0)}$
($k=2,L,A,I$), corresponding to the Born diagram (see
Fig.~\ref{Fg.2}a) are:
\begin{eqnarray}
\hat{\sigma}_{2,Q}^{(0)}(z,\lambda)&=&\hat{\sigma}_{B}(z)\sqrt{1+4\lambda z^{2}}\,\delta(1-z), \nonumber\\
\hat{\sigma}_{L,Q}^{(0)}(z,\lambda)&=&\hat{\sigma}_{B}(z)\frac{4\lambda
z^{2}}{\sqrt{1+4\lambda
z^{2}}}\,\delta(1-z), \label{22}\\
\hat{\sigma}_{A,Q}^{(0)}(z,\lambda)&=&\hat{\sigma}_{I,Q}^{(0)}(z,\lambda)=0,\nonumber
\end{eqnarray}
with
\begin{equation}\label{23}
\hat{\sigma}_{B}(z)=\frac{(2\pi)^2e_{Q}^{2}\alpha _{em}}{Q^{2}}\,z,
\end{equation}
where $e_{Q}$ is the quark charge in units of electromagnetic
coupling constant.

To take into account the NLO ${\cal O}(\alpha _{em}\alpha _{s})$
contributions, one needs to calculate the virtual corrections to the
Born process  (given in Fig.~\ref{Fg.2}c) as well as the real gluon
emission (see Fig.~\ref{Fg.2}b):
\begin{equation}
\gamma ^{*}(q)+Q(k_{Q})\rightarrow Q(p_{Q})+g(p_{g}).  \label{24}
\end{equation}

The NLO $\varphi$-dependent cross sections,
$\hat{\sigma}_{A,Q}^{(1)}$ and $\hat{\sigma}_{I,Q}^{(1)}$, are
described by the real gluon emission only. Corresponding
contributions are free of any type of singularities and the
quantities $\hat{\sigma}_{A,Q}^{(1)}$ and $\hat{\sigma}_{I,Q}^{(1)}$
can be calculated directly in four dimensions.

In the $\varphi$-independent case, $\hat{\sigma}_{2,Q}^{(1)}$ and
$\hat{\sigma}_{L,Q}^{(1)}$, we also work in four dimensions. The
virtual contribution (Fig.~\ref{Fg.2}c) contains ultraviolet (UV)
singularity that is removed using the on-mass-shell regularization
scheme. In particular, we calculate the absorptive part of the
Feynman diagram which has no UV divergences.  The real part is then
obtained by using the appropriate dispersion relations. As to the
infrared (IR) singularity, it is regularized with the help of an
infinitesimal gluon mass. This IR divergence is cancelled when we
add the bremsstrahlung contribution (Fig.~\ref{Fg.2}b). Some details
of our calculations are given in Appendix \ref{virt}.

The final (real+virtual) results for $\gamma ^{*}Q$ cross sections
can be cast into the following form:
\begin{eqnarray}
\hat{\sigma}_{2,Q}^{(1)}(z,\lambda)&=&\frac{\alpha_{s}}{2\pi}C_{F}\hat{\sigma}_{B}(1)
\sqrt{1+4\lambda}\,\delta(1-z)\Bigl\{-2+4\ln\lambda-\sqrt{1+4\lambda
}\,\ln r\nonumber\\
&&\qquad \qquad
+\frac{1+2\lambda}{\sqrt{1+4\lambda}}\Bigl[2\text{Li}_{2}(r^{2})+4\text{Li}_{2}(-r)
+3\ln^{2}(r)-4\ln r \nonumber\\
&&\qquad \qquad\qquad \qquad\qquad \qquad+4\ln r \ln(1+4\lambda)-2\ln r\ln\lambda\Bigr]\Bigr\} \nonumber \\
&+&\frac{\alpha_{s}}{4\pi}C_{F}\hat{\sigma}_{B}(z)\frac{1}{(1+4\lambda
z^{2})^{3/2}}\biggl\{
\frac{1}{\left[1-(1-\lambda)z\right]^{2}}\Bigl[1-3z-4z^{2}\nonumber\\
&&\qquad \qquad\qquad \qquad\qquad \qquad\qquad \qquad+6z^{3}+8z^{4}-8z^{5} \nonumber \\
&&\qquad \qquad\qquad \qquad+6\lambda z\left(3-18z+13z^{2}+10z^{3}-8z^{4}\right)  \nonumber \\
&&\qquad \qquad\qquad \qquad+4\lambda^{2}z^{2}\left(8-77z+65z^{2}-2z^{3}\right)\biggr.  \; \,  \label{25}\\
&&\qquad \qquad\qquad \qquad+16\lambda^{3}z^{3}\left(1-21z+12z^{2}\right)-128\lambda^{4}z^{5}\Bigr]  \; \,  \nonumber \\
&&+\frac{2\ln D(z,\lambda)}{\sqrt{1+4\lambda
z^{2}}}\Bigl[-\left(1+z+2z^{2}+2z^{3}\right)+2\lambda
z\left(2-11z-11z^{2}\right)\nonumber\\
&&\qquad \qquad\qquad \qquad\qquad \qquad \qquad \qquad\qquad +8\lambda^{2}z^{2}\left(1-9z\right)\Bigr] \nonumber \\
&&-\frac{8(1+4\lambda)^{2}z^{4}}{\left(1-z\right)_{+}}-
\frac{4(1+2\lambda)(1+4\lambda)^{2}z^{4}}{\sqrt{1+4\lambda
z^{2}}}\frac{\ln D(z,\lambda)}{\left(1-z\right)_{+}}\biggr\},
\nonumber
\end{eqnarray}
\begin{eqnarray}\label{26}
\hat{\sigma}_{L,Q}^{(1)}(z,\lambda)&=&\frac{\alpha_{s}}{\pi}C_{F}\hat{\sigma}_{B}(1)
\frac{2\lambda}{\sqrt{1+4\lambda}}\delta(1-z)\Bigl\{-2+4\ln\lambda\\
&&-\frac{4\lambda}{\sqrt{1+4\lambda}}\,\ln r\nonumber+\frac{1+2\lambda}{\sqrt{1+4\lambda}}\Bigl[2\text{Li}_{2}(r^{2})+4\text{Li}_{2}(-r)\nonumber\\
&&+3\ln^{2}(r)-4\ln r+4\ln r \ln(1+4\lambda)-2\ln r\ln\lambda\Bigr]\Bigr\} \nonumber \\
&+&\frac{\alpha_{s}}{\pi}C_{F}\hat{\sigma}_{B}(z)\frac{1}{(1+4\lambda
z^{2})^{3/2}}\biggl\{\frac{z}{\left[1-(1-\lambda)z\right]^{2}}\Bigl[(1-z)^{2} \nonumber \\
&&\qquad \qquad\qquad \qquad-\lambda z\left(13-19z-2z^{2}+8z^{3}\right)\Bigr.   \nonumber \\
&&\qquad \qquad\qquad \qquad-2\lambda^{2}z^{2}\left(31-39z+8z^{2}\right)\Bigr.  \nonumber \\
&&\qquad \qquad\qquad \qquad-8\lambda^{3}z^{3}\left(10-7z\right)-32\lambda^{4}z^{4}\Bigr]\Bigr.  \nonumber \\
&&-\frac{2\lambda z^{2}\ln D(z,\lambda)}{\sqrt{1+4\lambda
z^{2}}}\left[3+3z+16\lambda z\right]\Bigr.\nonumber \\
&&-\frac{8\lambda(1+4\lambda)z^{4}}{\left(1-z\right)_{+}}-
\frac{4\lambda(1+2\lambda)(1+4\lambda)z^{4}}{\sqrt{1+4\lambda
z^{2}}}\frac{\ln D(z,\lambda)}{\left(1-z\right)_{+}}\biggr\},
\nonumber
\end{eqnarray}
\begin{eqnarray}
\hat{\sigma}_{A,Q}^{(1)}(z,\lambda)=\frac{\alpha_{s}}{2\pi}C_{F}\hat{\sigma}_{B}(z)\frac{z(1-z)}{(1+4\lambda
z^{2})^{3/2}}\biggl\{
\frac{1}{\left[1-(1-\lambda)z\right]}\left[1+2\lambda(4-3z)+8\lambda^2
z\right]\nonumber \\
+\frac{2\lambda \ln D(z,\lambda)}{\sqrt{1+4\lambda
z^{2}}}\left[2+z+4\lambda z\right]\biggr\},\qquad\label{27}
\end{eqnarray}
\begin{eqnarray}
\hat{\sigma}_{I,Q}^{(1)}(z,\lambda)=\frac{\alpha_{s}}{8\sqrt{2}}C_{F}\hat{\sigma}_{B}(z)
\frac{1}{(1+4\lambda
z^{2})^{2}}\frac{\sqrt{z}}{\left[1-(1-\lambda)z\right]^{3/2}}\biggl\{
-(1-z)(1+2z)\nonumber\\
-4\lambda z\left(10-10z-z^{2}+2z^{3}\right) \nonumber \\
-8\lambda^{2}z^{2}\left(25-29z+8z^{2}\right)-96\lambda^{3}z^{3}\left(3-2z\right)-
128\lambda^{4}z^{4}\biggr.\;&& \label{28} \\
+8\sqrt{\lambda z\left[1-(1-\lambda)z\right]}\left[1-z^{2}+\lambda
z(13-11z)+4\lambda^{2}z^{2}(7-4z)+16\lambda^{3}z^{3}\right]\biggr\}.&&
\nonumber
\end{eqnarray}
In Eqs.~(\ref{25}-\ref{28}), $C_{F}=(N_{c}^{2}-1)/(2N_{c})$, where
$N_{c}$ is number of colors, while
\begin{equation}\label{29}
D(z,\lambda)=\frac{1+2\lambda z -\sqrt{1+4\lambda z^{2}}}{1+2\lambda
z +\sqrt{1+4\lambda z^{2}}},\qquad \qquad
r=\sqrt{D(z=1,\lambda)}=\frac{\sqrt{1+4\lambda}-1}{\sqrt{1+4\lambda}+1}.
\end{equation}
The so-called "plus" distributions are defined by
\begin{equation}\label{30}
\left[g(z)\right]_{+}=g(z)-\delta(1-z)\int\limits_{0}^{1}\text{d}\zeta\,g(\zeta).
\end{equation}
For any sufficiently regular test function $h(z)$, Eq.~(\ref{30})
gives
\begin{equation}\label{31}
\int\limits_{a}^{1}\text{d}z\,h(z)\left[\frac{\ln^{k}(1-z)}{1-z}\right]_{+}=
\int\limits_{a}^{1}\text{d}z\frac{\ln^{k}(1-z)}{1-z}\left[h(z)-h(1)\right]+
h(1)\frac{\ln^{k+1}(1-a)}{k+1}.
\end{equation}

To perform a numerical investigation of the inclusive partonic cross
sections, $\hat{\sigma}_{k,Q}$ ($k=T,L,A,I$),{\large \ } it is
convenient to introduce the dimensionless coefficient functions
$c_{k,Q}^{(n,l)}$,
\begin{equation}\label{32}
\hat{\sigma}_{k,Q}(\eta ,\lambda ,\mu ^{2})=\frac{e_{Q}^{2}\alpha
_{em}\alpha _{s}(\mu ^{2})}{m^{2}}\sum_{n=0}^{\infty }\left( 4\pi
\alpha _{s}(\mu ^{2})\right) ^{n}\sum_{l=0}^{n}c_{k,Q}^{(n,l)}(\eta
,\lambda )\ln ^{l}\left( \frac{\mu ^{2}}{m^{2}}\right) ,
\end{equation}
where $\mu$ is a factorization scale (we use $\mu=\mu_{F}=\mu_{R}$)
and the variable $\eta$ measures the distance to the partonic
threshold:
\begin{equation}\label{33}
\eta =\frac{s}{m^{2}}-1=\frac{1-z}{\lambda z}, \qquad \qquad s
=(q+k_{Q})^{2}.
\end{equation}
\begin{figure}
\begin{center}
\begin{tabular}{cc}
\mbox{\epsfig{file=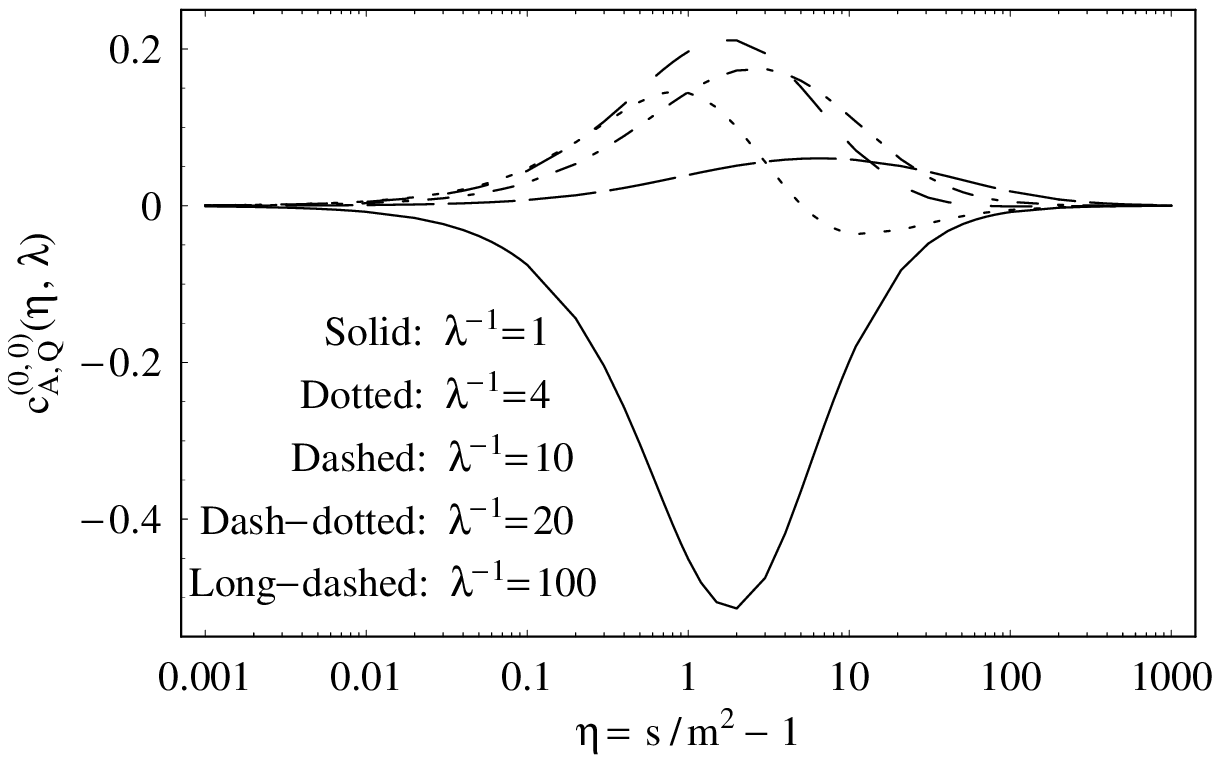,width=220pt}}
& \mbox{\epsfig{file=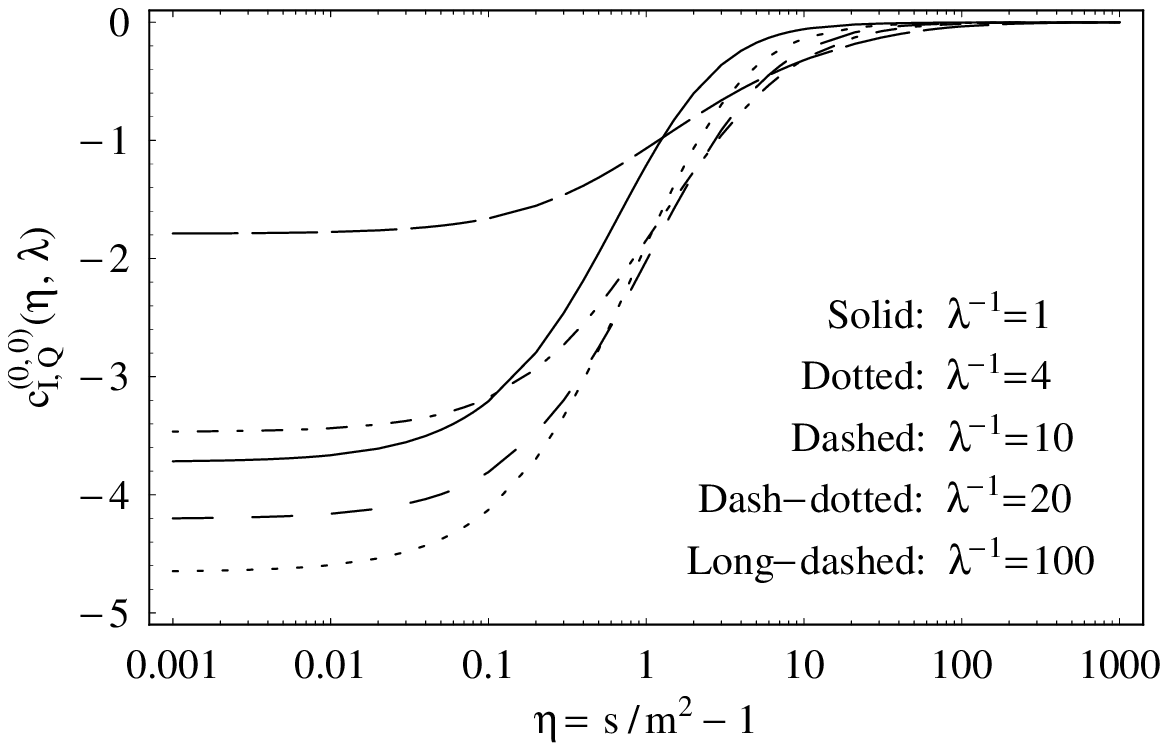,width=220pt}}\\
\end{tabular}
\caption[$c_{A,Q}^{(0,0)}(\eta,\lambda )$ and
$c_{I,Q}^{(0,0)}(\eta,\lambda )$ coefficient functions at several
values of $\lambda$.]{\label{Fg.3}\small
$c_{A,Q}^{(0,0)}(\eta,\lambda )$ and $c_{I,Q}^{(0,0)}(\eta,\lambda
)$ coefficient functions at several values of $\lambda$.}
\end{center}
\end{figure}
Our analysis of the quantity $c_{A,Q}^{(0,0)}(\eta ,\lambda)$ is
given in Fig.~\ref{Fg.3}. One can see that $c_{A,Q}^{(0,0)}$ is
negative at low $Q^{2}$ ($\lambda^{-1}\lesssim 1$) and positive at
high $Q^{2}$ ($\lambda^{-1}> 20$). For the intermediate values of
$Q^{2}$, $c_{A,Q}^{(0,0)}(\eta ,\lambda)$ is an alternating function
of $\eta$.

Our results for the coefficient function
$c_{I,Q}^{(0,0)}(\eta,\lambda )$ at several values of $\lambda$ are
presented in Fig.~\ref{Fg.3}. It is seen that $c_{I,Q}^{(0,0)}$ is
negative at all values of $\eta$ and $\lambda$. Note also the
threshold behavior of the coefficient function:
\begin{equation}\label{34}
c_{I,Q}^{(0,0)}(\eta\rightarrow 0,\lambda
)=-\sqrt{2}\pi^{2}C_{F}\frac{\sqrt{\lambda}}{1+4\lambda}+
{\cal{O}}(\eta).
\end{equation}
This quantity takes its minimum value at $\lambda_{m}=1/4$:
$$c_{I,Q}^{(0,0)}(\eta = 0,\lambda_{m})
=-\pi^{2}C_{F}/\left(2\sqrt{2}\right)$$.
 Let us analyze the
numerical significance of the $\cos\varphi$- and
$\cos2\varphi$-distributions for the QS component. It is difficult
to compare directly the $\hat{\sigma}^{(1)}_{A,Q}(z,\lambda)$ and
$\hat{\sigma}^{(1)}_{I,Q}(z,\lambda)$ cross sections given by the
usual functions (\ref{27}) and (\ref{28}) with the
$\varphi$-independent contributions
$\hat{\sigma}^{(0)}_{2,Q}(z,\lambda)$ and
$\hat{\sigma}^{(1)}_{2,Q}(z,\lambda)$ described by the generalized
functions (\ref{22}) and (\ref{25}). For this reason, we consider
the Mellin moments of the corresponding quantities defined as
\begin{equation}\label{37.4}
\hat{\sigma}_{i,Q}(N,\lambda)=\int\limits^{1}_{0}\hat{\sigma}_{i,Q}(z,\lambda)z^{N-1}\text{d}z,
\qquad \qquad \qquad (i=2,L,A,I).
\end{equation}
The Mellin transform of the Born level cross sections is trivial:
$\hat{\sigma}^{(0)}_{2,Q}(N,\lambda)=\hat{\sigma}_{B}(1)\sqrt{1+4\lambda}$.
The Mellin moments of the NLO results have been calculated
numerically. We use for $\alpha_{s}(\mu_{F})$ the one-loop
approximation with $\Lambda_{4}=326$ MeV,
$\mu_{F}=\sqrt{m^{2}+Q^{2}}$ and $m=1.3$ GeV.
\begin{figure}
\begin{center}
\begin{tabular}{cc}
\mbox{\epsfig{file=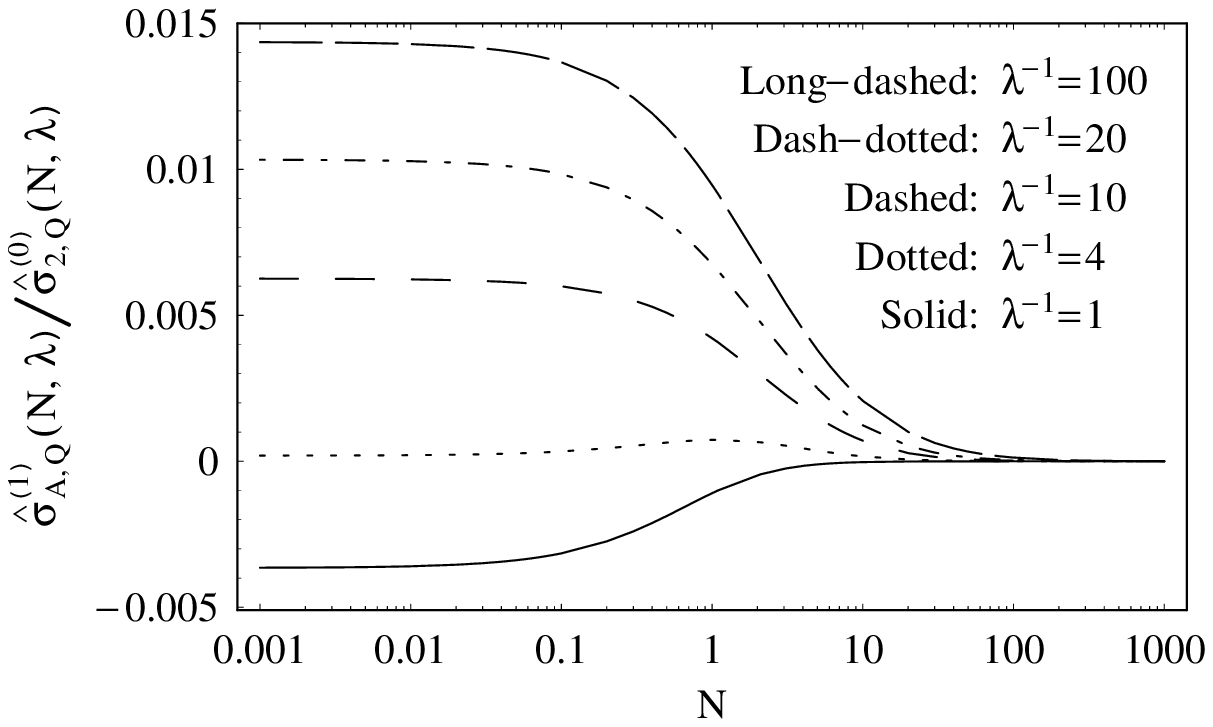,width=220pt}}
& \mbox{\epsfig{file=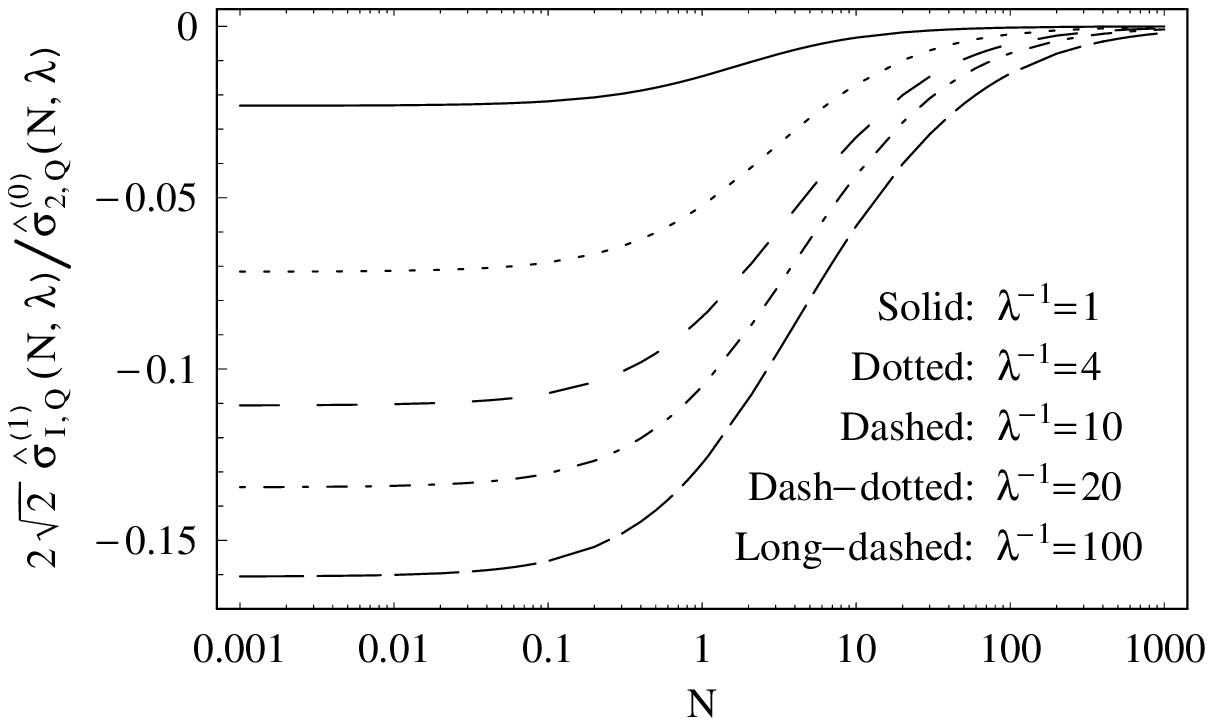,width=220pt}}\\
\end{tabular}
\caption[The quantities
$\hat{\sigma}^{(1)}_{A,Q}(N,\lambda)/\hat{\sigma}^{(0)}_{2,Q}(N,\lambda)$
and
$2\sqrt{2}\,\hat{\sigma}^{(1)}_{I,Q}(N,\lambda)/\hat{\sigma}^{(0)}_{2,Q}(N,\lambda)$
]{\label{Fg.4.0}\small The quantities
$\hat{\sigma}^{(1)}_{A,Q}(N,\lambda)/\hat{\sigma}^{(0)}_{2,Q}(N,\lambda)$
(\emph{left panel}) and
$2\sqrt{2}\,\hat{\sigma}^{(1)}_{I,Q}(N,\lambda)/\hat{\sigma}^{(0)}_{2,Q}(N,\lambda)$
(\emph{right panel}) at several values of $\lambda$.}
\end{center}
\end{figure}

The left panel of Fig.~\ref{Fg.4.0} presents the ratio
$\hat{\sigma}^{(1)}_{A,Q}(N,\lambda)/\hat{\sigma}^{(0)}_{2,Q}(N,\lambda)$
as a function of $N$ for several values of variable $\lambda$:
$\lambda^{-1}=1,4,10,20$ and 100. One can see that this ratio is
negligibly small (of the order of 1$\%$). Moreover, our analysis
shows that the ratio
$\hat{\sigma}^{(1)}_{A,Q}(N,\lambda)/\hat{\sigma}^{(0)}_{2,Q}(N,\lambda)$
is less than $1.5\%$ for all values of $\lambda$ and $N>0$. This
implies that the photon-quark scattering contribution is practically
$\cos2\varphi$-independent and, for this reason, we will neglect
 $\hat{\sigma}_{A,Q}(z,\lambda)$
cross section in our further analysis.

In the right panel of Fig.~\ref{Fg.4.0}, the $N$-dependence of the
ratio
$2\sqrt{2}\,\hat{\sigma}^{(1)}_{I,Q}(N,\lambda)/\hat{\sigma}^{(0)}_{2,Q}(N,\lambda)$
is given for the same values of $\lambda$. One can see that this
ratio is of the order of 10-15$\%$ at small $N$ and sufficiently
high $Q^{2}$. This fact indicates that the
$\cos\varphi$-distribution caused by the QS component may be
sizable.

\subsection{Comparison with Available Results}
For the first time, the NLO ${\cal O}(\alpha_{em}\alpha_{s})$
corrections to the $\varphi$-independent IC contribution have been
calculated a long time ago by Hoffmann and Moore (HM) \cite{HM}.
However, authors of Ref.~\cite{HM} don't give explicitly their
definition of the partonic cross sections that leads to a confusion
in interpretation of the original HM results. To clarify the
situation, we need first to derive the relation between the
lepton-quark DIS cross section, $\text{d}\hat{\sigma}_{lQ}$, and the
partonic cross sections, $\sigma^{(2)}$ and $\sigma^{(L)}$, used in
\cite{HM}. Using Eqs.~(C.1) and (C.5) in Ref.~\cite{HM}, one can
express the HM tensor $\sigma_{R}^{\mu\nu}$ in terms of "our" cross
sections $\hat{\sigma}_{2,Q}$ and $\hat{\sigma}_{L,Q}$ defined by
Eq.~(\ref{18}) in the present paper. Comparing the obtained results
with the corresponding definition of $\sigma_{R}^{\mu\nu}$ via the
HM cross sections $\sigma^{(2)}$ and $\sigma^{(L)}$ (given by
Eqs.~(C.16) and (C.17) in Ref.~\cite{HM}), we find that
\begin{eqnarray}
\hat{\sigma}_{2,Q}(z,\lambda)&\equiv
&\hat{\sigma}_{B}(z)\sqrt{1+4\lambda
z^{2}}\,\sigma^{(2)}(z,\lambda),  \label{35} \\
\hat{\sigma}_{L,Q}(z,\lambda)&\equiv
&\frac{2\hat{\sigma}_{B}(z)}{\sqrt{1+4\lambda
z^{2}}}\left[\sigma^{(L)}(z,\lambda)+2\lambda
z^{2}\sigma^{(2)}(z,\lambda)\right]. \label{36}
\end{eqnarray}
Now we are able to compare our results with original HM ones. It is easy to see that the LO cross
sections (defined by Eqs.~(37) in \cite{HM} and Eqs.~(\ref{22}) in our paper) obey both above
identities.  Comparing with each other the quantities $\sigma^{(2)}_{1}$ and
$\hat{\sigma}_{2,Q}^{(1)}$ (given by Eq.~(51) in \cite{HM} and Eq.~(\ref{25}) in this paper,
respectively), we find that identity (\ref{35}) is satisfied at NLO too. The situation with
longitudinal cross sections is more complicated. We have uncovered two misprints in the NLO
expression for $\sigma^{(L)}$ given by Eq.~(52) in \cite{HM}. First, the r.h.s. of this Eq. must be
multiplied by $z$. Second, the sign in front of the last term (proportional to $\delta (1-z)$) in
Eq.~(52) in Ref.~\cite{HM} must be changed \footnote{Note that this term originates from virtual
corrections and the virtual part of the longitudinal cross section given by Eq.~(39) in
Ref.~\cite{HM} also has wrong sign. See Appendix \ref{virt} for more details.}. Taking into account
these typos, we find that relation (\ref{36}) holds at NLO as well. So, our calculations of
$\hat{\sigma}_{2,Q}$ and $\hat{\sigma}_{L,Q}$ agree with the HM results.

Recently, the heavy quark initiated contributions to the
$\varphi$-independent DIS structure functions, $F_{2}$ and $F_{L}$,
have been calculated by Kretzer and Schienbein (KS) \cite{KS}. The
final KS results are expressed in terms of the parton level
structure functions $\hat{H}^{q}_{1}$ and $\hat{H}^{q}_{2}$. Using
the definition of $\hat{H}^{q}_{1}$ and $\hat{H}^{q}_{2}$ given by
Eqs.~(7, 8) in Ref.~\cite{KS}, we obtain that
\begin{eqnarray}\label{37}
\hat{\sigma}_{T,Q}(z,\lambda)\equiv
\frac{\alpha_{s}}{2\pi}\frac{\hat{\sigma}_{B}(z)}{\sqrt{1+4\lambda}}\frac{\hat{H}^{q}_{1}(\xi^{\prime},
\lambda)}{\sqrt{1+4\lambda z^{2}}},\\\nonumber
\hat{\sigma}_{2,Q}(z,\lambda)\equiv
\frac{\alpha_{s}}{2\pi}\hat{\sigma}_{B}(z)\sqrt{\frac{1+4\lambda}{1+4\lambda
z^{2}}}\,\hat{H}^{q}_{2}(\xi^{\prime},\lambda),
\end{eqnarray}
where $\hat{\sigma}_{T,Q}=\hat{\sigma}_{2,Q}-\hat{\sigma}_{L,Q}$ and $\hat{\sigma}_{L,Q}$ are
defined by Eq.~(\ref{18}) in our paper and
$\xi^{\prime}=z\left(1+\sqrt{1+4\lambda}\right)\left/\right.$ $\left(1+\sqrt{1+4\lambda
z^{2}}\right)$. To test identities (\ref{37}), one needs only to rewrite the NLO expressions for
the functions $\hat{H}^{q}_{1}(\xi^{\prime},\lambda)$ and $\hat{H}^{q}_{2}(\xi^{\prime},\lambda)$
(given in Appendix C in Ref.~\cite{KS}) in terms of variables $z$ and $\lambda$. Our analysis shows
that relations (\ref{37}) hold at both LO and NLO. Hence we coincide with the KS predictions for
the $\gamma^{*}Q$ cross sections.

However, we disagree with the conclusion of Ref.~\cite{KS} that there are errors in the NLO
expression for $\sigma^{(2)}$ given in Ref.~\cite{HM} \footnote{In detail, the KS point of view on
the HM results is presented in PhD thesis \cite{KS-thesis}, pp.~158-160.}. As explained above, a
correct interpretation of the quantities $\sigma^{(2)}$ and $\sigma^{(L)}$ used in \cite{HM} leads
to a complete agreement between the HM, KS and our results for $\varphi$-independent cross
sections.

As to the $\varphi$-dependent DIS, pQCD predictions for the
$\gamma^{*}Q$ cross sections $\hat{\sigma}_{A,Q}(z,\lambda)$ and
$\hat{\sigma}_{I,Q}(z,\lambda)$ in the case of arbitrary values of
$m^{2}$ and $Q^{2}$ are not, to our knowledge, available in the
literature. For this reason, we have performed several cross checks
of our results against well known calculations in two limits:
$m^{2}\rightarrow 0$ and $Q^{2}\rightarrow 0$. In particular, in the
chiral limit, we reproduce the original results of Georgi and
Politzer \cite{GP} and M\'{e}ndez \cite{Mendez} for
$\hat{\sigma}_{I,Q}(z,\lambda\rightarrow 0)$ and
$\hat{\sigma}_{A,Q}(z,\lambda\rightarrow 0)$. In the case of
$Q^{2}\rightarrow 0$, our predictions for
$\hat{\sigma}_{2,Q}(s,Q^{2}\rightarrow 0)$ and
$\hat{\sigma}_{A,Q}(s,Q^{2}\rightarrow 0)$ given by
Eqs.~(\ref{25},\ref{27}) reduce to the QED textbook results for the
Compton scattering of polarized photons \cite{Fano}.

\subsection{Photon-Gluon Fusion}
The gluon fusion component of the semi-inclusive DIS is the
following parton level interaction:
\begin{equation}
l(\ell )+g(k_{g})\rightarrow l(\ell
-q)+Q(p_{Q})+X[\overline{Q}](p_{X}). \label{d1}
\end{equation}
Corresponding lepton-gluon cross section,
$\text{d}\hat{\sigma}_{lg}$, has the following decomposition in
terms of the helicity $\gamma ^{*}g$ cross sections:
\begin{eqnarray}\nonumber
\frac{\text{d}^{3}\hat{\sigma}_{lg}}{\text{d}z\text{d}Q^{2}\text{d}\varphi
}=\frac{\alpha _{em}}{(2\pi )^{2}}\frac{1}{z
Q^{2}}\frac{y^2}{1-\varepsilon}\Bigl[\hat{\sigma}_{2,g}(z,\lambda)-
(1-\varepsilon)\hat{\sigma}_{L,g}(z,\lambda)\\
+\varepsilon\hat{\sigma}_{A,g}(z,\lambda)\cos 2\varphi+
2\sqrt{\varepsilon(1+\varepsilon)}\hat{\sigma}_{I,g}(z,\lambda)\cos
\varphi\Bigr],  \label{d2}
\end{eqnarray}
where the quantity $\varepsilon$ is defined by Eq.~(\ref{3}) with
$y=\left.(q\cdot k_{g})\right /(\ell\cdot k_{g})$.

At LO, ${\cal O}(\alpha_{em}\alpha_{s})$, the only gluon fusion
subprocess responsible for heavy flavor production is
\begin{equation}\label{38}
\gamma ^{*}(q)+g(k_{g})\rightarrow
Q(p_{Q})+\overline{Q}(p_{\stackrel{\_}{Q}}).
\end{equation}
\begin{figure}
\begin{center}
\mbox{\epsfig{file=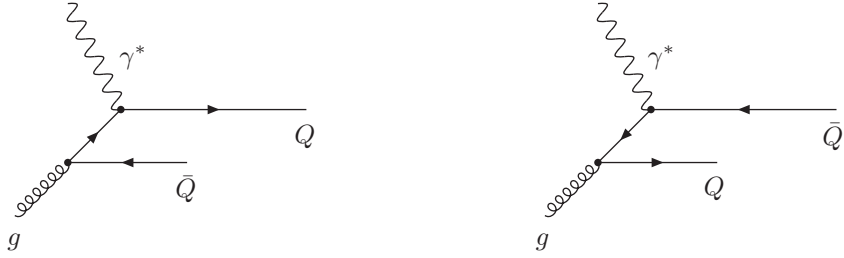,width=355pt}}
\end{center}
\caption[The LO photon-gluon fusion diagrams]{\label{Fg.4}\small The
LO photon-gluon fusion diagrams.}
\end{figure}
The $\gamma ^{*}g$ cross sections, $\hat{\sigma}_{k,g}^{(0)}$
($k=2,L,A,I$), corresponding to the Born diagrams given in
Fig.~\ref{Fg.4} have the form \cite{LW1,Watson}:
\begin{eqnarray}
\hat{\sigma}_{2,g}^{(0)}(z,\lambda)&=&\frac{\alpha_{s}}{2\pi}\hat{\sigma}_{B}(z)
\Bigl\{\left[(1-z)^{2}+z^{2}+4\lambda
z(1-3z)-8\lambda^{2}z^{2}\right] \ln\frac{1+\beta_{z}}{1-\beta_{z}}\\
&&\qquad \qquad\qquad \qquad\qquad \qquad -\left[1+4z(1-z)(\lambda-2)\right]\beta_{z}\Bigr\},\nonumber  \\
\hat{\sigma}_{L,g}^{(0)}(z,\lambda)&=&\frac{2\alpha_{s}}{\pi}\hat{\sigma}_{B}(z)z
\Bigl\{-2\lambda z\ln\frac{1+\beta_{z}}{1-\beta_{z}}+\left(1-z\right)\beta_{z}\Bigr\},  \label{39} \\
\hat{\sigma}_{A,g}^{(0)}(z,\lambda)&=&\frac{\alpha_{s}}{\pi}\hat{\sigma}_{B}(z)z
\Bigl\{2\lambda\left[1-2z(1+\lambda)\right]\ln\frac{1+\beta_{z}}{1-\beta_{z}}+
(1-2\lambda)(1-z)\beta_{z}\Bigr\},  \\
\hat{\sigma}_{I,g}^{(0)}(z,\lambda)&=&0,
\end{eqnarray}
where $\hat{\sigma}_{B}(z)$ is defined by Eq.~(\ref{23}) and the
following notations are used:
\begin{equation}\label{40}
z=\frac{Q^{2}}{2q\cdot k_{g}},\qquad \qquad\lambda
=\frac{m^{2}}{Q^{2}}, \qquad \qquad \beta_{z}=\sqrt{1-\frac{4\lambda
z}{1-z}}.
\end{equation}
Note that the $\cos \varphi $ dependence vanishes in the GF
mechanism due to the $Q\leftrightarrow \overline{Q}$ symmetry which,
at leading order, requires invariance under $\varphi \rightarrow
\varphi +\pi$ \cite{LW2}.

As to the NLO results, presently, only $\varphi$-independent
quantities $\hat{\sigma}_{2,g}^{(1)}$ and $\hat{\sigma}_{L,g}^{(1)}$
are known exactly \cite{LRSN}. For this reason, we will use in our
analysis the so-called soft-gluon approximation for the NLO $\gamma
^{*}g$ cross sections (see Appendix \ref{soft}). As shown in
Refs.~\cite{Laenen-Moch,we2,we4}, at energies not so far from the
production threshold, the soft-gluon radiation is the dominant
perturbative mechanism in the $\gamma ^{*}g$ interactions.

\section{\label{hadr}Hadron Level Results}
\subsection{\label{ha}Fixed Flavor Number Scheme and Nonperturbative Intrinsic Charm}
\hspace{14pt}In the fixed flavor number scheme \footnote{This approach is sometimes referred to as
the fixed-order perturbation theory (FOPT).}, the wave function of the proton consists of light
quarks $u,d,s$ and gluons $g$. Heavy flavor production in DIS is dominated by the gluon fusion
mechanism. Corresponding hadron level cross sections, $\sigma_{k,GF}(x,\lambda)$, have the form
\begin{eqnarray}
\sigma_{k,GF}(x,\lambda)&=&\int\limits_{\chi}^{1}\text{d}z\,g(z,\mu_{F})\,\hat{\sigma}_{k,g}
\!\left(x/z,\lambda,\mu_{F}\right), \qquad (k=2,L,A,I), \label{41}\\
\chi&=&x\left(1+4\lambda\right), \label{42}
\end{eqnarray}
where $g(z,\mu _{F})$ describes gluon density in the proton
evaluated at a factorization scale $\mu _{F}$. The lowest order GF
cross sections, $\hat{\sigma}_{k,g}^{(0)}$ ($k=2,L,A,I$), are given
by Eqs.~(\ref{39}). The NLO results, $\hat{\sigma}_{k,g}^{(1)}$, to
the next-to-leading logarithmic accuracy are presented in Appendix
\ref{soft}.

We neglect the $\gamma ^{*}q(\bar{q})$ fusion subprocesses. This is
justified as their contributions to heavy quark leptoproduction
vanish at LO and are small at NLO \cite{LRSN}.

In the FFNS, the intrinsic heavy flavor component of the proton wave
function is generated by $gg\rightarrow Q\bar{Q}$ fluctuations where
the gluons are coupled to different valence quarks. In the present
paper, this component is referred to as the nonperturbative
intrinsic charm (bottom). The probability of the corresponding
five-quark Fock state, $\left|uudQ\bar{Q}\right\rangle$, is of
higher twist since it scales as
$\Lambda_{QCD}^{2}\left/m^{2}\right.$ \cite{polyakov}. However,
since all of the quarks tend to travel coherently at same rapidity
in the $\left|uudQ\bar{Q}\right\rangle$ bound state, the heaviest
constituents carry the largest momentum fraction. For this reason,
the heavy flavor distribution function has a more "hard"
$z$-behavior than the light parton densities. Since all of the
densities vanish at $z\to 1$, the hardest PDF becomes dominant at
sufficiently large $z$ independently of normalization.

Convolution of PDFs with partonic cross sections does not violate
this observation. In particular, assuming a gluon density $g(z)\sim
(1-z)^n$ (where $n=3-5$), we obtain that the LO GF contribution to
$F_{2}$ scales as $(1-\chi)^{n+3/2}$ at $\chi\to 1$, where $\chi$ is
defined by Eq.~(\ref{42}). In the case of Hoffman and Moore charm
density (see below), the LO IC contribution is proportional to
$(1-\chi)$ at $\chi\to 1$. It is easy to see that, independently of
normalizations, the IC contribution to be dominate over the more
"soft" GF component at large enough $x$.

For the first time, the intrinsic charm momentum distribution in the
five-quark state $\left|uudc\bar{c}\right\rangle$ was derived by
Brodsky, Hoyer, Peterson and Sakai (BHPS) in the framework of a
light-cone model \cite{BHPS,BPS}. Neglecting the transverse motion
of constituents, they have obtained in the heavy quark limit that
\begin{equation}\label{43}
c(z)=\frac{N_{5}}{6}z^{2}\left[6z(1+z)\ln
z+(1-z)(1+10z+z^{2})\right],
\end{equation}
where $N_{5}=36$ corresponds to a $1\%$ probability for IC in the
nucleon: $\int^{1}_{0}c(z)\text{d}z=0.01$.

Hoffmann and Moore (HM) \cite{HM} incorporated mass effects in the
BHPS approach. They first introduced a mass scaling variable $\xi$,
\begin{equation}\label{44}
\xi=\frac{2ax}{1+\sqrt{1+4\lambda_{N}x^{2}}},\qquad
\qquad a=\frac{1+\sqrt{1+4\lambda}}{2},
\end{equation}
where $\lambda_{N}=m^{2}_{N}\left/Q^{2}\right.$. To provide correct
threshold behavior of the charm density, the constraint
$\xi\leq\gamma<1$ was imposed where
\begin{equation}\label{45}
\gamma=\frac{2a\hat{x}}{1+\sqrt{1+4\lambda_{N}\hat{x}^{2}}},
\qquad \qquad \hat{x}=\frac{1}{1+4\lambda-\lambda_{N}}.
\end{equation}
Resulting charm distribution function, $c(\xi,\gamma)$, has the
following form in the HM approach:
\begin{equation}\label{46}
c(\xi,\gamma)=\left\{%
\begin{array}{ll}
c(\xi)-\displaystyle{\frac{\xi}{\gamma}}c(\gamma),&\qquad \xi\leq\gamma\\
0,&\qquad \xi>\gamma
\end{array}%
\right.
\end{equation}
with $c(\xi)$ defined by Eq.~(\ref{43}). Corresponding hadron level
cross sections for the $(c+\bar{c})$ production,
$\sigma_{k,QS}(x,\lambda)$, due to the heavy quark scattering (QS)
mechanism, are
\begin{equation}\label{47}
\sigma_{k,QS}(x,\lambda)=\int\limits_{\xi}^{\gamma}\frac{\text{d}z}{\sqrt{1+4\lambda\xi^{2}/z^{2}}}
\,c_{+}(z,\gamma)\,\hat{\sigma}_{k,c}\!\left(\xi/z,\lambda\right), \qquad (k=2,L,A,I),
\end{equation}
where the charm density $c_{+}(z,\gamma)\equiv
c(z,\gamma)+\bar{c}(z,\gamma)$. The LO and NLO expressions for the
partonic cross sections $\hat{\sigma}_{k,c}(z,\lambda)$ are given by
Eqs.~(\ref{22}) and (\ref{25}-\ref{28}), respectively.

Note also that, in the FFNS, the full cross section for the charm
production, $\sigma_{k}(x,\lambda)$, is simply a sum of the GF and
IC components:
\begin{equation}\label{48}
\sigma_{k}(x,\lambda)=\sigma_{k,GF}(x,\lambda)+\sigma_{k,QS}(x,\lambda),
\qquad \qquad  (k=2,L,A,I).
\end{equation}

\begin{figure}
\begin{center}
\begin{tabular}{cc}
\mbox{\epsfig{file=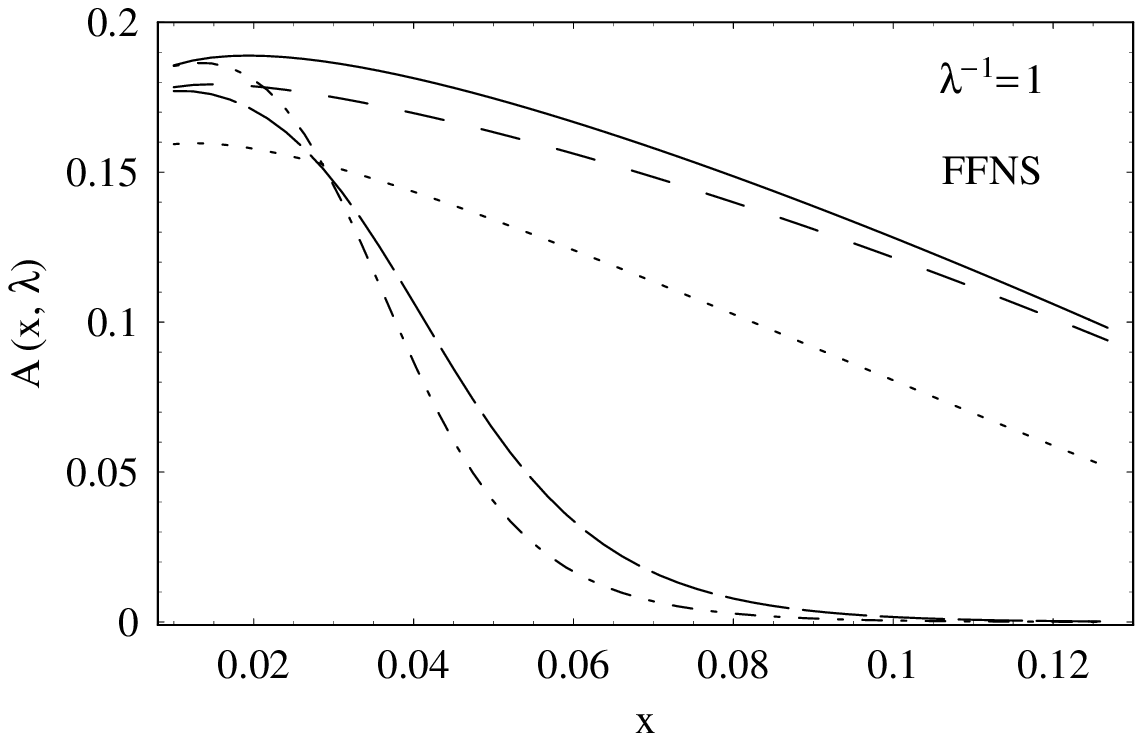,width=220pt}}
& \mbox{\epsfig{file=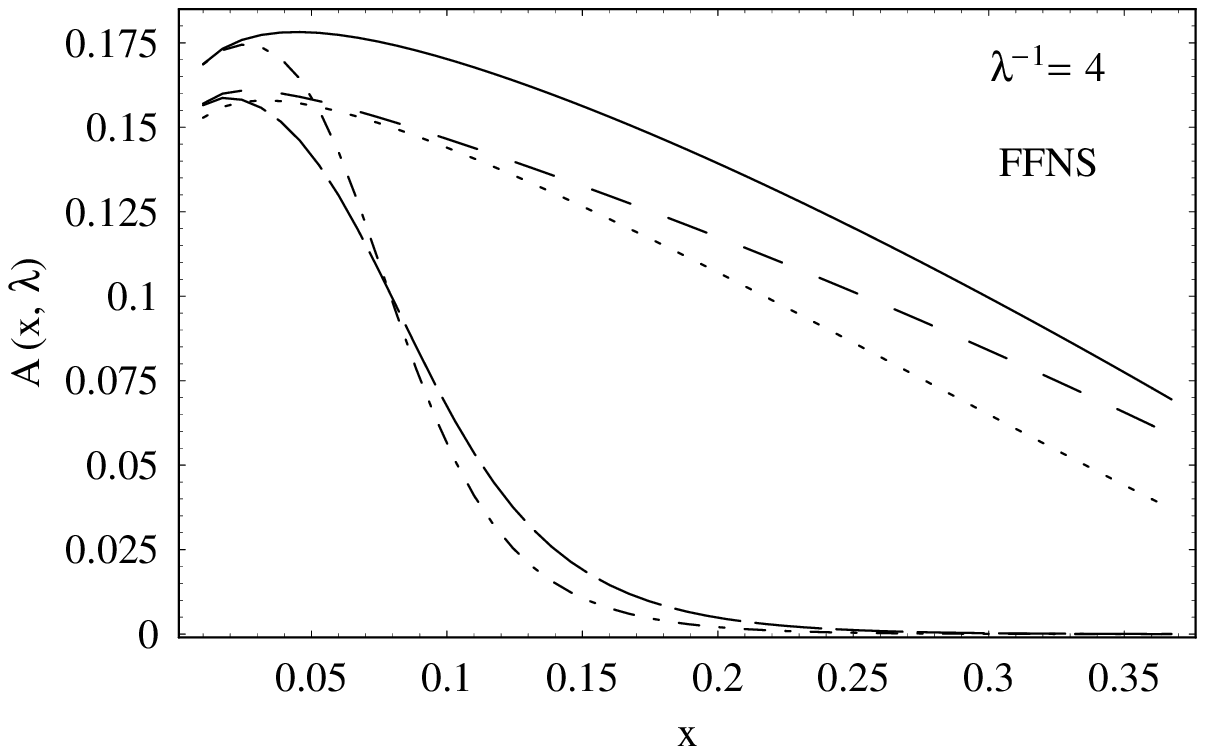,width=220pt}}\\
\mbox{\epsfig{file=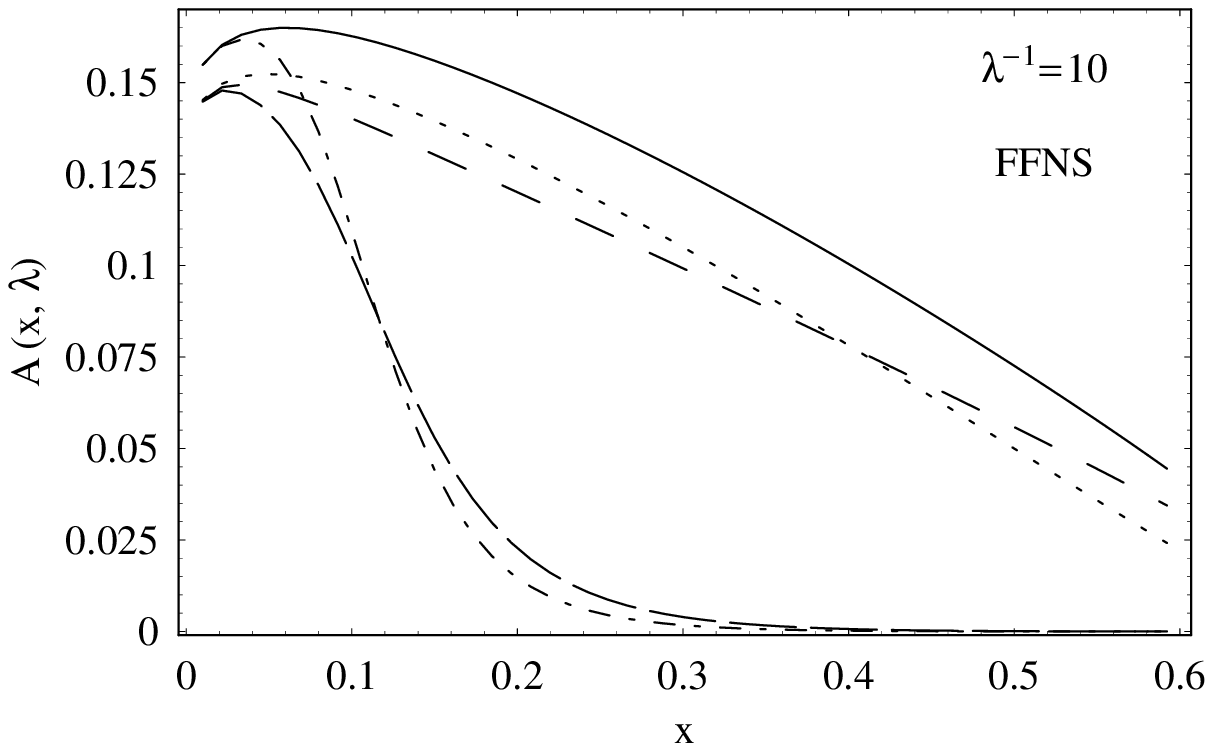,width=220pt}}
& \mbox{\epsfig{file=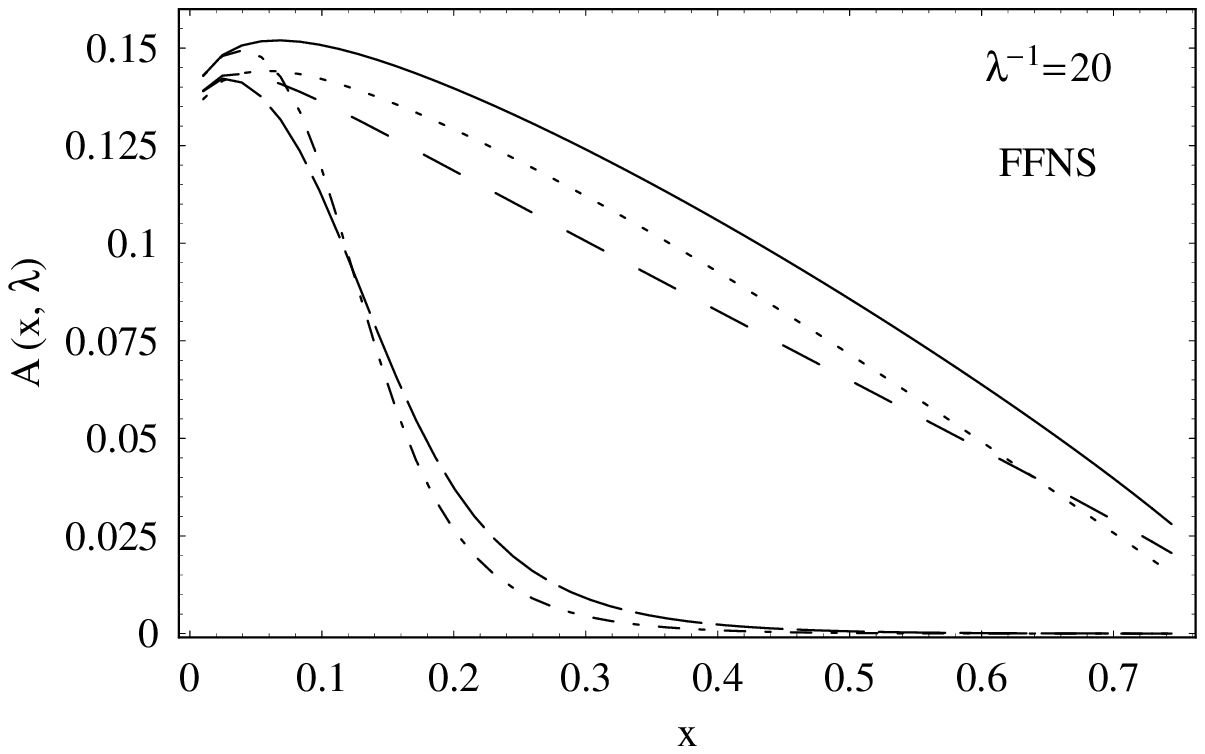,width=220pt}}\\
\mbox{\epsfig{file=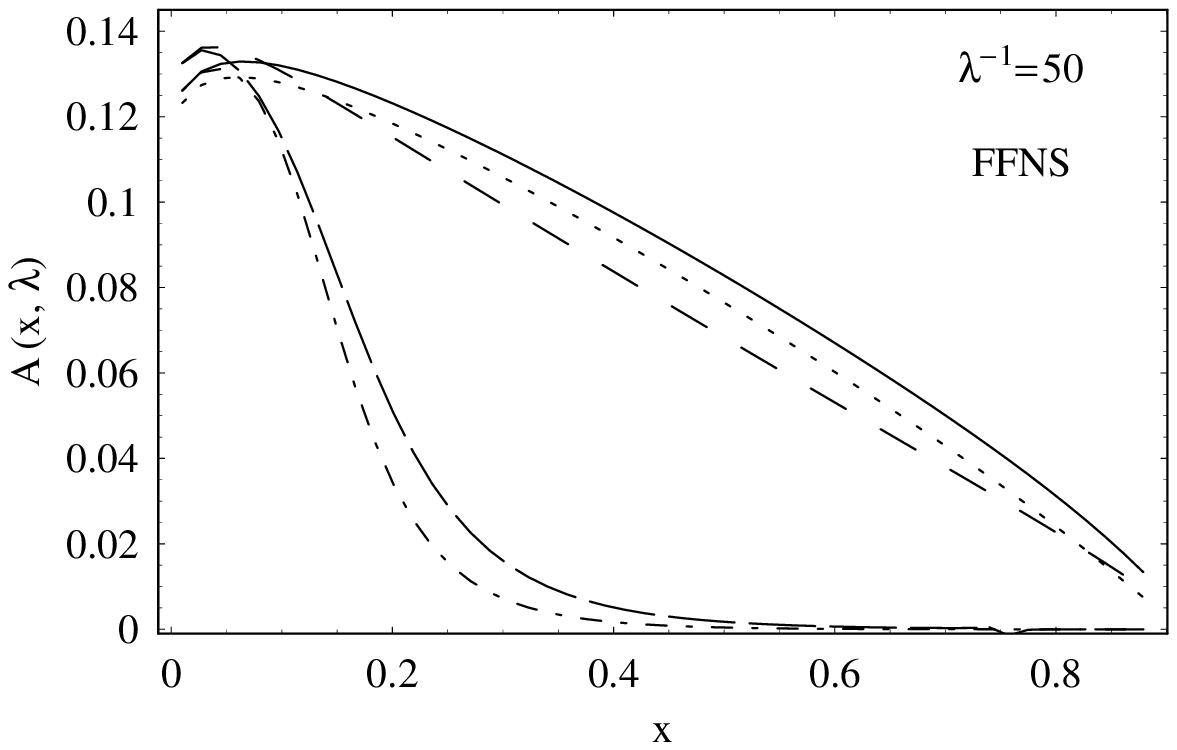,width=220pt}}
& \mbox{\epsfig{file=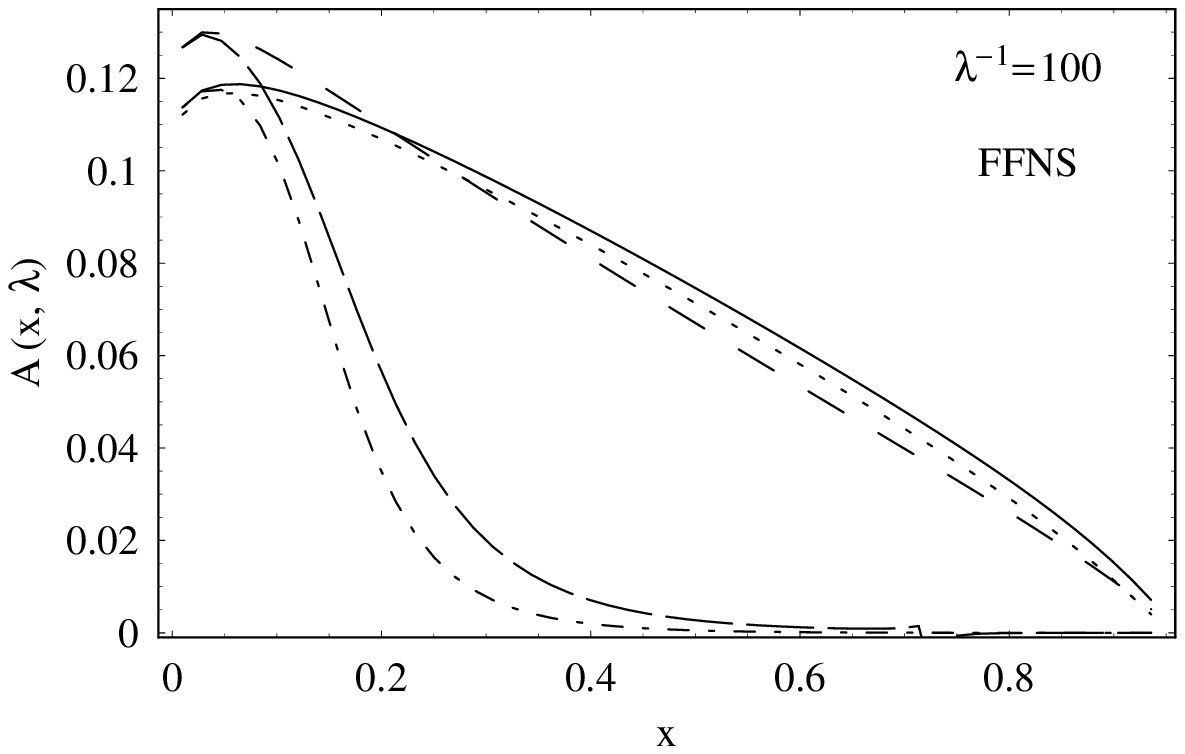,width=220pt}}\\
\end{tabular}
 \caption[Azimuthal asymmetry
parameter $A(x,\lambda)$ in the FFNS at several values of $\lambda$
in the case of $\int^{1}_{0}c(z)\text{d}z=1\%$.]{\label{Fg.6}\small
Azimuthal asymmetry parameter $A(x,\lambda)$ in the FFNS at several
values of $\lambda$ in the case of $\int^{1}_{0}c(z)\text{d}z=1\%$.
The following contributions are plotted: $\text{GF}^{\text{(LO)}}$
(solid lines), $\text{GF}^{\text{(LO)}}$+$k_{T}$-kick (dotted
lines), $\text{GF}^{\text{(NLO)}}$ (dashed lines),
$\text{GF}^{\text{(LO)}}$+$\text{QS}^{\text{(LO)}}$ (dash-dotted
lines) and $\text{GF}^{\text{(NLO)}}$+$\text{QS}^{\text{(NLO)}}$
(long-dashed lines).}
\end{center}
\end{figure}

Let us discuss the FFNS predictions for the hadron level asymmetry
parameter $A(x,Q^{2})$ defined by Eq.~(\ref{12}). Fig.~\ref{Fg.6}
shows $A(x,\lambda)$ as a function of $x$ for several values of
variable $\lambda$: $\lambda^{-1}=1,4,10,20,50$ and 100. We display
both LO and NLO predictions  of the GF mechanism as well as the
analogous results of the combined GF+QS contribution. The azimuthal
asymmetry due to the mere LO GF component is given by solid line.
The NLO GF predictions are plotted by dashed line. The LO and NLO
results of the total GF+QS contribution are given by dash-dotted and
long-dashed lines, respectively. In our calculations, the CTEQ5M
\cite{CTEQ5} parametrization of the gluon distribution function is
used and a $1\%$ probability for IC in the nucleon is assumed.
Throughout this paper, the value
$\mu_{F}=\mu_{R}=\sqrt{m^{2}+Q^{2}}$ for both factorization and
renormalization scales is chosen. In accordance with the CTEQ5M
parametrization, we use $m_{c}=1.3 $ GeV and $\Lambda_{4}=326$ MeV
\cite{CTEQ5}.

One can see from Fig.~\ref{Fg.6} the following basic features of the
azimuthal asymmetry, $A(x,\lambda)$, within the FFNS. First, as
expected, the nonperturbative IC contribution is practically
invisible at low $x$, but affects essentially the GF predictions at
large $x$. Since, contrary to the GF mechanism, the QS component is
practically $\cos2\varphi$-independent, the dominance of the IC
contribution at large $x$ leads to a more rapid (in comparison with
the GF predictions) decreasing of $A(x,\lambda)$ with growth of $x$.

The most remarkable property of the azimuthal asymmetry is its
perturbative stability. In Refs.~\cite{we2,we4}, the NLO soft-gluon
corrections to the GF predictions for the $\cos2\varphi$ asymmetry
in heavy quark photo- and leptoproduction was calculated. It was
shown that, contrary to the production cross sections, the quantity
$A(x,\lambda)$ is practically insensitive to soft radiation. One can
see from Fig.~\ref{Fg.6} in the present paper that the NLO
corrections to the LO GF predictions for $A(x,\lambda)$ are about
few percent at not large $x$.  This implies that large soft-gluon
corrections to $\sigma_{A,GF}^{(LO)}$ and $\sigma_{2,GF}^{(LO)}$
(increasing both cross sections by a factor of two) cancel each
other in the ratio
$\bigl(\sigma_{A,GF}^{(NLO)}\!\bigl/\sigma_{2,GF}^{(NLO)}\bigr)(x,\lambda)$
with a good accuracy. In terms of so-called $K$-factors,
$K_{k}(x,\lambda)=\bigl(\sigma_{k}^{(NLO)}\!\bigl/\sigma_{k}^{(LO)}\bigr)(x,\lambda)$
for $k=2,L,A,I$, perturbative stability of the GF predictions for
$A(x,\lambda)$ is provided by the fact that the corresponding
$K$-factors are approximately the same at not large $x$:
$K_{A,GF}(x,\lambda)\approx K_{2,GF}(x,\lambda)$.

Comparing with each other the dash-dotted and long-dashed curves in Fig.~\ref{Fg.6}, we see that
the NLO corrections to the combined GF+QS result for $A(x,\lambda)$ are also small. In this case,
three reasons are responsible for the closeness of the LO and NLO predictions. At small $x$, where
the nonperturbative IC contribution is negligible, perturbative stability of the asymmetry is
provided by the GF component. In the large-$x$ region, where the IC mechanism dominates, the
azimuthal asymmetry rapidly vanishes with growth of $x$ at both LO and NLO because the QS component
is practically $\cos2\varphi$-independent, $\hat{\sigma}^{(1)}_{A,c}(x,\lambda)\approx\hat{\sigma
}^{(0)}_{A,c}(x,\lambda)=0$ \footnote{Although the ratio $(A^{(NLO)}/A^{(LO)})(x,\lambda)$ is
sizeable at sufficiently large $x$, the absolute values of the quantities $A^{(LO)}(x,\lambda)$ and
$A^{(NLO)}(x,\lambda)$ become so small that it seems reasonable to consider the asymmetry as
equally negligible at both LO and NLO and treat the predictions as perturbatively stable.}. At
intermediate values of $x$, where both mechanisms are essential, perturbative stability of
$A(x,\lambda)$ is due to the similarity of the corresponding $K$-factors: $K_{2,GF}(x,\lambda)\sim
K_{2,QS}(x,\lambda)$ \footnote{Note however that this similarity takes only place at intermediate
values of $x$ where both GF and QS components are essential. In the low- and large-$x$ regions, the
factors $K_{2,GF}(x,\lambda)$ and $K_{2,QS}(x,\lambda)$ are strongly different.}.

Another remarkable property of the azimuthal asymmetry closely related to fast perturbative
convergence is its parametric stability \footnote{Of course, parametric stability of the fixed
order results does not imply a fast convergence of the corresponding series. However, a fast
convergent series must be parametrically stable. In particular, it must be $\mu _{R}$- and $\mu
_{F}$-independent.}. The analysis of Refs.~\cite{we1,we4} shows that the GF predictions for the
$\cos 2\varphi $ asymmetry are less sensitive to standard uncertainties in the QCD input parameters
($m,\mu _{R},\mu _{F},\Lambda_{QCD}$ and PDFs) than the corresponding ones for the production cross
sections. We have verified that the same situation takes also place for the combined GF+QS results.

Let us discuss how the GF predictions for the azimuthal asymmetry
are affected by nonperturbative contributions due to the intrinsic
transverse motion of the gluon in the target. Because of the
relatively low $c$-quark mass, these contributions are especially
important in the description of the cross sections for charmed
particle production.

To introduce $k_{T}$ degrees of freedom, $\vec{k}_{g}\simeq
\zeta\vec{p}+\vec{k}_{T}$, one extends the integral over the parton
distribution function in Eq. (\ref{41}) to $k_{T}$-space,
\begin{equation}  \label{49}
\text{d}\zeta\,g(\zeta,\mu _{F})\rightarrow
\text{d}\zeta\,\text{d}^{2}k_{T}f\big(\vec{k}_{T}\big) g(\zeta,\mu
_{F}).
\end{equation}
The transverse momentum distribution, $f\big( \vec{k}_{T}\big) $, is
usually taken to be a Gaussian:
\begin{equation}  \label{50}
f\big( \vec{k}_{T}\big) =\frac{{\rm {e}}^{-\vec{k}_{T}^{2}/\langle
k_{T}^{2}\rangle }}{\pi \langle k_{T}^{2}\rangle }.
\end{equation}
In practice, an analytic treatment of $k_{T}$ effects is usually
used. According to Ref.~\cite{kT}, the $k_{T}$-smeared differential
cross section of the process (\ref{1}) is a two-dimensional
convolution:
\begin{equation}  \label{51}
\frac{\text{d}^{4}\sigma _{lN}^{{\rm
{kick}}}}{\text{d}x\text{d}Q^{2}\text{d} p_{QT}\text{d}\varphi
}\left( \vec{p}_{QT}\right) =\int \text{d}^{2}k_{T} \frac{{\rm
{e}}^{-\vec{k}_{T}^{2}/\langle k_{T}^{2}\rangle }}{\pi \langle
k_{T}^{2}\rangle }\frac{\text{d}^{4}\sigma
_{lN}}{\text{d}x\text{d}Q^{2} \text{d}p_{QT}\text{d}\varphi }\Big(
\vec{p}_{QT}-\frac{1}{2}\vec{k} _{T}\Big) .
\end{equation}
The factor $\frac{1}{2}$ in front of $\vec{k}_{T}$ in the r.h.s. of
Eq.~(\ref {51}) reflects the fact that the heavy quark carries away
about one half of the initial energy in the reaction (\ref{1}).

Values of the $k_{T}$-kick corrections to the LO GF predictions for
the $\cos 2\varphi $ asymmetry in the charm production are shown in
Fig.~\ref{Fg.6} by dotted curves. Calculating the $k_{T}$-kick
effect we use $\langle k_{T}^{2}\rangle =0.5$ GeV$^{\text{2}}$. One
can see that $k_{T}$-smearing for $A(x,Q^{2})$ is about $20$-$25\%$
in the region of low $Q^{2}\lesssim m^{2}$ and rapidly decreases at
high $Q^{2}$.

In Fig.~\ref{Fg.6a}, the dependence of the asymmetry $A(x,\lambda)$
on the nonperturbative intrinsic charm content of the proton is
presented. We plot the LO predictions for $A(x,\lambda)$ as a
function of $x$ for several values of the variable $\lambda$ and
quantity $P_{c}=\int^{1}_{0}c(z)\text{d}z$ describing a probability
for IC in the nucleon. Dash-dotted curves describe the
$\text{GF}^{\text{(LO)}}$+$\text{QS}^{\text{(LO)}}$ contributions
with $P_{c}=5\%,~1\%,~0.1\%$ and $0.01\%$. Solid lines correspond to
the case when $P_{c}=0$. Comparing with each other Figs.~\ref{Fg.6}
and \ref{Fg.6a}, one can see that even a $0.1\%$ contribution of the
nonperturbative IC to the proton wave function could be extracted
from the $\cos 2\varphi $ asymmetry at large enough Bjorken $x$.
\begin{figure}
\begin{center}
\begin{tabular}{cc}
\mbox{\epsfig{file=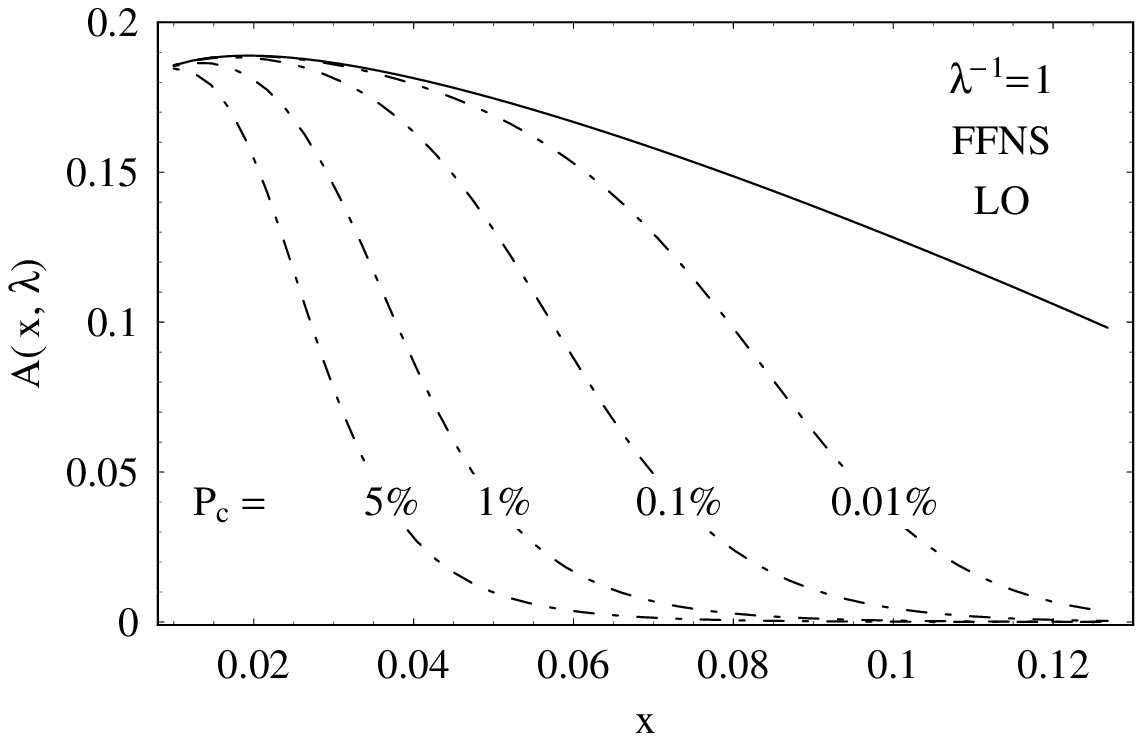,width=220pt}}
& \mbox{\epsfig{file=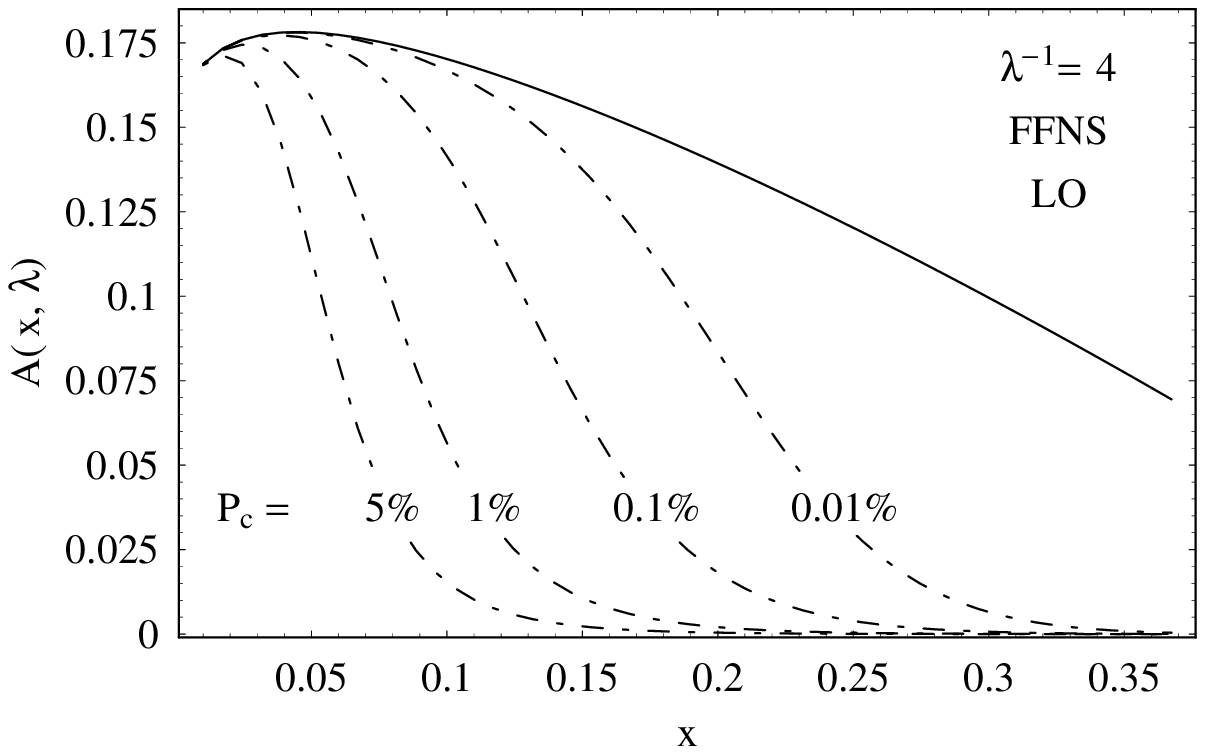,width=220pt}}\\
\mbox{\epsfig{file=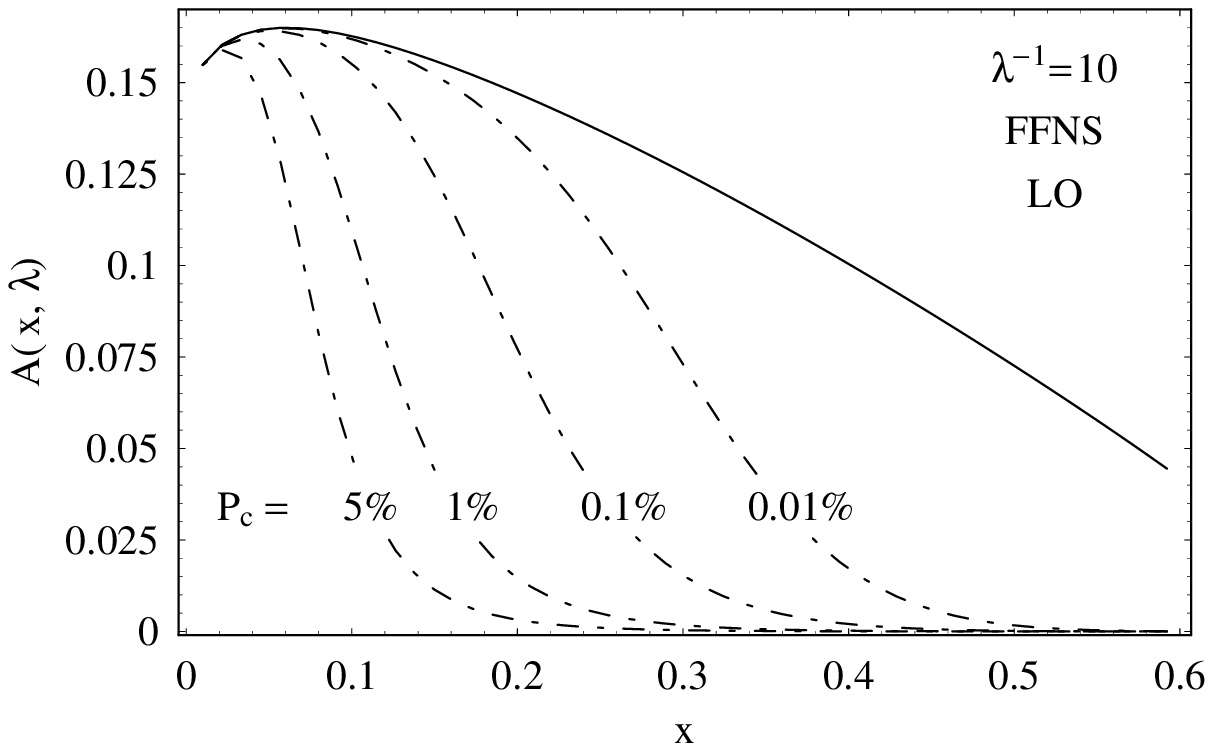,width=220pt}}
& \mbox{\epsfig{file=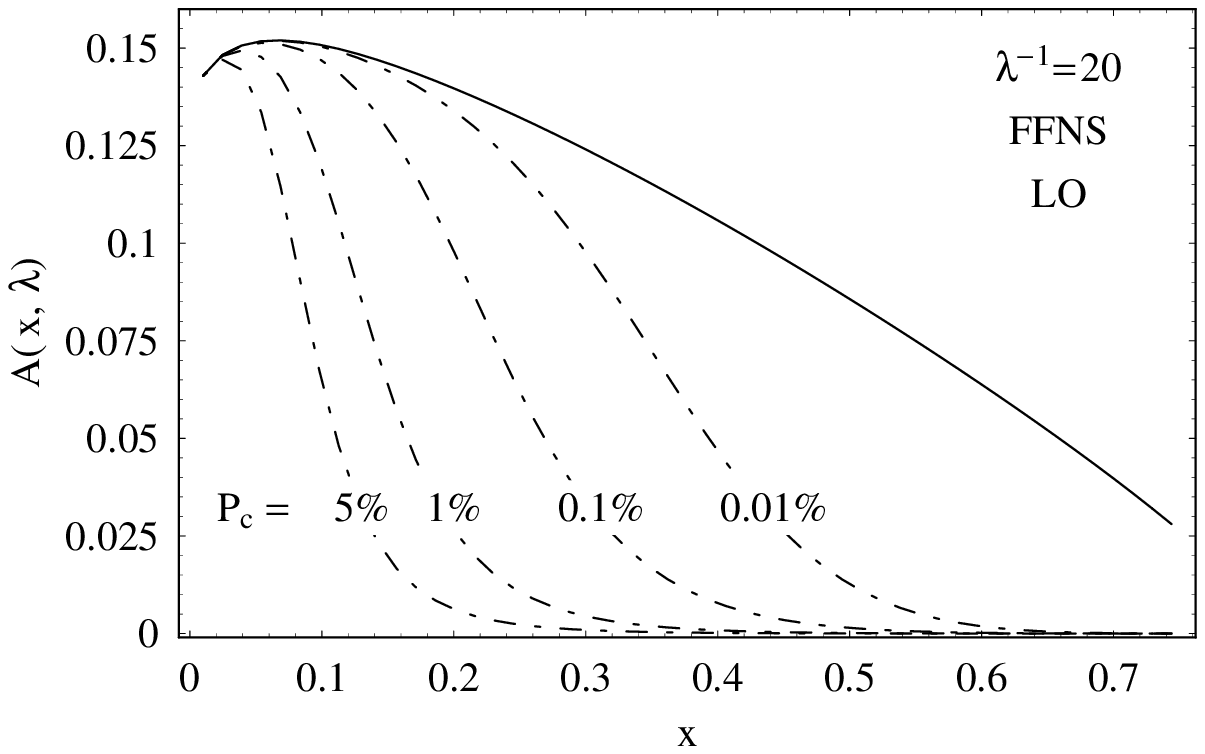,width=220pt}}\\
\mbox{\epsfig{file=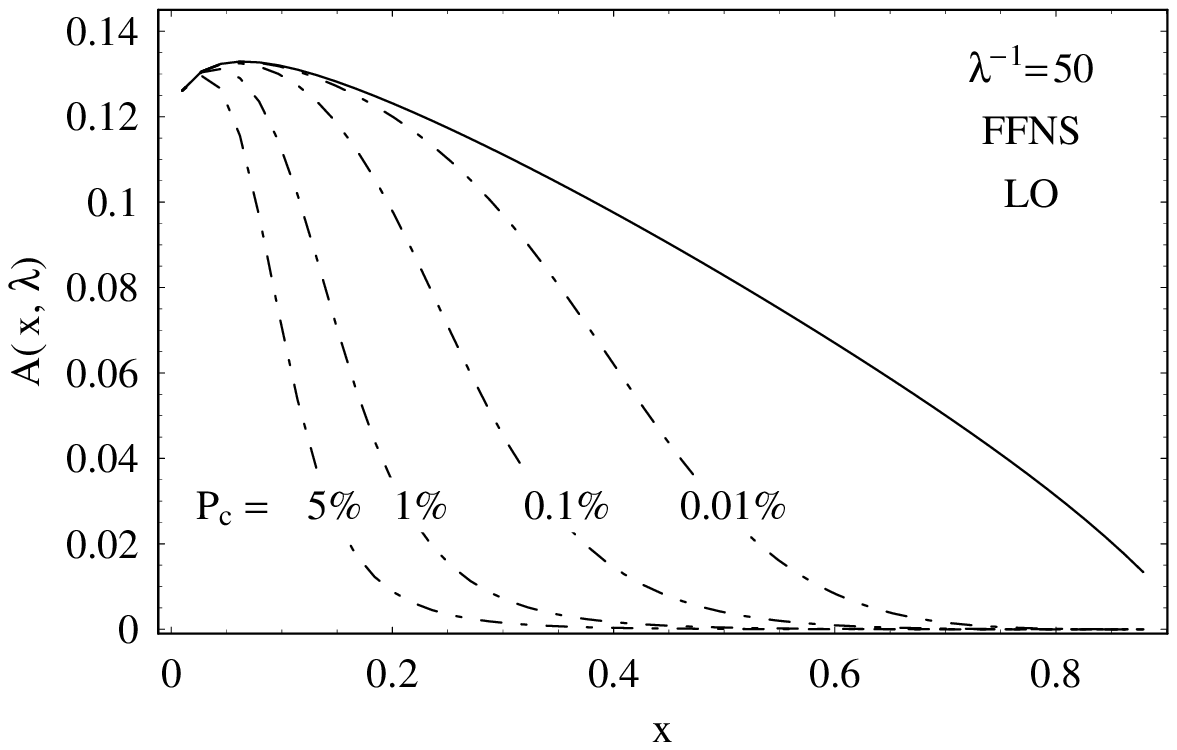,width=220pt}}
& \mbox{\epsfig{file=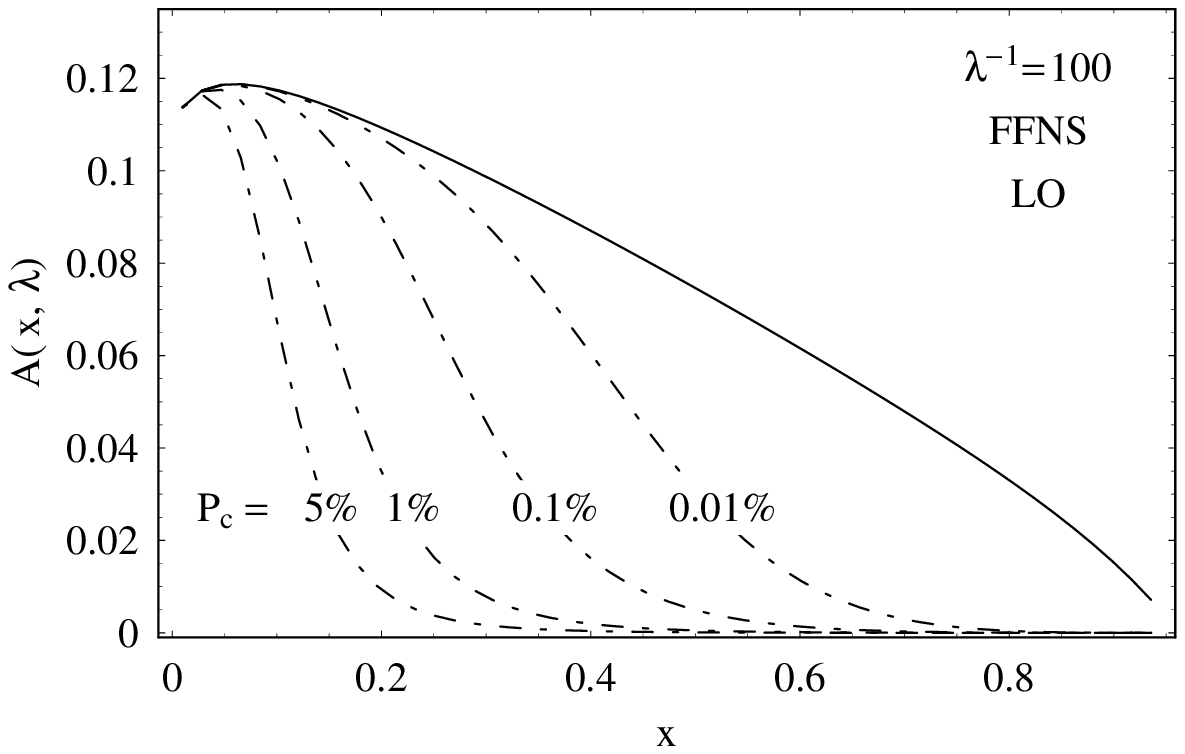,width=220pt}}\\
\end{tabular}
\caption[The LO predictions for $A(x,\lambda)$ in the FFNS at
several values of $\lambda$ and $P_{c}$.]{\label{Fg.6a}\small The LO
predictions for $A(x,\lambda)$ in the FFNS at several values of
$\lambda$ and $P_{c}=\int^{1}_{0}c(z)\text{d}z$. Dash-dotted curves
describe the $\text{GF}^{\text{(LO)}}$+$\text{QS}^{\text{(LO)}}$
contributions with $P_{c}=5\%,~1\%,~0.1\%$ and $0.01\%$. Solid lines
correspond to the case when $P_{c}=0$.}
\end{center}
\end{figure}

\subsection{\label{hb} Variable Flavor Number Scheme and Perturbative Intrinsic Charm}

One can see from Eqs.~(\ref{39}) that the GF cross section
$\hat{\sigma}_{2,g}^{(0)}(z,\lambda)$ contains potentially large
logarithm, $\ln (Q^{2}/m^{2})$. The same situation takes also place
for the QS cross section $\hat{\sigma}_{2,Q}^{(1)}(z,\lambda)$ given
by Eq.~(\ref{25}). At high energies, $Q^{2}\rightarrow \infty$, the
terms of the form $\alpha_{s}\ln (Q^{2}/m^{2})$ dominate the
production cross sections. To improve the convergence of the
perturbative series at high energies, the so-called variable flavor
number schemes (VFNS) have been proposed. Originally, this approach
was formulated by Aivazis, Collins, Olness and Tung (ACOT)
\cite{AOT,ACOT}.

In the VFNS, the mass logarithms of the type $\alpha_s^n\ln^n
(Q^{2}/m^{2})$ are resummed via the renormalization group equations.
In practice, the resummation procedure consists of two steps. First,
the mass logarithms have to be subtracted from the fixed order
predictions for the partonic cross sections in such a way that in
the asymptotic limit $Q^{2}\rightarrow \infty$ the well known
massless $\overline{\text{MS}}$ coefficient functions are recovered.
Instead, a charm parton density in the hadron, $c(x,Q^{2})$, has to
be introduced. This density obeys the usual massless NLO DGLAP
evolution equation with the boundary condition
$c(x,Q^{2}=Q_{0}^2)=0$ where $Q_{0}^2\sim m^{2}$. So, we may say
that, within the VFNS, the charm density arises perturbatively from
the $g\rightarrow c\bar{c}$ evolution.

In the VFNS, the treatment of the charm depends on the values chosen
for $Q^{2}$. At low $Q^{2}<Q_{0}^2$, the production cross sections
are described by the light parton contributions ($u,d,s$ and $g$).
The charm production is dominated by the GF process and its higher
order QCD corrections. At high $Q^{2}\gg m^{2}$, the charm is
treated in the same way as the other light quarks and it is
represented by a charm parton density in the hadron, which evolves
in $Q^{2}$. In the intermediate scale region, $Q^{2}\sim m^{2}$, one
has to make a smooth connection between the two different
prescriptions.

Strictly speaking, the perturbative charm density is well defined at
high $Q^2\gg m^2$ but does not have a clean interpretation at low
$Q^2$. Since the perturbative IC originates from resummation of the
mass logarithms of the type $\alpha_s^n\ln^n (Q^{2}/m^{2})$, it is
usually assumed that the corresponding PDF vanishes with these
logarithms, i.e. for $Q^{2}<Q_{0}^2\approx m^{2}$. On the other
hand, the threshold constraint $W^2=(q+p)^2=Q^2(1/x-1)>4m^2$ implies
that $Q_0$ is not a constant but "live" function of $x$. To avoid
this problem, several solutions have been proposed (see e.g.
Refs.~\cite{chi,SACOT}). In this paper, we use the so-called
ACOT($\chi$) prescription \cite{chi} which guarantees (at least at
$Q^2>m^2$) the correct threshold behavior of the
heavy-quark-initiated contributions.

Within the VFNS, the $\varphi$-independent charm production cross
sections have three pieces:
\begin{equation}\label{52}
\sigma_{2}(x,\lambda)=\sigma_{2,GF}(x,\lambda)-\sigma_{2,SUB}(x,\lambda)+\sigma_{2,QS}(x,\lambda),
\end{equation}
where the first and third terms on the right hand side describe the
usual (unsubtracted) GF and QS contributions while the second
(subtraction) term renders the total result infra-red safe in the
limit $m^{2}\rightarrow 0$. The only constraint imposed on the
subtraction term is to reproduce at high energies the familiar
$\overline{\text{MS}}$ partonic cross section:
\begin{equation}\label{53}
\lim_{\lambda\rightarrow 0}\left[\hat{\sigma}_{2,g}(z,\lambda)-
\hat{\sigma}_{2,SUB}(z,\lambda)\right]=\hat{\sigma}^{\overline{\text{MS}}}_{2,g}(z).
\end{equation}
Evidently, there is some freedom in the choice of finite mass terms
of the form $\lambda^{n}$ (with a positive $n$) in
$\hat{\sigma}_{2,SUB}(z,\lambda)$. For this reason, several
prescriptions have been proposed to fix the subtraction term. As
mentioned above, we use the so-called ACOT($\chi$) scheme
\cite{chi}.

According to the ACOT($\chi$) prescription, the lowest order
$\varphi$-independent cross section is
\begin{eqnarray}\label{54}
\sigma^{(LO)}_{2}(x,\lambda)=\int\limits_{\chi}^{1}\text{d}z\,g(z,\mu_{F})\left[\hat{\sigma}_{2,g}^{(0)}
\!\left(x/z,\lambda\right)-\frac{\alpha_{s}}{\pi}\ln\frac{\mu_{F}^{2}}{m^{2}}
\;\hat{\sigma}_{B}\left(x/z\right)P^{(0)}_{g\rightarrow
c}\left(\chi/z\right)\right]\\ \nonumber
+\hat{\sigma}_{B}(x)c_{+}(\chi,\mu_{F}),
\end{eqnarray}
where $P^{(0)}_{g\rightarrow c}$ is the LO gluon-quark splitting
function, $P^{(0)}_{g\rightarrow
c}(\zeta)=\left.\left[(1-\zeta)^{2}+\zeta^{2}\right]\right/2$, and
the LO GF cross section $\hat{\sigma}_{2,g}^{(0)}$ is given by
Eqs.~(\ref{39}). Remember also that $\chi=x(1+4\lambda)$ and
$c_{+}(\zeta,\mu_{F})=c(\zeta,\mu_{F})+\bar{c}(\zeta,\mu_{F})$.

The asymptotic behavior of the subtraction terms is fixed by the
parton level factorization theorem. This theorem implies that the
partonic cross sections d$\hat{\sigma}$ can be factorized into
process-dependent infra-red safe hard scattering cross sections
d$\tilde{\sigma}$, which are finite in the limit $m\rightarrow 0$,
and universal (process-independent) partonic PDFs $f_{a\rightarrow
i}$ and fragmentation functions $d_{n\rightarrow Q}$:
\begin{equation}\label{add1}
\text{d}\hat{\sigma}(\gamma^{*}+a\rightarrow
Q+X)=\sum_{i,n}f_{a\rightarrow
i}(\zeta)\otimes\text{d}\tilde{\sigma}(\gamma^{*}+i\rightarrow
n+X)\otimes d_{n\rightarrow Q}(z).
\end{equation}
In Eq.~(\ref{add1}), the symbol $\otimes$ denotes the usual
convolution integral, the indices $a,i,n$ and $Q$ denote partons,
$p_{i}=\zeta p_{a}$ and $p_{Q}=z p_{n}$. All the logarithms of the
heavy-quark mass (i.e., the singularities in the limit $m\rightarrow
0$) are contained in the PDFs $f_{a\rightarrow i}$ and fragmentation
functions $d_{n\rightarrow Q}$ while d$\tilde{\sigma}$ are IR-safe
(i.e., are free of the $\ln m^{2}$ terms). The expansion of
Eq.~(\ref{add1}) can be used to determine order by order the
subtraction terms. In particular, for the LO GF contribution to the
charm leptoproduction one finds \cite{ACOT}
\begin{equation}\label{add2}
\hat{\sigma}^{(0)}_{k,SUB}\left(z,\ln\,(\mu_{F}^{2}/m^{2})\right)=f^{(1)}_{g\rightarrow
c}\left(\zeta,\ln\,(\mu_{F}^{2}/m^{2})\right)\otimes\hat{\sigma}^{(0)}_{k,QS}(z/\zeta),
\qquad (k=2,L,A,I),
\end{equation}
where $f^{(1)}_{g\rightarrow
c}\left(\zeta,\ln\,(\mu_{F}^{2}/m^{2})\right)=\left(
\alpha_{s}/2\pi\right)\ln\,(\mu_{F}^{2}/m^{2})\,P^{(0)}_{g\rightarrow
c}\left(\zeta\right)$ describes the charm distribution in the gluon
within the $\overline{\text{MS}}$ factorization scheme.

One can see from Eq.~(\ref{add2}) that the azimuth-dependent GF
cross sections $\hat{\sigma}_{A,GF}$ and $\hat{\sigma}_{I,GF}$ don't
have subtraction terms at LO because the lowest order QS
contribution is $\varphi$-independent. For this reason, the
$\cos2\varphi$-dependence within the VFNS has the same form as in
the FFNS:
\begin{equation}\label{55}
\sigma^{(LO)}_{A}(x,\lambda)=\int\limits_{\chi}^{1}\text{d}z\,g(z,\mu_{F})\,\hat{\sigma}_{A,g}^{(0)}
\!\left(x/z,\lambda\right).
\end{equation}
\begin{figure}
\begin{center}
\begin{tabular}{cc}
\mbox{\epsfig{file=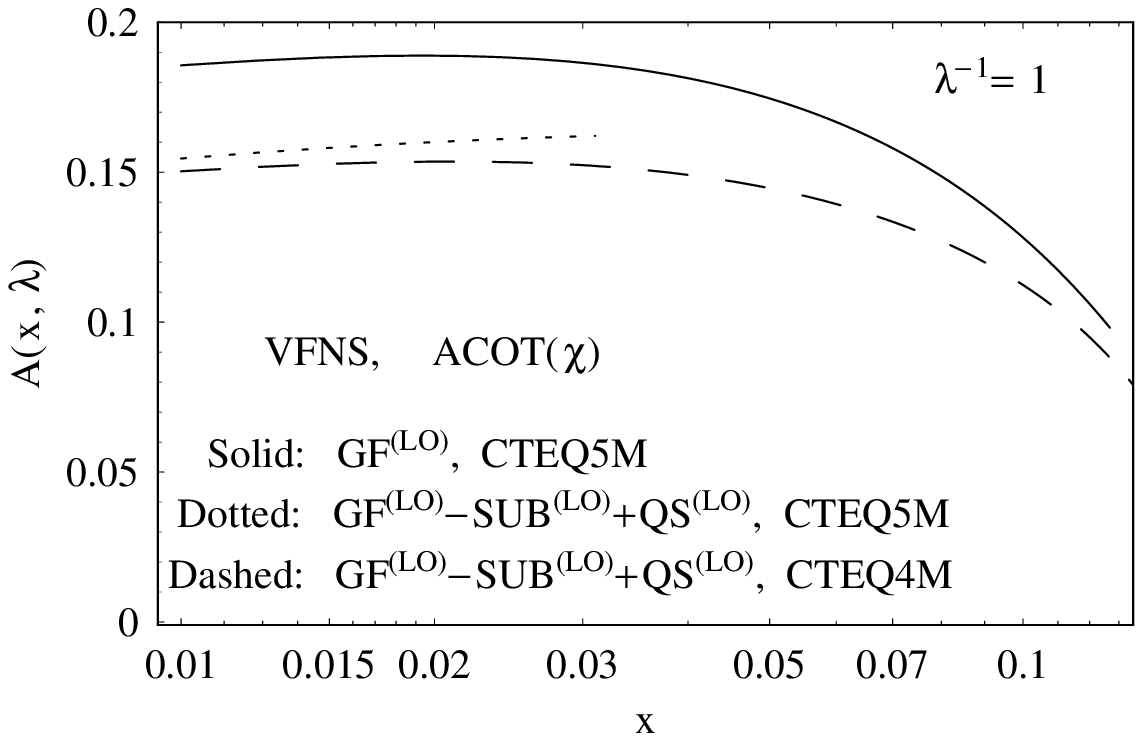,width=220pt}}
& \mbox{\epsfig{file=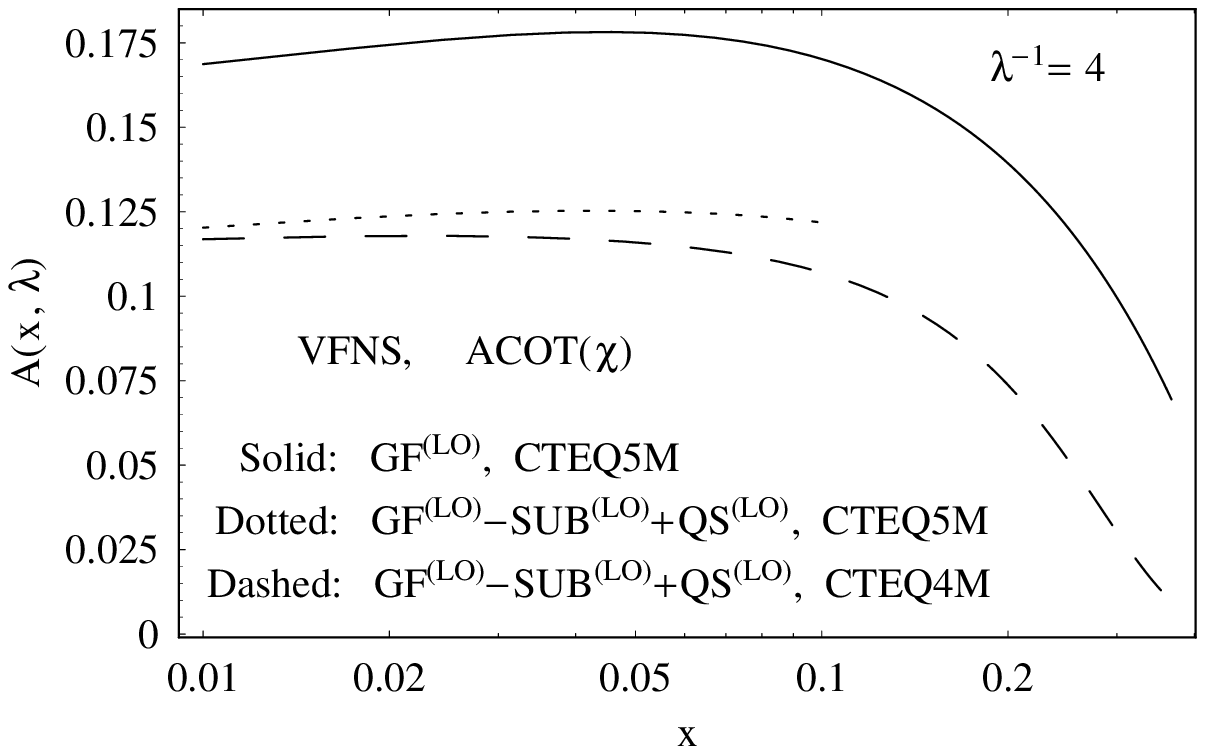,width=220pt}}\\
\mbox{\epsfig{file=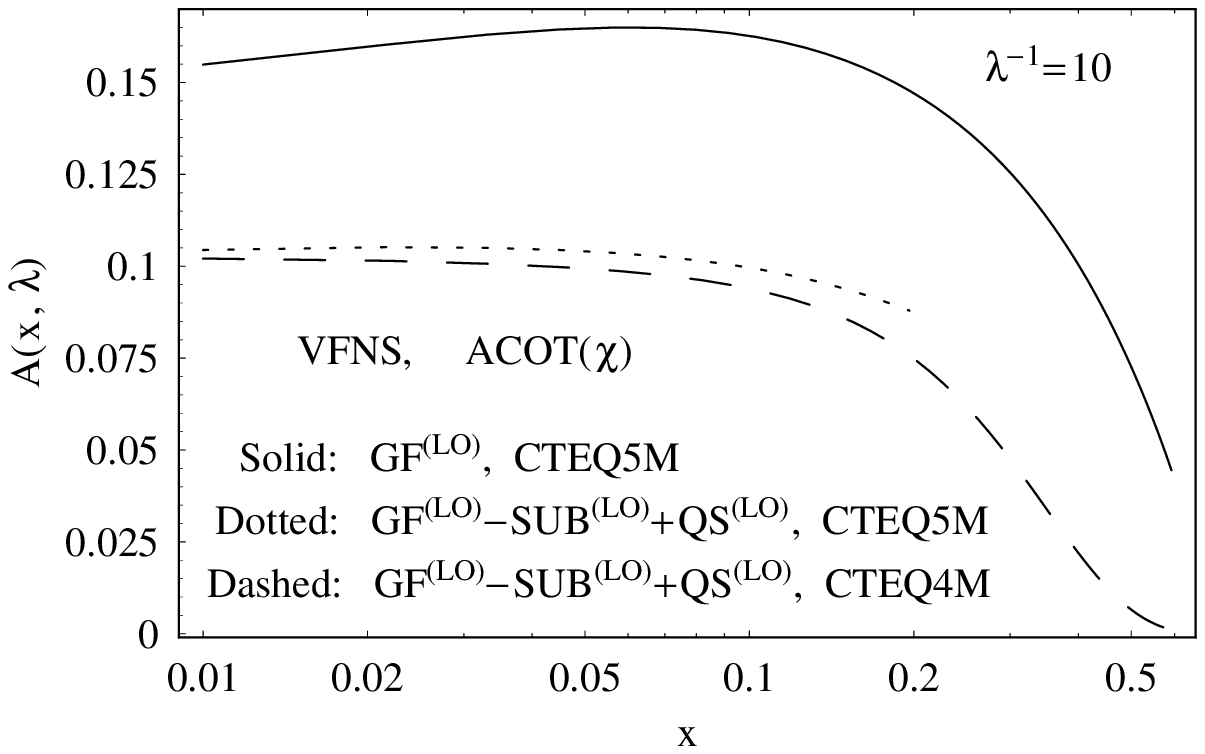,width=220pt}}
& \mbox{\epsfig{file=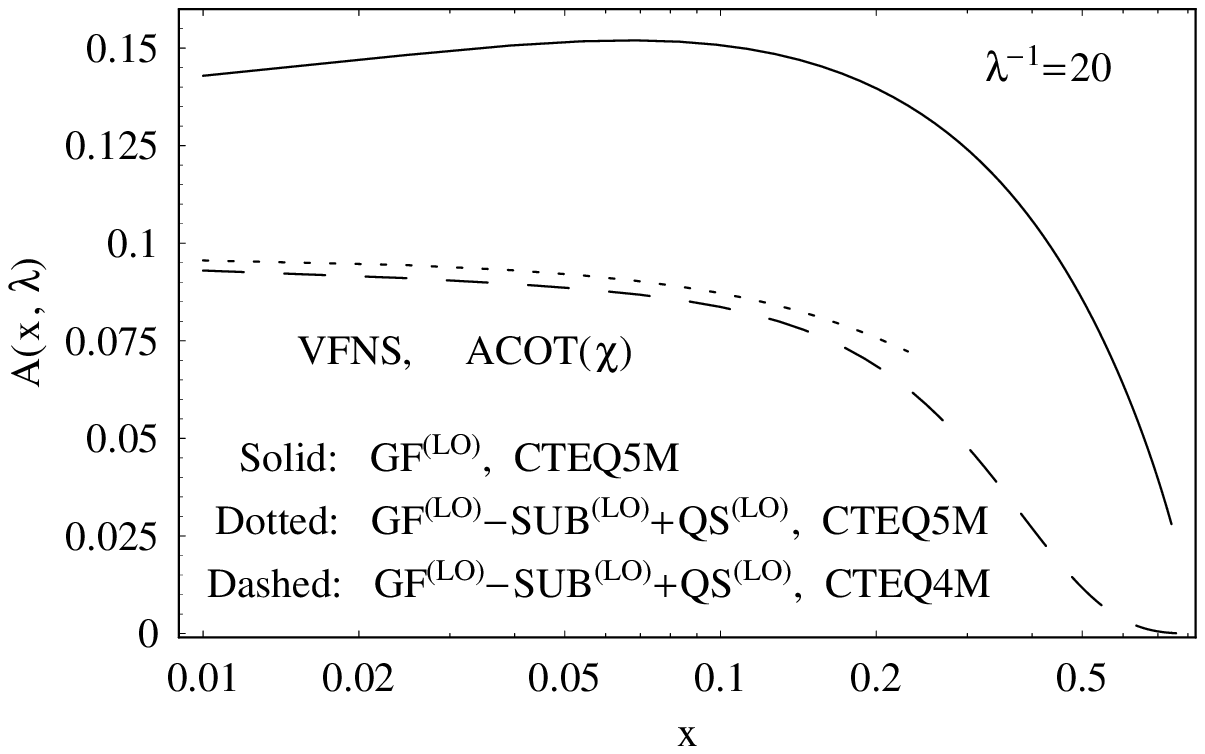,width=220pt}}\\
\mbox{\epsfig{file=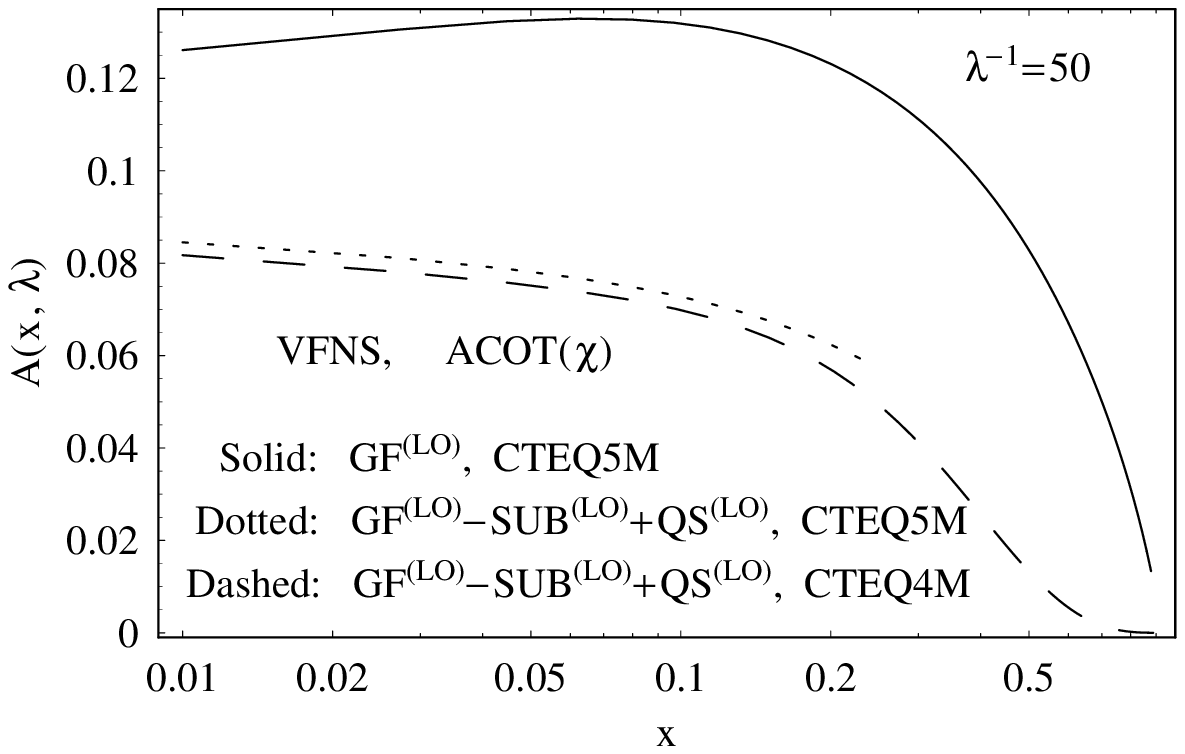,width=220pt}}
& \mbox{\epsfig{file=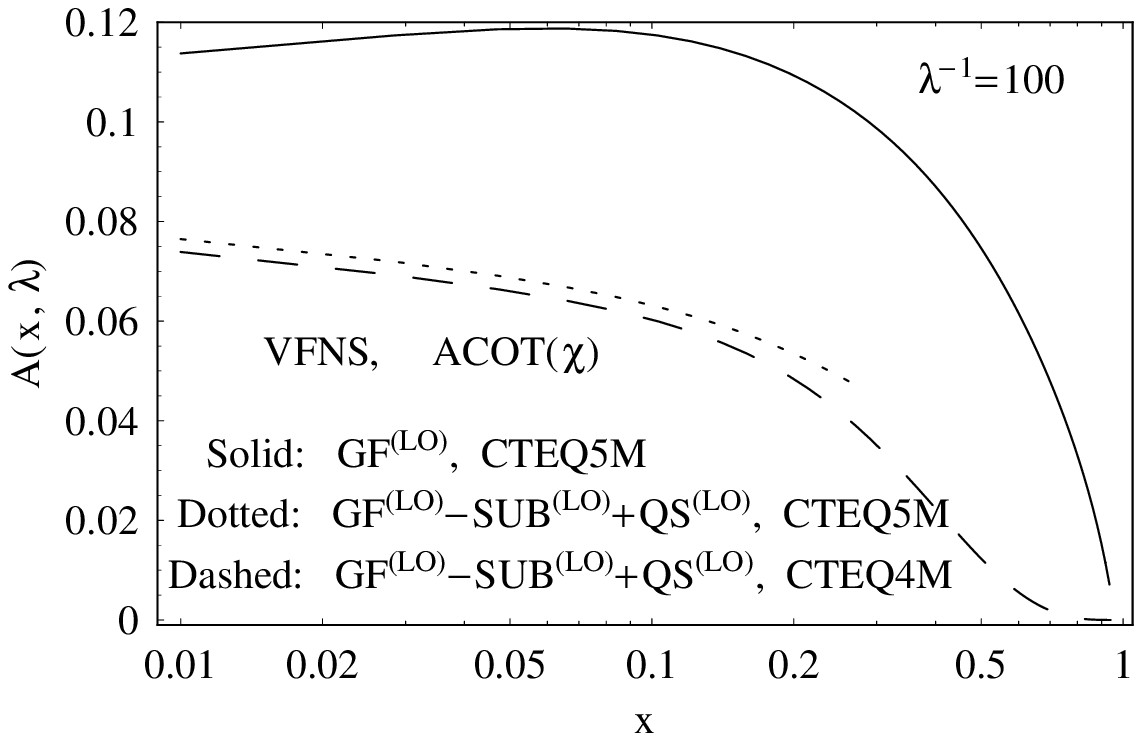,width=220pt}}\\
\end{tabular}
\caption[Azimuthal asymmetry parameter $A(x,\lambda)$ in the VFNS at
several values of $\lambda$.]{\label{Fg.7}\small Azimuthal asymmetry
parameter $A(x,\lambda)$ in the VFNS at several values of $\lambda$.
The following contributions are plotted: $\text{GF}^{\text{(LO)}}$
(solid curves),
$\text{GF}^{\text{(LO)}}$$-\text{SUB}^{\text{(LO)}}$+$\text{QS}^{\text{(LO)}}$
with the CTEQ5M set of PDFs (dotted curves) and
$\text{GF}^{\text{(LO)}}$$-\text{SUB}^{\text{(LO)}}$+
$\text{QS}^{\text{(LO)}}$ with the CTEQ4M set of PDFs (dashed
curves).}
\end{center}
\end{figure}

Fig.~\ref{Fg.7} shows the ACOT($\chi$) predictions for the asymmetry
parameter $A(x,\lambda)$ at several values of variable $\lambda$:
$\lambda^{-1}=1,4,10,20,50$ and 100. For comparison, we plot also
the LO GF predictions (solid curves). In the ACOT($\chi$) case, we
consider the CTEQ5M (dotted lines) and CTEQ4M (dashed curves)
parametrizations of the gluon and charm densities in the proton.
Corresponding values of the charm quark mass are $m_{c}=1.3 $ GeV
\cite{CTEQ5} (for the CTEQ5M PDFs) and $m_{c}=1.6 $ GeV \cite{CTEQ4}
(for the CTEQ4M PDFs). The default value of the factorization scale
is $\mu_{F}=\sqrt{m^{2}+Q^{2}}$.

One can see from Fig.~\ref{Fg.7} the following properties of the
azimuthal asymmetry, $A(x,\lambda)$, within the VFNS. Contrary to
the nonperturbative IC component, the perturbative one is
significant practically at all values of Bjorken $x$ and
$Q^{2}>m^{2}$. The perturbative charm contribution leads to a
sizeable decreasing of the GF predictions for the
$\cos2\varphi$-asymmetry. In the ACOT($\chi$) scheme, the IC
contribution reduces the GF results for $A(x,\lambda)$ by about
$30\%$. The origin of this reduction is straightforward: the QS
component is practically $\cos2\varphi$-independent.

The ACOT($\chi$) predictions for the asymmetry depend weakly on the
parton distribution functions we use. It is seen from
Fig.~\ref{Fg.7} that the CTEQ5M and CTEQ4M sets of PDFs lead to very
similar results  for $A(x,\lambda)$. Note that we give the CTEQ5M
predictions at low $x$ only because of irregularities in the CTEQ5M
charm density at large $x$.

We have also analyzed how the VFNS predictions depend on the choice
of subtraction prescription. In particular, the schemes proposed in
Refs.~\cite{KS,SACOT} have been considered. We find that,
sufficiently above the production threshold, these subtraction
prescriptions reduce the GF results for the asymmetry by
approximately $30\div 50 \%$.

One can conclude that impact of the perturbative IC on the
$\cos2\varphi$ asymmetry is essential in the whole region of Bjorken
$x$ and therefore can be tested experimentally.

\section{\label{virt}Appendix A: Virtual and Soft Contributions to
the Quark Scattering} In this Appendix we reproduce some results of
Hoffmann and Moore for the $\varphi$-independent QS cross sections,
and correct two misprints uncovered in Ref.~\cite{HM}. We work in
four dimensions, in the Feynman gauge and use the on-mass-shell
renormalization scheme. We compute the absorptive part of the
Feynman diagram (which is free of the UV divergences) and then
restore the real part using the appropriate dispersion relations.

In the on-mass-shell scheme, the renormalized fermion self-energy
vanishes like $(\hat{p}_{Q}-m)^{2}$ which means that the second and
third diagrams in Fig.~\ref{Fg.2}c do not contribute to the cross
section when the external quark legs are on-shell,
$\hat{p}_{Q}\rightarrow m$. The first graph in Fig.~\ref{Fg.2}c
describes the NLO corrections to the quark-photon vertex function:
\begin{equation}  \label{71}
\Lambda _{\mu }(q) =f\left( Q^{2}\right) \gamma _{\mu }-
\frac{g\left( Q^{2}\right) }{2m}\sigma _{\mu \nu }q^{\nu },
\end{equation}
where $\sigma _{\mu \nu }=\frac{1}{2}\left( \gamma _{\mu }\gamma
_{\nu }-\gamma _{\nu }\gamma _{\mu }\right) $ while $f\left(
Q^{2}\right) $ and $g\left( Q^{2}\right)$ are the quark
electromagnetic formfactors. At the lowest order $\Lambda _{\mu
}^{(0)}=\gamma _{\mu }$.

The virtual lepton-quark cross section, $\hat{\sigma}_{lQ}^{V}$, is
obtained from the interference term between the virtual and the Born
amplitude. The result can be written in terms of the electromagnetic
formfactors as:
\begin{equation} \label{72}
\frac{\text{d}^{\text{2}}\hat{\sigma}_{lQ}^{V}}{\text{d}z\text{d}Q^{2}}=
\frac{\alpha_{em}}{\pi }\frac{\hat{\sigma}_{B}(z)}{zQ^{2}}\delta
(1-z)\left\{ \left[ 1+(1-y)^{2}-2\lambda z^{2}y^{2}\right] f\left(
Q^{2}\right) +y^{2}g\left( Q^{2}\right) \right\}.
\end{equation}
Taking into account the definition of the HM cross sections,
$\sigma^{(2)}$ and $\sigma^{(L)}$, given by Eqs.~(\ref{35}),
(\ref{36}) and (\ref{18}), we find that corresponding virtual parts
are:
\begin{equation}\label{73}
\sigma_{1V}^{(2)}(z,Q^{2})=2\delta (1-z)f^{(1)}\left( Q^{2}\right),
\qquad \qquad \qquad \sigma _{1V}^{(L)}(z,Q^{2})=-\delta
(1-z)g^{(1)}\left( Q^{2}\right),
\end{equation}
where $f^{(1)}\left( Q^{2}\right)$ and $g^{(1)}\left( Q^{2}\right)$
are the NLO corrections to the electromagnetic formfactors. For the
NLO HM  cross sections, $\sigma_{1V}^{(2)}$ and $\sigma_{1V}^{(L)}$,
we use exactly the same notations as in Ref.~\cite{HM}.

In the on-mass-shell renormalization scheme, the renormalized vertex
correction vanishes as the photon virtuality goes to zero,
$f^{(1)}\left(0\right)=0$. This is a consequence of the Ward
identity and the fact that the real photon field (like the massive
fermion one) is unrenormalized in first order QCD. To satisfy the
condition $f^{(1)}\left(0\right)=0$ automatically, we should use for
$f^{(1)}\left(q^{2}\right)$ the dispersion relation with one
subtraction. The second formfactor, $g^{(1)}\left(q^{2}\right)$, has
no singularities. For this reason, we use for
$g^{(1)}\left(q^{2}\right)$ the dispersion relation without
subtractions:
\begin{equation}\label{74}
f^{(1)}\left( q^{2}\right) =\frac{q^{2}}{\pi
}\int\limits_{4m^{2}}^{\infty } \frac{\text{d}t\;\,\text{Im
}f^{(1)}(t)}{t(t-q^{2}-i0)}, \qquad \qquad g^{(1)}\left(
q^{2}\right) =\frac{1}{\pi }\int\limits_{4m^{2}}^{\infty }
\frac{\text{d}t\;\,\text{Im }g^{(1)}(t)}{t-q^{2}-i0}.
\end{equation}
Calculating the imaginary parts of the formafactors and restoring
their real parts with the help of Eqs.~(\ref{74}) yields
\begin{eqnarray}\label{75}
f^{(1)}\left( Q^{2}\right)&=&\frac{\alpha _{s}}{\pi }C_{F}\biggl\{
\left[ 1+ \frac{1+2\lambda
}{\sqrt{1+4\lambda }}\ln r\right]\left(\ln \frac{m}{m_{g}}-1\right) \\
&+&\frac{1+2\lambda}{\sqrt{1+4\lambda }}\left[ \text{Li}_{2}(-r)+
\frac{\pi ^{2}}{12}+ \frac{1}{4}\ln^{2}r+\frac{1}{2}\ln
r\ln\frac{1+4\lambda }{\lambda }\right]\nonumber
+\frac{1}{4}\frac{\ln r}{\sqrt{1+4\lambda }}\biggr\},\nonumber\\
g^{(1)}\left(
Q^{2}\right)&=&-\frac{\alpha_{s}}{\pi}C_{F}\frac{\lambda\ln
r}{\sqrt{1+4\lambda}}.\label{76}
\end{eqnarray}
In Eqs.~(\ref{75}) and (\ref{76}), $C_{F}=(N_{c}^{2}-1)/(2N_{c})$,
where $N_{c}$ is number of colors, and $r$ is defined by
Eq.~(\ref{29}). Taking into account that $\Lambda
^{(1)\mu}=\left(\alpha_{s}\left/\alpha_{em}\right.\right)C_{F}\,\Lambda_{QED}^{(1)\mu}$,
we see that Eqs.~(\ref{75}, \ref{76}) reproduce the textbook QED
results.

It is now straightforward to obtain the virtual contribution to the
longitudinal cross section. Combining Eqs.~(\ref{73}) and (\ref{76})
yields:
\begin{equation}\label{77}
\sigma _{1V}^{(L)}(z,Q^{2})=\frac{\alpha _{s}}{\pi }C_{F}\delta
(1-z)\frac{\lambda \ln r}{\sqrt{1+4\lambda }}.
\end{equation}
Comparing the above result with the corresponding one given by
Eq.~(39) in Ref.~\cite{HM}, we see that the HM expression for
$\sigma _{1V}^{(L)}$ has opposite sign. Note also that this typo
propagates into the final result for $\sigma _{1}^{(L)}$ given by
Eq.~(52) \cite{HM}.

Calculation of the bremsstrahlung contribution to the longitudinal
cross section, $\sigma _{1B}^{(L)}(z,Q^{2})$, is also
straightforward. We coincide with the HM result for $\sigma
_{1B}^{(L)}(z,Q^{2})$ given by Eq.~(49) in Ref.~\cite{HM}. However
there is one more misprint in the HM expression for $\sigma
_{1}^{(L)}$: the r.h.s of  Eq.~(52) \cite{HM} should be multiplied
by $z$.

In the case of $\sigma _{1}^{(2)}(z,Q^{2})$, the situation is
slightly more complicated due to the need to take into account the
IR singularities. One can see from Eq.~(\ref{75}) that
$f^{(1)}\left( Q^{2}\right)$ has an IR divergence which is
regularized with the help of an infinitesimal gluon mass $m_{g}$.
This singularity is cancelled when one adds the so-called soft
contribution originating from the real gluon emission. For this
purpose we introduce another infinitesimal parameter $\delta z$,
$\left(m_{g}\!\left/m\right.\right)\ll\delta z\ll 1$. The full
bremsstrahlung contribution, $\sigma _{1B}^{(2)}$, can then be
splitted into the soft and hard pieces as follows:
\begin{equation}\label{78}
\sigma _{1soft}^{(2)}(z,Q^{2})=\theta (z+\delta z-1)\sigma
_{1B}^{(2)}(z,Q^{2}), \qquad \sigma
_{1hard}^{(2)}(z,Q^{2})=\theta (1-z-\delta z)\sigma
_{1B}^{(2)}(z,Q^{2}),
\end{equation}
where $\theta (1-z-\delta z)$ is the Heaviside step function. The
soft cross section should be calculated in the eikonal
approximation, $\vec{p_{g}}\rightarrow 0$, taking into account the
infinitesimal gluon mass $m_{g}$. As a result, the sum of the
virtual and soft contributions is IR finite:
\begin{eqnarray}
\sigma _{1V}^{(2)}+\sigma _{1soft}^{(2)} =\frac{\alpha _{s}}{\pi }
C_{F}\delta (1-z)\biggl\{-2\ln (\delta z)\left[ 1+\frac{1+2\lambda
}{\sqrt{ 1+4\lambda }}\ln r\right] +2\ln \lambda
-1-\frac{\sqrt{1+4\lambda }}{2}\ln r \\\nonumber
+\frac{1+2\lambda}{\sqrt{1+4\lambda}}\Bigl[\text{Li}_{2}(r^{2})+2\text{Li}_{2}(-r)+\frac{3}{2}
\ln^{2}r -2\ln r-\ln r\ln\lambda+2\ln r\ln (1+4\lambda)\Bigr]
\biggr\}. \label{79}
\end{eqnarray}
Adding to the above expression the hard cross section $\sigma
_{1hard}^{(2)}$ defined by Eq.~(\ref{78}), we reproduce  in the
limit $\delta z\rightarrow 0$ the full result for $\sigma
_{1}^{(2)}$ given by Eq.~(51) in Ref.~\cite{HM}.

\section{\label{soft}Appendix B: NLO Soft-Gluon Corrections to the Photon-Gluon Fusion}
This Appendix provides an overview of the NLO soft-gluon
approximation for the photon-gluon fusion mechanism. We present the
final results for the parton level cross sections to the
next-to-leading logarithmic (NLL) accuracy. More details can be
found in Refs.~\cite{Laenen-Moch,we2,we4}.

To take into account the NLO contributions to the GF mechanism, one
needs to calculate the virtual ${\cal O}(\alpha _{em}\alpha
_{s}^{2})$ corrections to the Born process (\ref{38}) and the real
gluon emission:
\begin{equation}  \label{60}
\gamma ^{*}(q)+g(k_{g})\rightarrow
Q(p_{Q})+\overline{Q}(p_{\stackrel{\_}{Q}})+g(p_{g}).
\end{equation}
The partonic invariants describing the single-particle inclusive
(1PI) kinematics are
\begin{eqnarray}
s^{\prime }=2q\cdot k_{g}=s+Q^{2}=\zeta S^{\prime },\qquad
&&t_{1}=\left(
k_{g}-p_{Q}\right) ^{2}-m^{2}=\zeta T_{1},  \nonumber \\
s_{4}=s^{\prime }+t_{1}+u_{1},\qquad  &&u_{1}=\left(
q-p_{Q}\right) ^{2}-m^{2}=U_{1}, \label{61}
\end{eqnarray}
where $\zeta$ is defined by $\vec{k}_{g}= \zeta\vec{p}\,$ and
$s_{4}$ measures the inelasticity of the reaction (\ref{60}). The
corresponding 1PI hadron level variables describing the reaction
(\ref{1}) are
\begin{eqnarray}
S^{\prime }=2q\cdot p=S+Q^{2},\qquad \qquad &&T_{1}=\left(
p-p_{Q}\right)
^{2}-m^{2},  \nonumber \\
S_{4}=S^{\prime }+T_{1}+U_{1},\qquad \qquad &&U_{1}=\left(
q-p_{Q}\right) ^{2}-m^{2}.  \label{62}
\end{eqnarray}

The exact NLO calculations of the unpolarized heavy quark production
in $\gamma g$ \cite{Ellis-Nason,Smith-Neerven}, $\gamma ^{*}g$
\cite{LRSN}, and $gg$
\cite{Nason-D-E-1,Nason-D-E-2,Nason-D-E-3,BKNS} collisions show
that, near the partonic threshold, a strong logarithmic enhancement
of the cross sections takes place in the collinear, $\vec{p}_{g,T} $
$\rightarrow 0$, and soft, $\vec{p}_{g}\rightarrow 0$, limits. This
threshold (or soft-gluon) enhancement has universal nature in the
perturbation theory and originates from incomplete cancellation of
the soft and collinear singularities between the loop and the
bremsstrahlung contributions. Large leading and next-to-leading
threshold logarithms can be resummed to all orders of perturbative
expansion using the appropriate evolution equations
\cite{Contopanagos-L-S,Laenen-O-S,Kidonakis-O-S}. The analytic
results for the resummed cross sections are ill-defined due to the
Landau pole in the coupling strength $\alpha _{s}$. However, if one
considers the obtained expressions as generating functionals of the
perturbative theory and re-expands them at fixed order in $\alpha
_{s}$, no divergences associated with the Landau pole are
encountered.

Soft-gluon resummation for the photon-gluon fusion has been
performed in Ref.~\cite{Laenen-Moch} and checked in
Refs.~\cite{we2,we4}. To NLL accuracy, the perturbative expansion
for the partonic cross sections,
d$^{2}\hat{\sigma}_{k,g}/$d$t_{1}$d$u_{1}$ ($k=T,L,A,I$), can be
written in a factorized form as
\begin{eqnarray}  \label{63}
s^{\prime
2}\frac{\text{d}^{2}\hat{\sigma}_{k,g}}{\text{d}t_{1}\text{d}u_{1}}(
s^{\prime },t_{1},u_{1}) =B_{k,g}^{\text{{\rm Born}}}( s^{\prime
},t_{1},u_{1})\Bigl\{\delta (s^{\prime }+t_{1}+u_{1})
+\sum_{n=1}^{\infty }( \frac{\alpha _{s}C_{A}}{\pi})^{n}K^{(n)}(
s^{\prime },t_{1},u_{1})\Bigr\} ,
\end{eqnarray}
with the Born level distributions $B_{k,g}^{\text{{\rm Born}}}$
given by
\begin{eqnarray}
B_{T,g}^{\text{{\rm Born}}}( s^{\prime },t_{1},u_{1}) &=&\pi
e_{Q}^{2}\alpha _{em}\alpha _{s}\left[
\frac{t_{1}}{u_{1}}+\frac{u_{1}}{t_{1} }+4\left( \frac{s}{s^{\prime
}}-\frac{m^{2}s^{\prime }}{t_{1}u_{1}}\right) \left( \frac{s^{\prime
}(m^{2}-Q^{2}/2)}{t_{1}u_{1}}+\frac{Q^{2}}{s^{\prime}
}\right) \right] ,  \\
B_{L,g}^{\text{{\rm Born}}}( s^{\prime },t_{1},u_{1}) &=&\pi
e_{Q}^{2}\alpha _{em}\alpha _{s}\left[ \frac{8Q^{2}}{s^{\prime
}}\left( \frac{s}{s^{\prime }}-\frac{m^{2}s^{\prime
}}{t_{1}u_{1}}\right) \right] ,
 \\
B_{A,g}^{\text{{\rm Born}}}( s^{\prime },t_{1},u_{1}) &=&\pi
e_{Q}^{2}\alpha _{em}\alpha _{s}\left[ 4\left( \frac{s}{s^{\prime
}}-\frac{ m^{2}s^{\prime }}{t_{1}u_{1}}\right) \left(
\frac{m^{2}s^{\prime }}{
t_{1}u_{1}}+\frac{Q^{2}}{s^{\prime }}\right) \right] ,  \label{64} \\
B_{I,g}^{\text{{\rm Born}}}( s^{\prime },t_{1},u_{1}) &=&\pi
e_{Q}^{2}\alpha _{em}\alpha _{s}\left[ 4\sqrt{Q^{2}}\left(
\frac{t_{1}u_{1}s }{s^{\prime 2}}-m^{2}\right)
^{1/2}\frac{t_{1}-u_{1}}{t_{1}u_{1}}\left( 1-\frac{2Q^{2}}{s^{\prime
}}-\frac{2m^{2}s^{\prime }}{t_{1}u_{1}}\right) \right] .
\end{eqnarray}

Note that the functions $K^{(n)}( s^{\prime },t_{1},u_{1}) $ in
Eq.~(\ref{63}) originate from the collinear and soft limits.
Radiation of soft and collinear gluons does not affect the
transverse momentum of detected particles and therefore the
azimuthal angle $\varphi$. For this reason, the functions $
K^{(n)}(s^{\prime },t_{1},u_{1}) $ are the same for all helicity
cross sections $\hat{\sigma}_{k,g}$ ($k=T,L,A,I$). At NLO, the
soft-gluon corrections to NLL accuracy in the $\overline{\text{MS}}$
scheme are
\begin{eqnarray}
K^{(1)}( s^{\prime },t_{1},u_{1}) &=&-\left[ \frac{1}{s_{4}}\right]
_{+}\left\{ 1+\ln \left( \frac{u_{1}}{t_{1}}\right) -\left(
1-\frac{2C_{F}}{ C_{A}}\right) \left( 1+\text{Re}L_{\beta }\right)
+\ln \left( \frac{\mu ^{2} }{m^{2}}\right) \right\}   \\ \nonumber
&&+2\left[ \frac{\ln \left(s_{4}/m^{2}\right)
}{s_{4}}\right]_{+}+\delta ( s_{4}) \ln \left(
\frac{-u_{1}}{m^{2}}\right) \ln \left( \frac{\mu
^{2}}{m^{2}}\right), \label{65}
\end{eqnarray}
where we use $\mu =\mu _{F}=\mu_{R}$. In Eq.~(\ref{65}),
$C_{A}=N_{c}$ and $ C_{F}=(N_{c}^{2}-1)/(2N_{c})$, where $N_{c}$ is
number of colors, while $ L_{\beta }=(1-2m^{2}/s)\{\ln [(1-\beta
)/(1+\beta )]+$i$\pi\}$ with $\beta=\sqrt{1-4m^{2}/s}$. The
single-particle inclusive ''plus`` distributions are defined by
\begin{equation}  \label{66}
\left[\frac{\ln^{l}\left( s_{4}/m^{2}\right)
}{s_{4}}\right]_{+}=\lim_{\epsilon \rightarrow
0}\left\{\frac{\ln^{l}\left(s_{4}/m^{2}\right) }{s_{4}}\theta (
s_{4}-\epsilon)+\frac{1}{l+1}\ln ^{l+1}\left(\frac{\epsilon
}{m^{2}}\right) \delta ( s_{4})\right\}.
\end{equation}
For any sufficiently regular test function $h(s_{4})$,
Eq.~(\ref{66}) gives
\begin{eqnarray}\label{67}
\int\limits_{0}^{s_{4}^{\max }}\text{d}s_{4}\,h(s_{4})\left[
\frac{\ln ^{l}\left( s_{4}/m^{2}\right) }{s_{4}}\right]
_{+}=\int\limits_{0}^{s_{4}^{\max }}\text{d}s_{4}\left[
h(s_{4})-h(0)\right] \frac{\ln ^{l}\left( s_{4}/m^{2}\right)
}{s_{4}} +\frac{1}{l+1}h(0)\ln ^{l+1}\left( s_{4}^{\max
}/m^{2}\right) .
\end{eqnarray}

In Eq.~(\ref{65}), we have preserved the NLL terms for the
scale-dependent logarithms too. Note also that the results
(\ref{64}) and (\ref{65}) agree to NLL accuracy with the exact
${\cal O}(\alpha_{em}\alpha _{s}^{2})$ calculations of the
photon-gluon cross sections $\hat{\sigma}_{T,g}$ and
$\hat{\sigma}_{L,g}$ given in Ref.~\cite{LRSN}.

To investigate the scale dependence of the results
(\ref{63}$-$\ref{65}), it is convenient to introduce for the fully
inclusive (integrated over $t_{1}$ and $u_{1}$) cross sections,
$\hat{\sigma}_{k,g}$ ($k=T,L,A,I$),{\large \ }the dimensionless
coefficient functions $c_{k,g}^{(n,l)}$ defined by Eq.~(\ref{32}).
Concerning the NLO scale-independent coefficient functions, only $
c_{T,g}^{(1,0)}$ and $c_{L,g}^{(1,0)}$ are known exactly
\cite{LRSN,Harris-Smith}. As to the $\mu$-dependent coefficients,
they can by calculated explicitly using the evolution equation:
\begin{equation}  \label{68}
\frac{\text{d}\hat{\sigma}_{k,g}(z ,Q^{2},\mu ^{2})}{\text{d}\ln \mu
^{2}}=-\int\limits_{\zeta_{\min
}}^{1}\text{d}\zeta\,\hat{\sigma}_{k,g}(z/\zeta,Q^{2},\mu
^{2})P_{gg}(\zeta),
\end{equation}
where $z=Q^{2}/s^{\prime}$, $\zeta_{\min }=z(1+4\lambda)$,
$\hat{\sigma}_{k,g}(z,Q^{2},\mu )$ are the cross sections resummed
to all orders in $\alpha _{s}$ and $P_{gg}(\zeta)$ is the
corresponding (resummed) Altarelli-Parisi gluon-gluon splitting
function. Expanding Eq.~(\ref{68}) in $\alpha _{s}$, one can find
\cite{Laenen-Moch,we2}
\begin{equation}  \label{69}
c_{k,g}^{(1,1)}(z, \lambda)=\frac{1}{4\pi
^{2}}\int\limits_{\zeta_{\min}}^{1}\text{d}\zeta\left[ b_{2}\delta
(1-\zeta)-\,P_{gg}^{(0)}(\zeta)\right]
c_{k,g}^{(0,0)}(z/\zeta,\lambda),
\end{equation}
where $b_{2}=(11C_{A}-2n_{f})/12$ is the first coefficient of the
$\beta(\alpha _{s})$-function expansion and $n_{f}$ is the number of
active quark flavors. The one-loop gluon splitting function is:
\begin{eqnarray}  \label{70}
P_{gg}^{(0)}(\zeta)=\lim_{\epsilon \rightarrow 0}\left\{ \left(
\frac{\zeta}{1-\zeta}+ \frac{1-\zeta}{\zeta}+\zeta(1-\zeta)\right)
\theta(1-\zeta-\epsilon )+\delta(1-\zeta)\ln \epsilon \right\} C_{A}
+b_{2}\delta(1-\zeta).
\end{eqnarray}

With Eq.~(\ref{69}) in hand, it is possible to check the quality of
the NLL approximation against exact answers. As shown in
Ref.~\cite{we4}, the soft-gluon corrections reproduce satisfactorily
the threshold behavior of the available exact results for
$\lambda\sim1$. Since the gluon distribution function supports just
the threshold region, the soft-gluon contribution dominates the
photon-hadron cross sections $\sigma_{k,GF}$ ($k=T,L,A,I$) at
energies not so far from the production threshold and at relatively
low virtuality $Q^{2}\lesssim m^{2}$.

\section{\label{exp}Appendix C: Nonperturbative IC and Relevant Experimental Facts}
The most clean probe of the charm quark distribution function (both
perturbative and nonperturbative) is the semi-inclusive deep
inelastic lepton-proton scattering, $lp\rightarrow l'cX$. To measure
the nonperturbative IC contribution, one needs data on the charm
production at sufficiently large Bjorken $x$. The only experiment
which has investigated the large $x$ domain is the European Muon
Collaboration (EMC) \cite{EMC} where the decay lepton spectra have
been used to detect the produced charmed particles. In
Ref.~\cite{harris-emc}, a re-analysis of the EMC data on
$F_{2}^{c}(x,Q^{2})$ have been performed using the NLO results for
both GF and QS components. The analysis \cite{harris-emc} shows that
a nonperturbative intrinsic charm contribution to the proton wave
function of the order of $1\%$ is needed to fit the EMC data in the
large $x$ region. This value of the nonperturbative IC is consistent
with the estimates based on the operator product expansion
\cite{polyakov}. Note however that the EMC data are of limited
statistics and, for this reason, more accurate measurements of charm
leptoproduction at large $x$ are necessary.

It is also possible to extract useful information on the IC from
diffractive dissociation processes such as $p\rightarrow p J/\psi$
on a nuclear target. Comprehensive measurements of the
$pA\rightarrow J/\psi X$ and $\pi A \rightarrow J/\psi X$ cross
sections have been performed in the fixed target experiments NA3 at
CERN \cite{NA3} and E886 at FNAL \cite{E886}. According to the
arguments presented in
Refs.~\cite{brod-psi-1,hoyer-psi-1,brod-psi-2}, the IC contribution
is predicted to be strongly shadowed in the above reactions that is
in a complete agreement with the observed nuclear dependence of the
high Feynman $x_{F}$ component of the $J/\psi$ hadroproduction.

A non-vanishing five-quark Fock component $\left\vert
uudc\bar{c}\right\rangle$ leads to the production of open charm
states such as $\Lambda_{c}(cud)$ and $D^{-}(\bar{c}d)$ with large
Feynman $x_{F}$. This may occur either through a coalescence of the
valence and charm quarks which are moving with the same rapidity or
via hadronization of the produced $c$ and $\bar{c}$. As shown in
Refs.~\cite{barger-lead,brod-lead}, a model based on the
nonperturbative intrinsic charm naturally explains the leading
particle effect in the $pp\rightarrow DX$ and $pp\rightarrow
\Lambda_{c}X$ processes that has been observed at the ISR
\cite{ISR-lead} and Fermilab \cite{E791-lead-1,E791-lead-2}.

As to the high-$x_{F}$ hadroproduction of open bottom states like
$\Lambda_{b}(bud)$, corresponding cross sections are predicted to be
suppressed as $m_{c}^{2}/m_{b}^{2}\sim 1/10$ in comparison with the
case of charm production. Evidence for the forward $\Lambda_{b}$
production in the $pp$ collisions at the ISR energy was reported in
Refs.~\cite{ISR-B-1,ISR-B-2}.

Rare seven-quark fluctuations of the type $\left\vert
uudc\bar{c}c\bar{c}\right\rangle$ in the proton wave function can
lead to the production of two $J/\psi$ \cite{brod-7-quark} or a
double-charm baryon state at large $x_{F}$ and low $p_{T}$. Double
$J/\psi$ events with a high combined $x_{F}\geq 0.5$ have been
detected in the NA3 experiment \cite{NA3-7-quark}. An observation of
the double-charmed baryon $\Xi^{+}_{cc}(3520)$ with mean $\langle
x_{F}\rangle\simeq 0.33$ has been reported by the SELEX
collaboration at FNAL \cite{SELEX-7-quark}.

\chapter*{\label{Concl}Conclusion}
\addcontentsline{toc}{chapter}{Conclusion}
 \hspace{14pt}We conclude
by summarizing our main observations and results. In this thesis,
some spin effects in QCD and recurrence lattices with multi-site
exchanges are investigated. Our conclusions concerning the critical
phenomena in the recurrence spin models with multi-site exchanges
can be formulated as follows.

 We have studied the Yang-Lee complex zeros of the partition
function of the anti-ferromagnetic Potts model for biopolymer formulated on the recursive zigzag
ladder with three-spin interaction. Using the dynamical system approach of multi-dimensional
mapping, we have obtained the helix-coil pseudo-phase transitions in the thermodynamic limit.
Taking into account a non-classical helix-stabilizing interaction,
we describe the folding and unfolding processes  for proteins and
polypeptides. We have got also Arnold tongues with different winding
numbers in the considered microscopic theory \cite{arnold, Artuso}.


Using the dynamical system approach to solid and fluid $ ^3$He films
with MSE on the recurrent lattices, we have obtained magnetization
curves with plateaus (at $m=0, m=1/2, m=1/3$ and $m=2/3$) and one
period doubling~\cite{ar,lev2,lev3}. We have taken into account
two$-$, three$-$, four$-$,  five$-$, and six$-$spin exchanges.
Interaction parameters used in calculations were taken
        from experimental results.

We have also considered (on the Bethe lattice) the  model with cubic symmetry containing both
linear and quadratic spin-spin interactions. An expression for the free energy per
      spin in the thermodynamic limit  was found. Then we have identified the
       different thermodynamic phases of the system
       (disordered, partially ordered and completely ordered) in the
       ferromagnetic case ($J>0$, $K>0$). We have obtained the phase diagrams
        of the model which
       are found to be different for $Q\leq 2$ and $Q> 2 $. The
        case of $Q\leq 2$ contains three tricritical points. When $Q> 2$,
         one tricritical and one triple points there exist~\cite{vad}.

Next direction of our activity are spin effects in QCD. We consider
the azimuthal dependence in charm leptoproduction as a probe of the
IC content of the proton. Our analysis is based on the fact that the
GF and QS components have strongly different
$\cos2\varphi$-distributions~\cite{aniv}. This fact follows from the
NLO calculations of both parton level contributions. In the
framework of the FFNS, we justify the most remarkable property of
the hadron level azimuthal $\cos2\varphi$ asymmetry: the combined
GF+QS predictions for $A(x,Q^{2})$ are perturbatively and
parametrically stable. The nonperturbative IC contribution
(resulting from the five-quark $\left\vert uudc\bar{c}\right\rangle$
component  of the proton wave function) is practically invisible at
low $x$, but affects essentially the GF predictions for the
asymmetry at large Bjorken $x$. We conclude that measurements of the
$\cos2\varphi$ asymmetry at large $x$ could directly probe the
nonperturbative intrinsic charm~\cite{we6}.

Within the VFNS, charm density originates perturbatively from the
$g\rightarrow c\bar{c}$ process and obeys the DGLAP evolution
equation. Presently, charm densities are included practically in all
the global sets of PDFs like CTEQ and MRST. Our analysis shows that
these charm distribution functions reduce dramatically (by about
1/3) the GF predictions for $A(x,Q^{2})$ practically at all values
of $x$. For this reason, the perturbative IC contribution can easily
be measured using the azimuthal $\cos2\varphi$-distributions in
charm leptoproduction~\cite{we6}.


This dissertation is based on the material published in journals: Physics Letters A \cite{Artuso},
Physical Review B \cite{ar}, Physica A \cite{vad}, International Journal of Modern Physics B
\cite{lev2}, Proceedings of National Academy of Sciences of Armenia, Physics \cite{lev3}, Nuclear
Physics B \cite{we6}, Physical Review D \cite{aniv}. Our main results are reported on conferences:
Chaos and Supercomputers (Nor-Amberd, Armenia 2000), IX International Conference on Symmetry
Methods in Physics (Yerevan, Armenia 2001), Stat. Mech. and Dynamic Systems (Nor-Amberd, Armenia
2003), "Fizika-2000" Young Scientist's Republic Conference (Yerevan, Armenia 2004) , Selected
Topics in Theoretical Physics (Tbilisi, Georgia 2005), and discussed on seminars in Yerevan Physics
Institute and ICTP (Trieste, Italy).
\vskip 2cm
\begin{center}
\textbf{\LARGE{Acknowledgments}}
\end{center}
It is a pleasure to thank N.S.~Ananikian and N.Ya.~Ivanov for suggesting these very interesting
research topics to me. I am also grateful to my other co-authors, R.~Roger, V.~Ohanyan, T.
Arakelyan and R.~Artuso, for the fruitful collaboration. Finally, I would like express my gratitude
to all the members of the Theory Division of YerPhI for useful discussions.


\begin{thebibliography}{299}
\addcontentsline{toc}{chapter}{Bibliography}
\bibitem{uhlenbeck} S.A. Goudschmidt, G.H. Uhlenbeck, \emph{Spinning electrons and the structure of
spectra},\\
Nature \textbf{117}, pp. 264-265 (1926).
\bibitem{history} A.~Martin, \emph{History of spin and statistics},\\
hep-ph/0209068 pp. 1-27 (2002).
\bibitem{data} S.~Schael {\it et al.}, \emph{Precision electroweak measurements on the Z resonance},\\
Phys.\ Rept. {\bf 427}, 257-454 (2006).
\bibitem{Ellis} J.~Ellis, \emph{Polarization puts a New Spin on Physics},\\
 hep-ph/0701049 pp. 1-14 (2006).
\bibitem{Erler} J.~Erler, \emph{Low energy tests of the standard model with spin degrees of freedom},\\
hep-ph/0612030 pp.1-10 (2006).
\bibitem{Ising} E.~Ising, \emph{Beitrag zur theorie des ferromagnetismus},\\ Z.\ Physik {\bf 31}, pp. 253-258 (1925).

\bibitem{Onsager} L.~Onsager, \emph{Crystal statistics I. A two-dimensional model with an order disorder
transition},\\ Phys.\ Rev. {\bf 65}, pp. 117-149 (1944).
\bibitem{Vicari} A.~Pelissetto and E.~Vicari, \emph{Critical phenomena and renormalization-group
theory},\\ Phys.\ Rep. {\bf 368}, pp. 549-727 (2002).
\bibitem{glass1} T.~Castellani, A.~Cavagna, \emph{Spin-Glass Theory for Pedestrians}, \\
J.\ Stat. Mech. P05012 pp. 1-52 (2005).
\bibitem{glass2} M.~Mezard, G.~Parisi, M.~Virasoro, \emph{Spin Glass Theory and Beyond},\\
 World Scientific, Singapore (1987).
\bibitem{Wilson1} K.G.~Wilson, \emph{Renormalization group and critical phenomena. 1. Renormalization group and the Kadanoff scaling picture},\\
Phys.\ Rev. {\bf B4}, pp. 3174-3183 (1971).
\bibitem{Wilson2} K.G.~Wilson, J.~Kogut, \emph{The renormalization group and the $\epsilon$ expansion},\\
Phys.\ Rep. {\bf 12}  pp. 75-200 (1974).
\bibitem{Zinn-Justin} J.~Zinn-Justin, \emph{Quantum Field Theory and Critical Phenomena},\\
 3rd Edition, Clarendon Press, Oxford, 1996.
\bibitem{Wilson3} K.G.~Wilson, \emph{Confinement Of Quarks},\\
Phys.\ Rev. {\bf D10},  pp. 2445-2459 (1974).
\bibitem{Wilson4} K.G.~Wilson, \emph{Quarks and strings on a lattice}, in: A. Zichichi
(Ed.),\\
New Phenomena in  Subnuclear Physics, Plenum Press, New York, 1975.
\bibitem{Creutz}M.~Creutz, \emph{Quarks, Gluons and Lattices},\\
 Cambridge University Press, Cambridge, 1983.
\bibitem{Montvay}I.~Montvay, G.~M\"{u}nster, \emph{Quantum Fields on a Lattice},\\
 Cambridge University Press,
Cambridge, 1994.
\bibitem{Creutz2}M.~Creutz, L.~Jacobs, C.~Rebbi, \emph{Monte Carlo computations in lattice gauge theories},\\
Phys.\ Rep. {\bf 95}, pp. 201-282 (1983).
\bibitem{Mangano-N-R} M.~L.~Mangano, P.~Nason, and G.~Ridolfi, \emph{Heavy-quark correlations in hadron collisions at next-to-leading order}\\
Nucl.\ Phys.  {\bf B373}, pp. 295-345 (1992).
\bibitem{Frixione-M-N-R} S.~Frixione, M.~L.~Mangano, P.~Nason, and G.~Ridolfi, \emph{Heavy-quark correlations in photon-hadron collisions}\\
 Nucl.\ Phys.  {\bf B412}, pp. 225-259 (1994).
\bibitem{Ellis-Nason} R.~K.~Ellis and P.~Nason, \emph{QCD radiative corrections to the photoproduction of heavy quarks},\\
 Nucl.\ Phys.\ B {\bf 312}, pp. 551-570 (1989).
\bibitem{Smith-Neerven} J.~Smith and W.~L.~van Neerven, \emph{QCD corrections to heavy flavour photoproduction and
electroproduction},\\ Nucl.\ Phys.\ B {\bf 374}, pp. 36-82 (1992).
\bibitem{LRSN} E.~Laenen, S.~Riemersma, J.~Smith, and W.~L.~van Neerven, \emph{Complete $O(\alpha_S)$ corrections to heavy-flavour structure functions in electroproduction},
\\Nucl.\ Phys.\ B {\bf 392}, pp. 162-228 (1993).
\bibitem{Nason-D-E-1} P.~Nason, S.~Dawson, and R.~K.~Ellis, \emph{The total cross section for the production of heavy quarks in hadronic collisions},
\\ Nucl.\ Phys.\ B {\bf 303}, pp. 607-633 (1988).
\bibitem{Nason-D-E-2} P.~Nason, S.~Dawson, and R.~K.~Ellis, \emph{The one particle inclusive differential cross section for heavy quark production in hadronic
collisions},\\ Nucl.\ Phys.\ B {\bf 327}, pp. 49-92 (1989).
\bibitem{Nason-D-E-3} P.~Nason, S.~Dawson, and R.~K.~Ellis,
\emph{},\\
Nucl.\ Phys.\ B {\bf 335}, 260 (1990).
\bibitem{BKNS} W.~Beenakker, H.~Kuijf, W.~L.~van Neerven, and
J.~Smith, \emph{QCD Corrections to Heavy Quark Production In p
anti-p Collisions},\\ Phys.\ Rev.\ D {\bf 40}, pp. 54-82 (1989).
\bibitem{Contopanagos-L-S} H.~Contopanagos, E.~Laenen, and
G.~Sterman, \emph{Sudakov factorization and resummation},\\
 Nucl.\ Phys.\ B {\bf 484}, pp. 303-330 (1997).
\bibitem{Laenen-O-S} E.~Laenen, G.~Oderda, and G.~Sterman, \emph{Resummation of threshold corrections for single particle inclusive
cross-sections},\\ Phys.\ Lett.\ B {\bf 438}, 173-183 (1998).
\bibitem{Kidonakis-O-S} N.~Kidonakis, G.~Oderda, and G.~Sterman, \emph{Evolution of color exchange in QCD hard
scattering},\\ Nucl.\ Phys.\ B {\bf 531}, 365-402 (1998).
\bibitem{Laenen-Moch} E.~Laenen and S.~-O. Moch, \emph{Soft gluon resummation for heavy quark electroproduction},\\
Phys.\ Rev.\ D {\bf 59}, 034027 pp.1-18 (1999).
\bibitem{kid2} N.~Kidonakis,  \emph{Next-to-next-to-next-to-leading-order soft-gluon corrections in hard-scattering processes near threshold},\\
Phys.\ Rev.\ D {\bf 73}, 034001 pp.1-13 (2006).
\bibitem{kid1} N.~Kidonakis, \emph{High order corrections and subleading logarithms for top quark production},\\
Phys.\ Rev.\ D {\bf 64}, 014009 pp.1-21 (2001).
\bibitem{we1} N.~Ya.~Ivanov, A.~Capella, and A.~B.~Kaidalov, \emph{Single spin asymmetry in heavy flavor photoproduction as a test of
pQCD} \\
Nucl.\ Phys.\  {\bf B586}, pp. 382-396 (2000).
\bibitem{we2} N.~Ya.~Ivanov, \emph{Perturbative stability of the QCD predictions for single spin asymmetry in heavy quark photoproduction}, \\
 Nucl.\ Phys.\ B {\bf 615}, pp. 266-284 (2001).
\bibitem{we4} N.~Ya.~Ivanov, \emph{Azimuthal asymmetries in heavy quark leptoproduction as a test of pQCD},\\
 Nucl.\ Phys.\ B {\bf 666}, 88-104 (2003).
\bibitem{we3} N.~Ya.~Ivanov, P.~E.~Bosted, K.~Griffioen, and
S.~E.~Rock, \emph{Single spin asymmetry in open charm photoproduction and decay as a test of pQCD}\\
Nucl.\ Phys.\ B {\bf 650}, pp. 271-289 (2003).
\bibitem{HM} E.~Hoffmann and R.~Moore, \emph{Subleading Contributions To The Intrinsic Charm Of The Nucleon},\\ Z.\ Phys.\ C {\bf 20}, 71-82 (1983).
\bibitem{KS} S.~Kretzer and I.~Schienbein, \emph{Heavy quark initiated contributions to deep inelastic structure functions},\\
 Phys.\ Rev.\ D {\bf 58}, 094035 pp.1-12 (1998).
\bibitem{KS-thesis} I. Schienbein, \emph{Heavy Quark Production in CC and NC DIS and The Structure of Real and Virtual Photons in NLO QCD},\\
 hep-ph/0110292.
\bibitem{BHPS} S.~J.~Brodsky, P.~Hoyer, C.~Peterson, and N.~Sakai, \emph{The Intrinsic Charm Of The
Proton},\\ Phys.\ Lett.\ B {\bf 93}, pp. 451-455 (1980).
\bibitem{BPS} S.~J.~Brodsky, C.~Peterson, and N.~Sakai, \emph{Intrinsic Heavy Quark
States},\\ Phys.\ Rev.\ D {\bf 23}, pp. 2745-2757 (1981).
\bibitem{ACOT} M.~A.~G.~Aivazis, J.~C.~Collins, F.~I.~Olness, and
W.~-K.~Tung, \emph{Leptoproduction of heavy quarks. 2. A Unified QCD
formulation of charged and neutral current processes from fixed
target to collider energies},\\
 Phys.\ Rev.\ D {\bf 50}, pp. 3102-3118 (1994).
\bibitem{collins} J.~C.~Collins, \emph{Hard scattering factorization with heavy quarks: A General treatment},\\
Phys.\ Rev.\ D {\bf 58}, 094002 pp. 1-29 (1998).
\bibitem{CTEQ6} J.~Pumplin, D.~R.~Stump, J.~Huston, H.~L.~Lai, P.~Nadolsky,
and W.~K.~Tung, \emph{New generation of parton distributions with
uncertainties from global QCD analysis},\\ JHEP {\bf 07}, 012, pp.
1-47 (2002).
\bibitem{MRST2004} A.~D.~Martin, R.~G.~Roberts, W.~J.~Stirling, and
R.~S.~Thorne, \emph{Physical gluons and high E(T) jets},\\ Phys.\
Lett.\ B {\bf 604}, pp. 61-68 (2004).
\bibitem{eRHIC} A.~Deshpande, R.~Milner, R.~Venugopalan, and
W.~Vogelsang, \emph{Study of the fundamental structure of matter
with an electron-ion collider},\\ Ann.\ Rev.\ Nucl.\ Part.\ Sci.\
{\bf 55}, pp. 165-228 (2005).
\bibitem{EIC} See also http://www.bnl.gov/eic for information concernig
the eRHIC/EIC project.
\bibitem{LHeC} J.~B.~Dainton, M.~Klein, P.~Newman, E.~Perez, and
F.~Willeke, \emph{Deep Inelastic Electron-Nucleon Scattering at the LHC},\\
hep-ex/0603016.


\bibitem{yl1} C. N. Yang and T. D. Lee, \emph{Statistical Theory of Equations of State and Phase Transitions. I. Theory of
Condensation},\\
 \Journal{Phys. Rev}{87}{pp. 404-409}{1952}.
 \bibitem{yl2} T. D. Lee and C. N. Yang, \emph{Statistical Theory of Equations of State and Phase Transitions. II. Lattice Gas and Ising Model},\\
  \Journal {Phys.
 Rev.}{87}{pp. 410-419}{1952}.

\bibitem{kas} P. W. Kasteleyn and C.M Fortuin, \emph{Phase transitions in lattice systems with random local properties},\\
 \Journal {{ J. Phys. Soc. Japan}}{26}{(Suppl.)11}{1969};
\bibitem{kas2} C.M Fortuin and P. W. Kasteleyn, \emph{Random-cluster model. Introduction and relation to other models},\\
 \Journal {{Physica}}{57}{pp. 536-564}{1972};
\bibitem{salas} J.Salas and A. D. Sokal, \emph{Transfer matrices and partition-function zeros for antiferromagnetic Potts models},\\
 \Journal{{ J. Stat. Phys.}}{104}{pp. 609-699}{2001}.
 \bibitem{priez} J.G. Brankov, Vl.V. Papoyan, V.S. Poghosyan, V.B.
Priezzhev, \emph{The totally asymmetric exclusion process on a ring:
Exact
relaxation dynamics and associated model of clustering transition},\\
Physica A 368, pp. 471-480 (2006).
\bibitem{han}U.H.E. Hansmann,  Y. Okamoto, \emph{Finite-size scaling of helix–coil transitions in poly-alanine studied by multicanonical simulations},\\
  \Journal{{J. Chem. Phys.}}{110}{pp. 1267-1276}{1999}.
  \bibitem{han2}N. A. Alves, U.H.E. Hansmann, \emph{Partition Function Zeros and Finite Size Scaling of Helix-Coil Transitions in a Polypeptide},\\
 \Journal{{ Phys. Rev. Lett.}}{84}{pp. 1836-1839} {2000}.
\bibitem{han3} Y. Okamoto, U. H. E.Hansmann, \emph{Thermodynamics of Helix-Coil Transitions Studied by Multicanonical Algorithms} ,\\
 J. Phys. Chem. {\bf99} pp. 11276-11287 (1995).
\bibitem{han4} N. A. Alves and U.H.E. Hansmann, \emph{Yang-Lee zeros and the helix-coil transition in a continuum model of polyalanine},\\
Physica A{\bf 292}, pp. 509-518 (2001)


\bibitem{coom}S. Coombes and P. C. Bressloff, \emph{Mode locking and Arnold tongues in integrate-and-fire neural oscillators},\\
 \Journal{Phys. Rev}{E60}{pp. 2086-2096}{1999};\\
R. S. Mackay and C. Tresser, \emph{Transition to topological chaos for circle maps},\\
 \Journal{{ Physica}}{D19}{pp. 206-237}{1986}.
\bibitem{Pri} C. N. Pace and Ch. Tanford, \emph{Thermodynamics of the unfolding of beta-lactoglobulin A in aqueous urea solutions between 5 and 55
degrees},\\
 Biochemistry {\bf 7}, pp. 198-213 (1968);\\
  G. P. Privalov and P. L. Privalov, \emph{Problems and prospects in the microcalorimetry of biological macromolecules},\\
   Methods Enzymol. {\bf 323}, pp. 31-62 (2000).
\bibitem{bakk} A. Bakk, J. S. H{\o}ye, and A. Hassen, \emph{Apolar and polar solvation thermodynamics related to the protein unfolding process},\\
 Biophysical Journal {\bf 82} pp. 713-719 (2002);\\
O. Collet, \emph{Four-states phase diagram of proteins},\\
 Europhysics Letters {\bf 72}  pp. 301-307 (2005).
 \bibitem{collet} O. Collet, \emph{Warm and cold denaturation in the phase diagram of a protein lattice model},\\
 Europhysics Letters {\bf 53}  pp. 93-99 (2001).
 \bibitem{buzano} P.Bruscolini, C. Buzano, A. Pelizzola, M. Pretti, \emph{Bethe approximation for a model of polymer solvation},\\
  Phys. Rev. E {\bf 64} 050801(R) pp. 1-4 (2001).
\bibitem{don} D. J. Jacobs, S. Dallakyan, G. G. Wood, A.
Heckathorne, \emph{Network rigidity at finite temperature:
Relationships between thermodynamic stability, the nonadditivity of
entropy, and cooperativity in molecular systems},\\
\Journal{Phys. Rev.} {E68}{061109 pp. 1-22}{2003}.

\bibitem{maghis} D. Mattise, \emph{The theory of magnetism} (Harper and Row, New York) (1982).
\bibitem{supcon} J. G. Bednorz and K. A. M\"{u}ller, \emph{Possible high $T_c$ superconductivity in the Ba-La-Cu-O system}, \\
Z. Phys. {\bf B64}, pp. 189-193 (1986).
\bibitem{rog1}  M. Roger, J. H. Hetherington, J. M. Delrieu, \emph{Magnetism in solid $^3He$},\\
 Rev. Mod. Phys. \textbf{55}, pp. 1-64 (1983);\\
 H. Franco, R. Rapp and H. Godfrin, \emph{Nuclear Ferromagnetism of Two-Dimensional 3He},\\
Phys. Rev. Lett. \textbf{57}, pp. 1161-1164 (1986);\\
  H. Godfrin, R. Ruel and D. Osheroff, \emph{Experimental Observation of a Two-Dimensional Heisenberg Nuclear Ferromagnet},\\
 Phys. Rev. Lett. \textbf{60}, pp. 305-308 (1988).
 \bibitem{godfrin}
  H. Godfrin and R. E. Rapp, \emph{Two-dimensional nuclear magnets},\\
   Adv. Phys. \textbf{44}, pp. 113-186 (1995).
\bibitem{Bernu} M. Roger, C.B\"{a}uerle, Yu. M. Bunkov, A.-S. Chen and H.
Godfrin, \emph{Multiple-Spin Exchange on a Triangular Lattice: A
Quantitative Interpretation of Thermodynamic Properties of
Two-Dimensional Solid $^3He$},\\
 Phys. Rev. Lett. \textbf{80}, pp. 1308-1311 (1998).

\bibitem{God}  H. Godfrin and D. D. Osheroff,\emph{Multiple-spin-exchange calculation of the T=0 properties of solid
         3He},\\
        Phys. Rev. B  \textbf{38}, pp. 4492-4503 (1988).

\bibitem{Bet} N. S. Ananikan, A. R. Avakian, and N. Sh. Izmailian, \emph{Phase diagrams and tricritical effects in the beg model},\\
           Physica A 172, pp. 391-404 (1991).
 \bibitem{ana97} N.S. Ananikian, S.K. Dallakian, B. Hu, \emph{Chaotic Properties of the Q-state Potts Model on the Bethe Lattice: Q < 2},\\
 \Journal{{Complex Systems}}{11}{pp. 213-222}{1999};\\
 A.Z. Akheyan and  N.S. Ananikian, \emph{Global Bethe lattice consideration of the spin-1 Ising model},\\
 \Journal{J.Phys.}{A29}{pp. 721-731}{1996}.
\bibitem{ladderp} T. Sakai and M. Takahashi, \emph{Magnetization plateau in an S=3/2 antiferromagnetic Heisenberg chain with
anisotropy},\\
Phys. Rev. B \textbf{57}, R3201-R3204 (1998).
\bibitem{ladderp2}
D. C. Cabra, A. Honecker, and P. Pujo, \emph{Magnetization Curves of
Antiferromagnetic Heisenberg Spin- 1/2  Ladders},\\
Phys. Rev. Lett. \textbf{79}, 5126-5129 (1997).
\bibitem{ladderp3}
M. Maslen, M. T. Batchelor, and J. de Gier, \emph{Magnetization plateaux in Bethe ansatz solvable spin-S   ladders},\\
 Phys. Rev. B \textbf{68}, 024418 pp.1-8 (2003).
 \bibitem{ladderp4}
K. Okamoto, N. Okazaki, and T. Sakai, \emph{Magnetization Plateau of an S=1 Frustrated Spin Ladder},\\
  J. Phys. Soc. Jpn. \textbf{70}, 636-639 (2001).


\bibitem{arnold} N. S. Ananikian and L.N. Ananikyan,  ed. by S Rahvar, N Sadooghi and  F Shojai,
\emph{Arnold Tongues in One-, and Multi-Dimensional Mappings
       of Physical Systems}, World  Scientific; Singapure, 2005 pp. 21- 26.
\bibitem{Artuso} N. Ananikian, L. Ananikyan, and R.Artuso, \emph{Multi-dimensional Mapping and Folding Properties for  Non-classical
Helix-stabilizing},\\
 Phys. Lett. A. \textbf{360}, pp. 615-618 (2007).


\bibitem{ar} T.A. Arakelyan V.R Ohanyan, L.N. Ananikyan, N.S. Ananikian, and M.Roger, \emph{Multisite-interaction Ising model approach to the solid $^3He$ system on a triangular
lattice},\\
 Phys. Rev. B \textbf{67}, 024424 pp.1-13 (2003).
\bibitem{vad}V.R. Ohanyan, L.N. Ananikyan, N.S. Ananikyan,\emph{ An exact solution on the ferromagnetic face-cubic spin model on Bethe lattice},\\
 Physica A \textbf{377}  pp. 501-513 (2007).
\bibitem{lev2}L.N. Ananikyan, \emph{The hexagonal recursive approximation with multisite-interaction Ising model for the solid and fluid
         $^3$He system},\\
         Int. J. Mod. Phys. \textbf{B} 21, pp. 1-18 (2007).
\bibitem{lev3}L.N. Ananikyan, \emph{Magnetic Properties of  $^3He$ on the recursive
         lattices},\\
          Proceedings of National Academy of Sciences of Armenia, Physics, 42,
          pp. 17-25 (2007).
\bibitem{we6} L.~N.~Ananikyan and N.~Ya.~Ivanov, \emph{Azimuthal asymmetries in DIS as a probe of intrinsic charm content of the proton},\\
\emph{Nucl. Phys.} {\bf B762}, pp. 256-283 (2007).
\bibitem{aniv} L.~N.~Ananikyan and N.~Ya.~Ivanov, \emph{Azimuthal dependence of the heavy quark initiated contributions to DIS},\\
Phys. Rev {\bf D75}, 014010 pp. 1-9 (2007).

\bibitem{hu}P. D. Gujrati, \emph{Bethe or Bethe-like Lattice Calculations Are More Reliable Than Conventional Mean-Field Calculations},\\
 Phys. Rev. Lett. \textbf{74}, pp. 809-812 (1995).

\bibitem{mon91} J. L. Monroe, \emph{Phase diagrams of Ising models on Husimi trees. I. Pure multisite interaction systems},\\
J. Stat. Phys. \textbf{65}, pp. 255-268 (1991);\\
 J. L. Monroe, \emph{Phase-Diagrams Of Ising-Models On Husimi Trees. II. Pair And Multisite Interaction
Systems} J. Stat. Phys. 67, pp. 1185-1200 (1992).

\bibitem{ladder} M.Matsuma \emph{et al.}, \emph{Magnetic excitations and exchange interactions in the spin-1/2 two-leg ladder compound $La_6 Ca_8 Cu_{24}
O_{41}$},\\
 Phys. Rev. B \textbf{62}, pp. 8903-8908 (2000);\\
A. Lauchli, G. Schmid, and M. Troyer, \emph{Phase diagram of a spin
ladder with cyclic four-spin exchange},\\
   Phys. Rev. B \textbf{67}, 100409(R) pp. 1-4 (2003).
   \bibitem{ladder01}
 T. Hikihara, T.Momoi,and X. Hu, \emph{Spin-Chirality Duality in a Spin Ladder with Four-Spin Cyclic
 Exchange},\\
Phys. Rev. Lett. \textbf{90}, 087204 pp. 1-4 (2003).
\bibitem{ladder02}
 G. Japaridze and E. Pogosyan, \emph{Magnetization plateau in the S = 1/2 spin ladder
with alternating rung exchange},\\
J. Phys.: Condens. Matter \textbf{18}, pp. 9297-9306 (2006);\\
T. Vekua, G.I. Japaridze, and H.-J. Mikeska, \emph{Phase diagrams of
spin ladders with ferromagnetic legs},\\
Phys. Rev \textbf{B67}, 064419 pp. 1-11 (2003).

\bibitem{selke} W. Selke, \emph{The ANNNI model — Theoretical analysis and experimental application},\\
 Phys. Rep. \textbf{170}, pp. 213-264 (1988).


 \bibitem{bax} R. Baxter, \emph{Exactly Solved Models in Statistical Mechanics}\\
 (Academic press, New  York,1982), Chap.4.

\bibitem{Anf} C. B. Anfinsen, \emph{Principles that Govern Folding of Protein Chains},\\
 Science {\bf 181}, pp. 223-230 (1973).

\bibitem{Dill} K. A. Dill et. al., \emph{Principles of Protein-folding - a Perspective from Simple Exact Models},\\
 Protein Sci. {\bf 4} pp. 561-602 (1995).

\bibitem{one} J. D. Bryngelson and P. G. Wolynes, \emph{Spin-Glasses and the Statistical-Mechanics of Protein Folding},\\
 Proc. Natl. Acad. Sci. U.S.A. {\bf 84}, pp. 7524-7528 (1987).
 \bibitem{one2}
  V. S. Pande, A. Yu. Grosberg, and T. Tanaka, \emph{Freezing Transition of Random Heteropolymers Consisting of an Arbitrary Set of Monomers},\\
   Phys. Rev. E {\bf 51}, pp. 3381-3392 (1995).
   \bibitem{one3}
 A. Bakk and J. S. H{\o}ye, \emph{One-dimensional Ising model applied to protein folding},\\
    Physica A {\bf 323}, pp. 504-518 (2003).

\bibitem{two} M. S. Li and M. Cieplak, \emph{Folding in two-dimensional off-lattice models of proteins},\\
 Phys. Rev. E {\bf 59}, pp. 970-976 (1999).
 \bibitem{Salvi}  G. Salvi, S. M\"{o}lbert and P. Des Los Rios, \emph{Design of lattice proteins with explicit solvent},\\
 \Journal{Phys. Rev} {E66} {061911 pp. 1-5} {2002}.

\bibitem{three} E. Shakhvovich, \emph{Proteins with selected sequences fold into unique native conformation},\\
 Phys. Rev. Lett. {\bf 72} pp. 3907-3910 (1994).
 \bibitem{three2}
A. M. Gutin, V. L. Abkevich, and E. Shakhvovich, \emph{Chain Length
Scaling of Protein Folding Time},\\
 Phys. Rev. Lett. {\bf 77} pp. 5433-5436
(1996).

\bibitem{off_l} A. Irback, C. Peterson, and F. Pottast, \emph{Identification of amino acid sequences with good folding properties in an off-lattice
model},\\
 Phys. Rev. E {\bf 55} pp. 860-867 (1997).
\bibitem{off_l2}
  D. K. Klimov and D. Thirumalai, \emph{Viscosity Dependence of the Folding Rates of Proteins},\\
   Phys. Rev. Lett. {\bf 79} pp. 317-320 (1997).

\bibitem{off_lat} M. S. Li, M. Cieplak, and N. Sushko, \emph{Dynamical chaos and power spectra in toy models of heteropolymers and proteins},\\
 Phys. Rev. E {\bf 62}, pp. 4025-4031 (2000).


\bibitem{Bal} A. Chakrabartty and R. L. Baldwin, in \emph{Protein Folding: In
Vivo and In Vitro,} edited by J. Cleland and J. King (ACS.
Washington, D.C., 1993), pp. 166-177.

\bibitem{Maritan} H. S. Chan and K. A. Dill, \emph{The Protein Folding Problem},\\
Phys. Today {\bf46}, No. 2, pp. 24-32 (1993).
\bibitem{Maritan2}
 C. Micheletti, F. Seno, A. Maritan, and J. R. Banavar, \emph{Design of proteins with hydrophobic and polar amino acids},\\
  Proteins {\bf32}, pp. 80-87 (1998).

\bibitem{Kam} S. Kamtekar, J. M. Schiffer, H. Xiong, J. M. Babik, and M. H.
Hecht, \emph{Protein design by binary patterning of polar and nonpolar
amino-acids} ,\\
 Science {\bf 262}, pp. 1680-1685 (1993).

\bibitem{Dokh} G. Salvi and P. Des Los Rios, \emph{Effective Interactions Cannot Replace Solvent Effects in a Lattice Model of Proteins},\\
 Phys. Rev. Lett. {\bf 91} 258102 pp. 1-4 (2003).
 \bibitem{Dokh2}
 M. I. Marqu\'{e}s, J. M. Borreguero, H. E. Stanley, and N. V. Dokholyan, \emph{Possible Mechanism for Cold Denaturation of Proteins at High Pressure},\\
  Phys. Rev. Lett. {\bf 91} 138103 pp. 1-4 (2003).

\bibitem{zhou} H.-X. Zhou, \emph{Polymer models of protein stability, folding, and interactions},
 Biochemistry {\bf 43} pp. 2141-2154 (2004);\\
V.N. Morozov, S.G. Gevorkian, \emph{Low-temperature glass-transition in proteins},\\
Biopolymers {\bf 24} pp. 1785-1799 (1985);\\
S.G. Gevorkian, D.S. Gevorgyan, E.H. Karagyan, \emph{How elastic are biopolymers?},\\
Eur. Biophys. J. {\bf 34} pp. 539-610 (2005).



\bibitem{Sch} H. Qian and J. A. Schellman, \emph{Helix-coil theories: a comparative study for finite length polypeptides},\\
 \Journal{J. Phys. Chem.}{96}{pp. 3987-3994}{1992}.




\bibitem{Shi} Z. Shi, C. A. Olson, A. J. Bell Jr., N. R. Kallenbach,\emph{ Non-classical helix-stabilizing interactions: C-H...O H-bonding between Phe and Glu side chains in alpha-helical peptides},\\
 Biophysical Chemistry {\bf 101-102}, pp. 267-279 (2002).

\bibitem{hboundbook} G.R. Desiraju and T. Steiner, The Weak Hydrogen \emph{Bound in Structural Chemistry and Biology},\\
 Oxford university
press (1999).

\bibitem{TIBS} M.C. Wahl and M. Sundaralingam, \emph{C-H center dot center dot center dot O hydrogen bonding in biology},\\
 TIBS {\bf 22}, pp. 97-102 (1997).

\bibitem{Kar} S. Scheiner, T. Kar, and Y. Gu, \emph{Strength of the $C_\alpha H\cdot\cdot O$ Hydrogen Bond of Amino Acid Residues},\\
 J. Biol. Chem. 276, pp. 9832-9837 (2001).

\bibitem{bella} J. Bella and H. M. Berman, \emph{Crystallographic Evidence for $C_\alpha - H \cdot\cdot\cdot O=C$ Hydrogen
Bonds in a Collagen Triple Helix},\\
 JMB 264, 734 (1996).

 \bibitem{zimbrag} B.H.Zimm and I.K.Bragg, \emph{Theory of the Phase Transition between Helix and Random Coil in Polypeptide Chains},\\
\Journal{J.Chem.Phys.}{31}{pp. 526-535}{1959};\\
 S. Lifson, A. Roig, \emph{On the Theory of Helix—Coil Transition in Polypeptides},\\
 \Journal{J.Chem.Phys.}{34}{pp. 1963-1974}{1961}.

\bibitem{ra} G.N. Ramachandran, C. Ramakrishnan, V. Sasisekharan, \emph{Stereochemistry of polypeptide. Chain configurations},\\
  J. Mol. Biol.{\bf 7} 95 (1963);\\
 J.T. Edsall, P.J. Flory, J.C. Kendrew, A.M. Liquory, G. Nementhy, G.N. Ramachandran, H.A.
Scheraga, \emph{A proposal of standard conventions and nomenclature
for description of polypeptide conformations},\\
 J. Mol. Biol. {\bf 15} pp. 399-429 (1966).

\bibitem{hay1}N.S. Ananikyan, Sh.A. Hajryan,  E.Sh. Mamasakhlisov,
V. F. Morozov, \emph{Helix-Coil transition in polypeptides: A
microscopical approach},\\
\Journal{Biopolymers}{30}{pp. 357-367}{1990}.

\bibitem{yokoi} C.S.O. Yokoi, M.J. de Oliveira, and S.R.Salinas, \emph{Strange Attractor In The Ising-Model With Competing Interactions On The Cayley Tree},\\
 Phys. Rev. Lett. {\bf 54} pp. 163-166 (1985);\\
  M.H.R. Tragtenberg, C.S.O. Yokoi, \emph{Field behavior of an ising-model with competing interactions on the bethe lattice},\\
  Phys. Rev. E {\bf 52} pp. 2187-2197 (1995).


\bibitem {sl}
A. Alahverdian, N.  Ananikian, S.Dallakian, \emph{Singularities at a dense set of temperature singularities in the Husimi tree},\\
 \Journal{Phys. Rev E}{57}{pp. 2452-2454}{1998};\\
R. G. Ghulghazaryan, N.S.Ananikian, \emph{Partition function zeros
of the one-dimensional Potts model: the recursive method},\\
\Journal{J. Phys.}{A36}{pp. 6297-6312}{2003}.



\bibitem{rog2} H. Jichu and K. Kuroda, \emph{Theory Of The Surface-Induced Magnetism In Liquid $^3$He},\\
 Prog. Theor. Phys. \textbf{67}, pp. 715-725 (1982);\\
 R. A. Gayer,\emph{$^3 He$ films and the Ruderman-Kittel-Kasuya-Yosida interaction},\\
  Phys. Rev. Lett. \textbf{64}, pp. 1919-1923 (1990);\\
   M. Roger, \emph{High-temperature series expansions with cyclic exchanges on a triangular lattice: Application to two-dimensional solid
   $^3He$},\\
  Phys. Rev. B. \textbf{56}, pp. R2928 - R2931 (1997).

\bibitem{exp} M. Siquera, J. Ny\'{e}ki, B. Cowan and J. Saunders, \emph{Heat Capacity Study of the Quantum Antiferromagnetism of a
3He Monolayer},\\
Phys. Rev. Lett. \textbf{76}, pp. 1884-1887 (1996) ;\\
K. Ishida, M. Morishita, K. Yawata and H. Fukuyama, \emph{Low Temperature Heat-Capacity Anomalies in Two-Dimensional Solid $^3He$},\\
Phys. Rev. Lett. \textbf{79}, pp. 3451 - 3454 (1997).


\bibitem{theor}J. M. Delrieu, M. Roger, and J. H. Hetherington, \emph{Exchange and magnetic order in HCP 3He and adsorbed 3He with triangular
lattice},\\
J. Low Temp. Phys. \textbf{40}, pp. 71-87 (1980);\\
 M. Roger, \emph{Multiple exchange in $^3He$ and in the Wigner solid}, \\
Phys. Rev. B \textbf{30}, pp. 6432-6457 (1984); \\

\bibitem{ladder2} G. Misguich, B. Bernu, C. Lhuillier, and C.
Waldtmann, \emph{Spin Liquid in the Multiple-Spin Exchange Model on
the Triangular Lattice: $^3He$ on Graphite},\\
 Phys. Rev. Lett. \textbf{81}, pp. 1098-1101 (1998).

\bibitem{hida} K.Hida, \emph{Magnetic Properties of the Spin-1/2 Ferromagnetic-Ferromagnetic-Antiferromagnetic Trimerized Heisenberg Chain},\\
J. Phys. Soc. Jpn.  \textbf{63}, pp. 2359-2364 (1994).

\bibitem{pl} M.Oshikawa, M.Yamanaka and I.Affleck, \emph{Magnetization Plateaus in Spin Chains: "Haldane Gap" for Half-Integer Spins},\\
 Phys. Rev. Lett. \textbf{78}, pp. 1984-1987 (1997).

\bibitem{pl2} H. Nishimori and S. Miyashita, \emph{Magnetization Process of the Spin-1/2 Antiferromagnetic Ising-Like Heisenberg Model on the Triangular Lattice},
\\ J. Phys. Soc. Jpn. \textbf{55}, pp. 4448-4455 (1986);\\
A. Honecker, \emph{A comparative study of the magnetization process of two-dimensional antiferromagnets}, \\
 J. Phys. Cond. Matt. \textbf{11}, pp. 4697-4713 (1999).

\bibitem{kag} C. Zeng and V. Elser, \emph{Quantum dimer calculations on the spin-1/2 kagom\'{e} Heisenberg
antiferromagnet}\\
 Phys. Rev. B \textbf{51}, pp. 8318-8324 (1995);\\
P. W. Leung and V. Elser, \emph{Numerical studies of a 36-site
kagome\'{e} antiferromagnet},\\
 Phys. Rev. B \textbf{47}, pp. 5459-5462 (1993);\\
 Ch. Waldtmann \emph{et al.}, \emph{First excitations of the spin 1/2 Heisenberg antiferromagnet on the kagome lattice},\\
  Eur. Phys. J B \textbf{2}, pp. 501-507 (1998).





\bibitem{aha74} A. Aharony and A. D. Bruce, \emph{Polycritical Points and Floplike Displacive Transitions in
Perovskites},\\
 Phys. Rev. Lett. \textbf{33}, pp. 427-430 (1974).

\bibitem{bru75} A. D. Bruce and A. Aharony, \emph{Coupled order parameters, symmetry-breaking irrelevant scaling fields, and tetracritical points},\\
Phys. Rev. B \textbf{11}, pp. 478-499 (1975).

\bibitem{aha76}A. Aharony, \textit{Phase Transition and Critical
Phenomena}, ed. by C. Domb and J. Lebowitz\\
 Academic Pess, New York, 1976, vol. 6, pp. 357-403.



\bibitem{bar89} N. C. Bartelt, T. L. Einstein, and L. T. Wille, \emph{Phase diagram and critical properties of a two-dimensional lattice-gas model of oxygen ordering in
$YBa_2Cu_3O_z$},\\
 Phys. Rev. B \textbf{40}, pp. 10759-10765  (1989);\\
 T. Aukrust, M. A. Novotny, P. A. Rikvold, and D. P. Landau, \emph{Numerical investigation of a model for oxygen ordering in $YBa_2Cu_3O_{6+x}$}, \\
 Phys. Rev. B \textbf{41}, pp. 8772-8791 (1990).

\bibitem{cho93} T. Chou and D. R. Nelson, \emph{Buckling instabilities of a confined colloid crystal
layer},\\
 Phys. Rev. E \textbf{48}, pp. 4611-4621 (1993).

\bibitem{bek00} S. Bekhechi, A. Benyoussef, and N. Moussa, \emph{Numerical study of a lattice-gas model for micellar binary
solutions},\\
 Phys. Rev. B \textbf{61}, pp. 3362-3371 (2000).

 \bibitem{cal02} P. Calabrese and A. Celi, \emph{Critical behavior of the two-dimensional N-component Landau-Ginzburg Hamiltonian with cubic
anisotropy},\\
 Phys. Rev. B \textbf{66}, 184410 pp. 1-13 (2002).

\bibitem{cal04} P. Calabrese, E. V. Orlov, D. V. Pakhnin, and A. I.
Sokolov, \emph{Critical behavior of two-dimensional cubic and MN
models in the five-loop renormalization group approximation},\\
 Phys. Rev. B \textbf{70}, 094425 pp. 1-15 (2004).

\bibitem{kle97} H. Kleinert, S. Thoms, and V. Schulte-Frohlinde, \emph{Stability of a three-dimensional cubic fixed point in the two-coupling-constant
$\varphi^4$ theory},\\
Phys. Rev. B \textbf{56}, pp. 14428-14434 (1997).


\bibitem{pak00} D. V. Pakhnin and A. I. Sokolov, \emph{Five-loop renormalization-group expansions for the three-dimensional n-vector cubic model and critical exponents for impure Ising
systems},\\
 Phys. Rev. B \textbf{61}, pp. 15130-15135 (2000).

\bibitem{pak01} D. V. Pakhnin and A. I. Sokolov, \emph{Renormalization group and nonlinear susceptibilities of cubic ferromagnets at
criticality},\\
 Phys. Rev. B \textbf{64}, 094407 pp. 1-6 (2001).

\bibitem{cal02a}  P. Calabrese, A. Pelissetto and E. Vicary, \emph{Critical behavior of vector models with cubic symmetry},\\
 Acta Phys. Slov. \textbf{52} , pp. 311-316 (2002).

\bibitem{usa04} Z. Usatenko and J. Spa\l ek, \emph{Surface critical behaviour of semi-infinite systems with cubic anisotropy at the ordinary transition},\\
 J. Phys. A \textbf{37}, pp. 7113-7125 (2004).

\bibitem{kac06} H. Kachkachi and E. Bonet, \emph{Surface-induced cubic anisotropy in nanomagnets},\\
Phys. Rev. B \textbf{73}, 224402 pp. 1-7 (2006).

\bibitem{kim75} D. Kim, P. M. Levy and L. F. Uffer, \emph{Cubic rare-earth compounds: Variants of the three-state Potts
model},\\
Phys. Rev. B \textbf{12}, pp. 989-1004 (1975).

\bibitem{kim75d} D. Kim and P. M. Levy, \emph{Critical behavior of the cubic model}, \\
Phys. Rev. B \textbf{12}, pp. 5105-5111 (1975).

\bibitem{kim76} D. Kim, P. M. Levy, and J. J. Sudano, \emph{Applicability of the cubic model to the critical behavior of real systems},\\
Phys. Rev. B \textbf{13}, pp. 2054-2065 (1976).

\bibitem{aha77} A. Aharony, \emph{Critical behaviour of the discrete spin cubic model},\\
J. Phys. A \textbf{10}, pp. 389-398  (1977).

\bibitem{rie81} E. K. Riedel, \emph{ The Potts and cubic models in two dimensions: A renormalization-group description},\\
Physica A \textbf{106}, pp. 110-121 (1981).

\bibitem{nie83} B. Nienhuis, E. K. Riedel, and M. Schick, \emph{Critical behavior of the n-component cubic model and the Ashkin-Teller fixed
line},\\
 Phys. Rev. B \textbf{27}, pp. 5625-5643 (1983).



\bibitem{egg74} T. P. Eggarter, \emph{Cayley trees, the Ising problem, and the thermodynamic limit} ,\\
Phys. Rev. B. \textbf{9}, pp. 2989-2992 (1974).

\bibitem{mul74} E. M\"{u}ller--Hartmann and J. Zittartz, \emph{New Type of Phase
Transition},\\
 Phys. Rev. Lett. \textbf{33}, pp. 893-897 (1974).

\bibitem{tho82} C. J. Thompson, \emph{Local properties of an Ising model on a Cayley tree},\\
 J. Stat. Phys. \textbf{27}, pp. 441-456 (1982).

\bibitem{sto03} T. Sto\v{s}i\'{c}, B. D. Sto\v{s}i\'{c} and I.P.Fittipaldi, \emph{Anomalous behavior of the zero field
susceptibility of the Ising model on the Cayley tree},\\
Physica A \textbf{320}, pp. 443-448 (2003);\\
 T. Sto\v{s}i\'{c}, B. D. Sto\v{s}i\'{c} and I. P. Fittipaldi,\emph{Thermodynamic limit for the Ising model on the Cayley tree},\\
Physica A \textbf{355}, pp. 346-354 (2005).

\bibitem{che74} M.-S. Chen, L. Onsager, J. Bonner, and J. Nagle, \emph{Hopping of ions in ice},\\
 J. Chem. Phys. \textbf{60}, pp. 405-419 (1974).


\bibitem{van81} J. Vannimenus, Z. Phys. B \textbf{43}, \emph{Modulated phase of an Ising system with competing interactions on a Cayley tree},\\
 141 (1981).


\bibitem{guj84} P. D. Gujrati, \emph{Ordering Field, Order Parameter, and Self-Avoiding
Walks},\\
 Phys. Rev. Lett. \textbf{53}, pp. 2453-2456 (1984);\\
P. D. Gujrati, \emph{Thermal and percolative transitions and the
need for independent symmetry breakings in branched polymers on a
Bethe lattice},\\
J. Chem. Phys. \textbf{98}, pp. 1613-1634 (1993).

\bibitem{pap95} Vl. V. Papoyan and R. R. Scherbakov, \emph{Abelian sandpile model on the Husimi lattice of square plaquettes},\\ J. Phys.
A \textbf{28}, 6099-6107 (1995).


\bibitem{whe70} J. C. Wheeler and B. Widom,
\emph{Phase Equilibrium and Critical Behavior in a Two-Component Bethe-Lattice Gas or Three-Component Bethe-Lattice Solution,
 }\\
 J. Chem. Phys. \textbf{52}, pp. 5334-5343 (1970).

\bibitem{erd06} A. Erdi\c{c}, O. Canko and A. Albayrak, \emph{Multicritical behaviors of the antiferromagnetic Blume-Emery-Griffiths model with the external magnetic field on the Bethe
lattice},\\ J. Magn. Magn. Mater. \textbf{303}, pp. 185-190 (2006).

\bibitem{mon96} J. L. Monroe, \emph{Potts models with period doubling cascades, chaos, etc},\\
 J. Phys. A \textbf{29}, 5421 (1996).

\bibitem{eck05} M. Eckstein, M. Kollar, K. Byczuk, and D.
Vollhardt, \emph{Hopping on the Bethe lattice: Exact results for
densities of states and dynamical mean-field theory},\\ Phys. Rev. B
\textbf{71}, 235119 pp. 1-13 (2005).


\bibitem{eki05}
 C. Ekiz, \emph{Tricritical behavior in the mixed spin-1/2 and spin-1 Ising ferromagnetic system on the two-fold Cayley tree},\\
  Physica \textbf{A347}, pp. 353-362 (2005);\\
   C. Ekiz, \emph{Mixed spin-1/2 and spin-S Ising ferrimagnets with a crystal field},\\
   Physica A \textbf{353}, 286-296 (2005).

\bibitem{ala98} N. S. Ananikian, R. G. Ghulghazaryan, \emph{Yang-Lee and Fisher zeros of multisite interaction Ising models on the Cayley-type lattices},\\
  Phys. Lett. A \textbf{277}, 249 (2000).

\bibitem{wu} F. Y. Wu, \emph{The Potts-model},\\
Rev. Mod. Phys. \textbf{54}, pp. 235-268 (1982).

\bibitem{graph} B. Balobas, \textit{Modern Graph Theory} (Springer, New York,
1998).

\bibitem{brod1} S.~J.~Brodsky, \emph{"Light-front QCD"}, hep-ph/0412101 pp. 1-66.
\bibitem{brod2} S.~J.~Brodsky, \emph{New results in light-front phenomenology}, Few Body Syst. {\bf 36}, pp. 35-52 (2005).
\bibitem{polyakov} M.~Franz, V.~Polyakov, and K.~Goeke, \emph{Heavy quark mass expansion and intrinsic charm in light hadrons}\\
Phys.\ Rev. {\bf D} 62, 074024 pp. 1-20 (2000).
\bibitem{grib-lip} V.~N.~Gribov and L.~N.~Lipatov, \emph{Deep inelastic e p scattering in perturbation theory},\\ Sov.\ J.\ Nucl.\ Phys. {\bf 15}, pp. 438-450 (1972).
\bibitem{dokshitzer} Y.~L.~Dokshitzer, \emph{Calculation Of The Structure Functions For Deep Inelastic Scattering And E+ E- Annihilation By Perturbation Theory In Quantum Chromodynamics},\\ Sov.\ Phys.\ JETP {\bf 46}, pp. 641-653 (1977).
\bibitem{alt-par} G.~Altarelli and G.~Parisi, \emph{Asymptotic freedom in parton language},
\\ \emph{Nucl.\ Phys.}\  {\bf B126}, pp. 298-331 (1977).
\bibitem{pumplin} J.~Pumplin, \emph{Light-cone models for intrinsic charm and bottom},\\
 Phys.\ Rev.\ D {\bf 73}, 114015 pp. 1-15 (2006).

\bibitem{brod-higgs} S.~J.~Brodsky, B.~Kopeliovich, I.~Schmidt, and J.~Soffer, \emph{Diffractive Higgs production from intrinsic heavy flavors in the
proton},\\
 Phys.\ Rev.\ D {\bf 73}, 113005 pp. 1-30 (2006).
\bibitem{E161} SLAC E161 (2000), http://www.slac.stanford.edu/exp/e160.
\bibitem{dombey} N.~Dombey, \emph{Scattering of polarized leptons at high energy},\\ Rev.\ Mod.\ Phys.\ {\bf 41}, pp. 236-246 (1969).
\bibitem{GP} H.~Georgi and H.~D.~Politzer, \emph{Clean tests of QCD in mu p scattering},\\ Phys.\ Rev.\ Lett.\ {\bf 40}, pp. 3-7 (1978).
\bibitem{Mendez} A.~M\'{e}ndez, \emph{QCD predictions for semi-inclusive and inclusive leptoproduction},\\
 Nucl.\ Phys.\ B {\bf 145}, pp. 199-232 (1978).
\bibitem{Fano} U.~Fano, \emph{A Stokes-Parameter Technique for the Treatment of Polarization in Quantum Mechanics}\\
Phys.\ Rev.\ {\bf 93}, pp. 121-123 (1954).
\bibitem{LW1} J.~P.~Leveille and T.~Weiler, \emph{Azimuthal dependence of diffractive psi and d anti-d electroproduction and a test of gluon spin, parity and k-transverse},\\ Phys.\ Rev.\ D {\bf 24}, pp. 1789-1801 (1981).
\bibitem{Watson} A.~D.~Watson, \emph{Spin Asymmetries In Inclusive Muon Proton Charm Production},\\Z.\ Phys.\ C {\bf 12}, pp. 123-133 (1982).
\bibitem{LW2} J.~P.~Leveille and T.~Weiler, \emph{Characteristics of heavy quark leptoproduction in QCD},\\ Nucl.\ Phys.\ B {\bf 147}, pp. 147-200 (1979).
\bibitem{CTEQ5} H.~L.~Lai et al., \emph{Global QCD analysis of parton structure of the nucleon: CTEQ5 parton distributions},\\ Eur.\ Phys.\ J.\ C {\bf 12}, pp. 375-392 (2000).
\bibitem{kT} L.~Apanasevich et al., \emph{k(T) effects in direct photon production},\\ Phys.\ Rev.\ D {\bf 59}, 074007 pp. 1-37 (1999).
\bibitem{AOT} M.~A.~G.~Aivazis, F.~I.~Olness, and W.~-K.~Tung, \emph{Leptoproduction of heavy quarks. 1. General formalism and kinematics of charged current and neutral current production
processes},\\ Phys.\ Rev.\ D {\bf 50}, pp. 3085-3101 (1994).
\bibitem{chi} W.~-K. Tung, S.~Kretzer, and C.~Schmidt, \emph{Open heavy flavor production in QCD: Conceptual framework and implementation issues}, J.\ Phys.\ G {\bf 28}, pp. 983-996 (2002).
\bibitem{CTEQ4} H.~L.~Lai et al., \emph{Improved parton distributions from global analysis of recent deep inelastic scattering and inclusive jet data},\\
Phys.\ Rev.\ D {\bf 55} pp. 1280-1296 (1997).
\bibitem{SACOT} M.~Kramer, F.~I.~Olness, and D.~E.~Soper, \emph{Treatment of heavy quarks in deeply inelastic
scattering},\\ Phys.\ Rev.\ D {\bf 62}, 096007 pp. 1-15 (2000).
\bibitem{neerven} W.~L.~van Neerven, \emph{Production of heavy quarks in deep-inelastic lepton-hadron scattering},\\hep-ph/0107193.
\bibitem{BMSN} M.~Buza, Y.~Matiounine, J.~Smith, and W.~L.~van
Neerven, \emph{Charm electroproduction viewed in the variable flavor
number scheme versus fixed order perturbation theory},\\Eur.\ Phys.\
J.\ C {\bf 1}, pp. 301-320 (1998).

\bibitem{Harris-Smith} B.~W.~Harris and J.~Smith, \emph{Heavy-quark correlations in deep-inelastic electroproduction},\\ Nucl.\ Phys.\ B {\bf 452}, pp. 109-160 (1995).
\bibitem{EMC} J.~J.~Aubert et al., Production of charmed particles in 250 GeV $\mu^+$-iron interactions,\\ Nucl.\ Phys.\ B {\bf 213}, pp. 31-86 (1983).
\bibitem{harris-emc} B.~W.~Harris, J.~Smith, and R.~Vogt, \emph{Reanalysis of the EMC charm production data with extrinsic and intrinsic charm at
NLO},\\ Nucl.\ Phys.\ B {\bf 461}, pp. 181-196 (1996).
\bibitem{NA3} J.~Badier et al., \emph{Experimental J/psi hadronic production from 150-GeV/c to 280-GeV/c},\\ Z.\ Phys.\ C {\bf 20}, pp. 101-123 (1983).
\bibitem{E886} M.~J.~Leitch et al., \emph{Measurement of J/psi and psi-prime suppression in p-A collisions at 800-GeV/c},\\ Phys.\ Rev.\ Lett.\ {\bf 84}, pp. 3256-3260 (2000).
\bibitem{brod-psi-1} S.~J.~Brodsky and P.~Hoyer, \emph{The Nucleus As A Color Filter In Qcd Decays: Hadroproduction In Nuclei},\\
Phys.\ Rev.\ Lett.\ {\bf 63}, pp. 1566-1570 (1989).
\bibitem{hoyer-psi-1} P.~Hoyer, M.~Vanttinen, and U.~Sukhatme, \emph{Violation of factorization in charm
hadroproduction},\\ Phys.\ Lett.\ B {\bf 246}, pp. 217-220 (1990).
\bibitem{brod-psi-2} S.~J.~Brodsky, P.~Hoyer, A.~H.~Mueller, and
W.~K.~Tang, \emph{New QCD production mechanisms for hard processes
at large x},\\ Nucl.\ Phys.\ B {\bf 369}, pp. 519-542 (1992).
\bibitem{barger-lead} V.~D.~Barger, F.~Halzen, and W.~Y.~Keung, T\emph{he Central And Diffractive Components Of Charm
Production},\\ Phys.\ Rev.\ D {\bf 25}, pp. 112-133 (1982).
\bibitem{brod-lead} R.~Vogt and S.~J.~Brodsky, \emph{Charmed hadron asymmetries in the intrinsic charm coalescence model},\\ Nucl.\ Phys.\ B {\bf 478}, pp. 311-334 (1996).
\bibitem{ISR-lead} P.~Chauvat et al., \emph{Production of lambda(c) with large x(f) at the isr},\\ Phys.\ Lett.\ B {\bf 199}, pp. 304-315 (1987).
\bibitem{E791-lead-1} E.~M.~Aitala et al., \emph{Asymmetries in the production of Lambda(c)+ and Lambda(c)- baryons in 500-GeV/c pi- nucleon interactions},\\ Phys.\ Lett.\ B {\bf 495}, pp. 42-48 (2000).
\bibitem{E791-lead-2} E.~M.~Aitala et al., \emph{Differential cross-sections, charge production asymmetry, and spin density matrix elements for D*+-(2010) produced in 500-GeV/c pi- nucleon interactions},\\ Phys.\ Lett.\ B {\bf 539}, pp. 218-226 (2002).
\bibitem{ISR-B-1} M.~Basile et al., \emph{The leading baryon effect in lambda(b)0 production in proton proton interactions at s**(1/2) = 62-GeV},\\ Nuovo\ Cim.\ A {\bf 65}, pp. 408-416 (1981).
\bibitem{ISR-B-2} G.~Bari et al., \emph{The Lambda/b0 beauty baryon production in proton proton interactions at s**(1/2) = 62-GeV: A Second observation},\\Nuovo\ Cim.\ A {\bf 104}, pp. 1787-1800 (1991).
\bibitem{brod-7-quark} R.~Vogt and S.~J.~Brodsky, \emph{Intrinsic charm contribution to double quarkonium hadroproduction},\\ Phys.\ Lett.\ B {\bf 349}, pp. 569-575 (1995).
\bibitem{NA3-7-quark} J.~Badier et al., \emph{Evidence for psi psi production in pi- Interactions at 150-GeV/c and 280-GeV/c,}\\ Phys.\ Lett.\ B {\bf 114}, pp. 457-468 (1982).
\bibitem{SELEX-7-quark} A.~Ocherashvili et al., \emph{Confirmation of the double charm baryon Xi+(cc)(3520) via its decay to p D+
K-}, \\
 Phys.\ Lett.\ B {\bf 628}, pp. 18-24 (2005).

\end{thebibliography}
\end{document}